\documentclass[journal]{IEEEtran}
\usepackage{amsmath,amsfonts}
\usepackage{mathrsfs}
\usepackage{algorithmicx}
\usepackage{algorithm}
\usepackage{algpseudocode}
\usepackage{array}
\usepackage[caption=false,font=normalsize,labelfont=sf,textfont=sf]{subfig}
\usepackage{textcomp}
\usepackage{stfloats}
\usepackage{url}
\usepackage{verbatim}
\usepackage{graphicx}
\usepackage{cite}
\usepackage{balance}
\usepackage{enumitem}

\makeatletter
\def\submodules#1{\expandafter\@submodules\csname c@#1\endcsname}
\def\@submodules#1{\ifcase#1\or o\or i\fi}
\makeatother
\AddEnumerateCounter{\submodules}{\@submodules}{i}

\newcommand{\nonameditem}{\item[]\addtocounter{enumi}{1}}
\newcommand{\A}{\mathrm{A}}
\newcommand{\B}{\mathrm{B}}
\newcommand{\x}{\mathrm{x}}
\newcommand{\z}{\mathrm{z}}
\newcommand{\peq}{\overset{\mathrm{p}}{=}}
\newcommand{\pto}{\overset{\mathrm{p}}{\to}}
\newcommand{\plto}{\overset{\mathrm{PL}(2)}{\to}}

\newtheorem{assumption}{Assumption}
\newtheorem{theorem}{Theorem}
\newtheorem{lemma}{Lemma}
\newtheorem{definition}{Definition}
\newtheorem{proposition}{Proposition}
\newtheorem{remark}{Remark}
\newtheorem{corollary}{Corollary}

\begin{document}

\title{Generalized Orthogonal Approximate Message-Passing for Sublinear Sparsity}

\author{Keigo~Takeuchi,~\IEEEmembership{Member,~IEEE}
\thanks{
This work was supported in part by the Grant-in-Aid for Scientific 
Research~(B) (Japan Society for the Promotion of Science (JSPS) KAKENHI) 
under Grant 26K00948. An earlier version of this paper will be presented 
in part to the 2026 IEEE International Conference on Acoustics, Speech 
and Signal Processing.
}
\thanks{K.~Takeuchi is with the Department of Electrical and Electronic Information Engineering, Toyohashi University of Technology, Toyohashi 441-8580, Japan (e-mail: takeuchi@ee.tut.ac.jp).}
}

\markboth{submitted to IEEE transactions on information theory}%
{Takeuchi: Generalized Orthogonal Approximate Message-Passing for Sublinear Sparsity}

\IEEEpubid{0000--0000/00\$00.00~\copyright~xxxx IEEE}

\maketitle

\begin{abstract}
This paper addresses the reconstruction of sparse signals from generalized linear measurements. Signal sparsity is assumed to be sublinear in the signal dimension while it was proportional to the signal dimension in conventional research. Approximate message-passing (AMP) has poor convergence properties for sensing matrices beyond standard Gaussian matrices. To solve this convergence issue, generalized orthogonal AMP (GOAMP) is proposed for signals with sublinear sparsity. The main feature of GOAMP is the so-called Onsager correction to realize asymptotic Gaussianity of estimation errors. The Onsager correction in GOAMP is designed via state evolution for orthogonally invariant sensing matrices in the sublinear sparsity limit, where the signal sparsity and measurement dimension tend to infinity at sublinear speed in the signal dimension. When the support of non-zero signals does not contain a neighborhood of the origin, GOAMP using Bayesian denoisers is proved to achieve error-free signal reconstruction for linear measurements if and only if the measurement dimension is larger than a threshold, which is equal to that of AMP for standard Gaussian sensing matrices. Numerical simulations are also presented for linear measurements and 1-bit compressed sensing. When ill-conditioned sensing matrices are used, GOAMP for sublinear sparsity is shown to outperform existing reconstruction algorithms, including generalized AMP for sublinear sparsity.  
\end{abstract}


\section{Introduction}
\subsection{Sparse Signal Reconstruction}
\IEEEPARstart{T}{his} paper considers reconstruction of unknown signals with 
sublinear sparsity from generalized linear 
measurements~\cite{Donoho06,Candes06,Wegkamp07,Geer08}. Let 
$\boldsymbol{x}$ denote an $N$-dimensional signal vector with sparsity 
$k\in\{1,\ldots,N\}$, which is the number of non-zero elements in 
$\boldsymbol{x}$. An $M$-dimensional measurement vector 
$\boldsymbol{y}\in\mathbb{R}^{M}$ is given by 
\begin{equation} \label{measurement}
\boldsymbol{y} = g(\boldsymbol{z}, \boldsymbol{w}), \quad 
\boldsymbol{z} = \boldsymbol{A}\boldsymbol{x}.
\end{equation}
In (\ref{measurement}), $\boldsymbol{A}\in\mathbb{R}^{M\times N}$ and 
$\boldsymbol{w}\in\mathbb{R}^{M}$ denote a sensing matrix and a noise 
vector, respectively. The signal vector $\boldsymbol{x}$ depends on the 
measurement vector $\boldsymbol{y}$ only through the linear transform 
$\boldsymbol{z}=\boldsymbol{A}\boldsymbol{x}$. The random variables 
$\{\boldsymbol{A}, \boldsymbol{x}, \boldsymbol{w}\}$ are independent. 

The measurement function $g: \mathbb{R}^{2}\to\mathbb{R}$ in 
(\ref{measurement}) represents nonlinearity in measurements.  
For instance, $g(z, w)=z + w$ corresponds to the linear measurement. 
1-bit compressed sensing~\cite{Boufounos08} and 
phase retrieval~\cite{Gerchberg72,Fienup82} postulate the sign function 
$g(z, w)=\mathrm{sgn}(z + w)$ and $g(z, w)= |z| + w$, respectively. 

Information-theoretic analysis~\cite{Wainwright09,Fletcher09,Aeron10,Scarlett17,Gamarnik17,Reeves20} revealed that the optimum sample complexity is 
$M={\cal O}(k\log(N/k))$. In particular, the prefactor in the optimum sample 
complexity is known for constant non-zero signals: 
$M=C_{\mathrm{AWGN}}^{-1}k\log(N/k)$~\cite{Reeves20}, 
with $C_{\mathrm{AWGN}}$~[nat/s/Hz] denoting 
the capacity of the additive white Gaussian noise (AWGN) channel.   

Two asymptotic situations have been considered: 
One situation is linear sparsity, for which $N$ and $k$ tend to infinity 
with the ratio $\rho=k/N\in(0, 1]$ fixed. In this case, the optimum sample 
complexity is linear in $N$. In the other situation, 
$\log k/\log N$ tends to $\gamma\in[0, 1)$, i.e.\ $k={\cal O}(N^{\gamma})$. 
For this sublinear sparsity, the optimum sample complexity is sublinear 
in $N$. The sublinear sparsity--- $k\to\infty$ is required 
in message-passing algorithms---is the main focus of this paper. 

\subsection{Lasso-Type Algorithms}
For the linear measurement, least absolute shrinkage and selection operator 
(Lasso)~\cite{Tibshirani96} is a popular approach to signal reconstruction 
for the linear and sublinear sparsity. More generally, the 
following optimization problem is considered: 
\begin{equation} \label{Lasso}
\min_{\boldsymbol{x}\in\mathbb{R}^{N}}\left\{
 \frac{1}{2M}\|\boldsymbol{y} - \boldsymbol{A}\boldsymbol{x}\|_{2}^{2}
 + \lambda\|\boldsymbol{x}\|_{p}
\right\},
\end{equation}
with some $\lambda>0$. Lasso uses the $\ell_{1}$ 
norm $p=1$. In this paper, algorithms to solve (\ref{Lasso}) are 
called Lasso-type algorithms. 

Classical optimization techniques have been used to propose several 
Lasso-type algorithms for $\ell_{1}$ regularization. 
Iterative shrinkage thresholding algorithm 
(ISTA)~\cite{Daubechies04} or fast ISTA 
(FISTA)~\cite{Beck09} is an instance of proximal gradient. 
The other algorithms include an interior-point method~\cite{Koh07}, gradient projection~\cite{Figueiredo07}, Bregman iteration~\cite{Yin08}, and alternating direction method of multipliers (ADMM)~\cite{Esser09,Yang11}. 
Iterative hard thresholding (IHT)~\cite{Blumensath09} is an alternative 
of ISTA to solve the problem~(\ref{Lasso}) with $\ell_{0}$ regularization. 

\IEEEpubidadjcol

Lasso-type algorithms can be generalized to 1-bit compressed sensing. 
The Lasso problem~(\ref{Lasso}) with $\ell_{1}$ regularization was generalized 
to a generalized Lasso (GLasso) problem~\cite{Plan13,Plan16}. Since GLasso is 
essentially equivalent to Lasso for the linear 
measurement~\cite{Thrampoulidis15}, existing Lasso-type algorithms can be 
used to solve the GLasso problem. For $\ell_{0}$ regularization, IHT was 
generalized to binary IHT (BIHT)~\cite{Jacques13,Matsumoto22}.

\subsection{Message-Passing Algorithms}
Message-passing is another promising approach to signal reconstruction mainly 
for linear sparsity. Approximate message-passing (AMP)~\cite{Donoho09} and 
generalized AMP (GAMP)~\cite{Rangan11} are iterative algorithms for signal 
reconstruction from the linear and generalized linear measurements, 
respectively. AMP is regarded as an exact approximation of loopy belief 
propagation~\cite{Kabashima03}. See \cite{Kamilov12} and 
\cite{Schniter14,Mondelli22} for applications of GAMP to 1-bit compressed 
sensing and phase retrieval, respectively. 

State evolution~\cite{Bayati11,Javanmard13,Takeuchi242}---inspired by 
Bolthausen's conditioning technique~\cite{Bolthausen14}---is a rigorous 
technique to analyze the dynamics of GAMP for linear sparsity. 
In the conditioning technique, statistical properties of the current messages 
are evaluated via the conditional distribution of the sensing matrix 
given all previous messages. 
State evolution for GAMP requires the assumption of zero-mean independent and 
identically distributed (i.i.d.) Gaussian sensing matrices. In particular, 
GAMP using Bayes-optimal denoisers---called Bayesian GAMP---can achieve the 
Bayes-optimal performance~\cite{Reeves191,Barbier201,Barbier19} when 
state evolution recursion has a unique fixed point. 

GAMP has poor convergence properties when the sensing 
matrix is beyond zero-mean i.i.d. Gaussian 
matrices~\cite{Caltagirone14,Rangan191}. To relax this limitation, 
several message-passing algorithms have been proposed, such as 
GAMP with mean removal~\cite{Vila15}, swept AMP~\cite{Manoel15}, orthogonal 
AMP (OAMP)~\cite{Ma17}, vector AMP (VAMP)~\cite{Rangan192}, and AMP with unitary transformation~\cite{Yuan21}. 

OAMP or equivalently VAMP is a promising algorithm for solving the limitation 
of AMP. OAMP/VAMP using Bayes-optimal denoisers---called Bayesian 
OAMP/VAMP---is equivalent to \cite[Appendix~D]{Opper05} and regarded as an 
approximation of expectation propagation~\cite{Cespedes14,Takeuchi20}. 
State evolution~\cite{Rangan192,Takeuchi20} revealed that Bayesian OAMP/VAMP 
can achieve the Bayes-optimal performance~\cite{Barbier18,Li24} for 
right-orthogonally invariant sensing matrices when state evolution recursion 
has a unique fixed point. See \cite{Wang24,Dudeja24} for a practical 
alternative of 
right-orthogonally invariant matrices. VAMP was extended to generalized VAMP 
(GVAMP)~\cite{Schniter16} for the generalized linear 
measurement, which was analyzed via state evolution~\cite{Fletcher18}. 

OAMP/VAMP has higher complexity than AMP, 
because of matrix inversion or singular value decomposition (SVD). 
Long-memory message-passing aims to circumvent this complexity issue, 
such as rotationally invariant AMP~\cite{Opper16,Fan22,Venkataramanan21}, 
convolutional AMP~\cite{Takeuchi21}, memory AMP~\cite{Liu22}, conjugate 
gradient OAMP/VAMP~\cite{Takeuchi17,Skuratovs22}. These algorithms utilize 
messages in all previous iterations to update the current messages. 
Since the complexity issue is outside the scope of this paper, 
long-memory message-passing is not discussed anymore. 

Message-passing algorithms for sublinear sparsity are limited. 
Only a few papers have proposed GAMP for sublinear 
sparsity~\cite{Barbier202,Truong23,Takeuchi251}. In particular, the 
so-called all-or-nothing phenomenon was proved to occur in the sublinear 
sparsity limit~\cite{Takeuchi251}---all $k$, $M$, and $N$ tend to infinity 
while $\log k/\log N\to\gamma\in[0, 1)$ and $M/\{k\log(N/k)\}\to\delta>0$ 
are satisfied: Bayesian GAMP can realize asymptotically 
error-free signal reconstruction if and only if $\delta$ is 
larger than a reconstruction threshold. The additional assumption $k\to\infty$ 
is required to justify the so-called Onsager correction in GAMP for sublinear 
sparsity. However, to the best of author's knowledge, 
OAMP/VAMP has not yet been proposed for reconstruction of signals with 
sublinear sparsity.   

\subsection{Contributions}
This paper proposes generalized OAMP/VAMP for sublinear sparsity---called 
GOAMP simply. The main contributions of this paper are fourfold: A first 
contribution is the establishment of a unified state evolution framework 
for all orthogonally invariant sensing matrices in the sublinear sparsity 
limit (Theorem~\ref{theorem_SE_tech}). The framework is utilized to 
design the Onsager correction in GOAMP, which realizes asymptotic Gaussianity 
for estimation errors. Since it is presented in a general form, the framework 
can be used to design long-memory message-passing for sublinear sparsity 
in future research.   

A second contribution is the proposal of GOAMP for sublinear sparsity. 
The Onsager correction in GOAMP is designed in order for the state 
evolution framework to contain an error model of GOAMP. The proposed GOAMP 
has essentially the same Onsager correction for mean messages as conventional 
GVAMP~\cite{Schniter16} for linear sparsity. However, variance messages in 
GOAMP are different from those in GVAMP~\cite{Schniter16}. Under a strong 
assumption on a denoiser in GOAMP (Assumption~\ref{assumption_inner}), 
the asymptotic dynamics of GOAMP can be described with four 
discrete-time dynamical systems---called state evolution recursion 
in this paper (Theorem~\ref{theorem_SE}).  

A third contribution is theoretical comparison of GOAMP with 
GAMP~\cite{Takeuchi251} for sublinear sparsity. 
The strong assumption on the denoiser is justified for GOAMP using Bayesian 
denoisers---called Bayesian GOAMP (Theorem~\ref{theorem_SE_Bayes}). 
For the linear measurement, the state evolution recursion for Bayesian 
OAMP is simplified to two dynamical systems 
(Corollary~\ref{corollary_SE_Bayes_linear}). When the support of non-zero 
signals does not contain a neighborhood of the origin, Bayesian OAMP is 
proved to achieve asymptotically error-free signal reconstruction 
if and only if the prefactor $\delta$ in the optimum sample complexity is 
larger than an optimized threshold over sensing 
matrices (Theorem~\ref{theorem_linear}), which is equal to that of Bayesian 
AMP for standard Gaussian sensing matrices~\cite{Takeuchi251}. 

The last contribution is numerical comparison of Bayesian GOAMP with existing 
reconstruction algorithms, including Bayesian GAMP~\cite{Takeuchi251} for 
sublinear sparsity. The linear measurement and 1-bit compressed sensing are 
considered to test Bayesian GOAMP for sublinear sparsity. For both cases, 
numerical simulations show that Bayesian GOAMP is significantly superior to 
the existing reconstruction algorithms especially for ill-conditioned 
sensing matrices.  

Part of these contributions were presented in \cite{Takeuchi26}. More 
precisely, \cite{Takeuchi26} restricted the measurement 
model~(\ref{measurement}) to the linear 
measurement. Furthermore, it presented Bayesian OAMP for sublinear sparsity, 
the third contribution in this paper, and the linear measurement part in the 
last contribution.

\subsection{Organization}
The remainder of this paper is organized as follows: After presenting the 
notation used in this paper, GOAMP for sublinear sparsity is proposed 
in Section~\ref{sec2}. Section~\ref{sec3} presents the main results of this 
paper. Bayesian GOAMP is numerically compared to existing reconstruction 
algorithms in Section~\ref{sec4}. Section~\ref{sec5} concludes this paper. 
Theoretical results in Section~\ref{sec3} are proved in the appendices. 

\subsection{Notation}
Throughout this paper, $x_{n,\mathcal{I}}$ denotes the $n$th element of a 
vector $\boldsymbol{x}_{\mathcal{I}}$ with a set of indices $\mathcal{I}$. 
The notation $\boldsymbol{o}(N)$ represents a vector of which the elements 
are all $o(N)$. The $\ell_{p}$ norm is denoted by $\|\cdot\|_{p}$ while 
$\|\cdot\|_{\mathrm{F}}$ represents the Frobenius norm of a matrix. The notation 
$\boldsymbol{1}$ represents a vector of which the elements are all $1$ while 
$1(\cdot)$ denotes the indicator function. For a vector 
$\boldsymbol{v}\in\mathbb{R}^{N}$, the notation 
$\mathrm{diag}\{\boldsymbol{v}\}$ denotes the diagonal matrix of which the 
$n$th diagonal element is $v_{n}$. The arithmetic mean is written as 
$\langle\boldsymbol{v}\rangle=N^{-1}\sum_{n=1}^{N}v_{n}$.  

For a function $f: \mathbb{R}^{t}\to\mathbb{R}$ and vectors
$\{\boldsymbol{x}_{\tau}\in\mathbb{R}^{N}\}_{\tau=0}^{t-1}$, the notation 
$f(\boldsymbol{x}_{0},\ldots,\boldsymbol{x}_{t-1})$ represents the
element-wise application of $f$, 
i.e.\ $[f(\boldsymbol{x}_{0},\ldots,\boldsymbol{x}_{t-1})]_{n}
=f([\boldsymbol{x}_{0}]_{n},\ldots,[\boldsymbol{x}_{t-1}]_{n})$. The notation 
$\partial_{\tau}f$ denotes the partial derivative of $f$ with respect 
to the $\tau$th variable. The divergence of $f$ with respect to 
$\boldsymbol{x}_{\tau}$ is represented as 
$\mathrm{div}_{\boldsymbol{x}_{\tau}}f=\sum_{n=1}^{N}\partial_{\tau}[f]_{n}$.  

The notation $\mathcal{N}(\boldsymbol{\mu}, \boldsymbol{\Sigma})$ denotes 
the Gaussian distribution with mean $\boldsymbol{\mu}$ and 
covariance $\boldsymbol{\Sigma}$. For a sequence of random variables 
$\{X_{n}\}$, the convergence in probability of $X_{n}$ to some $X$ is 
represented as $X_{n}\pto X$ while $\peq$ denotes the equivalence in 
probability. Thus, $X_{n}\pto X$ is equivalent to $X_{n}\peq X + o(1)$. 

For square and symmetric matrices $\boldsymbol{S}$, the 
trace and minimum eigenvalue are represented as 
$\mathrm{Tr}(\boldsymbol{S})$ and $\lambda_{\mathrm{min}}(\boldsymbol{S})$, 
respectively. The transpose of a matrix $\boldsymbol{M}$ is denoted
by $\boldsymbol{M}^{\mathrm{T}}$. The SVD of a tall matrix $\boldsymbol{M}$ is 
represented as $\boldsymbol{M}=\boldsymbol{\Phi}_{\boldsymbol{M}}
\boldsymbol{\Sigma}_{\boldsymbol{M}}\boldsymbol{\Psi}_{\boldsymbol{M}}^{\mathrm{T}}$. 
The matrix $\boldsymbol{\Phi}_{\boldsymbol{M}}^{\parallel}$ consists of the left 
singular vectors of $\boldsymbol{M}$ associated with positive 
singular values while $\boldsymbol{\Phi}_{\boldsymbol{M}}^{\perp}$ consists of 
those associated with zero singular values. This paper uses the projection 
matrices $\boldsymbol{P}_{\boldsymbol{M}}^{\parallel}
=\boldsymbol{\Phi}_{\boldsymbol{M}}^{\parallel}
(\boldsymbol{\Phi}_{\boldsymbol{M}}^{\parallel})^{\mathrm{T}}$ and 
$\boldsymbol{P}_{\boldsymbol{M}}^{\perp}
=\boldsymbol{\Phi}_{\boldsymbol{M}}^{\perp}
(\boldsymbol{\Phi}_{\boldsymbol{M}}^{\perp})^{\mathrm{T}}$. 
When $\boldsymbol{M}$ has full rank, 
$\boldsymbol{P}_{\boldsymbol{M}}^{\parallel}=\boldsymbol{M}
(\boldsymbol{M}^{\mathrm{T}}\boldsymbol{M})^{\mathrm{T}}\boldsymbol{M}^{\mathrm{T}}$ 
and $\boldsymbol{P}_{\boldsymbol{M}}^{\perp}=\boldsymbol{I} 
- \boldsymbol{P}_{\boldsymbol{M}}^{\parallel}$ hold.

\section{Generalized OAMP} \label{sec2}
\subsection{Overview}
GOAMP is composed of two modules---called modules A and B in this paper.  
Module~A computes the linear minimum mean-square error (LMMSE) 
estimators of both $\boldsymbol{x}$ and $\boldsymbol{z}$ 
while module~B exploits the signal prior and the measurement vector 
$\boldsymbol{y}$ in (\ref{measurement}) to refine them. Each module 
has two sub-modules---called inner and outer sub-modules. The inner sub-module 
estimates the signal vector $\boldsymbol{x}$ while the outer sub-module 
estimates the outer signal vector 
$\boldsymbol{z} = \boldsymbol{A}\boldsymbol{x}$. The inner and outer 
sub-modules in module~A utilize the feedback from both inner and outer 
sub-modules in module~B. On the other hand, the inner and outer sub-modules 
in module~B operate independently: The inner and outer sub-modules exploit 
messages sent from the inner and outer sub-modules in module~A, respectively. 

Each sub-module Onsager-corrects the estimation of $\boldsymbol{x}$ or 
$\boldsymbol{z}$. In general, the original estimation errors follow 
intractable distributions. The Onsager correction is performed to realize 
asymptotic Gaussianity for the estimation errors. 

Table~\ref{table1} lists differences between existing 
GVAMP~\cite{Schniter16} and GOAMP for sublinear sparsity. 
The outer sub-modules are essentially the same as those in 
GVAMP~\cite{Schniter16} while variance messages may be updated 
with consistent estimators. 
The effects of sublinear sparsity are encapsulated within the inner 
sub-modules as long as normalization is appropriately designed in the inner 
sub-module of module~A. 
Thus, readers can skip the definitions of the outer sub-modules if they 
are familiar with GVAMP~\cite{Schniter16}. 

\begin{table}[t]
\caption{Sublinear sparsity versus linear sparsity}
\label{table1}
\centering
\begin{tabular}{|c|c|c|}
\hline
& Linear & Sublinear \\
\hline\hline
Sparsity $k$ & ${\cal O}(N)$ & $o(N)$ \\
\hline 
\#Measurements $M$ & ${\cal O}(N)$ & ${\cal O}(k\log(N/k))$ \\
\hline 
Power & $\mathbb{E}[\|\boldsymbol{x}\|_{2}^{2}]={\cal O}(N)$ & 
$\mathbb{E}[\|\boldsymbol{x}\|_{2}^{2}] = {\cal O}(1)$ \\
normalization & $\mathbb{E}[\|\boldsymbol{A}\|_{\mathrm{F}}^{2}]
={\cal O}(N)$ & 
$\mathbb{E}[\|\boldsymbol{A}\|_{\mathrm{F}}^{2}]={\cal O}(MN)$ \\
\hline 
Signal prior & i.i.d.\ spike and slab & Uniformly distributed \\
& & support sets of size $k$\\ 
\hline\hline 
Outer sub-modules & Essentially same & Essentially same \\
\hline 
Inner sub-module  & Square error  & Square error\\
in module~A & divided by $N$ & divided by $N/M$  \\
\hline 
Inner sub-module  & Square error  & Unnormalized \\
in module~B & divided by $N$ & square error\\
\hline\hline
$\boldsymbol{x}_{\A}^{t}$ in module~A & $\boldsymbol{x}_{\B \to \A}^{t} 
+ \widehat{\Delta \boldsymbol{x}^{t}}$ & 
$\boldsymbol{x}_{\B \to \A}^{t} 
+ (N/M)\widehat{\Delta \boldsymbol{x}^{t}}$ \\
\hline 
$v_{\A \to \B}^{\mathrm{x}, t}$ in module~A & 
$\frac{\xi_{\A, t}^{\mathrm{x}} v_{\B \to \A}^{\mathrm{x}, t}}
{1 - \xi_{\A, t}^{\mathrm{x}}}$ & $\frac{M \| \boldsymbol{x}_{\A}^{t} 
- \boldsymbol{x}_{\B \to \A}^{t} \|_{2}^{2}}{N (1 - \xi_{\A, t}^{\mathrm{x}})^{2}}$ \\
\hline 
$\boldsymbol{\omega}_{t}$ in (\ref{virtual_AWGN}) & 
$\mathcal{N}(\boldsymbol{0}, \bar{v}_{\A\to\B}^{\x,t}\boldsymbol{I}_{N})$ &
$\mathcal{N}(\boldsymbol{0}, 
M^{-1}\bar{v}_{\A\to\B}^{\x,t}\boldsymbol{I}_{N})$ \\
\hline
Other messages & & \\
in inner modules  & Same$^{\mathrm{a}}$ & Same \\
\hline
\multicolumn{3}{l}{$^{\mathrm{a}}$$N^{-1}v_{\B \to \A}^{\mathrm{x}, t}$ and 
$(M/N)\xi_{\B,t}^{\mathrm{x}}$ should be replaced by $v_{\B \to \A}^{\mathrm{x}, t}$ 
and $\xi_{\B,t}^{\mathrm{x}}$.}
\end{tabular}
\end{table}

\subsection{Modules}
Suppose that module~A in GOAMP has received messages 
$\boldsymbol{x}_{\B\to\A}^{t}\in\mathbb{R}^{N}$ and 
$\boldsymbol{z}_{\B\to\A}^{t}\in\mathbb{R}^{M}$ in iteration~$t$ from module~B, 
as well as messages $v_{\B\to\A}^{\x,t}>0$ and $v_{\B\to\A}^{\z,t}>0$. The messages  
$\boldsymbol{x}_{\B\to\A}^{t}\in\mathbb{R}^{N}$ and 
$\boldsymbol{z}_{\B\to\A}^{t}\in\mathbb{R}^{M}$ are estimators of 
$\boldsymbol{x}$ and $\boldsymbol{z}$ in (\ref{measurement}), 
respectively. The messages $v_{\B\to\A}^{\x,t}$ and $v_{\B\to\A}^{\z,t}$ correspond 
to estimators of the unnormalized square error 
$\|\boldsymbol{x}_{\B\to\A}^{t} - \boldsymbol{x}\|_{2}^{2}$ and the mean-square 
error (MSE) $M^{-1}\|\boldsymbol{z}_{\B\to\A}^{t} - \boldsymbol{z}\|_{2}^{2}$. 

The outer sub-module in module~A computes the following Onsager-corrected 
messages $\boldsymbol{z}_{\A\to\B}^{t}$ and $v_{\A\to\B}^{\z, t}$, and 
sends them to the outer sub-module in module~B: 
\begin{equation} \label{moduleA_z_denoiser}
\boldsymbol{z}_{\A}^{t} = f_{\A}^{\z}(\boldsymbol{z}_{\B\to\A}^{t}, 
\boldsymbol{A}\boldsymbol{x}_{\B\to\A}^{t}; 
v_{\B\to\A}^{\z, t}, v_{\B\to\A}^{\x, t}), 
\end{equation}
\begin{equation} \label{moduleA_z_mean}
\boldsymbol{z}_{\A\to\B}^{t} 
= \frac{\boldsymbol{z}_{\A}^{t} - \xi_{\A,t}^{\z}\boldsymbol{z}_{\B\to\A}^{t}}
{1 - \xi_{\A,t}^{\z}}, \quad 
v_{\A\to\B}^{\z, t} = \frac{\xi_{\A,t}^{\z}v_{\B\to\A}^{\z,t}}
{1 - \xi_{\A,t}^{\z}}, 
\end{equation}
with
\begin{equation} \label{xi_A_z}
\xi_{\A,t}^{\z} 
= M^{-1}\mathrm{div}_{\boldsymbol{z}_{\B\to\A}^{t}}
\boldsymbol{z}_{\A}^{t}. 
\end{equation}
In (\ref{moduleA_z_denoiser}), $f_{\A}^{\z}$  
denotes an outer denoiser in module~A, 
which computes an estimator of $\boldsymbol{z}$. 
The message $v_{\A\to\B}^{\z,t}$ is an estimator of the MSE
$M^{-1}\|\boldsymbol{z}_{\A\to\B}^{t} - \boldsymbol{z}\|_{2}^{2}$.

Similarly, the inner sub-module in module~A computes the following 
Onsager-corrected messages 
$\boldsymbol{x}_{\A\to\B}^{t}$ and $v_{\A\to\B}^{\x, t}$, and sends them to the 
inner sub-module in module~B:
\begin{equation} \label{moduleA_x_denoiser}
\boldsymbol{x}_{\A}^{t} 
= f_{\A}^{\x}(\boldsymbol{x}_{\B\to\A}^{t}, \boldsymbol{z}_{\B\to\A}^{t}; 
v_{\B\to\A}^{\x, t}, v_{\B\to\A}^{\z, t}), 
\end{equation} 
\begin{equation} 
\boldsymbol{x}_{\A\to\B}^{t} 
= \frac{\boldsymbol{x}_{\A}^{t}  - \xi_{\A,t}^{\x}\boldsymbol{x}_{\B\to\A}^{t}}
{1 - \xi_{\A,t}^{\x}}, \quad
v_{\A\to\B}^{\x, t} = \frac{M\|\boldsymbol{x}_{\A}^{t} 
- \boldsymbol{x}_{\B\to\A}^{t}\|_{2}^{2}}{N(1 - \xi_{\A,t}^{\x})^{2}}, 
\label{moduleA_x_mean}
\end{equation}
with 
\begin{equation} \label{xi_A_x}
\xi_{\A,t}^{\x} = N^{-1}\mathrm{div}_{\boldsymbol{x}_{\B\to\A}^{t}}
\boldsymbol{x}_{\A}^{t}.
\end{equation}
In (\ref{moduleA_x_denoiser}), $f_{\A}^{\x}$  
denotes an inner denoiser in module~A, which computes an estimator of 
$\boldsymbol{x}$. The message $M^{-1}v_{\A\to\B}^{\x,t}$ is an estimator of the MSE 
$N^{-1}\|\boldsymbol{x}_{\A\to\B}^{t} - \boldsymbol{x}\|_{2}^{2}$.  

The outer sub-module in module~B uses the messages 
$\boldsymbol{z}_{\A\to\B}^{t}$ and $v_{\A\to\B}^{\z, t}$ to feeds 
$\boldsymbol{z}_{\B\to\A}^{t+1}$ and $v_{\B\to\A}^{\z, t+1}$ back to module~A: 
\begin{equation} \label{moduleB_z_denoiser}
\boldsymbol{z}_{\B}^{t+1} 
= f_{\B}^{\z}(\boldsymbol{z}_{\A\to\B}^{t}, \boldsymbol{y}; v_{\A\to\B}^{\z, t}), 
\end{equation}
\begin{equation} \label{moduleB_z_mean}
\boldsymbol{z}_{\B\to\A}^{t+1}
= \frac{\boldsymbol{z}_{\B}^{t+1} - \xi_{\B,t}^{\z}\boldsymbol{z}_{\A\to\B}^{t}}
{1 - \xi_{\B,t}^{\z} - \eta_{\B,t}}, 
v_{\B\to\A}^{\z, t+1} = \frac{\|\boldsymbol{z}_{\B\to\A}^{t+1}\|_{2}^{2}}{M}
- \mathbb{E}[\|\boldsymbol{x}\|_{2}^{2}], 
\end{equation}
with 
\begin{equation} \label{xi_B_z}
\xi_{\B,t}^{\z} 
= M^{-1}\mathrm{div}_{\boldsymbol{z}_{\A\to\B}^{t}}
\boldsymbol{z}_{\B}^{t+1},   
\end{equation}
\begin{equation} \label{eta_B}
\eta_{\B,t} = 1 - \xi_{\B,t}^{\z} - M^{-1}\mathrm{div}_{\boldsymbol{z}}
f_{\B}^{\z}(\boldsymbol{z}_{\A\to\B}^{t}, 
g(\boldsymbol{z}, \boldsymbol{w}); v_{\A\to\B}^{\z,t}). 
\end{equation}
In (\ref{moduleB_z_denoiser}), $f_{\B}^{\z}$ denotes a separable outer 
denoiser in module~B. The message $v_{\B\to\A}^{\z,t+1}$ is an estimator of 
the MSE $M^{-1}\|\boldsymbol{z}_{\B\to\A}^{t+1} - \boldsymbol{z}\|_{2}^{2}$. 
The message $\eta_{\B,t}$ is introduced to realize asymptotic Gaussianity 
for the estimation error $\boldsymbol{z}_{\B\to\A}^{t+1} - \boldsymbol{z}$. 
As proved shortly, $\eta_{\B,t}=0$ is satisfied for the Bayes-optimal 
outer denoiser $f_{\B}^{\z}$. Thus, the outer sub-module in module~B is 
equivalent to that in \cite{Schniter16} as long as the Bayes-optimal outer 
denoiser is used.  

The message $\eta_{\B,t}$ is computable when the 
last term in (\ref{eta_B}) depends on $\boldsymbol{z}$ and 
$\boldsymbol{w}$ only through 
$\boldsymbol{y}=g(\boldsymbol{z}, \boldsymbol{w})$. However, 
$\eta_{\B,t}$ depends on unobservable $\boldsymbol{z}$ and 
$\boldsymbol{w}$ in general, so that it is not computable. When $\eta_{\B,t}$ is 
not computable, it should be replaced with its asymptotic alternative. 

Similarly, the inner sub-module in module~B computes the messages 
$\boldsymbol{x}_{\B\to\A}^{t+1}$ and $v_{\B\to\A}^{\x, t+1}$ to feed them 
back to module~A. 

\begin{equation} \label{moduleB_x_denoiser}
\boldsymbol{x}_{\B}^{t+1}
= f_{\B}^{\x}(\boldsymbol{x}_{\A\to\B}^{t}; v_{\A\to\B}^{\x, t}), 
\end{equation}
\begin{equation} \label{moduleB_x_mean}
\boldsymbol{x}_{\B\to\A}^{t+1}
= \frac{\boldsymbol{x}_{\B}^{t+1} - (M/N)\xi_{\B,t}^{\x}\boldsymbol{x}_{\A\to\B}^{t}}
{1 - (M/N)\xi_{\B,t}^{\x}},
\end{equation}
\begin{equation} \label{moduleB_x_var}
v_{\B\to\A}^{\x,t+1} = \frac{v_{\B}^{\x,t+1} - (M/N)(\xi_{\B,t}^{\x})^{2}v_{\A\to\B}^{\x,t}}
{\{1 - (M/N)\xi_{\B,t}^{\x}\}^{2}}, 
\end{equation}
with
\begin{equation} \label{xi_B_x}
\xi_{\B,t}^{\x} = M^{-1}\mathrm{div}_{\boldsymbol{x}_{\A\to\B}^{t}}
\boldsymbol{x}_{\B}^{t+1}. 
\end{equation}
In (\ref{moduleB_x_denoiser}), $f_{\B}^{\x}$ 
denotes a separable inner denoiser in module~B. 
In particular, $\boldsymbol{x}_{\B}^{t+1}$ is the estimator 
of $\boldsymbol{x}$ for GOAMP in iteration~$t+1$. 
The message $v_{\B\to\A}^{\x,t+1}$ is an estimator of the unnormalized square 
error $\|\boldsymbol{x}_{\B\to\A}^{t+1} - \boldsymbol{x}\|_{2}^{2}$. 

The inner sub-module needs to compute a message $v_{\B}^{\x,t+1}$ in 
(\ref{moduleB_x_var}) that 
corresponds to a consistent estimator of the unnormalized square error 
$\|\boldsymbol{x}_{\B}^{t+1} - \boldsymbol{x}\|_{2}^{2}$. In particular, 
posterior variance of $\boldsymbol{x}$ can be used as $v_{\B}^{\x,t+1}$ 
for Bayesian denoisers. 

The Onsager correction in (\ref{moduleB_x_mean}) is required to realize 
asymptotic Gaussianity. As shown in state evolution, however, the quantity 
$(M/N)\xi_{\B,t}^{\x}$ is $o(1)$ in the sublinear sparsity limit. Thus, 
the variance message $v_{\B\to\A}^{\x,t+1}$ in (\ref{moduleB_x_var}) is equal to 
$v_{\B}^{\x,t+1}$ in the sublinear sparsity limit. Nonetheless, 
numerical simulations showed that $v_{\B\to\A}^{\x,t+1}$ in 
(\ref{moduleB_x_var}) was a better estimator than $v_{\B}^{\x,t+1}$ itself 
for finite-sized systems. 

\subsection{Initialization}
Module~A needs initial messages 
$(\boldsymbol{z}_{\B\to\A}^{0}, v_{\B\to\A}^{\z, 0})$ and 
$(\boldsymbol{x}_{\B\to\A}^{0}, v_{\B\to\A}^{\x, 0})$ for $t=0$.  
The initial message for the inner sub-module is set to 
$(\boldsymbol{x}_{\B\to\A}^{0}, v_{\B\to\A}^{\x, 0})
= (\boldsymbol{0}, \mathbb{E}[\|\boldsymbol{x}\|_{2}^{2}])$.  
For the outer sub-module we use the initial messages  
\begin{equation} \label{initial_condition_z}
\boldsymbol{z}_{\B\to\A}^{0} 
= \frac{f_{\B}^{\z}(\boldsymbol{0}, \boldsymbol{y}; v_{\A\to\B}^{\z,-1})}
{1 - \eta_{\B,-1}},\; 
v_{\B\to\A}^{\z, 0} = \frac{\|\boldsymbol{z}_{\B\to\A}^{0}\|_{2}^{2}}{M} 
- \mathbb{E}[\| \boldsymbol{x}\|_{2}^{2}],
\end{equation} 
with $v_{\A\to\B}^{\z,-1}=M^{-1}\mathbb{E}[\|\boldsymbol{z}\|_{2}^{2}]$ and 
\begin{equation} \label{eta_B_init}
\eta_{\B,-1} = 1 - M^{-1}\mathrm{div}_{\boldsymbol{z}}
f_{\B}^{\z}\left(
 \boldsymbol{0}, g(\boldsymbol{z}, \boldsymbol{w}); v_{\A\to\B}^{\z,-1}
\right).
\end{equation}
 
This initialization is used to realize 
$M^{-1}\boldsymbol{z}^{\mathrm{T}}(\boldsymbol{z}_{\B\to\A}^{0}
- \boldsymbol{z})\pto 0$. As shown shortly, for the Bayes-optimal outer 
denoiser $f_{\B}^{\z}$ we have $\eta_{\B,-1}\peq \xi_{\B,-1}^{\z}+o(1)$, with  
\begin{equation}
\xi_{\B,-1}^{\z} 
= \left\langle
 \partial_{0}f_{\B}^{\z}\left(
  \boldsymbol{0}, g(\boldsymbol{z}, \boldsymbol{w}); v_{\A\to\B}^{\z,-1}
 \right)
\right\rangle.    
\end{equation}
Thus, the initialization~(\ref{initial_condition_z}) is essentially 
equivalent to that in \cite{Schniter16} for the Bayes-optimal outer 
denoiser, with a minor exception of $v_{\B\to\A}^{\z,-1}=\infty$ 
in \cite{Schniter16}. 

\subsection{Denoisers in module~A} 
Denoisers in module~A are designed in this section while state evolution in 
the next section is utilized to design denoisers in module~B. 
In designing the inner denoiser $f_{\A}^{\x}$ in module~A, the signal vector 
$\boldsymbol{x}\sim\mathcal{N}(\boldsymbol{x}_{\B\to\A}^{t},  
N^{-1}v_{\B\to\A}^{\x,t}\boldsymbol{I}_{N})$ is regarded as a Gaussian vector. 
The definition of the variance $N^{-1}v_{\B\to\A}^{\x,t}$ is consistent with the 
fact that $v_{\B\to\A}^{\x,t}$ is an estimator of the unnormalized square error 
$\|\boldsymbol{x}_{\B\to\A}^{t} - \boldsymbol{x}\|_{2}^{2}$. 
Similarly, the estimation error $\boldsymbol{q}_{\z,t}
=\boldsymbol{z}_{\B\to\A}^{t} - \boldsymbol{z}
\sim\mathcal{N}(\boldsymbol{0}, v_{\B\to\A}^{\z,t}\boldsymbol{I}_{M})$ is assumed 
to be Gaussian. These assumptions are postulated to reduce the complexity 
in module~A. 

Under these Gaussian assumptions, we know that the LMMSE estimator 
of $\boldsymbol{x}$ is Bayes-optimal. It is equivalent to that from the 
following linear measurement:  
\begin{equation}
\boldsymbol{z}_{\B\to\A}^{t} - \boldsymbol{A}\boldsymbol{x}_{\B\to\A}^{t}
= \boldsymbol{q}_{\z,t} 
+ \boldsymbol{A}(\boldsymbol{x} - \boldsymbol{x}_{\B\to\A}^{t}), 
\end{equation}
where we have used $\boldsymbol{z}=\boldsymbol{A}\boldsymbol{x}$ in 
(\ref{measurement}). 
Since $\Delta\boldsymbol{x}^{t}=\boldsymbol{x} - \boldsymbol{x}_{\B\to\A}^{t}
\sim\mathcal{N}(\boldsymbol{0}, N^{-1}v_{\B\to\A}^{\x,t}\boldsymbol{I}_{N})$ and 
$\boldsymbol{q}_{\z,t}\sim\mathcal{N}(\boldsymbol{0}, 
v_{\B\to\A}^{\z,t}\boldsymbol{I}_{M})$ are assumed, 
the LMMSE estimator of $\Delta\boldsymbol{x}^{t}$ is represented as 
\begin{equation} \label{Delta_x}
\widehat{\Delta\boldsymbol{x}^{t}}
= \boldsymbol{A}^{\mathrm{T}}
\left(
 \frac{Nv_{\B\to\A}^{\z,t}}{v_{\B\to\A}^{\x,t}}\boldsymbol{I}_{M} 
 + \boldsymbol{A}\boldsymbol{A}^{\mathrm{T}}
\right)^{-1}
(\boldsymbol{z}_{\B\to\A}^{t} - \boldsymbol{A}\boldsymbol{x}_{\B\to\A}^{t}). 
\end{equation}

The definition $\boldsymbol{x}= \boldsymbol{x}_{\B\to\A}^{t} 
+ \Delta\boldsymbol{x}^{t}$ implies that the naive LMMSE estimator of 
$\boldsymbol{x}$ would be $\boldsymbol{x}_{\B\to\A}^{t} 
+ \widehat{\Delta\boldsymbol{x}^{t}}$. However, small variance 
$N^{-1}v_{\B\to\A}^{\x,t}$ results in 
$\mathbb{E}[\|\boldsymbol{x}_{\B\to\A}^{t} - \boldsymbol{x}\|_{2}^{2}]
\gg\mathbb{E}[\|\widehat{\Delta\boldsymbol{x}^{t}}\|_{2}^{2}]$ 
in the sublinear sparsity limit. In other words, the naive LMMSE estimator 
cannot refine the estimator $\boldsymbol{x}_{\B\to\A}^{t}$ in module~B. 
To reduce the influence of $\boldsymbol{x}_{\B\to\A}^{t}$, this paper 
proposes the following scaled LMMSE estimator:   
\begin{equation}
f_{\A}^{\x}(\boldsymbol{x}_{\B\to\A}^{t}, \boldsymbol{z}_{\B\to\A}^{t}; 
v_{\B\to\A}^{\x,t}, v_{\B\to\A}^{\z,t}) 
= \boldsymbol{x}_{\B\to\A}^{t} 
+ \frac{N}{M}\widehat{\Delta\boldsymbol{x}^{t}}, 
\label{LMMSE_x}
\end{equation}
where the coefficient $N/M$ in the second term has been multiplied to keep 
$\boldsymbol{x}_{\B\to\A}^{t} - \boldsymbol{x}$ and 
$(N/M)\widehat{\Delta\boldsymbol{x}^{t}}$ be the same order in the sublinear 
sparsity limit. This heuristic modification is a crucial point in GOAMP 
and motivated by state evolution.  

The outer denoiser $f_{\A}^{\z}$ is designed as the true LMMSE estimator 
of $\boldsymbol{z}$. From $\boldsymbol{z}=\boldsymbol{A}\boldsymbol{x}$, 
we use the true LMMSE estimator $\boldsymbol{x}_{\B\to\A}^{t} 
+ \widehat{\Delta\boldsymbol{x}^{t}}$ of $\boldsymbol{x}$ to obtain 
\begin{equation}
f_{\A}^{\z}(\boldsymbol{z}_{\B\to\A}^{t}, 
\boldsymbol{A}\boldsymbol{x}_{\B\to\A}^{t}; v_{\B\to\A}^{\z,t}, v_{\B\to\A}^{\x,t})
= \boldsymbol{A}\boldsymbol{x}_{\B\to\A}^{t} 
+ \boldsymbol{A}\widehat{\Delta\boldsymbol{x}^{t}}.  
\label{LMMSE_z}
\end{equation}
Since $M^{-1}\mathbb{E}[\|\boldsymbol{A}\boldsymbol{x}_{\B\to\A}^{t} 
- \boldsymbol{z}\|_{2}^{2}]$ and 
$M^{-1}\mathbb{E}[\|\boldsymbol{A}\widehat{\Delta\boldsymbol{x}^{t}}\|_{2}^{2}]$ 
are the same order in the sublinear sparsity limit, we need not introduce 
any scaling coefficient in the outer denoiser $f_{\A}^{\z}$. Throughout this 
paper, (\ref{LMMSE_x}) and (\ref{LMMSE_z}) are used in GOAMP.

\subsection{Orthogonal Approximate Message-Passing} \label{sec_OAMP}
To reduce GOAMP to OAMP, consider the linear measurement $g(z, w)=z+w$ and 
a linear outer denoiser in module~B
$f_{\B}^{\z}(\boldsymbol{z}_{\A\to\B}^{t}, \boldsymbol{y}; v_{\A\to\B}^{\z, t})
= a_{t}\boldsymbol{z}_{\A\to\B}^{t} + b_{t}\boldsymbol{y}$
for any $a_{t}\in\mathbb{R}$ and $b_{t}\in\mathbb{R}$. 
In this case,
$\xi_{\B\,t}^{\z}$ and $\eta_{\B,t}$ in (\ref{xi_B_z}) and (\ref{eta_B}) reduce
to $\xi_{\B,t}^{\z}=a_{t}$ and $\eta_{\B,t}=1 - \xi_{\B,t}^{\z} - b_{t}$.
Substituting these results into $\boldsymbol{z}_{\B\to\A}^{t+1}$ in
(\ref{moduleB_z_mean}) yields $\boldsymbol{z}_{\B\to\A}^{t+1}=\boldsymbol{y}$.
Using the consistency $v_{\B\to\A}^{\z,t+1}
- M^{-1}\mathbb{E}[\|\boldsymbol{z}_{\B\to\A}^{t+1} - \boldsymbol{z}\|_{2}^{2}]
\pto 0$ proved via state evolution, as well as 
$M^{-1}\mathbb{E}[\|\boldsymbol{z}_{\B\to\A}^{t+1} - \boldsymbol{z}\|_{2}^{2}]
= M^{-1}\mathbb{E}[\|\boldsymbol{w}\|_{2}^{2}]\to\sigma^{2}$ obtained from 
$\boldsymbol{z}_{\B\to\A}^{t+1}=\boldsymbol{y}$, 
we obtain $v_{\B\to\A}^{\z,t+1}\pto\sigma^{2}$.

We reduce GOAMP to OAMP. Applying  
$\boldsymbol{z}_{\B\to\A}^{t+1} = \boldsymbol{y}$ and $v_{\B\to\A}^{\z,t+1}
\pto\sigma^{2}$ to $\boldsymbol{x}_{\A}^{t}$ in (\ref{moduleA_x_denoiser}) 
with $f_{\A}^{\x}$ in (\ref{LMMSE_x}), we arrive at 
\begin{equation}
\boldsymbol{x}_{\A}^{t} 
\peq \boldsymbol{x}_{\B\to\A}^{t} 
+ \frac{N}{M}\boldsymbol{A}^{\mathrm{T}}
\left(
 \frac{N\sigma^{2}}{v_{\B\to\A}^{\x,t}}\boldsymbol{I}_{M} 
 + \boldsymbol{A}\boldsymbol{A}^{\mathrm{T}}
\right)^{-1}
(\boldsymbol{y} - \boldsymbol{A}\boldsymbol{x}_{\B\to\A}^{t}). 
\end{equation}
This allows us to iterate the inner sub-modules in modules~A 
and B without updating the outer sub-modules. This simplified GOAMP 
is called OAMP and presented in \cite{Takeuchi26}. 

\section{Main Results} \label{sec3}
\subsection{State Evolution}
The dynamics of GOAMP is analyzed via state evolution in the sublinear 
sparsity limit. To present state evolution results, technical definitions and 
assumptions are summarized. 

\begin{definition}[Pseudo-Lipschitz]
A function $f: \mathbb{R}^{t}\to\mathbb{R}$ is said to be $j$th-order 
pseudo-Lipschitz if there is some $L>0$ such that 
the following inequality holds for all 
$\boldsymbol{x}, \boldsymbol{y}\in\mathbb{R}^{t}$:
\begin{equation}
|f(\boldsymbol{x}) - f(\boldsymbol{y})|\leq 
L(1 + \|\boldsymbol{x}\|_{2}^{j-1} + \|\boldsymbol{y}\|_{2}^{j-1})
\|\boldsymbol{x} - \boldsymbol{y}\|_{2}. 
\end{equation}
\end{definition}

\begin{definition}[$\mathrm{PL}(j)$ Convergence]
We say that a random vector $\boldsymbol{v}$ converges empirically to a 
scalar random variable $V$ in the sense of $j$th-order pseudo-Lipschitz 
if $\langle f(\boldsymbol{v}) \rangle$ converges in probability to 
$\mathbb{E}[f(V)]$ for all almost everywhere $j$th-order pseudo-Lipschitz 
functions $f$. This convergence is called $\mathrm{PL}(j)$ convergence 
and denoted by $\boldsymbol{v}\overset{\mathrm{PL}(j)}{\to} V$.  
\end{definition}

The $\mathrm{PL}(2)$ convergence is used to represent statistical properties 
of the noise vector $\boldsymbol{w}$. 

\begin{assumption} \label{assumption_w}
The noise vector $\boldsymbol{w}$ satisfies the $\mathrm{PL}(2)$ convergence 
$\boldsymbol{w}\plto W$ for some absolutely 
continuous random variable $W\in\mathbb{R}$ with variance $\sigma^{2}>0$. 
\end{assumption}

The Gaussian noise vector $\boldsymbol{w}\sim\mathcal{N}(\boldsymbol{0}, 
\sigma^{2}\boldsymbol{I}_{M})$ satisfies Assumption~\ref{assumption_w}. 
The absolute continuity in $W$ is required to postulate a piecewise 
Lipschitz-continuous denoiser. It can be eliminated when 
everywhere Lipschitz continuity is postulated. 

\begin{assumption} \label{assumption_x}
The support $\mathcal{S}=\{n\in\{1,\ldots,N\}: x_{n}\neq0\}$ of the 
$k$-sparse signal vector $\boldsymbol{x}$ is a uniform and random 
sample of size~$k$ from $\{1,\ldots, N\}$ without replacement. 
Let $\boldsymbol{u}=\sqrt{k}\boldsymbol{x}$. The scaled 
non-zero elements $\{u_{n}: n\in\mathcal{S}\}$ are i.i.d.\ 
random variables. The common distribution of $u_{n}$ is represented by a 
random variable $U$, i.e.\ $u_{n}\sim U$, which has no probability mass at 
the origin and power $\mathbb{E}[U^{2}]=P>0$. The cumulative 
distribution of $U$ is everywhere left-continuously or right-continuously 
differentiable and almost everywhere continuously differentiable. 
\end{assumption}

The support condition in Assumption~\ref{assumption_x} is equivalent to that in 
\cite{Reeves20,Takeuchi251} for sublinear sparsity. 
The mild conditions on $U$ indicate that $U$ is not sampled from any singular 
distribution. They are required to utilize existing results for 
a separable Bayesian inner denoiser $f_{\B}^{\x}$ proposed in 
\cite{Takeuchi251}. 

State evolution for linear 
sparsity~\cite{Bayati11,Rangan192,Takeuchi20,Fletcher18} postulated 
a signal vector satisfying the $\mathrm{PL}(2)$ convergence 
$\boldsymbol{x}\pto X$ for some random variable $X$. For instance, 
the $\mathrm{PL}(2)$ convergence is satisfied for 
$\boldsymbol{x}$ with i.i.d.\ elements $x_{n}\sim X$, in which $X$ takes 
non-zero values with probability $k/N$. In this case the number of 
non-zero elements is Poisson-distributed with mean $k$ while it is exactly 
equal to $k$ in Assumption~\ref{assumption_x}.     

\begin{assumption} \label{assumption_A}
The sensing matrix $\boldsymbol{A}$ is orthogonally invariant: The SVD 
$\boldsymbol{A}=\boldsymbol{U}\boldsymbol{\Sigma}\boldsymbol{V}^{\mathrm{T}}$ 
is composed of three independent matrices $\{\boldsymbol{U}, 
\boldsymbol{\Sigma}, \boldsymbol{V}\}$. The orthogonal matrices 
$\boldsymbol{U}$ and $\boldsymbol{V}$ are Haar-distributed. Furthermore, 
the empirical eigenvalue distribution of 
$\boldsymbol{\Lambda}=N^{-1}\boldsymbol{\Sigma}\boldsymbol{\Sigma}^{\mathrm{T}}$ 
converges in probability to a deterministic distribution with 
a compact support---represented with a positive random variable 
$\Lambda$---in the sublinear sparsity limit. In particular, 
the first moment $\mu_{1}=M^{-1}\mathrm{Tr}(\boldsymbol{\Lambda})$ converges 
in mean to $\bar{\mu}_{1}=1$ in the sublinear sparsity limit. 
\end{assumption}

Assumption~\ref{assumption_A} is an important assumption to analyze the 
dynamics of GOAMP via state evolution. It is satisfied 
for standard Gaussian sensing matrices considered in 
\cite{Reeves20,Takeuchi251}. The convergence in distribution is too weak 
to imply that the $j$th moment $\mu_{j}$ of the empirical eigenvalue 
distribution converges in probability to a constant 
$\bar{\mu}_{j}=\mathbb{E}[\Lambda^{j}]>0$ in the sublinear sparsity limit: 
$\mu_{j} = M^{-1}\mathrm{Tr}(\boldsymbol{\Lambda}^{j})\pto \bar{\mu}_{j}$. 
While additional assumptions on $\boldsymbol{\Lambda}$ are postulated 
in state evolution, we prove that Assumption~\ref{assumption_A} implies 
them for GOAMP. 

OAMP does not need left-orthogonal invariance: In the SVD 
$\boldsymbol{A}=\boldsymbol{U}\boldsymbol{\Sigma}\boldsymbol{V}^{\mathrm{T}}$,  
only $\boldsymbol{V}$ should be assumed to be a Haar-distributed orthogonal 
matrix and independent of $\boldsymbol{U}\boldsymbol{\Sigma}$. 
This is because OAMP does not update the outer sub-modules.   

\begin{assumption} \label{assumption_denoisers}
For any fixed $v>0$ the composition $f_{\B}^{\z}(\cdot, g(\cdot, \cdot); v)$ is 
nonlinear with respect to the first variable and piecewise 
Lipschitz-continuous. 
\end{assumption}

When $f_{\B}^{\z}(\cdot, g(\cdot, \cdot); v)$ is linear with respect to the 
first variable, $\boldsymbol{z}_{\B\to\A}^{t+1}$ in (\ref{moduleB_z_mean}) does 
not depend on $\boldsymbol{z}_{\A\to\B}^{t}$. To circumvent this issue, 
the nonlinearity is assumed. 
The Lipschitz continuity is the standard assumption in 
state evolution~\cite{Bayati11,Rangan192,Takeuchi20}. The piecewiseness 
is assumed to treat practical measurements, such as 
clipping~\cite{Takeuchi242}. Note that OAMP does not need 
Assumption~\ref{assumption_denoisers} because it has no outer sub-modules. 

To present state evolution recursion for GOAMP, we first define Gaussian 
$Z\sim\mathcal{N}(0, P)$ that is an asymptotic alternative of 
$\boldsymbol{z}$ in (\ref{measurement}) and independent of $W$ 
in Assumption~\ref{assumption_w}. We start with 
the initial conditions $\bar{v}_{\B\to\A}^{\x,0}=P$ and 
\begin{equation} \label{moduleB_z_var0_bar}
\bar{v}_{\B\to\A}^{\z,0} 
= \frac{\mathbb{E}[\{f_{\B}^{\z}(0, g(Z,W);P) - Z\}^{2}]
- \bar{\eta}_{\B,-1}^{2}P}
{(1 - \bar{\eta}_{\B,-1})^{2}},
\end{equation}
with 
\begin{equation} \label{eta_B0_bar}
\bar{\eta}_{\B,-1} = 1 - \mathbb{E}\left[
 \left.
  \frac{\partial}{\partial z}f_{\B}^{\z}(0, g(z, W);P)
 \right|_{z=Z}
\right]. 
\end{equation}
The variables $\bar{v}_{\B\to\A}^{\x,0}$, $\bar{v}_{\B\to\A}^{\z,0}$, and 
$\bar{\eta}_{\B,-1}$ correspond to the messages 
$v_{\B\to\A}^{\x,0}$, $v_{\B\to\A}^{\z,0}$, and $\eta_{\B,-1}$ in GOAMP, respectively. 

To describe the dynamics of the outer sub-module in module~A, 
we define asymptotic alternative of $\xi_{\A,t}^{\z}$ in (\ref{xi_A_z}) as 
\begin{equation} \label{xi_A_z_bar}
\bar{\xi}_{\A,t}^{\z} = \mathbb{E}\left[
 \frac{\bar{v}_{\B\to\A}^{\x,t}\Lambda} 
 {\bar{v}_{\B\to\A}^{\z,t} + \bar{v}_{\B\to\A}^{\x,t}\Lambda}
\right], 
\end{equation}
where $\Lambda$ follows the asymptotic eigenvalue 
distribution in Assumption~\ref{assumption_A}. Using $\bar{\xi}_{\A,t}^{\z}$, 
we define $\bar{v}_{\A\to\B}^{\z,t}$ and $\bar{v}_{\A\to\B}^{\z,*,t}$ as 
\begin{equation} \label{moduleA_z_var_bar}
\bar{v}_{\A\to\B}^{\z,t} 
= \frac{\bar{\xi}_{\A,t}^{\z}\bar{v}_{\B\to\A}^{\z,t}}{1 - \bar{\xi}_{\A,t}^{\z}},  
\quad 
\bar{v}_{\A\to\B}^{\z,*,t} 
= \frac{\bar{v}_{\B\to\A}^{\x,0,t}}{\bar{v}_{\B\to\A}^{\x,t}}\bar{v}_{\A\to\B}^{\z,t}, 
\end{equation}
with $\bar{v}_{\B\to\A}^{\x,0,t}$ defined shortly, 
which correspond to $v_{\A\to\B}^{\z,t}$ in (\ref{moduleA_z_mean}) and 
$-M^{-1}\boldsymbol{z}^{\mathrm{T}}(\boldsymbol{z}_{\A\to\B}^{t} - \boldsymbol{z})$, 
respectively. 

Similarly, we represent the dynamics of the inner sub-module in module~A 
with 
\begin{equation}\label{moduleA_x_var_bar}
\bar{v}_{\A\to\B}^{\x,\tau,t} = \frac{\bar{v}_{\B\to\A}^{\x,t}}
{1 - \bar{\xi}_{\A,t}^{\x}} 
\end{equation}
for all $\tau\in\{0,\ldots,t\}$, where $\bar{\xi}_{\A,t}^{\x}$ is given by 
\begin{equation} \label{xi_A_x_bar}
\bar{\xi}_{\A,t}^{\x} = 1 - \bar{\xi}_{\A,t}^{\z}. 
\end{equation}
The variable $\bar{v}_{\A\to\B}^{\x,\tau,t}$ is an asymptotic alternative of 
the error covariance 
$(M/N)(\boldsymbol{x}_{\A\to\B}^{\tau} - \boldsymbol{x})^{\mathrm{T}}
(\boldsymbol{x}_{\A\to\B}^{t} - \boldsymbol{x})$  
while $\bar{\xi}_{\A,t}^{\x}$ corresponds to $\xi_{\A,t}^{\x}$ in (\ref{xi_A_x}). 
We write $\bar{v}_{\A\to\B}^{\x,t,t}$ as 
$\bar{v}_{\A\to\B}^{\x,t}$ simply. 

To present state evolution recursion for the outer sub-module in 
module~B, we define a zero-mean Gaussian random variable $H_{\z,t}$ that is 
an asymptotic alternative of the estimation 
error $\boldsymbol{z}_{\A\to\B}^{t} - \boldsymbol{z}$ and is independent of 
$W$. The random variables $Z$ and $H_{\z,t}$ are jointly Gaussian-distributed 
with covariance $\mathbb{E}[H_{\z,t}^{2}]=\bar{v}_{\A\to\B}^{\z,t}$ and 
$\mathbb{E}[ZH_{\z,t}] = -\bar{v}_{\A\to\B}^{\z,*,t}$. Let 
\begin{equation} \label{xi_B_z_bar}
\bar{\xi}_{\B,t}^{\z} = \mathbb{E}\left[
 \partial_{0}f_{\B}^{\z}(Z + H_{\z,t}, g(Z, W); \bar{v}_{\A\to\B}^{\z,t}) 
\right], 
\end{equation}
\begin{equation} \label{eta_B_bar}
\bar{\eta}_{\B,t}
= 1 - \bar{\xi}_{\B,t}^{\z}
- \mathbb{E}\left[
 \left.
  \frac{\partial}{\partial z}f_{\B}^{\z}(Z + H_{\z,t}, g(z, W); 
  \bar{v}_{\A\to\B}^{\z,t}) 
 \right|_{z=Z}
\right],
\end{equation}
which correspond to $\xi_{\B,t}^{\z}$ and $\eta_{\B,t}$ in (\ref{xi_B_z}) and 
(\ref{eta_B}), respectively.  
Using these definitions, we define an 
asymptotic alternative of $v_{\B\to\A}^{\z,t+1}$ in (\ref{moduleB_z_mean}) as 
\begin{align} 
&(1 - \bar{\xi}_{\B,t}^{\z} - \bar{\eta}_{\B,t})^{2}\bar{v}_{\B\to\A}^{\z,t+1} 
=  - \bar{\eta}_{\B,t}^{2}P
- 2\bar{\xi}_{\B,t}^{\z}\bar{\eta}_{\B,t}\bar{v}_{\A\to\B}^{\z,*,t}
\nonumber \\
&+ \mathbb{E}[\{f_{\B}^{\z}(Z + H_{\z,t}, g(Z, W); \bar{v}_{\A\to\B}^{\z,t}) 
- Z\}^{2}]
- (\bar{\xi}_{\B,t}^{\z})^{2}\bar{v}_{\A\to\B}^{\z,t}. 
\label{moduleB_z_var_bar}
\end{align}

Finally, we define state evolution recursion for the inner sub-module 
in module~B. Let \{$\boldsymbol{\omega}_{\tau}\in\mathbb{R}^{N}\}_{\tau=0}^{t}$ 
denote zero-mean Gaussian random vectors that correspond to the estimation 
errors $\{\boldsymbol{x}_{\A\to\B}^{\tau} - \boldsymbol{x}\}$ and are 
independent of $\boldsymbol{x}$. They have the covariance 
$\mathbb{E}[\boldsymbol{\omega}_{\tau}\boldsymbol{\omega}_{t}^{\mathrm{T}}]
= M^{-1}\bar{v}_{\A\to\B}^{\x,\tau,t}\boldsymbol{I}_{N}$. 
We define asymptotic alternatives of $v_{\B\to\A}^{\x,t}$, 
$-\boldsymbol{x}^{\mathrm{T}}(\boldsymbol{x}_{\B\to\A}^{t+1} - \boldsymbol{x})$,  
and $\xi_{\B,t}^{\x}$ as 
\begin{equation} \label{moduleB_x_var_bar}
\bar{v}_{\B\to\A}^{\x,t+1} 
= \lim_{N\to\infty}\mathbb{E}\left[
 \|f_{\B}^{\x}(\boldsymbol{x} + \boldsymbol{\omega}_{t}; \bar{v}_{\A\to\B}^{\x,t}) 
 - \boldsymbol{x}\|_{2}^{2}
\right],
\end{equation}
\begin{equation} \label{moduleB_x_cov_bar}
\bar{v}_{\B\to\A}^{\x,0,t+1}
= -\lim_{N\to\infty}\mathbb{E}\left[
 \boldsymbol{x}^{\mathrm{T}}\{f_{\B}^{\x}(\boldsymbol{x} 
 + \boldsymbol{\omega}_{t}; \bar{v}_{\A\to\B}^{\x,t}) 
 - \boldsymbol{x}\}
\right],  
\end{equation}
\begin{equation} \label{xi_B_x_bar}
\bar{\xi}_{\B,t}^{\x}
= \lim_{N\to\infty}\frac{1}{M}\mathbb{E}\left[
 \boldsymbol{1}^{\mathrm{T}}
 \partial_{0}f_{\B}^{\x}(\boldsymbol{x} + \boldsymbol{\omega}_{t}; 
 \bar{v}_{\A\to\B}^{\x,t})
\right], 
\end{equation}
respectively, 
where the limits denote the sublinear sparsity limit. Note that the limits  
exist for a Bayesian estimator in \cite{Takeuchi251}. 
The state evolution recursion~(\ref{moduleA_z_var_bar}), 
(\ref{moduleA_x_var_bar}), (\ref{moduleB_z_var_bar}), and  
(\ref{moduleB_x_var_bar}) describes the asymptotic dynamics of GOAMP. 

To prove the main theorem of this paper, we need additional technical 
assumptions. 
\begin{assumption} \label{assumption_inner}
Let $\boldsymbol{\Omega}_{t+1,N}=[\boldsymbol{\omega}_{0,N},\ldots, 
\boldsymbol{\omega}_{t,N}]$ denote a zero-mean 
Gaussian random matrix that is independent of the signal vector 
$\boldsymbol{x}$ and 
has the covariance $\mathbb{E}[\boldsymbol{\omega}_{\tau,N}
\boldsymbol{\omega}_{t,N}^{\mathrm{T}}]=v_{\tau,t,N}\boldsymbol{I}_{N}$ with 
$v_{\tau,t,N}=N^{-1}(\boldsymbol{x}_{\A\to\B}^{\tau}-\boldsymbol{x})^{\mathrm{T}}
(\boldsymbol{x}_{\A\to\B}^{t}-\boldsymbol{x})$. 
Suppose that a random vector $\boldsymbol{c}\in\mathbb{R}^{N}$ satisfies 
the boundedness in probability of $\|\boldsymbol{c}\|_{2}^{2}$ 
in the sublinear sparsity limit. 
\begin{itemize}
\item The inner denoiser $f_{\B}^{\x}(\cdot; v)$ in module~B is 
almost everywhere differentiable with respect to the first variable. 
Furthermore, the sublinear sparsity limit in (\ref{xi_B_x_bar}) exists. 

\item $\{f_{\B}^{\x}(\boldsymbol{x} + \boldsymbol{\omega}_{\tau}; 
\bar{v}_{\A\to\B}^{\x,\tau}) - \boldsymbol{x}\}^{\mathrm{T}}
\{f_{\B}^{\x}(\boldsymbol{x} + \boldsymbol{\omega}_{t}; \bar{v}_{\A\to\B}^{\x,t}) 
- \boldsymbol{x}\}$ and $\boldsymbol{x}^{\mathrm{T}}
\{f_{\B}^{\x}(\boldsymbol{x} + \boldsymbol{\omega}_{t}; \bar{v}_{\A\to\B}^{\x,t}) 
- \boldsymbol{x}\}$ converge 
in probability to their expectation in the sublinear sparsity limit. 

\item 
$\|f_{\B}^{\x}(o(1)\boldsymbol{c} + \boldsymbol{x} 
+ \boldsymbol{\omega}_{t,N};v_{\A\to\B}^{\x,t})
- f_{\B}^{\x}(\boldsymbol{x} + \boldsymbol{\omega}_{t}; 
\bar{v}_{\A\to\B}^{\x,t})\|_{2}\pto0$ holds in the sublinear sparsity limit. 

\item $\boldsymbol{\omega}_{\tau}^{\mathrm{T}}
f_{\B}^{\x}(o(1)\boldsymbol{c} + \boldsymbol{x} 
+ \boldsymbol{\omega}_{t};\bar{v}_{\A\to\B}^{\x,t}) 
- \mathbb{E}[\boldsymbol{\omega}_{\tau}^{\mathrm{T}}
f_{\B}^{\x}(\boldsymbol{x} + \boldsymbol{\omega}_{t};\bar{v}_{\A\to\B}^{\x,t})]$
$\pto0$ holds in the sublinear sparsity limit. 
\end{itemize}
\end{assumption}

Assumption~\ref{assumption_inner} contains strong assumptions for the inner 
denoiser $f_{\B}^{\x}$ in module~B. It needs to be 
justified for each inner denoiser in module~B. For a Bayesian estimator 
$f_{\B}^{\x}$ of $\boldsymbol{x}$ proposed in \cite{Takeuchi251},  
\cite[Lemmas~1--4]{Takeuchi251} justifies Assumption~\ref{assumption_inner}. 

\begin{assumption} \label{assumption_SE}
There is some $T\in\mathbb{N}$ such that 
$\bar{v}_{\A\to\B}^{\z,t}>0$, $\bar{v}_{\A\to\B}^{\x,t}>0$, 
$\bar{v}_{\B\to\A}^{\z,t+1}>0$, and $\bar{v}_{\B\to\A}^{\x,t+1}>0$ in 
(\ref{moduleA_z_var_bar}), (\ref{moduleA_x_var_bar}), 
(\ref{moduleB_z_var_bar}), and (\ref{moduleB_x_var_bar}) hold for all 
$t\in\{0,\ldots,T\}$. 
\end{assumption}

Assumption~\ref{assumption_SE} is a technical assumption to terminate state 
evolution just before achieving error-free signal reconstruction in the 
sublinear sparsity limit. Note that Assumption~\ref{assumption_SE} postulates 
GOAMP for nonlinear measurements. When OAMP is considered for the linear 
measurement, $\bar{v}_{\A\to\B}^{\z,t}>0$ and 
$\bar{v}_{\B\to\A}^{\z,t+1}>0$---not updated in OAMP---can be eliminated. 

To simplify state evolution for GOAMP in the sublinear sparsity limit, 
we consider modified GOAMP with $\xi_{\B,t}^{\x}$ in (\ref{xi_B_x}) replaced 
by $\bar{\xi}_{\B,t}^{\x}$ in (\ref{xi_B_x_bar}). 
This replacement allows us to omit 
the proof of the convergence in probability 
$\xi_{\B,t}^{\x}\pto\bar{\xi}_{\B,t}^{\x}$.  
Modified GOAMP is considered only for theoretical analysis. GOAMP 
with $\xi_{\B,t}^{\x}$ in (\ref{xi_B_x}) is used in numerical simulations.  

\begin{theorem} \label{theorem_SE}
Suppose that Assumptions~\ref{assumption_w}--\ref{assumption_SE} are 
satisfied. Assume that $v_{\B}^{\x,t+1}$ in the inner sub-module for 
module~B is a consistent estimator of 
$\|\boldsymbol{x}_{\B}^{t+1} - \boldsymbol{x}\|_{2}^{2}$ in the sublinear 
sparsity limit. Then, the unnormalized square error 
$\|\boldsymbol{x}_{\B}^{t+1} - \boldsymbol{x}\|_{2}^{2}$ for modified GOAMP 
converges in probability to $\bar{v}_{\B\to\A}^{\x,t+1}$---given via 
the state evolution recursion~(\ref{moduleA_z_var_bar}), 
(\ref{moduleA_x_var_bar}), (\ref{moduleB_z_var_bar}), and  
(\ref{moduleB_x_var_bar})---in the sublinear sparsity limit 
for all $t\in\{0,\ldots,T\}$. 
\end{theorem}
\begin{IEEEproof}
See Appendix~\ref{proof_theorem_SE}. 
\end{IEEEproof}

Note that both $\|\boldsymbol{x}_{\B}^{t+1} - \boldsymbol{x}\|_{2}^{2}$ and 
$\|\boldsymbol{x}_{\B\to\A}^{t+1} - \boldsymbol{x}\|_{2}^{2}$ converge 
in probability to the identical value $\bar{v}_{\B\to\A}^{\x,t+1}$ in 
the sublinear sparsity limit. 
Theorem~\ref{theorem_SE} implies asymptotic Gaussianity for the estimation 
error $\boldsymbol{x}_{\B}^{t+1} - \boldsymbol{x}$: The unnormalized square 
error $\|\boldsymbol{x}_{\B}^{t+1} - \boldsymbol{x}\|_{2}^{2}$ can be described 
with $\bar{v}_{\B\to\A}^{\x,t+1}$ in (\ref{moduleB_x_var_bar}), which is 
the unnormalized square error for the estimator  
$f_{\B}^{\x}(\boldsymbol{x} + \boldsymbol{\omega}_{t}; \bar{v}_{\A\to\B}^{\x,t})$ 
of $\boldsymbol{x}$ from the virtual Gaussian measurement 
$\boldsymbol{x} + \boldsymbol{\omega}_{t}$.   

The meaning of the asymptotic Gaussianity is restrictive for sublinear 
sparsity. The strong assumptions in Assumption~\ref{assumption_inner} are 
required to prove Theorem~\ref{theorem_SE} while the asymptotic Gaussianity 
in linear sparsity was proved for all Lipschitz-continuous inner 
denoisers~\cite{Bayati11,Rangan11,Javanmard13}.  

\subsection{Bayesian GOAMP} 
We first design a Bayesian inner denoiser $f_{\B}^{\x}$ in module~B. 
Consider the following virtual Gaussian measurements: 
\begin{equation} \label{virtual_AWGN}
\boldsymbol{x}_{t} = \boldsymbol{x} + \boldsymbol{\omega}_{t}, \quad 
\boldsymbol{\omega}_{t}\sim\mathcal{N}(\boldsymbol{0}, 
M^{-1}\bar{v}_{\A\to\B}^{\x,t}\boldsymbol{I}_{N}). 
\end{equation}
The posterior mean estimator $f_{\B}^{\x}(\boldsymbol{x}_{t}; 
\bar{v}_{\A\to\B}^{\x,t})=\mathbb{E}[\boldsymbol{x} | \boldsymbol{x}_{t}]$ is 
Bayes-optimal in terms of the unnormalized square error 
$\bar{v}_{\B\to\A}^{\x,t+1}$. Unfortunately, this Bayes-optimal inner denoiser 
requires high complexity, because of Assumption~\ref{assumption_x}, so that 
this paper follows \cite{Takeuchi251} to use the suboptimal Bayesian inner 
denoiser $f_{\B}^{\x}(x_{n,t}; \bar{v}_{\A\to\B}^{\x,t})=\mathbb{E}[x_{n} | x_{n,t}]$ 
and the following posterior variance for the inner sub-module in module~B:
\begin{equation} \label{posterior_variance_Bayes} 
v_{\B}^{\x,t+1} = \sum_{n=1}^{N}\mathbb{E}\left[ 
 \left.
  \left\{
   x_{n} - f_{\B}^{\x}(x_{n,t}; v_{\A\to\B}^{\x,t})
  \right\}^{2}
 \right| x_{n,t} 
\right]. 
\end{equation}
In this case, we use the well-known identity $v_{\B}^{\x,t+1}
=v_{\A\to\B}^{\x,t}\xi_{\B,t}^{\x}$~\cite{Rangan11} to find that $v_{\B\to\A}^{\x,t+1}$ 
in (\ref{moduleB_x_var}) reduces to 
\begin{equation} \label{moduleB_x_var_modify}
v_{\B\to\A}^{\x,t+1} = \frac{v_{\A\to\B}^{\x,t}\xi_{\B,t}^{\x}}{1 - (M/N)\xi_{\B,t}^{\x}}. 
\end{equation}

For the Bayesian inner denoiser, we use \cite[Lemma~1]{Takeuchi251} to confirm  
that $\bar{v}_{\B\to\A}^{\x,t+1}$ in (\ref{moduleB_x_var_bar}) is given by 
\begin{equation} \label{moduleB_x_var_bar_Bayes}
\bar{v}_{\B\to\A}^{\x,t+1} 
= \mathbb{E}\left[
 U^{2}1\left(
  U^{2} < \frac{2\bar{v}_{\A\to\B}^{\x,t}}{\delta}
 \right)
\right]
\end{equation}
for $M/\{k\log(N/k)\}\to\delta$, with $U$ in Assumption~\ref{assumption_x}. 
Using the well-known uncorrelation property 
$\mathbb{E}[\mathbb{E}[x_{n} | x_{n,t}](
\mathbb{E}[x_{n} | x_{n,t}] - x_{n})]=0$, 
we find that (\ref{moduleB_x_cov_bar}) reduces to (\ref{moduleB_x_var_bar}):
\begin{equation}
\bar{v}_{\B\to\A}^{\x,0,t+1}
= -\lim_{N\to\infty}\sum_{n=1}^{N}\mathbb{E}\left[
 x_{n}\{f_{\B}^{\x}(x_{n} + \omega_{n,t}; \bar{v}_{\A\to\B}^{\x,t}) 
 - x_{n}\}
\right] \nonumber
\end{equation}
\begin{equation}
= \lim_{N\to\infty}\sum_{n=1}^{N}\mathbb{E}\left[
 \{f_{\B}^{\x}(x_{n} + \omega_{n,t}; \bar{v}_{\A\to\B}^{\x,t}) 
 - x_{n}\}^{2}
\right]
= \bar{v}_{\B\to\A}^{\x,t+1}. 
\end{equation}
This identity implies that $\bar{v}_{\A\to\B}^{\z,*,t}$ in 
(\ref{moduleA_z_var_bar}) is equal to $\bar{v}_{\A\to\B}^{\z,t}$. 
Thus, the covariance variables are not needed 
to solve the state evolution recursion for Bayesian GOAMP. 
 
We next define the Bayes-optimal outer denoiser $f_{\B}^{\z}$. 
Consider virtual scalar measurements,  
\begin{equation} \label{virtual_outer_measurement_model}
Y = g(Z, W), \quad Z_{t} = Z + H_{\z,t}, 
\end{equation}
where $Z$ and $H_{\z,t}$ are jointly zero-mean Gaussian-distributed with 
the covariance $\mathbb{E}[Z^{2}]=P$, 
$\mathbb{E}[ZH_{\z,t}]=-\bar{v}_{\A\to\B}^{\z,t}$, and 
$\mathbb{E}[H_{\z,t}^{2}]=\bar{v}_{\A\to\B}^{\z,t}$ 
under the assumption $\bar{v}_{\A\to\B}^{\z,*,t}=\bar{v}_{\A\to\B}^{\z,t}$ while 
$W$ in Assumption~\ref{assumption_w} is independent of $Z$ and $H_{\z,t}$. 
We know that the Bayes-optimal outer denoiser 
$f_{\B}^{\z}(Z_{t}, Y; \bar{v}_{\A\to\B}^{\z,t})=\mathbb{E}[Z | Y, Z_{t}]$ minimizes 
the MSE $\mathbb{E}[(f_{\B}^{\z} - Z)^{2}]$ in (\ref{moduleB_z_var_bar}). 

We evaluate the Bayes-optimal outer denoiser. We follow 
\cite[p.~7405]{Takeuchi242} to represent $Z$ as 
$Z\sim Z_{t} + N_{\z,t}$ with $N_{\z,t}\sim\mathcal{N}(0, \bar{v}_{\A\to\B}^{\z,t})$ 
independent of $Z_{t}$ and $W$. This representation implies that the 
Bayes-optimal outer denoiser is given by 
\begin{equation} \label{outer_denoiser_Bayes} 
f_{\B}^{\z}(z_{t}, y; \bar{v}_{\A\to\B}^{\z,t}) 
= \frac{\int z\mathbb{P}_{Y|Z}(y | z)e^{-\frac{(z - z_{t})^{2}}{2\bar{v}_{\A\to\B}^{\z,t}}}dz}
{\int \mathbb{P}_{Y|Z}(y | z)e^{-\frac{(z - z_{t})^{2}}{2\bar{v}_{\A\to\B}^{\z,t}}}dz}, 
\end{equation}
where $\mathbb{P}_{Y|Z}$ denotes the conditional probability mass function of 
discrete $Y$ given $Z$ induced from the randomness of $W$. For absolutely 
continuous $Y$, $\mathbb{P}_{Y|Z}$ should be replaced with the conditional 
probability density function $p_{Y|Z}$. 

This paper refers to GOAMP using the suboptimal Bayesian inner denoiser with 
the posterior variance $v_{\B}^{\x,t+1}$ in (\ref{posterior_variance_Bayes}) 
and the Bayes-optimal outer denoiser as Bayesian GOAMP. For Bayesian GOAMP, 
we follow \cite{Takeuchi251} to replace Assumption~\ref{assumption_inner} 
with the following more practical assumption: 
\begin{assumption} \label{assumption_inner_Bayes}
For $i\in\{1, 2\}$, let $f_{U}^{(i)}(y; kM^{-1}\bar{v}_{\A\to\B}^{\x,t})
=\mathbb{E}[U^{i} | U + \sqrt{k}\omega_{n,t} = y]$ denote the posterior mean 
of $U^{i}$ in Assumption~\ref{assumption_x} for $n\in\mathcal{S}$ when 
the support $\mathcal{S}$ of $\boldsymbol{x}$ is known. 
\begin{itemize}
\item The function $f_{U}^{(1)}$ is uniformly bounded by a linear 
function: There are some $L>0$ and $C>0$ such that 
$|f_{U}^{(1)}(y ;v)| < L|y| + C$ holds for all $y\in\mathbb{R}$ and $v>0$. 

\item The function $f_{U}^{(1)}$ is consistent in the following sense: 
\begin{equation}
\left|
 y - f_{U}^{(1)}(y ;v)
\right| \leq o(|y|)
\quad \hbox{as $v\to0$}
\end{equation} 
for all $y\in\mathbb{R}$. 
\item The function $f_{U}^{(2)}(\cdot; v)$ is uniformly bounded on any bounded 
interval $\mathcal{D}$ for all $v>0$: There is some $C>0$ such that 
$|f_{U}^{(2)}(y; v)|<C$ holds 
for all $y\in\mathcal{D}$ and $v>0$. 
\item The probability $\mathbb{P}(U^{2}=2\bar{v}_{\A\to\B}^{\x,t}/\delta)=0$ 
is satisfied for all $t\in\{0,\ldots,T\}$. 
\end{itemize}
\end{assumption} 

\begin{remark}
The uniform Lipschitz-continuity for $f_{U}^{(1)}(\cdot; v)$ was assumed 
in \cite{Takeuchi251}, instead of the first assumption in 
Assumption~\ref{assumption_inner_Bayes}. However, the uniform 
Lipschitz-continuity was used in \cite{Takeuchi251} only to justify 
the first assumption. Thus, the first assumption is directly assumed 
in this paper. 
\end{remark}

For Gaussian $U\sim\mathcal{N}(0, P)$,  
we have $f_{U}^{(1)}(y; v)=Py/(P + v)$ and $f_{U}^{(2)}(y; v) = Pv/(P+v) 
+ \{f_{U}^{(1)}(y; v)\}^{2}$~\cite{Takeuchi251}. Thus, the first three conditions 
in Assumption~\ref{assumption_inner_Bayes} are satisfied. The last condition 
is a technical assumption to justify Assumption~\ref{assumption_inner} 
for the Bayesian inner denoiser in module~B. 

For discrete $U\in\mathcal{U}=\{u_{q}\neq 0: q\in\{1,\ldots, Q\}\}$, the first three assumptions are 
satisfied~\cite[Appendix A-A]{Takeuchi262}. The last assumption is difficult 
to prove in advance. After solving the state evolution recursion, 
one should confirm that the trajectory does not pass through 
$u_{q}^{2}=2\bar{v}_{\A\to\B}^{\x,t}/\delta$ for all $q$ and $t$. 

To present state evolution recursion for Bayesian GOAMP, define  
$\bar{\xi}_{\B,-1}^{\z} = \mathbb{E}[
\partial_{0}f_{\B}^{\z}(0, Y; \bar{v}_{\A\to\B}^{\z,-1})]$. 
For the Bayes-optimal outer denoiser~(\ref{outer_denoiser_Bayes}), we have 
$\bar{v}_{\A\to\B}^{\z,-1}=P$, 
\begin{equation}
\bar{v}_{\B\to\A}^{\z,t+1} 
= \frac{\bar{\xi}_{\B,t}^{\z}\bar{v}_{\A\to\B}^{\z,t}}
{1 - \bar{\xi}_{\B,t}^{\z}} 
\label{moduleB_z_var_bar_Bayes}
\end{equation}
for all $t\geq-1$, 
$\bar{\eta}_{\B,-1}=\bar{\xi}_{\B,-1}$, and $\bar{\eta}_{\B,t}=0$ for all $t\geq0$. 

As considered in Theorem~\ref{theorem_SE}, for theoretical analysis 
we consider modified Bayesian GOAMP with $\xi_{\B,t}^{\x}$ in (\ref{xi_B_x}) 
replaced by $\bar{\xi}_{\B,t}^{\x}$ in (\ref{xi_B_x_bar}). 
The following theorem provides state evolution results for 
modified Bayesian GOAMP: 

\begin{theorem} \label{theorem_SE_Bayes}
Suppose that Assumptions~\ref{assumption_w}, \ref{assumption_x}, 
\ref{assumption_A}, \ref{assumption_denoisers}, \ref{assumption_SE}, and 
\ref{assumption_inner_Bayes} hold. Then, the unnormalized square error 
$\|\boldsymbol{x}_{\B}^{t+1} - \boldsymbol{x}\|_{2}^{2}$ for modified 
Bayesian GOAMP converges in probability to $\bar{v}_{\B\to\A}^{\x,t+1}$---given 
via the state evolution recursion~(\ref{moduleA_z_var_bar}), 
(\ref{moduleA_x_var_bar}), (\ref{moduleB_x_var_bar_Bayes}), and  
(\ref{moduleB_z_var_bar_Bayes})---in the sublinear sparsity limit 
for all $t\in\{0,\ldots,T\}$.  
\end{theorem}
\begin{IEEEproof}
See Appendix~\ref{proof_theorem_SE_Bayes}. 
\end{IEEEproof}

GVAMP~\cite{Schniter16} used $\eta_{\B,t}=0$ in (\ref{eta_B}) and 
\begin{equation} \label{moduleB_z_var_Bayes}
v_{\B\to\A}^{\z,t+1} = \frac{\xi_{\B,t}^{\z}v_{\A\to\B}^{\z,t}}{1 - \xi_{\B,t}^{\z}}
\end{equation}
instead of $v_{\B\to\A}^{\z,t+1}$ in (\ref{moduleB_z_mean}). Thus, 
Theorem~\ref{theorem_SE_Bayes} implies that GVAMP~\cite{Schniter16} requires 
the Bayes-optimal outer denoiser~(\ref{outer_denoiser_Bayes}). 

To simplify Bayesian GOAMP, we replace the update rule~(\ref{moduleB_z_mean}) 
for $v_{\B\to\A}^{\z,t+1}$ with (\ref{moduleB_z_var_Bayes}). Furthermore, 
numerical simulations showed that the update rule~(\ref{moduleA_z_mean}) for 
$v_{\A\to\B}^{\z,t}$ was not necessarily an accurate estimator for 
$M^{-1}\|\boldsymbol{z}_{\A\to\B}^{t} - \boldsymbol{z}\|_{2}^{2}$. 
Thus, we use the following modification: Let 
\begin{equation} \label{moduleA_z_var_modify}
v_{\A\to\B}^{\z,t} = P
- M^{-1}\|\boldsymbol{z}_{\A\to\B}^{t}\|_{2}^{2}.
\end{equation}
We replace $v_{\A\to\B}^{\z,t}$ in (\ref{moduleA_z_mean}) with 
(\ref{moduleA_z_var_modify}) if (\ref{moduleA_z_var_modify}) is positive. 
The estimator~(\ref{moduleA_z_var_modify}) can be proved to be consistent 
for Bayesian GOAMP in the sublinear sparsity limit. 
Bayesian GOAMP with these replacements is referred to simply 
as Bayesian GOAMP. See Appendix~\ref{appen_pseudocodes} for pseudocodes 
of Bayesian GOAMP and its state evolution recursion.  

\subsection{Linear Measurement} 
We focus on Bayesian GOAMP for the linear measurement 
$g(\boldsymbol{z}, \boldsymbol{w})=\boldsymbol{z}+\boldsymbol{w}$ with 
Gaussian $\boldsymbol{w}\sim\mathcal{N}(\boldsymbol{0}, 
\sigma^{2}\boldsymbol{I}_{M})$. In this case, as discussed in 
Section~\ref{sec_OAMP}, Bayesian GOAMP reduces to Bayesian OAMP with 
$\boldsymbol{z}_{\B\to\A}^{t}=\boldsymbol{y}$ and $v_{\B\to\A}^{\z,t}=\sigma^{2}$. 
Furthermore, the state evolution recursion reduces to 
(\ref{moduleB_x_var_bar_Bayes}) and 
\begin{equation} \label{moduleA_x_var_bar_linear} 
\bar{v}_{\A\to\B}^{\x,t} = \left\{
 \mathbb{E}\left[
  \frac{\Lambda}{\sigma^{2} + \bar{v}_{\B\to\A}^{\x,t}\Lambda}
 \right]
\right\}^{-1},  
\end{equation}
which is obtained from (\ref{xi_A_z_bar}), 
(\ref{moduleA_x_var_bar}), and (\ref{xi_A_x_bar}). Thus, for 
Bayesian OAMP with $\xi_{\B,t}^{\x}$ in (\ref{xi_B_x}) replaced by 
$\bar{\xi}_{\B,t}^{\x}$ in (\ref{xi_B_x_bar})---called modified Bayesian 
OAMP---we arrive at the following corollary as a special case of 
Theorem~\ref{theorem_SE_Bayes}:
\begin{corollary} \label{corollary_SE_Bayes_linear}
Consider the linear measurement with $\boldsymbol{w}\sim\mathcal{N}
(\boldsymbol{0}, \sigma^{2}\boldsymbol{I}_{M})$. 
Suppose that Assumptions~\ref{assumption_x}, \ref{assumption_A}, 
\ref{assumption_SE}, and \ref{assumption_inner_Bayes} hold.  
Then, the unnormalized square error 
$\|\boldsymbol{x}_{\B}^{t+1} - \boldsymbol{x}\|_{2}^{2}$ for modified 
Bayesian OAMP converges in probability to $\bar{v}_{\B\to\A}^{\x,t+1}$---given 
via the state evolution recursion~(\ref{moduleB_x_var_bar_Bayes}) and 
(\ref{moduleA_x_var_bar_linear})---in the sublinear sparsity limit 
for all $t\in\{0,\ldots,T\}$.  
\end{corollary}

See Appendix~\ref{appen_pseudocodes} for a pseudocode of the state evolution 
recursion of Bayesian OAMP. 
Since (\ref{moduleB_x_var_bar_Bayes}) and (\ref{moduleA_x_var_bar_linear}) 
are monotonically non-decreasing, the state evolution recursion converges 
to a fixed point as $t\to\infty$. By investigating properties of fixed 
points, the following theorem is obtained: 

\begin{theorem} \label{theorem_linear}
Let $u_{\mathrm{min}}\geq0$ denote the essential minimum of the non-zero signal 
amplitude $|U|$, i.e.\ the supremum of $u\geq0$ such that 
$\mathbb{P}(|U|>u) = 1$ is satisfied. If $u_{\mathrm{min}}>0$ holds, then 
$\bar{v}_{\B\to\A}^{\x,t}$---given 
via the state evolution recursion~(\ref{moduleB_x_var_bar_Bayes}) and 
(\ref{moduleA_x_var_bar_linear})---converges to zero for any distribution of 
$U$ as $t\to\infty$ if and only if $\delta$ is larger than $\delta_{*}$, 
given by  
\begin{equation} \label{delta_threshold}
\delta_{*} = 2\left\{
 \mathbb{E}\left[
  \frac{u_{\mathrm{min}}^{2}\Lambda}{\sigma^{2} + u_{\mathrm{min}}^{2}\Lambda}
 \right]
\right\}^{-1}. 
\end{equation}
\end{theorem}
\begin{IEEEproof}
See Appendix~\ref{proof_theorem_linear}. 
\end{IEEEproof}

Theorem~\ref{theorem_linear} provides the reconstruction 
threshold~(\ref{delta_threshold}) of Bayesian OAMP for the linear 
measurement: Bayesian OAMP achieves error-free signal reconstruction 
in the sublinear sparsity limit if and only if $M/\{k\log(N/k)\}$ 
tends to $\delta>\delta_{*}$. 

To compare the reconstruction threshold~(\ref{delta_threshold}) with that for 
Bayesian AMP~\cite{Takeuchi251}, as well as the information-theoretic 
result~\cite{Reeves20}, Jensen's inequality is used to optimize 
the reconstruction threshold~(\ref{delta_threshold}) over $\Lambda$ as 
\begin{equation}
\delta_{*} \geq 2\left( 
 \frac{u_{\mathrm{min}}^{2}\mathbb{E}[\Lambda]}
 {\sigma^{2} + u_{\mathrm{min}}^{2}\mathbb{E}[\Lambda]}
\right)^{-1} = 2\left(
 1 + \frac{\sigma^{2}}{u_{\mathrm{min}}^{2}}
\right),  
\end{equation}
where we have used $\mathbb{E}[\Lambda]=1$ in Assumption~\ref{assumption_A}. 
Furthermore, the equality in the 
inequality is satisfied only for $\Lambda=1$ with probability~$1$. 
This optimized reconstruction threshold is equal to that of  
Bayesian AMP~\cite{Takeuchi251} for standard Gaussian sensing matrices. 
A similar result was also observed in \cite{Takeuchi241}. 

To compare the optimized threshold with the 
information-theoretic result~\cite{Reeves20}, assume the constant 
non-zero signals, i.e.\ $U=\sqrt{P}$ with probability~$1$. In this case, 
we have $u_{\mathrm{min}}^{2}=P$, so that the optimized reconstruction threshold 
is strictly larger than the information-theoretic threshold 
$2/\log(1 + P/\sigma^{2})$~\cite{Reeves20}. 

\section{Numerical Results} \label{sec4}
\subsection{Numerical Conditions}
In all numerical results, $k$-sparse signals in Assumption~\ref{assumption_x} 
were simulated for Gaussian $U\sim\mathcal{N}(0, P)$ with $P=1$. 
The Gaussian noise vector 
$\boldsymbol{w}\sim\mathcal{N}(\boldsymbol{0}, \sigma^{2}\boldsymbol{I}_{M})$ 
was considered in the generalized linear measurement~(\ref{measurement}). 

As an approximation of $\boldsymbol{A}$ satisfying 
Assumption~\ref{assumption_A}, the SVD 
structure $\boldsymbol{A}=\boldsymbol{U}_{\mathrm{DCT}}
\boldsymbol{P}_{1}\boldsymbol{\Sigma}
\boldsymbol{P}_{2}\boldsymbol{V}_{\mathrm{DCT}}$ was assumed. In this definition, 
the orthogonal matrices $\boldsymbol{U}_{\mathrm{DCT}}$ and 
$\boldsymbol{V}_{\mathrm{DCT}}$ denote $M$-point and $N$-point discrete cosine 
transforms (DCTs), respectively. The matrices $\boldsymbol{P}_{1}$ and 
$\boldsymbol{P}_{2}$ are independent random permutation matrices. The 
diagonal matrix $\boldsymbol{\Sigma}\in\mathbb{R}^{M\times N}$ has $M$ 
geometric positive singular values considered in \cite{Takeuchi21}. 
In this model, all singular values are uniquely determined once the 
condition number $\kappa\geq1$ of $\boldsymbol{A}$---the ratio of 
the maximum singular value to the minimum singular value---is given. 

\subsection{Linear Measurement} 
For the linear measurement, Bayesian OAMP is compared to Bayesian 
AMP~\cite{Takeuchi251}, FISTA~\cite{Beck09}, and orthogonal matching 
pursuit (OMP)~\cite{Tropp07}.   
To improve the convergence property of Bayesian AMP, 
damping~\cite{Rangan191} was inserted just after the inner denoising. 
The common damping factor over all iterations was optimized via exhaustive 
search for each $\kappa$. Note that damping was not used for Bayesian OAMP. 
FISTA was implemented with 
backtracking~\cite{Beck09} and gradient-based restart~\cite{O'Donoghue15} 
to solve the Lasso problem~(\ref{Lasso}) with $p=1$, in which  
$\lambda$ was optimized via exhaustive search for each $\kappa$. 

\begin{figure}[t]
\centering
\includegraphics[width=\hsize]{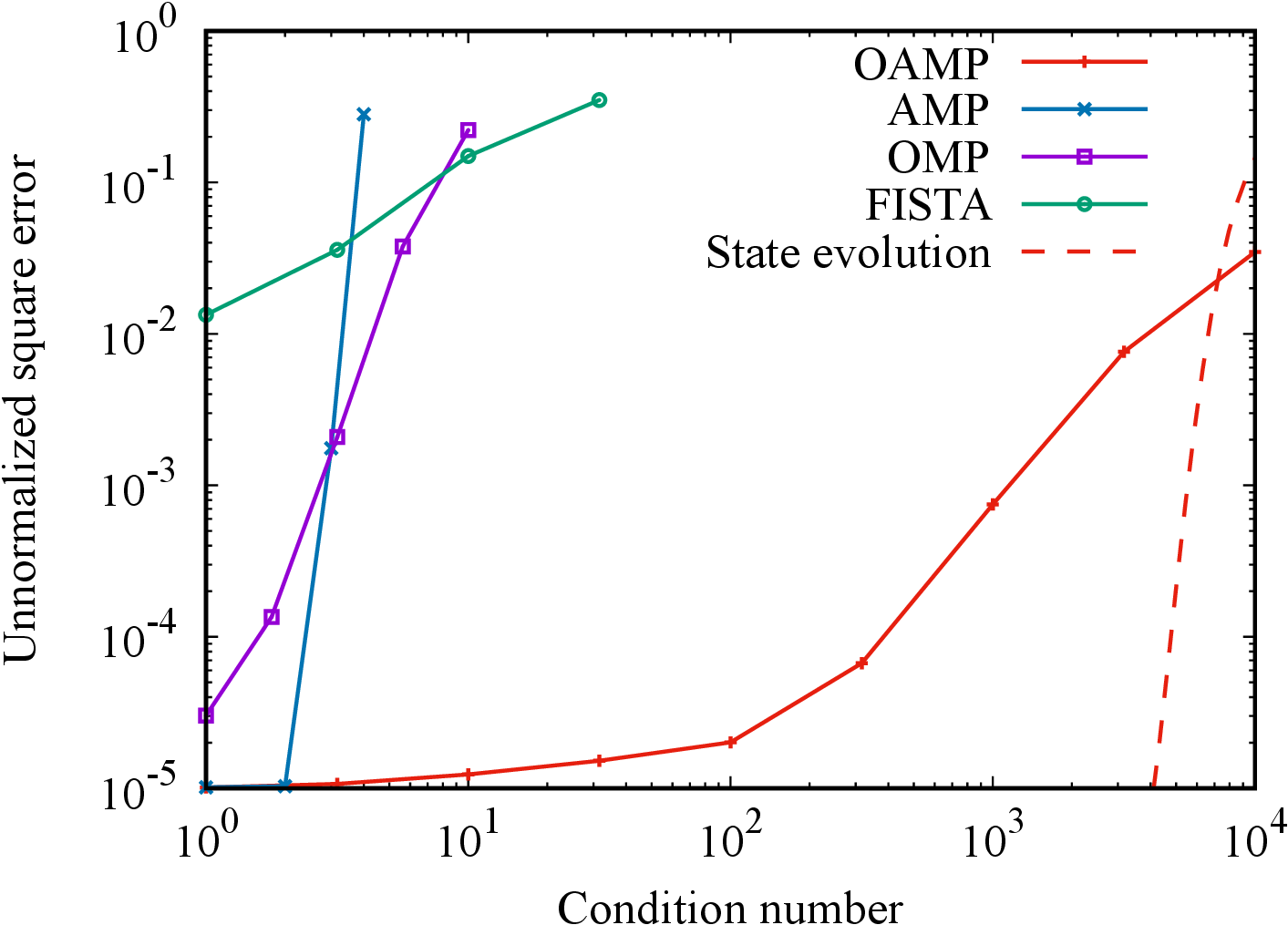}
\caption{Unnormalized square error versus the condition number $\kappa$ for 
the linear measurement, $N=2^{16}$, $k=16$, $M=200$, and $P/\sigma^{2}=40$~dB. 
Bayesian OAMP with $20$ iterations is compared to Bayesian 
AMP~\cite{Takeuchi251}, FISTA~\cite{Beck09}, and OMP~\cite{Tropp07} with $20$, 
$10^{3}$, and $k$ iterations, respectively. $10^{5}$ independent trials were 
simulated for all algorithms. The dashed curve shows the state evolution 
of Bayesian OAMP for $\delta=M/\{k\log(N/k)\}\approx1.50$.}
\label{fig1}
\end{figure}

Figure~\ref{fig1} shows the unnormalized square errors for the four 
algorithms. Bayesian OAMP can achieve good reconstruction performance up to 
condition number $\kappa\approx100$ and the best performance among the four 
algorithms for all $\kappa$. Bayesian AMP~\cite{Takeuchi251} is comparable 
to Bayesian OAMP only for $\kappa\leq2$. As conjectured from 
\cite{Rangan191}, the performance of Bayesian AMP degrades rapidly 
as $\kappa$ increases. FISTA has the worst performance among the four 
algorithms for small $\kappa$ while it outperforms Bayesian AMP and OMP 
for moderate $\kappa$. This is because $10^{3}$ iterations are too few for 
FISTA to converge. FISTA could achieve better performance if it used 
much more iterations, which result in high complexity.  

The state evolution result for Bayesian OAMP provides a reasonably accurate 
prediction of $\kappa$ at which the unnormalized square error changes rapidly.  
Since the sublinear sparsity limit is too far, as shown in 
\cite[Fig.~1]{Takeuchi251}, we cannot expect that the state evolution 
prediction shows a quantitative agreement with numerical results for 
all $\kappa$. To improve the prediction, we need higher-order analysis 
for the Bayesian inner denoiser in module~B, which is outside the scope of 
this paper. 

\begin{figure}[t]
\centering
\includegraphics[width=\hsize]{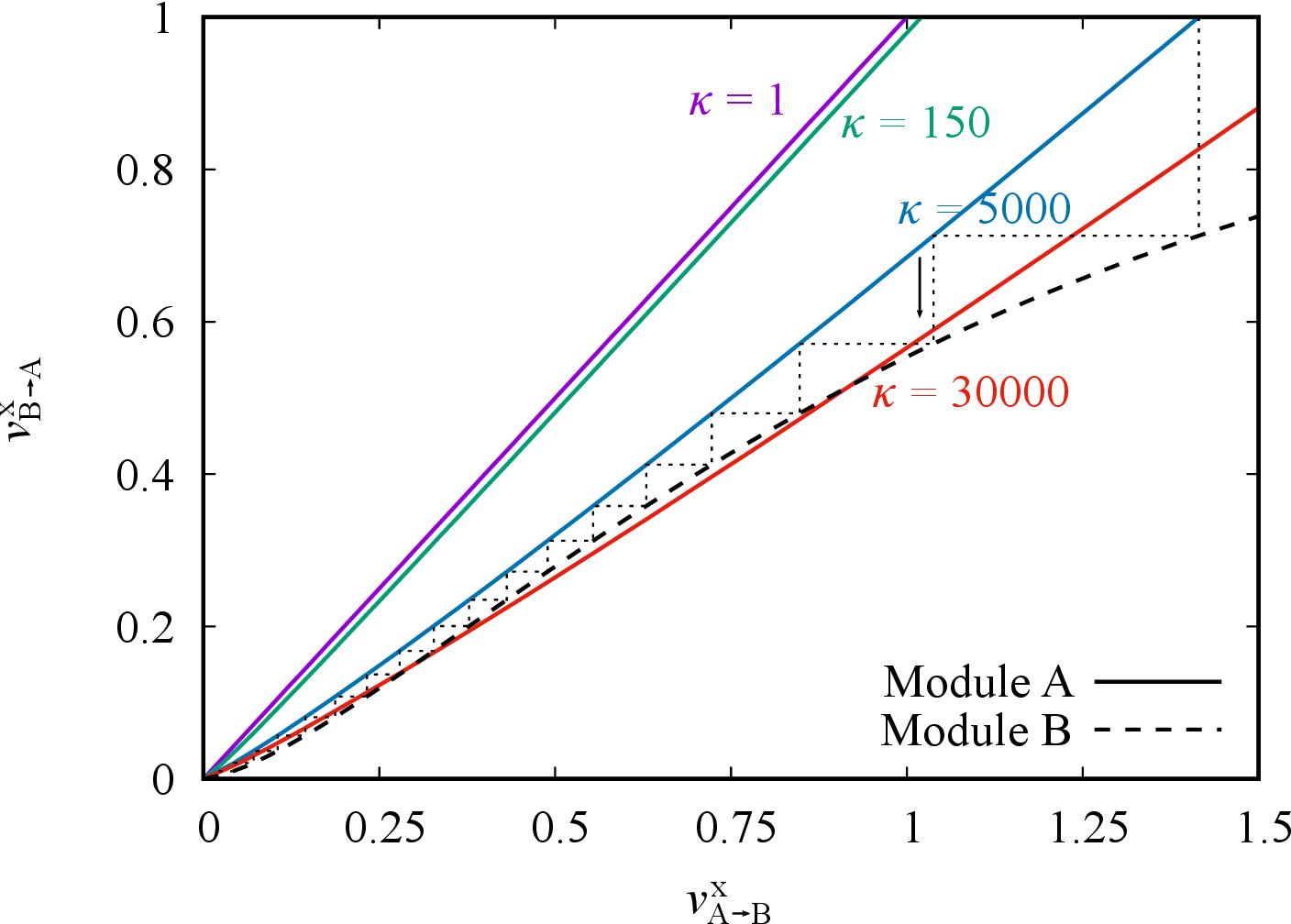}
\caption{State evolution chart of Bayesian OAMP for the linear measurement, 
$\delta=1.5$, and $P/\sigma^{2}=40$~dB. Module~A shows 
(\ref{moduleA_x_var_bar_linear}) while module~B represents 
(\ref{moduleB_x_var_bar_Bayes}). The zigzag line shows the trajectory of 
Bayesian OAMP for $\kappa=5000$. The state evolution recursion does not depend 
on $\gamma$ in $k=N^{\gamma}$.}
\label{fig2}
\end{figure}
 
To understand the qualitative influence of the condition number, we draw a 
chart of Bayesian OAMP based on the state evolution 
recursion~(\ref{moduleB_x_var_bar_Bayes}) and 
(\ref{moduleA_x_var_bar_linear}) in Fig.~\ref{fig2}. When the two kinds of 
curves have a unique 
intersection, as shown for $\kappa=1$, $150$, and $5000$, the state evolution 
recursion converges to the fixed point. Otherwise, it converges to a fixed 
point with the largest unnormalized square error. As the condition number 
increases from $\kappa=1$ to $\kappa=150$, (\ref{moduleA_x_var_bar_linear}) 
for module~A hardly changes. It changes significantly when $\kappa$ gets 
even larger. These observations explain the flat unnormalized square errors 
of Bayesian OAMP for low-to-moderate condition numbers in Fig.~\ref{fig1} 
and the rapid change for large condition numbers.  

\begin{figure}[t]
\centering
\includegraphics[width=\hsize]{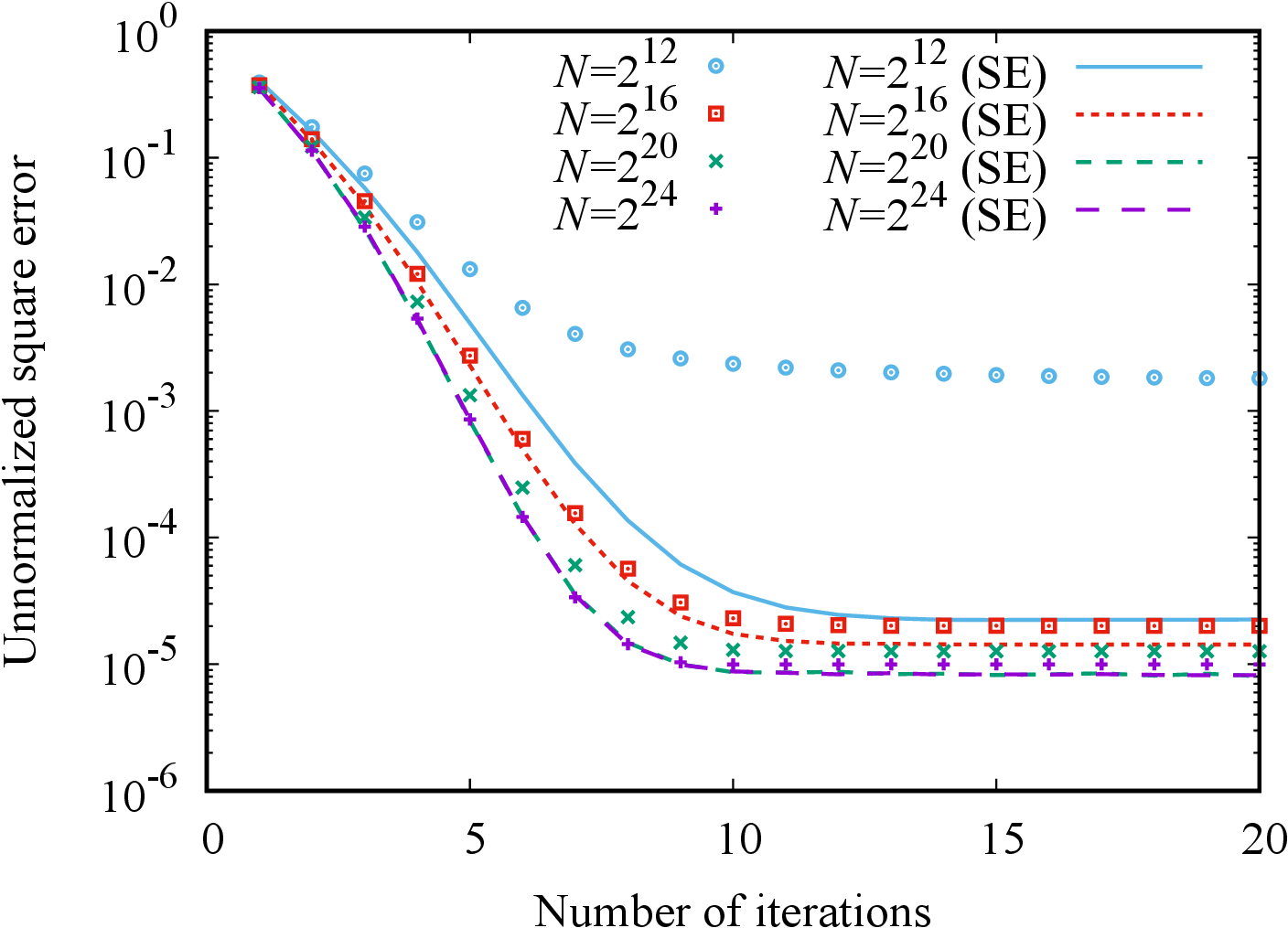}
\caption{Unnormalized square error of Bayesian OAMP versus the number of 
iterations for the linear measurement, $\gamma=1/4$, $\delta\approx 1.50$, 
$\kappa=100$, $P/\sigma^{2}=40$~dB, and $10^{5}$ independent trials. 
SE shows state evolution results with (\ref{moduleB_x_var_bar_Bayes}) 
replaced by (\ref{moduleB_x_var_bar}), which was approximately evaluated 
via numerical sampling for each finite $N$.  
} 
\label{fig3}
\end{figure}

We investigate the influence of the signal dimension $N$ in Fig.~\ref{fig3}. 
To improve the accuracy of state evolution for finite $N$, 
the state evolution recursion~(\ref{moduleB_x_var_bar_Bayes}) was replaced 
with (\ref{moduleB_x_var_bar}), in which the sublinear sparsity limit was 
approximately evaluated via numerical sampling for each $N$. 
The state evolution result for $N=2^{20}$ is indistinguishable from that 
for $N=2^{24}$. Numerical results for $N=2^{24}$ are also close to the 
corresponding state evolution results. These results imply that state 
evolution can accurately predict numerical simulations for large systems 
if finite-length analysis of (\ref{moduleB_x_var_bar}) is available. 

\subsection{1-Bit Compressed Sensing}
For 1-bit compressed sensing, Bayesian GOAMP is compared to Bayesian 
GAMP~\cite{Takeuchi251}, BIHT~\cite{Matsumoto22}, and 
GLasso~\cite{Plan13, Thrampoulidis15}. To improve the convergence property 
of Bayesian GOAMP, damping was inserted at the end of module~B,
\begin{equation}
\boldsymbol{x}_{\B\to\A}^{t+1}:= \theta_{\x}\boldsymbol{x}_{\B\to\A}^{t+1} 
+ (1 - \theta_{\x})\boldsymbol{x}_{\B\to\A}^{t}, 
\end{equation}
\begin{equation}
v_{\B\to\A}^{\x,t+1} := \theta_{\x}v_{\B\to\A}^{\x,t+1} 
+ (1 - \theta_{\x})v_{\B\to\A}^{\x,t},
\end{equation} 
\begin{equation}
\boldsymbol{z}_{\B\to\A}^{t+1}:= \theta_{\z}\boldsymbol{z}_{\B\to\A}^{t+1} 
+ (1 - \theta_{\z})\boldsymbol{z}_{\B\to\A}^{t}, 
\end{equation}
\begin{equation}
v_{\B\to\A}^{\z,t+1} := \theta_{\z}v_{\B\to\A}^{\z,t+1} 
+ (1 - \theta_{\z})v_{\B\to\A}^{\z,t} 
\end{equation} 
for $t>0$, with damping factors $\theta_{\x}\in(0, 1]$ and 
$\theta_{\z}\in(0, 1]$. For Bayesian GAMP~\cite{Takeuchi251}, similarly, 
damping was also inserted just after the inner denoising. The common 
damping factors over all iterations were optimized via exhaustive search 
for each $\kappa$. GLasso~\cite{Plan13, Thrampoulidis15} was implemented 
by using FISTA with backtracking~\cite{Beck09}, gradient-based 
restart~\cite{O'Donoghue15}, and $\lambda$ optimized via exhaustive search 
for each $\kappa$.  

\begin{figure}[t]
\centering
\includegraphics[width=\hsize]{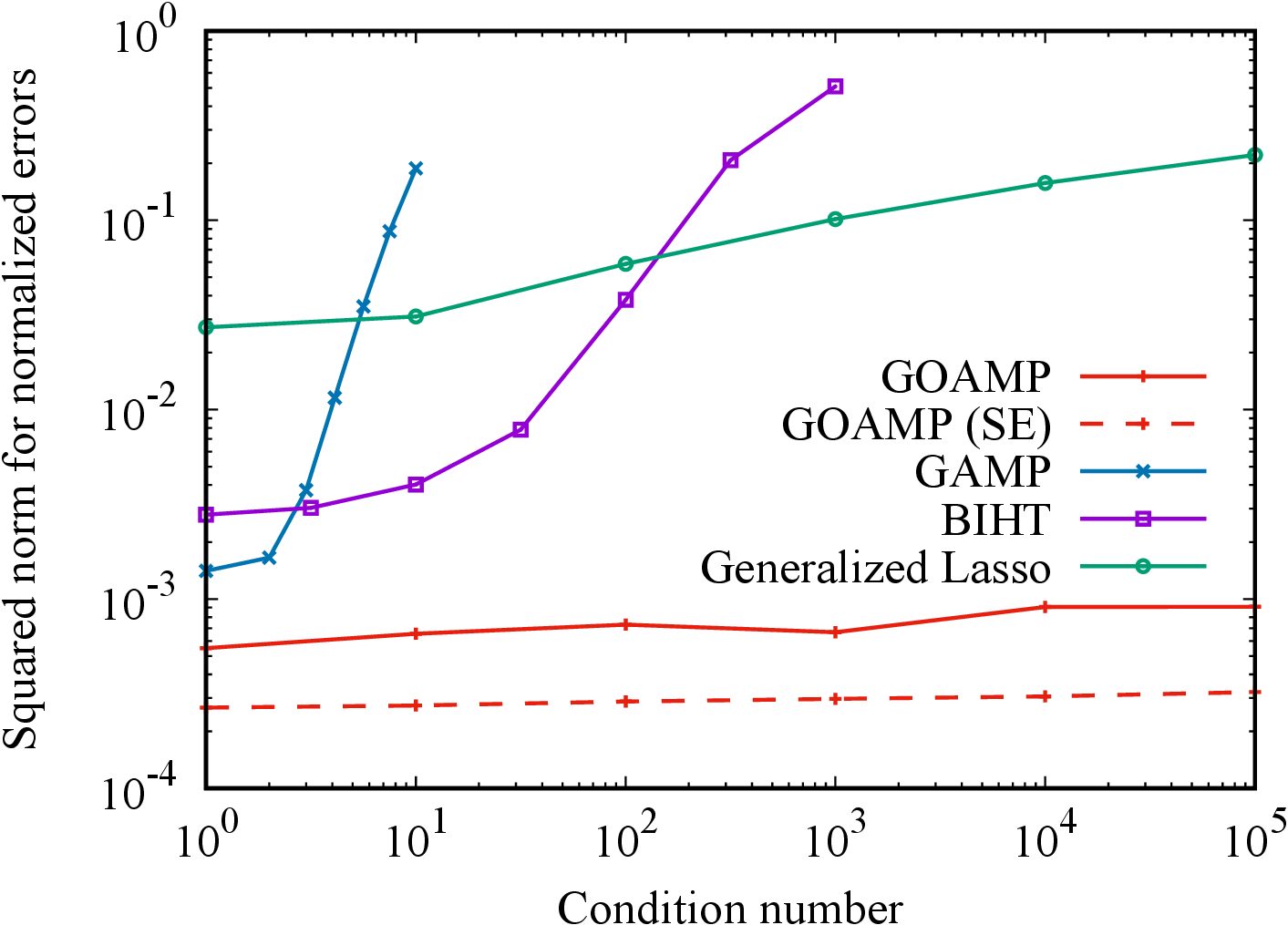}
\caption{Squared norm for the normalized errors versus the condition number 
$\kappa$ for 1-bit compressed sensing, $N=2^{16}$, $k=16$, $M=2000$, and 
$\sigma^{2}=0$. The damping factors in Bayesian GOAMP were set to 
$(\theta_{\x}, \theta_{\z})=(0.6, 0.6)$ for $\kappa>1$ while 
$(\theta_{\x}, \theta_{\z})=(0.7, 1)$ was used for $\kappa=1$.  
Bayesian GOAMP with $20$ iterations is compared to Bayesian 
GAMP~\cite{Takeuchi251}, BIHT~\cite{Matsumoto22}, and 
GLasso~\cite{Plan13, Thrampoulidis15} with $20$, $20$, and $50$ iterations, 
respectively. $10^{4}$ independent trials were simulated for all algorithms. 
SE shows state evolution results of Bayesian GOAMP 
with (\ref{moduleB_x_var_bar_Bayes}) 
replaced by (\ref{moduleB_x_var_bar}), which was approximately evaluated 
via numerical sampling for $N=2^{16}$ and $k=16$.
}
\label{fig4}
\end{figure}

In 1-bit compressed sensing, the measurement vector $\boldsymbol{y}$ is 
independent of the signal power $\|\boldsymbol{x}\|_{2}^{2}$. Thus, 
BIHT and GLasso cannot estimate the signal power while Bayesian GOAMP and 
GAMP can estimate it with the prior information on the signal vector. 
For a fair comparison, the performance of the four algorithms is evaluated 
with the squared norm $\|\boldsymbol{x}/\|\boldsymbol{x}\|_{2} 
- \hat{\boldsymbol{x}}/\|\hat{\boldsymbol{x}}\|_{2}\|_{2}^{2}$ 
for normalized errors.  

Figure~\ref{fig4} shows the squared norm for the normalized errors  of 
the four algorithms. The noiseless case $\sigma=0$ was considered to simulate 
BIHT~\cite{Matsumoto22}. Bayesian GOAMP outperforms Bayesian 
GAMP~\cite{Takeuchi251} even for small condition number $\kappa$ and is robust 
against the increase of $\kappa$. The performance of Bayesian 
GAMP~\cite{Takeuchi251} degrades rapidly as the condition number increases.  
BIHT~\cite{Matsumoto22} is also sensitive to the increase of the condition 
number. GLasso has the worst performance for $\kappa=1$ while it converged  
after $50$ iterations. On the other hand, GLasso is robust 
against the increase of the condition number. As a result, Bayesian GOAMP 
achieves the best performance among the four algorithms for all $\kappa$. 

\begin{figure}[t]
\centering
\includegraphics[width=\hsize]{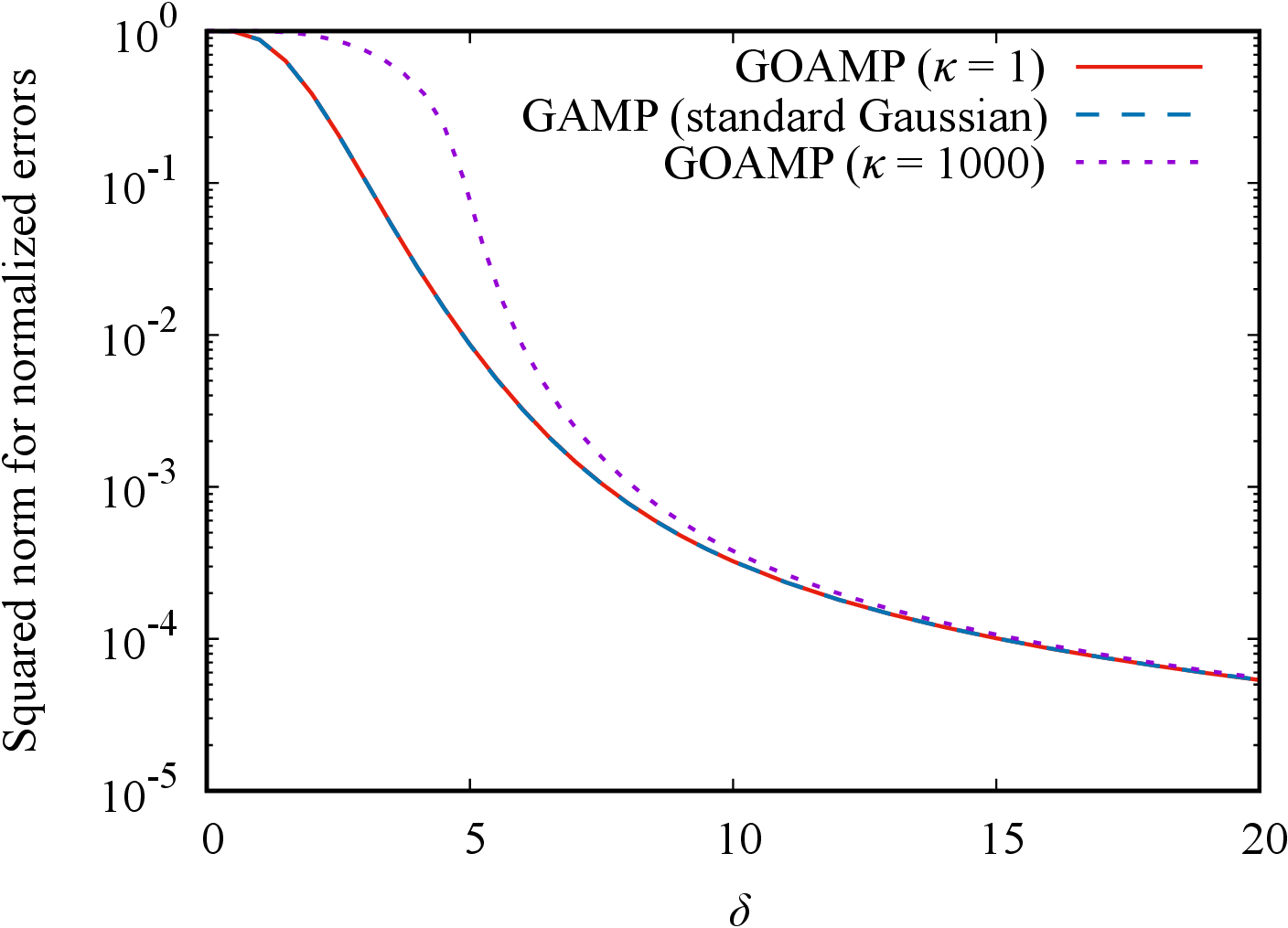}
\caption{Squared norm for the normalized errors versus 
$\delta=M/\{k\log(N/k)\}$ for 1-bit compressed sensing with 
$P/\sigma^{2}=40$~dB in the sublinear sparsity limit. For Bayesian GOAMP, 
the state evolution recursion~(\ref{moduleA_z_var_bar}), 
(\ref{moduleA_x_var_bar}), (\ref{moduleB_x_var_bar_Bayes}), and  
(\ref{moduleB_z_var_bar_Bayes}) was solved with $50$ iterations. 
For Bayesian GAMP, state evolution 
recursion~\cite[Eqs.~(78) and (79)]{Takeuchi251} with $50$ 
iterations was evaluated for standard Gaussian sensing matrices. 
The state evolution results do not depend on $\gamma$.}
\label{fig5}
\end{figure}

It is interesting to observe that Bayesian GAMP is strictly inferior to 
Bayesian GOAMP for $\kappa=1$. To elucidate this reason, Bayesian GOAMP 
is compared to Bayesian GAMP in terms of state evolution. As shown in 
Fig.~\ref{fig5}, the asymptotic performance of Bayesian GOAMP for $\kappa=1$ 
is indistinguishable from that of Bayesian GAMP for standard Gaussian 
sensing matrices. Since standard Gaussian matrices correspond to orthogonally 
invariant matrices with orthogonal rows (i.e.\ $\kappa=1$) in the sublinear 
sparsity limit, we should interpret the gap between Bayesian GOAMP 
and GAMP for $\kappa=1$ in Fig.~\ref{fig4} as finite size effect. 

It is possible to understand why the performance of Bayesian GOAMP hardly 
changes in Fig.~\ref{fig4} as the condition number increases. For the 
numerical conditions in Fig.~\ref{fig4}, we have $\delta\approx 15.0$. For 
this $\delta$, Bayesian GOAMP for $\kappa=1$ is indistinguishable from that 
for $\kappa=1000$ in Fig.~\ref{fig5} 
while there is a significant gap for small $\delta$. This observation is consistent with the robustness of Bayesian GOAMP against the increase of the condition number in Fig.~\ref{fig4}. 

\section{Conclusion} \label{sec5}
This paper has proposed GOAMP for reconstruction of an unknown signal vector 
with sublinear sparsity from the generalized linear measurement. State 
evolution has been utilized to design the Onsager correction in GOAMP for 
sublinear sparsity. For the linear measurement, the optimized reconstruction 
threshold of Bayesian OAMP has been proved to be equal to that of 
Bayesian AMP for standard Gaussian sensing matrices. However, there is a gap 
between the reconstruction threshold and the information-theoretic threshold. 
For the linear measurement and 1-bit compressed sensing, numerical simulations 
have shown that Bayesian GOAMP outperforms existing reconstruction 
algorithms---including Bayesian GAMP---especially for ill-conditioned 
sensing matrices.   

\appendices

\section{Proof of Theorem~\ref{theorem_SE}}
\label{proof_theorem_SE}
\subsection{Overview}
Modified GOAMP with $\xi_{\B,t}^{\x}$ in (\ref{xi_B_x}) replaced 
by $\bar{\xi}_{\B,t}^{\x}$ in (\ref{xi_B_x_bar}) is simply called GOAMP. 
The proof of Theorem~\ref{theorem_SE} consists of four steps. A first step 
is the establishment of a general error model that contains an error model 
of GOAMP. The definition of the general error model is the main challenge 
in state evolution. If the general error model were defined inappropriately, 
existing state evolution for linear sparsity could not be extended to that 
for sublinear sparsity. The general error model is defined so that existing 
state evolution is naturally generalized to that for sublinear sparsity. 
In addition, it is defined as an error model with long-term memory so as to 
analyze the dynamics of long-memory message-passing.  

A second step is proofs of technical results to analyze the general error 
model via state evolution. Normalized quantities are evaluated for the inner 
sub-module in module~A and outer sub-modules while unnormalized quantities 
are considered for the inner sub-module in module~B. Owing to this 
difference in normalization, two types of lemmas are proved to evaluate the 
normalized and unnormalized quantities, respectively. More precisely, 
Lemma~\ref{lemma_elimination} is proved in a unified form that treats both 
normalization cases. The other technical lemmas are essentially equivalent 
to those used for linear sparsity. 

A third step is state evolution of the general error model. Since the 
general error model has been defined appropriately, standard state evolution 
can be used to analyze the dynamics of the general error model. A difference 
between the linear sparsity and sublinear sparsity is in the meaning of 
empirical convergence. In the linear 
sparsity~\cite{Bayati11,Rangan192,Takeuchi20,Fletcher18}, 
the $\mathrm{PL}(2)$ convergence was considered. For the sublinear sparsity, this paper focuses on the product of two piecewise Lipschitz-continuous functions, which is a special instance of piecewise second-order pseudo-Lipschitz functions. 

The last step is the reduction of a general theorem to the GOAMP case. The assumptions in Theorem~\ref{theorem_SE} are utilized to justify non-trivial assumptions in the general theorem.

\subsection{Error Model}
An error model for GOAMP is first defined. 
Let $\boldsymbol{q}_{\z,t+1}=\boldsymbol{z}_{\B\to\A}^{t+1} - \boldsymbol{z}$ and 
$\boldsymbol{q}_{\x,t+1}=\boldsymbol{x}_{\B\to\A}^{t+1} - \boldsymbol{x}$ denote 
the estimation errors just after the Onsager correction for outer and inner 
denoising in module~B, 
respectively. Similarly, we write the estimation errors just after the Onsager 
correction for outer and inner denoising in module~A 
as $\boldsymbol{h}_{\x,t}=\boldsymbol{x}_{\A\to\B}^{t} - \boldsymbol{x}$ and   
$\boldsymbol{h}_{\z,t}=\boldsymbol{z}_{\A\to\B}^{t} - \boldsymbol{z}$. 


We formulate the dynamics of the estimation errors for GOAMP. 
Let $\boldsymbol{\lambda}\in\mathbb{R}^{M}$ denote a vector 
that is composed of the diagonal elements of $\boldsymbol{\Lambda}
=N^{-1}\boldsymbol{\Sigma}\boldsymbol{\Sigma}^{\mathrm{T}}$ 
in Assumption~\ref{assumption_A}, i.e.\ $\boldsymbol{\Lambda}
=\mathrm{diag}\{\boldsymbol{\lambda}\}$. For module~A, define 
\begin{equation}
\phi_{\z,t}^{\mathrm{GO}}(\boldsymbol{b}_{\z,t}, 
\boldsymbol{\Sigma}\boldsymbol{b}_{\x,t}, 
\boldsymbol{\lambda})
= \boldsymbol{\Sigma}\boldsymbol{b}_{\x,t}
+ \boldsymbol{\Lambda}\phi_{t}^{\mathrm{GO}}
(\boldsymbol{\Sigma}\boldsymbol{b}_{\x,t}, 
\boldsymbol{b}_{\z,t}, \boldsymbol{\lambda}), 
\label{phi_z_GOAMP}
\end{equation}
\begin{equation}
\phi_{\x,t}^{\mathrm{GO}}(\boldsymbol{b}_{\x,t}, \boldsymbol{b}_{\z,t},  
\boldsymbol{\Sigma})
= \boldsymbol{b}_{\x,t} + \frac{1}{M}\boldsymbol{\Sigma}^{\mathrm{T}}
\phi_{t}^{\mathrm{GO}}(\boldsymbol{\Sigma}\boldsymbol{b}_{\x,t}, 
\boldsymbol{b}_{\z,t}, \boldsymbol{\lambda}), 
\label{phi_x_GOAMP}
\end{equation}
with
\begin{equation} \label{phi_t_GOAMP}
\phi_{t}^{\mathrm{GO}}(\boldsymbol{\Sigma}\boldsymbol{b}_{\x,t}, 
\boldsymbol{b}_{\z,t}, \boldsymbol{\lambda})
= \left(
 \frac{v_{\B\to\A}^{\z,t}}{v_{\B\to\A}^{\x,t}}\boldsymbol{I}_{M} 
 + \boldsymbol{\Lambda}
\right)^{-1}
(\boldsymbol{b}_{\z,t} - \boldsymbol{\Sigma}\boldsymbol{b}_{\x,t}). 
\end{equation}

For module~B, let 
\begin{equation} 
\psi_{\z,t}^{\mathrm{GO}}(\boldsymbol{h}_{\z,t}, \boldsymbol{z}, 
\boldsymbol{w})
= f_{\B}^{\z}(\boldsymbol{z} + \boldsymbol{h}_{\z,t}, 
g(\boldsymbol{z}, \boldsymbol{w}); v_{\A\to\B}^{\z,t}) 
- \boldsymbol{z},
\label{psi_z_GOAMP}
\end{equation}
\begin{equation} 
\psi_{\x,t}^{\mathrm{GO}}(\boldsymbol{h}_{\x,t}, \boldsymbol{x}) 
= f_{\B}^{\x}(\boldsymbol{x} + \boldsymbol{h}_{\x,t}; v_{\A\to\B}^{\x,t}) 
- \boldsymbol{x},
\label{psi_x_GOAMP}
\end{equation}
with the measurement $g(\boldsymbol{z}, \boldsymbol{w})$ in 
(\ref{measurement}). 
In these definitions, the superscripts ``GO'' mean GOAMP. 

\begin{lemma} \label{lemma_error_model} 
The estimation errors of GOAMP satisfy the following dynamical system: 
\begin{equation} \label{b_GOAMP}
(\boldsymbol{b}_{\z,t}, \boldsymbol{b}_{\x,t}) 
= \left(
 \boldsymbol{U}^{\mathrm{T}}\boldsymbol{q}_{\z,t}, 
 \boldsymbol{V}^{\mathrm{T}}\boldsymbol{q}_{\x,t}
\right), 
\end{equation}
\begin{equation} \label{mz_GOAMP}
(\boldsymbol{m}_{\z,t}, \boldsymbol{m}_{\x,t})
= \left(
 \frac{\phi_{\z,t}^{\mathrm{GO}} - \xi_{\A,t}^{\z}\boldsymbol{b}_{\z,t}}
 {1 - \xi_{\A,t}^{\z}}, 
 \frac{\phi_{\x,t}^{\mathrm{GO}} - \xi_{\A,t}^{\x}\boldsymbol{b}_{\x,t}}
 {1 - \xi_{\A,t}^{\x}}
\right), 
\end{equation}
\begin{equation} \label{h_GOAMP}
(\boldsymbol{h}_{\z,t},\boldsymbol{h}_{\x,t}) 
= \left(
 \boldsymbol{U}\boldsymbol{m}_{\z,t}, 
 \boldsymbol{V}\boldsymbol{m}_{\x,t}
\right),
\end{equation}
\begin{align} 
&(\boldsymbol{q}_{\z,t+1}, \boldsymbol{q}_{\x,t+1}) 
\nonumber \\
&= \left(
 \frac{\psi_{\z,t}^{\mathrm{GO}} - \xi_{\B,t}^{\z}\boldsymbol{h}_{\z,t} 
 + \eta_{\B,t}\boldsymbol{z}} {1 - \xi_{\B,t}^{\z} - \eta_{\B,t}},
 \frac{\psi_{\x,t}^{\mathrm{GO}} - (M/N)\xi_{\B,t}^{\x}\boldsymbol{h}_{\x,t}}
 {1 - (M/N)\xi_{\B,t}^{\x}}
\right),\label{q_GOAMP}
\end{align}
with the initial condition
\begin{equation} \label{q_0_GOAMP}
(\boldsymbol{q}_{\z,0}, \boldsymbol{q}_{\x,0}) 
= \left(
 \frac{\psi_{\z,-1}^{\mathrm{GO}}(\boldsymbol{z},\boldsymbol{w})
 + \eta_{\B,-1}\boldsymbol{z}}{1 - \eta_{\B,-1}},
 - \boldsymbol{x}
\right)
\end{equation} 
for $\psi_{\z,-1}^{\mathrm{GO}}(z,w) = f_{\B}^{\z}(0, g(z, w); 
v_{\A\to\B}^{\z,-1}) - z$. 
In the error model, $\xi_{\A,t}^{a}$, $\xi_{\B,t}^{a}$, and $\eta_{\B,t}$  
for $a\in\{\x,\z\}$ are defined as  
\begin{equation} \label{xi_A_z_GOAMP}
\xi_{\A,t}^{\z} = \left\langle
 \partial_{0}\phi_{\z,t}^{\mathrm{GO}}(\boldsymbol{b}_{\z,t}, 
 \boldsymbol{\Sigma}\boldsymbol{b}_{\x,t}, \boldsymbol{\lambda})
\right\rangle, 
\end{equation}
\begin{equation} \label{xi_A_x_GOAMP}
\xi_{\A,t}^{\x} = \left\langle
 \partial_{0}\phi_{\x,t}^{\mathrm{GO}}(\boldsymbol{b}_{\x,t}, 
 \boldsymbol{b}_{\z,t}, \boldsymbol{\Sigma})
\right\rangle,
\end{equation}
\begin{equation} \label{xi_B_z_GOAMP}
\xi_{\B,t}^{\z} = \langle \partial_{0}\psi_{\z,t}^{\mathrm{GO}}(\boldsymbol{h}_{\z,t}, 
\boldsymbol{z}, \boldsymbol{w}) \rangle,
\end{equation}
\begin{equation} \label{eta_B_GOAMP}
\eta_{\B,t}
= - \left\langle
 \partial_{1}\psi_{\z,t}^{\mathrm{GO}}(\boldsymbol{h}_{\z,t}, \boldsymbol{z},
 \boldsymbol{w})
\right\rangle,
\end{equation}
\begin{equation} \label{xi_B_x_GOAMP}
\xi_{\B,t}^{\x} = M^{-1}\boldsymbol{1}^{\mathrm{T}}
\partial_{0}\psi_{\x,t}^{\mathrm{GO}}(\boldsymbol{h}_{\x,t}, \boldsymbol{x}), 
\end{equation}
with $\eta_{\B,-1} = -\langle \partial_{0}\psi_{\z,-1}^{\mathrm{GO}}
(\boldsymbol{z}, \boldsymbol{w}) \rangle$. 
\end{lemma}

\begin{IEEEproof}
We first confirm (\ref{q_GOAMP}). Using 
$\boldsymbol{x}_{\B\to\A}^{t+1}$ in (\ref{moduleB_x_mean}) yields 
\begin{equation}
\boldsymbol{q}_{\x,t+1} 
= \frac{f_{\B}^{\x} - (M/N)\xi_{\B,t}^{\x}(\boldsymbol{x} 
+ \boldsymbol{h}_{\x,t})}
{1 - (M/N)\xi_{\B,t}^{\x}} - \boldsymbol{x}, 
\end{equation}
which is equivalent to $\boldsymbol{q}_{\x,t+1}$ in (\ref{q_GOAMP}) 
with $\psi_{\x,t}^{\mathrm{GO}}$ in (\ref{psi_x_GOAMP}). Similarly, 
the expression of $\boldsymbol{q}_{\z,t+1}$ 
in (\ref{q_GOAMP}) can be confirmed:
\begin{equation}
\boldsymbol{q}_{\z,t+1} 
= \frac{f_{\B}^{\z} - \xi_{\B,t}^{\z}
(\boldsymbol{z} + \boldsymbol{h}_{\z,t})}{1 - \xi_{\B,t}^{\z} - \eta_{\B,t}}
- \boldsymbol{z},
\end{equation}
which is equivalent to $\boldsymbol{q}_{\z,t+1}$ in (\ref{q_GOAMP}) 
with $\psi_{\z,t}^{\mathrm{GO}}$ in 
(\ref{psi_z_GOAMP}) and $\boldsymbol{y}=g(\boldsymbol{z},\boldsymbol{w})$ 
in (\ref{measurement}). 
Furthermore, we use the initial conditions 
$\boldsymbol{x}_{\B\to\A}^{0}=\boldsymbol{0}$ and (\ref{initial_condition_z}) 
to obtain the initial condition~(\ref{q_0_GOAMP}). 

We next prove $\boldsymbol{m}_{\x,t}$ in (\ref{mz_GOAMP}). Using 
$\boldsymbol{x}_{\A\to\B}^{t}$ in (\ref{moduleA_x_mean}) yields 
\begin{equation} 
\boldsymbol{h}_{\x,t} 
= \frac{f_{\A}^{\x} 
- \xi_{\A,t}^{\x}\boldsymbol{x}_{\B\to\A}^{t}}{1 - \xi_{\A,t}^{\x}} 
- \boldsymbol{x} 
= \frac{f_{\A}^{\x} - \boldsymbol{x} 
- \xi_{\A,t}^{\x}\boldsymbol{q}_{\x,t}}{1 - \xi_{\A,t}^{\x}}. 
\label{h_x_tmp}
\end{equation}
From the definition of $f_{\A}^{\x}$ in (\ref{LMMSE_x}) and 
$\boldsymbol{z}=\boldsymbol{A}\boldsymbol{x}$ we have 
\begin{align}
&f_{\A}^{\x}(\boldsymbol{x} + \boldsymbol{q}_{\x,t}, 
\boldsymbol{z} + \boldsymbol{q}_{\z,t}; 
v_{\B\to\A}^{\x,t}, v_{\B\to\A}^{\z,t}) - \boldsymbol{x} 
\nonumber \\
&= \boldsymbol{q}_{\x,t} 
+ \frac{N}{M}\boldsymbol{A}^{\mathrm{T}}
\left(
 \frac{Nv_{\B\to\A}^{\z,t}}{v_{\B\to\A}^{\x,t}}\boldsymbol{I}_{M} 
 + \boldsymbol{A}\boldsymbol{A}^{\mathrm{T}}
\right)^{-1}
(\boldsymbol{q}_{\z,t} - \boldsymbol{A}\boldsymbol{q}_{\x,t})
\nonumber \\
&= \boldsymbol{V}\phi_{\x,t}^{\mathrm{GO}}(\boldsymbol{b}_{\x,t}, 
\boldsymbol{b}_{\z,t}, \boldsymbol{\Sigma}), \label{mx_GOAMP_proof}
\end{align}
with $\phi_{\x,t}^{\mathrm{GO}}$ in (\ref{phi_x_GOAMP}), where the last equality 
follows from the SVD $\boldsymbol{A}=\boldsymbol{U}\boldsymbol{\Sigma}
\boldsymbol{V}^{\mathrm{T}}$ and the definition of $(\boldsymbol{b}_{\z,t}, 
\boldsymbol{b}_{\x,t})$ in (\ref{b_GOAMP}). Substituting this expression into 
(\ref{h_x_tmp}) and using the definitions of $\boldsymbol{b}_{\x,t}$ and 
$\boldsymbol{h}_{\x,t}$ in (\ref{b_GOAMP}) and (\ref{h_GOAMP}), 
we arrive at $\boldsymbol{m}_{\x,t}$ in (\ref{mz_GOAMP}). 

Similarly, the expression~(\ref{mz_GOAMP}) for $\boldsymbol{m}_{\z,t}$ 
can be proved: 
\begin{equation}
\boldsymbol{h}_{\z,t} 
= \frac{f_{\A}^{\z} - \xi_{\A,t}^{\z}\boldsymbol{z}_{\B\to\A}^{t}}
{1 - \xi_{\A,t}^{\z}} - \boldsymbol{z}
= \frac{f_{\A}^{\z} - \boldsymbol{z} 
- \xi_{\A,t}^{\z}\boldsymbol{q}_{\z,t}}{1 - \xi_{\A,t}^{\z}}. 
\end{equation}
Using the following expression:
\begin{align}
&f_{\A}^{\z}(\boldsymbol{z} + \boldsymbol{q}_{\z,t}, 
\boldsymbol{A}(\boldsymbol{x} + \boldsymbol{q}_{\x,t}); 
v_{\B\to\A}^{\z,t}, v_{\B\to\A}^{\x, t})
- \boldsymbol{z} 
\nonumber \\
&= \boldsymbol{A}\boldsymbol{q}_{\x,t} 
+ \boldsymbol{A}\boldsymbol{A}^{\mathrm{T}}
\left(
 \frac{Nv_{\B\to\A}^{\z,t}}{v_{\B\to\A}^{\x,t}}\boldsymbol{I}_{M} 
 + \boldsymbol{A}\boldsymbol{A}^{\mathrm{T}}
\right)^{-1}
(\boldsymbol{q}_{\z,t} - \boldsymbol{A}\boldsymbol{q}_{\x,t})
\nonumber \\
&= \boldsymbol{U}\phi_{\z,t}^{\mathrm{GO}}(\boldsymbol{b}_{\z,t}, 
\boldsymbol{\Sigma}\boldsymbol{b}_{\x,t}, \boldsymbol{\lambda}), 
\label{phi_z_GOAMP_tmp}
\end{align}
with $\phi_{\z,t}^{\mathrm{GO}}$ in (\ref{phi_z_GOAMP}), we arrive at 
$\boldsymbol{m}_{\z,t}$ in (\ref{mz_GOAMP}). 

Finally, we confirm (\ref{xi_A_z_GOAMP})--(\ref{xi_B_x_GOAMP}). 
Applying (\ref{phi_z_GOAMP_tmp}) to the right-hand side (RHS) on  
(\ref{xi_A_z_GOAMP}) and using $\boldsymbol{b}_{\z,t}$ in (\ref{b_GOAMP}) and 
the linearity of $f_{\A}^{\z}$ in (\ref{LMMSE_z}), we find that 
(\ref{xi_A_z_GOAMP}) is equivalent to (\ref{xi_A_z}). Similarly, we can 
confirm that (\ref{xi_A_x_GOAMP})--(\ref{xi_B_x_GOAMP})
are equivalent to (\ref{xi_A_x}), (\ref{xi_B_z}), (\ref{eta_B}), and 
(\ref{xi_B_x}), respectively. Thus, Lemma~\ref{lemma_error_model} holds. 
\end{IEEEproof}

The error model for GOAMP in Lemma~\ref{lemma_error_model} contains 
the Onsager correction for $\boldsymbol{z}$ in (\ref{q_GOAMP}) while GOAMP 
does not perform such Onsager correction in (\ref{moduleB_z_mean}). 
This definition of the error model simplifies state evolution. 

To unify the notation, we define $\tilde{\boldsymbol{h}}=-\boldsymbol{z}$ and $\tilde{\boldsymbol{m}}=\boldsymbol{U}^{\mathrm{T}}\tilde{\boldsymbol{h}}$. Using $\boldsymbol{z}=\boldsymbol{A}\boldsymbol{x}$ in (\ref{measurement}), the SVD $\boldsymbol{A}=\boldsymbol{U}\boldsymbol{\Sigma}\boldsymbol{V}^{\mathrm{T}}$, $\boldsymbol{q}_{\x,0}= -\boldsymbol{x}$, and $\boldsymbol{b}_{\x,0}$ in (\ref{b_GOAMP}), we obtain $\tilde{\boldsymbol{m}}=\boldsymbol{\Sigma}\boldsymbol{b}_{\x,0}$. 

We propose a general error model that contains the error model for GOAMP. 
In the general error model, the denoisers~(\ref{phi_z_GOAMP}), 
(\ref{phi_x_GOAMP}), (\ref{psi_z_GOAMP}), and (\ref{psi_x_GOAMP}) are 
generalized to those with long-term memory. 
The general error model is available as a framework of state evolution 
for long-memory message-passing in the sublinear sparsity limit. 

We first introduce several notations used in the general error model. 
Let $\boldsymbol{B}_{\z,t}=[\boldsymbol{b}_{\z,0},\ldots,
\boldsymbol{b}_{\z,t-1}]$. Similarly, we define other matrices, 
such as $\boldsymbol{B}_{\x,t}$, $\boldsymbol{H}_{\z,t}$, and 
$\boldsymbol{H}_{\x,t}$. 
In the proposed general error model, we replace $\phi_{\z,t}^{\mathrm{GO}}$, 
$\phi_{\x,t}^{\mathrm{GO}}$, $\psi_{\z,t}^{\mathrm{GO}}$, and $\psi_{\x,t}^{\mathrm{GO}}$ 
in (\ref{phi_z_GOAMP}), (\ref{phi_x_GOAMP}), 
(\ref{psi_z_GOAMP}), and (\ref{psi_x_GOAMP}) with more general functions 
\begin{equation} \label{phi_z}
\phi_{\z,t} = \phi_{\z,t}(\boldsymbol{B}_{\z,t+1}, 
\boldsymbol{\Sigma}\boldsymbol{B}_{\x,t+1}, \boldsymbol{\lambda}), 
\end{equation}
\begin{equation} \label{phi_x}
\phi_{\x,t} = \boldsymbol{b}_{\x,t} 
+ \frac{1}{M}\boldsymbol{\Sigma}^{\mathrm{T}}
\tilde{\phi}_{\x,t}(\boldsymbol{\Sigma}\boldsymbol{B}_{\x,t+1}, 
\boldsymbol{B}_{\z,t+1}, \boldsymbol{\lambda}), 
\end{equation}
\begin{equation} \label{psi_z}
\psi_{\z,t} 
= \psi_{\z,t}(\boldsymbol{H}_{\z,t+1}, \tilde{\boldsymbol{h}}, 
\boldsymbol{w}), 
\end{equation}
\begin{equation} \label{psi_x}
\psi_{\x,t} = \psi_{\x,t}(\boldsymbol{H}_{\x,t+1}, \boldsymbol{x}), 
\end{equation}
with $\phi_{\z,t}: \mathbb{R}^{2t+3}\to\mathbb{R}$, 
$\tilde{\phi}_{\x,t}: \mathbb{R}^{2t+3}\to\mathbb{R}$, 
$\psi_{\z,t}: \mathbb{R}^{t+3}\to\mathbb{R}$, and 
$\psi_{\x,t}: \mathbb{R}^{t+2}\to\mathbb{R}$. 

\begin{assumption} \label{assumption_denoisers_general}
The functions $\phi_{\z,t}$ and $\tilde{\phi}_{\x,t}$ in (\ref{phi_z}) and 
(\ref{phi_x}) are piecewise Lipschitz-continuous with respect to the first 
$2(t+1)$ variables. 
The function $\psi_{\z,t}$ in (\ref{psi_z}) is piecewise 
Lipschitz-continuous with respect to all variables. 
\end{assumption}

The Lipschitz continuity of denoisers is the standard assumption in 
state evolution~\cite{Bayati11,Rangan192,Takeuchi20}. The piecewiseness 
provides no essential changes in state evolution as long as asymptotic 
Gaussianity holds with ``piecewise'' replaced by ``everywhere.'' 
Assumptions for $\psi_{\x,t}$ in (\ref{psi_x}) are defined shortly.

The general error model is given by 
\begin{equation} \label{b}
(\boldsymbol{b}_{\z,t}, \boldsymbol{b}_{\x,t}) 
= \left(
 \boldsymbol{U}^{\mathrm{T}}\boldsymbol{q}_{\z,t}, 
 \boldsymbol{V}^{\mathrm{T}}\boldsymbol{q}_{\x,t}
\right), 
\end{equation}
\begin{equation} \label{mz}
\boldsymbol{m}_{\z,t}
= \phi_{\z,t} - \sum_{\tau=0}^{t}\xi_{\A,t,\tau}^{\z}\boldsymbol{b}_{\z,\tau}, 
\end{equation}
\begin{equation} \label{mx}
\boldsymbol{m}_{\x,t}
= \phi_{\x,t} - \sum_{\tau=0}^{t}\xi_{\A,t,\tau}^{\x}\boldsymbol{b}_{\x,\tau}, 
\end{equation}
\begin{equation} \label{h}
(\boldsymbol{h}_{\z,t},\boldsymbol{h}_{\x,t}) 
= \left(
 \boldsymbol{U}\boldsymbol{m}_{\z,t}, 
 \boldsymbol{V}\boldsymbol{m}_{\x,t}
\right),
\end{equation}
\begin{equation} \label{qz}
\boldsymbol{q}_{\z,t+1}
= \psi_{\z,t} - \sum_{\tau=0}^{t}\xi_{\B,t,\tau}^{\z}\boldsymbol{h}_{\z,\tau}
- \eta_{\B,t}\tilde{\boldsymbol{h}}, 
\end{equation}
\begin{equation} \label{qx}
\boldsymbol{q}_{\x,t+1}
= \psi_{\x,t} - \frac{M}{N}\sum_{\tau=0}^{t}\xi_{\B,t,\tau}^{\x}
\boldsymbol{h}_{\x,\tau}, 
\end{equation}
with the initial condition
\begin{equation} \label{initial_condition_general}
(\boldsymbol{q}_{\z,0}, \boldsymbol{q}_{\x,0}) 
= \left(
 \psi_{\z,-1}(\tilde{\boldsymbol{h}}, \boldsymbol{w}) 
 - \eta_{\B,-1}\tilde{\boldsymbol{h}},
 - \boldsymbol{x}
\right). 
\end{equation}
In the general error model, the variables $\xi_{\A,t,\tau}^{a}$, 
$\xi_{\B,t,\tau}^{a}$, and $\eta_{\B,t}$ for $a\in\{\x, \z\}$ are given by 
\begin{equation} \label{xi_A_z_general}
\xi_{\A,t,\tau}^{\z}
= \left\langle
 \partial_{\tau}\phi_{\z,t}(\boldsymbol{B}_{\z,t+1}, 
 \boldsymbol{\Sigma}\boldsymbol{B}_{\x,t+1}, \boldsymbol{\lambda})
\right\rangle,
\end{equation}
\begin{equation} \label{xi_A_x_general}
\xi_{\A,t,\tau}^{\x} 
= \delta_{t,\tau} + \langle\boldsymbol{\Lambda}
\partial_{\tau}\tilde{\phi}_{\x,t}(\boldsymbol{\Sigma}
\boldsymbol{B}_{\x,t+1}, \boldsymbol{B}_{\z,t+1}, \boldsymbol{\lambda})
\rangle, 
\end{equation}
\begin{equation} \label{xi_B_z_general}
\xi_{\B,t,\tau}^{\z} = \langle \partial_{\tau}\psi_{\z,t}(\boldsymbol{H}_{\z,t+1}, 
\tilde{\boldsymbol{h}}, \boldsymbol{w}) \rangle,  
\end{equation}
\begin{equation} \label{eta_B_general}
\eta_{\B,t} 
= \left\langle
 \partial_{t+1}\psi_{\z,t}(\boldsymbol{H}_{\z,t+1}, \tilde{\boldsymbol{h}}, 
 \boldsymbol{w}) 
\right\rangle, 
\end{equation}
\begin{equation} \label{xi_B_x_general}
\xi_{\B,t,\tau}^{\x} = \frac{1}{M}\boldsymbol{1}^{\mathrm{T}}
\partial_{\tau}\psi_{\x,t}(\boldsymbol{H}_{\x,t+1}, \boldsymbol{x}),
\end{equation}
with $\eta_{\B,-1} = \langle\partial_{0}\psi_{\z,-1}
(\tilde{\boldsymbol{h}}, \boldsymbol{w}) \rangle$. The variable 
$\xi_{\A,t,\tau}^{\x}$ in (\ref{xi_A_x_general}) is equal to the 
arithmetic mean for the partial derivatives of $\phi_{\x,t}$ 
in (\ref{phi_x}) with respect to the elements of $\boldsymbol{b}_{\z,\tau}$. 
The partial derivative in $\eta_{\B,t}$ is taken with 
respect to the elements of $\tilde{\boldsymbol{h}}$. 

The Onsager correction in the general error model utilizes messages 
in all previous iterations, as shown in (\ref{mz}), (\ref{mx}), (\ref{qz}), 
and (\ref{qx}). It realizes asymptotic joint Gaussianity for 
$\boldsymbol{B}_{\z,t}$, $\boldsymbol{B}_{\x,t}$,   
$\boldsymbol{H}_{\z,t}$, $\tilde{\boldsymbol{h}}$, and $\boldsymbol{H}_{\x,t}$. 

\subsection{Ensemble of Sensing Matrices}
Assumption~\ref{assumption_A} only postulates the convergence in distribution 
of the empirical eigenvalue distribution for $\boldsymbol{\Lambda}
=N^{-1}\boldsymbol{\Sigma}\boldsymbol{\Sigma}^{\mathrm{T}}$. 
This convergence is too weak to 
justify the convergence in probability of empirical average with respect to 
$\boldsymbol{\lambda}$. We define the ensemble of sensing matrices assumed 
in state evolution via the piecewise Lipschitz-continuous functions 
$\phi_{\z,t}$ and $\tilde{\phi}_{\x,t}$ in (\ref{phi_z}) and (\ref{phi_x}) 
under Assumption~\ref{assumption_denoisers_general}. 
 
For fixed $\lambda\in[0,\infty)$, 
let $L_{\z,t,i}(\lambda)>0$ denote a Lipschitz constant 
of $\phi_{\z,t}$ for the $i$th piece $\mathcal{D}_{\z,t,i}\subset\mathbb{R}^{2t+2}$ 
in the domain of $\phi_{\z,t}(\cdots, \lambda)$. Similarly, we write a piece for 
$\tilde{\phi}_{\x,t}(\cdots, \lambda)$ and the corresponding Lipschitz 
constant as $\mathcal{D}_{\x,t,i}\subset\mathbb{R}^{2t+2}$ and 
$L_{\x,t,i}(\lambda)>0$, respectively. 

\begin{assumption} \label{assumption_A_addition}
Under Assumptions~\ref{assumption_A} and \ref{assumption_denoisers_general}, 
the Lipschitz constants $L_{\z,t,i}(\lambda)$ and $L_{\x,t,i}(\lambda)$ 
for $\phi_{\z,t}$ and $\tilde{\phi}_{\x,t}$ are 
almost everywhere continuous functions of $\lambda$
with respect to the empirical eigenvalue distribution of 
$\boldsymbol{\Lambda}$. Furthermore, 
for all $t$, $i$, and $a\in\{\z, \x\}$,  
the empirical average $M^{-1}\sum_{m=1}^{M}\lambda_{m}L_{a,t,i}^{2}(\lambda_{m})$ 
is uniformly integrable in the sublinear sparsity limit. 
\end{assumption}

Assumption~\ref{assumption_A_addition} imposes additional conditions on 
the sensing matrix $\boldsymbol{A}$. 
For fixed $\{\phi_{\z,t}\}$ and $\{\tilde{\phi}_{\x,t}\}$, 
Assumptions~\ref{assumption_A} and \ref{assumption_A_addition} define 
an ensemble of sensing matrices. We write this ensemble induced from 
$\{\phi_{\z,t}\}$ and $\{\tilde{\phi}_{\x,t}\}$ as 
$\mathcal{A}(\{\phi_{\z,t}\}, \{\tilde{\phi}_{\x,t}\})$. 

Conversely, $\mathcal{A}(\{\phi_{\z,t}\}, \{\tilde{\phi}_{\x,t}\})$ defines 
a subset of piecewise Lipschitz-continuous functions 
$\phi:\mathbb{R}^{\tau}\to\mathbb{R}$ with respect to 
the first $(\tau-1)$ variables. For fixed $\lambda$, let 
$L_{\phi,i}(\lambda)>0$ denote a Lipschitz constant for piece~$i$ in the 
domain of $\phi(\cdots, \lambda)$. 

\begin{definition} \label{definition_integrability}
The set $\mathcal{L}_{\tau}(\{\phi_{\z,t}\}, \{\tilde{\phi}_{\x,t}\})$ consists 
of all possible piecewise Lipschitz-continuous functions 
$\phi:\mathbb{R}^{\tau}\to\mathbb{R}$ with respect to the first $(\tau-1)$ 
variables such that, for any piece $i$, the Lipschitz constant 
$L_{\phi,i}(\lambda)$ is almost everywhere continuous with respect 
to the empirical eigenvalue distribution of $\boldsymbol{\Lambda}$ and that 
$M^{-1}\sum_{m=1}^{M}\lambda_{m}L_{\phi,i}^{2}(\lambda_{m})$ is uniformly 
integrable under the ensemble $\mathcal{A}(\{\phi_{\z,t}\}, 
\{\tilde{\phi}_{\x,t}\})$. 
\end{definition}

Since $\{\phi_{\z,t}\}$ and $\{\tilde{\phi}_{\x,t}\}$ are fixed throughout 
state evolution, we abbreviate $\mathcal{L}_{\tau}
(\{\phi_{\z,t}\}, \{\tilde{\phi}_{\x,t}\})$ as $\mathcal{L}_{\tau}$.  

In state evolution we encounter empirical average with respect to 
$\boldsymbol{\lambda}$. Consider $\langle f(\boldsymbol{\lambda}) \rangle$  
for $f(\lambda)=\lambda L_{\phi,i}(\lambda)L_{\psi,i'}(\lambda)\geq0$ for 
$\phi, \psi\in\mathcal{L}_{\tau}$ under 
Assumption~\ref{assumption_A}. By definition, $f$ is almost 
everywhere continuous. Using the Cauchy-Schwarz inequality 
for $f(\lambda)=\sqrt{\lambda}L_{\phi,i}(\lambda)\sqrt{\lambda}
L_{\psi,i'}(\lambda)$, we have 
\begin{equation}
\langle f(\boldsymbol{\lambda}) \rangle
\leq \left(
 \frac{1}{M}\sum_{m=1}^{M}\lambda_{m}L_{\phi,i}^{2}(\lambda_{m})
 \frac{1}{M}\sum_{m=1}^{M}\lambda_{m}L_{\psi,i'}^{2}(\lambda_{m})
\right)^{1/2},
\end{equation}
which is uniformly integrable, because of $\sqrt{xy}\leq(x + y)/2$ and 
$\phi, \psi\in\mathcal{L}_{\tau}$ in Definition~\ref{definition_integrability}. 
From the continuity of $f$ and the uniform integrability of $\langle f(\boldsymbol{\lambda}) \rangle$, the convergence in distribution under  Assumption~\ref{assumption_A} implies $\langle f(\boldsymbol{\lambda}) \rangle \pto \mathbb{E}[f(\Lambda)]$. 

\subsection{Technical Results}
Technical results are presented to analyze the general error model via state evolution. 

\begin{lemma}[\cite{Rangan192,Takeuchi20}] \label{lemma_conditioning}
Suppose that $\boldsymbol{V}$ is an $N\times N$ Haar-distributed orthogonal 
matrix. For $t<N$, let $\boldsymbol{X}\in\mathbb{R}^{N\times t}$ denote a 
random matrix with full rank that is independent of $\boldsymbol{V}$. 
Consider the noiseless and compressed measurement 
$\boldsymbol{Y}\in\mathbb{R}^{N\times t}$ of $\boldsymbol{V}$ given by 
\begin{equation} \label{constraint}
\boldsymbol{Y} = \boldsymbol{V}\boldsymbol{X}. 
\end{equation}
Then, the conditional distribution of 
$\boldsymbol{V}$ given $\{\boldsymbol{X}, \boldsymbol{Y}\}$ satisfies 
\begin{equation} \label{V_conditional_distribution} 
\boldsymbol{V}\sim \boldsymbol{Y}(\boldsymbol{Y}^{\mathrm{T}}\boldsymbol{Y})^{-1}
\boldsymbol{X}^{\mathrm{T}} + \boldsymbol{\Phi}_{\boldsymbol{Y}}^{\perp}
\tilde{\boldsymbol{V}}(\boldsymbol{\Phi}_{\boldsymbol{X}}^{\perp})^{\mathrm{T}}, 
\end{equation}
where $\tilde{\boldsymbol{V}}$ is an $(N-t)\times(N-t)$ Haar-distributed 
orthogonal matrix independent of $\{\boldsymbol{X}, \boldsymbol{Y}\}$. 
\end{lemma}

Lemma~\ref{lemma_conditioning} is a key lemma in Bolthausen's conditioning 
technique~\cite{Bolthausen14}. It depends heavily on the Haar-distribution 
assumption of $\boldsymbol{V}$. The usage of Lemma~\ref{lemma_conditioning} 
is the same as that for linear sparsity. However, we need to evaluate 
negligibly small errors due to this conditioning technique more carefully 
for sublinear sparsity than for linear sparsity.  

\begin{lemma} \label{lemma_bias_elimination}
For a separable function $\boldsymbol{f}: \mathbb{R}^{M\times t}
\to\mathbb{R}^{M}$, suppose that the $m$th function $f_{m}$ is 
second-order pseudo-Lipschitz with a Lipschitz constant $L_{m}>0$.
Let $\delta_{M}\in\mathbb{R}$ denote a random variable satisfying   
$\delta_{M}\pto0$ and consider 
\begin{equation}
\Delta\boldsymbol{f} 
= \boldsymbol{f}(\boldsymbol{a}_{1}, \ldots, \boldsymbol{a}_{t-1},
\delta_{M}\boldsymbol{a}_{0} + \boldsymbol{a}_{t}) 
- \boldsymbol{f}(\boldsymbol{a}_{1},\ldots, \boldsymbol{a}_{t})
\end{equation}
for random vectors $\boldsymbol{a}_{\tau}\in\mathbb{R}^{M}$  
with $\tau\in\{0, \ldots, t\}$. 
Then, $\langle\Delta\boldsymbol{f}\rangle$ 
converges in probability to zero as $M\to\infty$ 
if the following conditions hold for some $C>0$:
\begin{equation}
\lim_{M\to\infty}\frac{1}{M}\sum_{m=1}^{M}L_{m} < \infty, 
\end{equation} 
\begin{equation}
\lim_{M\to\infty}\mathbb{P}\left(
 \frac{1}{M}\sum_{m=1}^{M}L_{m}a_{m,\tau}^{2} < C
\right) = 1 \quad \hbox{for all $\tau$.} 
\end{equation}
\end{lemma}
\begin{IEEEproof}
See Appendix~\ref{proof_lemma_bias_elimination}. 
\end{IEEEproof}

\begin{lemma}[\cite{Takeuchi251}] \label{lemma_bias_elimination_inner} 
For random vectors $\boldsymbol{\phi}$, $\tilde{\boldsymbol{\phi}}$, 
$\boldsymbol{\psi}$, and $\tilde{\boldsymbol{\psi}}$,  
let $\Delta\boldsymbol{\phi} = \tilde{\boldsymbol{\phi}}
- \boldsymbol{\phi}$ and $\Delta\boldsymbol{\psi} 
= \tilde{\boldsymbol{\psi}} - \boldsymbol{\psi}$. 
If $\|\boldsymbol{\phi}\|_{2}$ and $\|\boldsymbol{\psi}\|_{2}$
are bounded in probability and if $\|\Delta\boldsymbol{\phi}\|_{2}$ and 
$\|\Delta\boldsymbol{\psi}\|_{2}$ converge in probability to zero, 
then $\tilde{\boldsymbol{\phi}}^{\mathrm{T}}
\tilde{\boldsymbol{\psi}} - \boldsymbol{\phi}^{\mathrm{T}}
\boldsymbol{\psi}\pto0$ holds.
\end{lemma}
\begin{IEEEproof}
Repeat the proof of \cite[Lemma~9]{Takeuchi251}. 
\end{IEEEproof}

Lemmas~\ref{lemma_bias_elimination} and \ref{lemma_bias_elimination_inner} 
are utilized to eliminate negligibly 
small terms that originate from Bolthausen's conditioning 
technique~\cite{Bolthausen14}. 

\begin{lemma}[\cite{Rangan192,Takeuchi20}] \label{lemma_representation}
Suppose that $\boldsymbol{V}$ is an $N\times N$ Haar-distributed orthogonal 
matrix. Let $\boldsymbol{a}\in\mathbb{R}^{N}$ denote a random vector 
independent of $\boldsymbol{V}$. Then, we have 
\begin{equation} 
\boldsymbol{V}\boldsymbol{a} 
\sim \frac{\|\boldsymbol{a}\|_{2}}
{\|\boldsymbol{\omega}\|_{2}}\boldsymbol{\omega},
\end{equation}
where $\boldsymbol{\omega}\sim\mathcal{N}(\boldsymbol{0},\boldsymbol{I}_{N})$ 
is independent of $\boldsymbol{a}$. 
\end{lemma}
\begin{lemma} \label{lemma_modified_representation}
For $t<N$, suppose that $\boldsymbol{V}$ is an $(N-t)\times(N-t)$ 
Haar-distributed orthogonal matrix. Let $\boldsymbol{a}\in\mathbb{R}^{N-t}$ and 
$\boldsymbol{M}\in\mathbb{R}^{N\times t}$ denote a random vector and matrix that 
are independent of $\boldsymbol{V}$. Then, we have 
\begin{equation} \label{modified_representation}
\boldsymbol{\Phi}_{\boldsymbol{M}}^{\perp}
\boldsymbol{V}\boldsymbol{a}
\sim \boldsymbol{M}\boldsymbol{v} 
+ \{1 + o(1)\}\frac{\|\boldsymbol{a}\|_{2}}{\|\boldsymbol{\omega}\|_{2}}
\boldsymbol{\omega}
\end{equation}
as $N\to\infty$. In (\ref{modified_representation}),  
$\boldsymbol{\omega}\sim\mathcal{N}(\boldsymbol{0}, \boldsymbol{I}_{N})$ 
is independent of $\{\boldsymbol{a}, \boldsymbol{M}\}$. 
The random vector $\boldsymbol{v}\in\mathbb{R}^{t}$ satisfies 
\begin{equation}
\mathbb{E}\left[
 \left.
  \|\boldsymbol{v}\|_{2}^{2}
 \right| \boldsymbol{M}, \boldsymbol{a} 
\right]
\leq \frac{t\|\boldsymbol{a}\|_{2}^{2}}
{N\lambda_{\mathrm{min}}
(\boldsymbol{M}^{\mathrm{T}}\boldsymbol{M})} \quad 
\hbox{in probability.}  
\end{equation}
In particular, $\boldsymbol{v}=\boldsymbol{o}(1)$ holds if 
$\|\boldsymbol{a}\|_{2}^{2}/
\{N\lambda_{\mathrm{min}}(\boldsymbol{M}^{\mathrm{T}}\boldsymbol{M})\}\pto0$ 
is satisfied. 
\end{lemma}
\begin{IEEEproof}
See Appendix~\ref{proof_lemma_modified_representation}. 
\end{IEEEproof}

Lemma~\ref{lemma_modified_representation} implies that 
$\boldsymbol{\Phi}_{\boldsymbol{M}}^{\perp}\boldsymbol{V}\boldsymbol{a}$ is  
zero-mean Gaussian with the amplitude correction 
$\|\boldsymbol{a}\|_{2}/\|\boldsymbol{\omega}\|_{2}$ when the negligibly 
small terms are ignored.  
When $\|\boldsymbol{a}\|_{2}/\|\boldsymbol{\omega}\|_{2}$ converges in 
probability to a constant, all finite sets of elements in 
$\boldsymbol{\Phi}_{\boldsymbol{M}}^{\perp}\boldsymbol{V}\boldsymbol{a}$ are 
jointly Gaussian-distributed. This property is utilized to prove asymptotic 
Gaussianity in the general error model. 

\begin{proposition} \label{proposition1}
For $\boldsymbol{w}\sim\mathcal{N}(\boldsymbol{0}, \boldsymbol{I}_{N})$ and 
any $p\geq 0$ we have 
\begin{equation}
\mathbb{E}\left[
 \left|
  \frac{1}{\|\boldsymbol{w}\|_{2}} - \frac{1}{\sqrt{N}}
 \right|^{p}
\right] = {\cal O}(N^{-p}). 
\end{equation}
\end{proposition}
\begin{IEEEproof}
See Appendix~\ref{proof_proposition1}. 
\end{IEEEproof}

Proposition~\ref{proposition1} is used to evaluate the last term in 
(\ref{modified_representation}) as 
$\|\boldsymbol{a}\|_{2}\|\boldsymbol{w}\|_{2}^{-1}\boldsymbol{w}
\peq N^{-1/2}\|\boldsymbol{a}\|_{2}\boldsymbol{w} 
+ {\cal O}(N^{-1}\|\boldsymbol{a}\|_{2})\boldsymbol{w}$.  

\begin{lemma} \label{lemma_elimination}
Define 
$\boldsymbol{\omega}\sim\mathcal{N}(\boldsymbol{0}, \boldsymbol{I}_{N})$. 
Let $a_{N}\in\mathbb{R}$ denote a random variable that is independent of 
$\boldsymbol{\omega}$. For $\tilde{a}_{N}=a_{N}/\{\sqrt{N}f(N)\}$ with 
some function $f$ of $N$, assume $\tilde{a}_{N}\pto a$ for some 
$a\in\mathbb{R}$. For two separable functions $\boldsymbol{\phi}: 
\mathbb{R}^{N}\to\mathbb{R}^{N}$ and $\boldsymbol{\psi}: 
\mathbb{R}^{N}\to\mathbb{R}^{N}$, suppose that 
$\|\boldsymbol{\phi}(af(N)\boldsymbol{\omega})\|_{2}$ and 
$\|\boldsymbol{\psi}(af(N)\boldsymbol{\omega})\|_{2}$ are 
bounded in probability. Define 
\begin{equation}
\Delta\boldsymbol{\phi}(\boldsymbol{\omega}) 
=\boldsymbol{\phi}\left(
  \frac{a_{N}}{\|\boldsymbol{\omega}\|_{2}}\boldsymbol{\omega}
 \right)
- \boldsymbol{\phi}\left(
 af(N)\boldsymbol{\omega}
\right),
\end{equation}
\begin{equation}
\Delta\boldsymbol{\psi}(\boldsymbol{\omega}) 
=\boldsymbol{\psi}\left(
  \frac{a_{N}}{\|\boldsymbol{\omega}\|_{2}}\boldsymbol{\omega}
 \right)
- \boldsymbol{\psi}\left(
 af(N)\boldsymbol{\omega}
\right).
\end{equation}
If the $n$th functions $\phi_{n}$ and $\psi_{n}$ are 
Lipschitz-continuous with a common Lipschitz constant $L_{n}>0$ satisfying 
the condition $\lim_{N\to\infty}f^{2}(N)\sum_{n=1}^{N}L_{n}^{2}<\infty$,  
then the difference $\boldsymbol{\phi}^{\mathrm{T}}(a_{N}\boldsymbol{\omega}
/\|\boldsymbol{\omega}\|_{2})\boldsymbol{\psi}(a_{N}\boldsymbol{\omega}
/\|\boldsymbol{\omega}\|_{2})
- \boldsymbol{\phi}^{\mathrm{T}}(af(N)\boldsymbol{\omega})
\boldsymbol{\psi}(af(N)\boldsymbol{\omega})$
converges in probability to zero as $N\to\infty$. 
\end{lemma}
\begin{IEEEproof}
See Appendix~\ref{proof_lemma_elimination}. 
\end{IEEEproof}

Lemma~\ref{lemma_elimination} is utilized to prove asymptotic Gaussianity 
for module~A and the outer denoiser in module~B. 

\begin{lemma}[Stein's Lemma~\cite{Stein72,Takeuchi21}] \label{lemma_Stein}
Suppose that $\{Z_{\tau}\}_{\tau=1}^{t}$ are zero-mean Gaussian random variables. 
Then, for a function $f:\mathbb{R}^{t}\to\mathbb{R}$ we have 
\begin{equation} \label{Stein}
\mathbb{E}[Z_{t'}f(Z_{1},\ldots,Z_{t})] = 
\sum_{\tau=1}^{t}\mathbb{E}[Z_{t'}Z_{\tau}]\mathbb{E}\left[
 \frac{\partial f}{\partial Z_{\tau}}(Z_{1},\ldots,Z_{t})
\right] 
\end{equation}
if $f$ is piecewise Lipschitz-continuous, or if $f$ is almost everywhere 
differentiable and both sides in (\ref{Stein}) are bounded. 
\end{lemma}
\begin{IEEEproof}
The piecewise Lipschitz-continuity of $f$ implies the latter conditions: 
the almost everywhere differentiability of $f$ and the boundedness of both 
sides in (\ref{Stein}), as shown in \cite[Lemma 2]{Takeuchi21}. 
See \cite[Lemma 2]{Takeuchi21} for the proof of 
(\ref{Stein}) under the latter conditions. 
\end{IEEEproof}

Lemma~\ref{lemma_Stein} is used to justify the Onsager correction in the 
general error model. Since the Lipschitz continuity is not assumed for the 
inner denoiser in module~B, the latter conditions in Lemma~\ref{lemma_Stein} 
are presented to utilize Stein's lemma for the inner denoiser in module~B.  

We present a technical result to generalize the class of denoisers. 
For that purpose, we start with definitions. 
For a random matrix $\boldsymbol{X}_{M}=\{X_{m,\tau,M}\}\in\mathbb{R}^{M\times t}$, 
define the empirical distribution of 
$\boldsymbol{X}_{M}$ with respect to the rows as 
\begin{equation} \label{empirical_distribution}
\rho_{M}(\mathcal{A}) = \frac{1}{M}\sum_{m=1}^{M}1\left(
 \{X_{m,\tau,M}\}_{\tau=1}^{t}\in\mathcal{A}
\right) 
\end{equation}
for measurable $\mathcal{A}\subset\mathbb{R}^{t}$. The empirical 
distribution $\rho_{M}$ is a random function that depends on 
$\boldsymbol{X}_{M}$. 
Assume the convergence in probability $\rho_{M}\pto\rho_{*}$ for some 
distribution $\rho_{*}$. 
Let $\tilde{\boldsymbol{X}}_{M}\in\mathbb{R}^{t}$ and 
$\tilde{\boldsymbol{X}}_{*}\in\mathbb{R}^{t}$ denote random vectors that 
follow the empirical distribution $\rho_{M}$ and the limiting distribution 
$\rho_{*}$, respectively. 
For some $t_{0}\in\{1,\ldots,t\}$, we consider the decomposition 
$\tilde{\boldsymbol{X}}_{M}=[\tilde{\boldsymbol{X}}_{1,M}^{\mathrm{T}}, 
\tilde{\boldsymbol{X}}_{2,M}^{\mathrm{T}}]^{\mathrm{T}}$ for 
$\tilde{\boldsymbol{X}}_{1,M}\in\mathbb{R}^{t_{0}}$ and 
$\tilde{\boldsymbol{X}}_{2,M}\in\mathbb{R}^{t-t_{0}}$. Similarly, we define 
$\tilde{\boldsymbol{X}}_{*}
=[\tilde{\boldsymbol{X}}_{1,*}^{\mathrm{T}}, 
\tilde{\boldsymbol{X}}_{2,*}^{\mathrm{T}}]^{\mathrm{T}}$. 

\begin{lemma} \label{lemma_generalization}
Suppose that the empirical distribution $\rho_{M}$ converges in probability to 
some deterministic distribution $\rho_{*}$ in the sense of the weak convergence 
as $M\to\infty$: For all bounded and Lipschitz-continuous functions 
$\phi: \mathbb{R}^{t}\to\mathbb{R}$, 
$\lim_{M\to\infty}\mathbb{E}[\phi(\tilde{\boldsymbol{X}}_{M}) 
| \boldsymbol{X}_{M}]\peq\mathbb{E}[\phi(\tilde{\boldsymbol{X}}_{*})]$ holds. 
For some $t_{0}\in\{1,\ldots,t\}$, consider the decomposition 
$\boldsymbol{X}_{M}=[\boldsymbol{X}_{1,M}, \boldsymbol{X}_{2,M}]$ for 
$\boldsymbol{X}_{1,M}\in\mathbb{R}^{M\times t_{0}}$ and 
$\boldsymbol{X}_{2,M}\in\mathbb{R}^{M\times(t-t_{0})}$. Suppose that
$f: \mathbb{R}^{t}\to \mathbb{R}$ has some positive function
$L(\boldsymbol{x}_{2})>0$ of
$\boldsymbol{x}_{2}=[x_{t_{0}+1},\ldots,x_{t}]^{\mathrm{T}}$
such that $|f(\boldsymbol{x})| \leq L(\boldsymbol{x}_{2})
(1 + \|\boldsymbol{x}_{1}\|_{2}^{2})$ holds for all $\boldsymbol{x}
=[\boldsymbol{x}_{1}^{\mathrm{T}},\boldsymbol{x}_{2}^{\mathrm{T}}]^{\mathrm{T}}
\in\mathbb{R}^{t}$. Then, we have 
\begin{equation}
\lim_{M\to\infty}\mathbb{E}[f(\tilde{\boldsymbol{X}}_{M}) | \boldsymbol{X}_{M}]
\peq \mathbb{E}[f(\tilde{\boldsymbol{X}}_{*})] 
\end{equation}
for $f: \mathbb{R}^{t}\to \mathbb{R}$ 
if the following conditions are satisfied: 
\begin{itemize}
\item Let $\mathcal{C}_{f}(\boldsymbol{x}_{2})\subset\mathbb{R}^{t_{0}}$ denote 
the set of all continuous points in $f(\cdot, \boldsymbol{x}_{2})$ for given 
$\boldsymbol{x}_{2}$. The null-set condition 
$\mathbb{P}(\tilde{\boldsymbol{X}}_{1,*}\notin\mathcal{C}_{f}
(\tilde{\boldsymbol{X}}_{2,M}) | \boldsymbol{X}_{M}) = 0$ holds.

\item $\mathbb{E}[\|\tilde{\boldsymbol{X}}_{1,M}\|_{2}^{2}
| \boldsymbol{X}_{M}]\pto 
\mathbb{E}[\|\tilde{\boldsymbol{X}}_{1,*}\|_{2}^{2}]<\infty$ holds. 

\item $\mathbb{E}[f(\tilde{\boldsymbol{X}}_{1,*}, 
\tilde{\boldsymbol{X}}_{2,M}) | \boldsymbol{X}_{M}] \pto 
\mathbb{E}[f(\tilde{\boldsymbol{X}}_{*})]$ holds. 

\item $L(\tilde{\boldsymbol{X}}_{2,M})$ given 
$\boldsymbol{X}_{M}$ is uniformly integrable as $M\to\infty$. 
\end{itemize}
\end{lemma}
\begin{IEEEproof}
See Appendix~\ref{proof_lemma_generalization}. 
\end{IEEEproof}

From the definition of $\rho_{M}$, the identity 
$\mathbb{E}[f(\tilde{\boldsymbol{X}}_{M}) | \boldsymbol{X}_{M}]
=M^{-1}\sum_{m=1}^{M}f(X_{m,1,M},\ldots,X_{m,t,M})$ holds. 
Lemma~\ref{lemma_generalization} is utilized to prove the weak law of 
large numbers for this average. The class of bounded and Lipschitz-continuous 
functions in the weak convergence is extended to that of more general 
functions $f$ that satisfy the assumptions in 
Lemma~\ref{lemma_generalization}. 

The condition $\mathbb{P}(\tilde{\boldsymbol{X}}_{1,*}\in
\mathcal{C}_{f}(\tilde{\boldsymbol{X}}_{2,M}) | \boldsymbol{X}_{M})=1$ means 
that $f(\cdot, \tilde{\boldsymbol{X}}_{2,M})$ is $\rho_{*}$-almost everywhere 
continuous. 
In particular, this condition is satisfied for almost everywhere continuous 
$f$ with respect to the first $t_{0}$ variables 
under the Lebesgue measure when $\rho_{*}$ is absolutely 
continuous with respect to the first $t_{0}$ variables under 
the Lebesgue measure. 

In state evolution, the last condition is justified for 
$\boldsymbol{X}_{2,M}=\boldsymbol{0}$ or 
$\boldsymbol{X}_{2,M}=\boldsymbol{\lambda}$ under 
Assumptions~\ref{assumption_A} and \ref{assumption_A_addition}.  

\begin{lemma} \label{lemma_positive_definite}
For a random vector $\boldsymbol{X}_{t+1}=[X_{0},\ldots, X_{t}]^{\mathrm{T}}
\in\mathbb{R}^{t+1}$, suppose that $\mathbb{E}[\boldsymbol{X}_{t}
\boldsymbol{X}_{t}^{\mathrm{T}}]$ is invertible. Then, 
we have 
\begin{align}
&\mathbb{E}[X_{t}^{2}] 
- \mathbb{E}[X_{t}\boldsymbol{X}_{t}^{\mathrm{T}}]
(\mathbb{E}[\boldsymbol{X}_{t}\boldsymbol{X}_{t}^{\mathrm{T}}])^{-1}
\mathbb{E}[\boldsymbol{X}_{t}X_{t}] 
\nonumber \\
&\geq \mathbb{E}\left[
 (X_{t} - \mathbb{E}[X_{t} | \boldsymbol{X}_{t}])^{2}
\right]. 
\end{align}
\end{lemma}
\begin{IEEEproof}
See Appendix~\ref{proof_lemma_positive_definite}. 
\end{IEEEproof}

Lemma~\ref{lemma_positive_definite} is utilized to prove the full-rank 
condition in Lemma~\ref{lemma_conditioning}. 
It implies that the Schur complement of 
$\mathbb{E}[\boldsymbol{X}_{t+1}\boldsymbol{X}_{t+1}^{\mathrm{T}}]$ is bounded 
from below by the conditional variance of $X_{t}$ given $\boldsymbol{X}_{t}$.   
Thus, $\mathbb{E}[\boldsymbol{X}_{t+1}\boldsymbol{X}_{t+1}^{\mathrm{T}}]$ has 
full rank if $X_{t}$ given $\boldsymbol{X}_{t}$ is random and if 
$\mathbb{E}[\boldsymbol{X}_{t}\boldsymbol{X}_{t}^{\mathrm{T}}]$ has full rank.

\subsection{State Evolution}
In state evolution, the dynamics of the general error model is evaluated via 
the conditional distributions of $\boldsymbol{U}$ and $\boldsymbol{V}$ 
in the SVD $\boldsymbol{A}=\boldsymbol{U}\boldsymbol{\Sigma}
\boldsymbol{V}^{\mathrm{T}}$. We write random variables given throughout 
state evolution as 
$\Theta=\{\boldsymbol{x}, \boldsymbol{\Sigma}, \boldsymbol{w}\}$. Let 
$\mathfrak{E}_{t,t'}^{\x}=\{\boldsymbol{Q}_{\x,t'}, \boldsymbol{B}_{\x,t'}, 
\boldsymbol{M}_{\x,t}, \boldsymbol{H}_{\x,t}\}$, defined in the same manner as 
for $\boldsymbol{B}_{\z,t}$. The set $\mathfrak{E}_{t,t}^{\x}$ contains 
all $\boldsymbol{V}$-dependent messages just before updating 
$\boldsymbol{b}_{\x,t}=\boldsymbol{V}^{\mathrm{T}}\boldsymbol{q}_{\x,t}$ in 
(\ref{b}) while $\mathfrak{E}_{t,t+1}^{\x}$ includes them just before updating 
$\boldsymbol{h}_{\x,t}=\boldsymbol{V}\boldsymbol{m}_{\x,t}$ in (\ref{h}). 

For the outer modules, we define  
$\tilde{\boldsymbol{M}}_{\z,t}=[\boldsymbol{M}_{\z,t}, \tilde{\boldsymbol{m}}]$ 
and $\tilde{\boldsymbol{H}}_{\z,t}=[\boldsymbol{H}_{\z,t}, 
\tilde{\boldsymbol{h}}]$ if $\tilde{\boldsymbol{m}}$ and 
$\boldsymbol{m}_{\z, 0}$ are linearly independent. Otherwise, we let  
$\tilde{\boldsymbol{M}}_{\z,t}=\boldsymbol{M}_{\z,t}$ 
and $\tilde{\boldsymbol{H}}_{\z,t}=\boldsymbol{H}_{\z,t}$. Define 
$\mathfrak{E}_{t,t'}^{\z}=\{ \boldsymbol{Q}_{\z,t'}, 
\boldsymbol{B}_{\z,t'}, \tilde{\boldsymbol{M}}_{\z,t}, 
\tilde{\boldsymbol{H}}_{\z,t}\}$. 
The sets $\mathfrak{E}_{t,t}^{\z}$ and $\mathfrak{E}_{t,t+1}^{\z}$ contain 
all $\boldsymbol{U}$-dependent messages just before updating 
$\boldsymbol{b}_{\z,t}=\boldsymbol{U}^{\mathrm{T}}\boldsymbol{q}_{\z,t}$ and 
$\boldsymbol{h}_{\z,t}=\boldsymbol{U}\boldsymbol{m}_{\z,t}$, respectively. 
Note that $\boldsymbol{U}$ and $\boldsymbol{V}$ given 
$\mathfrak{E}_{t,t'}^{\x}$ and $\mathfrak{E}_{t,t'}^{\z}$ are independent 
since $\boldsymbol{U}$ is assumed to be independent of 
$\boldsymbol{V}$ in Assumption~\ref{assumption_A}. 
Thus, we can evaluate the conditional distributions of $\boldsymbol{U}$ and 
$\boldsymbol{V}$ separately.  

To present state evolution results, we first define random variables that 
represent the asymptotic dynamics of the general error model. 
Let $\tilde{H}\sim\mathcal{N}(0, P)$ denote a zero-mean Gaussian random 
variable that is an asymptotic alternative of $\tilde{\boldsymbol{h}}
=-\boldsymbol{z}$ in (\ref{measurement}) and independent of $W$ 
in Assumption~\ref{assumption_w}. Furthermore, define 
$Q_{\z,0} = \psi_{\z,-1}(\tilde{H}, W) - \bar{\eta}_{\B,-1}\tilde{H}$ 
and $\psi_{\x,-1}(\boldsymbol{x})=-\boldsymbol{x}$ that correspond to 
$\boldsymbol{q}_{\z,0}$ and $\boldsymbol{q}_{\x,0}$ in 
(\ref{initial_condition_general}), respectively, with 
\begin{equation} \label{eta_B_0}
\bar{\eta}_{\B,-1} = \mathbb{E}\left[
 \partial_{0}\psi_{\z,-1}(\tilde{H}, W)
\right],  
\end{equation}
which is an asymptotic alternative of $\eta_{\B,-1}$. 
Assumption~\ref{assumption_x} implies the convergence in 
probability, 
\begin{equation} \label{Qx_0}
\|\psi_{\x,-1}(\boldsymbol{x})\|_{2}^{2} 
= \|\boldsymbol{x}\|_{2}^{2} 
= \sum_{n\in\mathcal{S}}x_{n}^{2}
\pto P \equiv Q_{\x,0,0}, 
\end{equation}
where the convergence in probability follows from the weak law of large 
numbers. 

We next define a sequence of zero-mean Gaussian random variables 
$\mathcal{B}_{\z,t+1}=\{B_{\z,\tau}\}_{\tau=0}^{t}$, which are asymptotic 
alternatives of $\{\boldsymbol{b}_{\z,\tau}\}$. They are 
independent of $\Lambda$ in Assumption~\ref{assumption_A} and have the 
covariance $\mathbb{E}[B_{\z,t}B_{\z,\tau}]=\mathbb{E}[Q_{\z,t}Q_{\z,\tau}]$, 
with $Q_{\z,t}$ defined shortly. Similarly,  
let $\mathcal{B}_{\x,t+1}=\{B_{\x,\tau}\}_{\tau=0}^{t}$ denote zero-mean Gaussian 
random variables that are independent of 
$\{\Lambda, \mathcal{B}_{\z,t+1}\}$ and have 
the covariance $\mathbb{E}[B_{\x,t}B_{\x,\tau}]
=\lim_{N\to\infty}\mathbb{E}[\psi_{\x,t-1}^{\mathrm{T}}(\boldsymbol{\Omega}_{t}, 
\boldsymbol{x})\psi_{\x,\tau-1}(\boldsymbol{\Omega}_{\tau}, \boldsymbol{x})]
\equiv Q_{\x,t,\tau}$ in the sublinear sparsity limit, 
with $\boldsymbol{\Omega}_{t}\in\mathbb{R}^{N\times t}$ 
defined shortly.  

To present an asymptotic alternative of $\boldsymbol{m}_{\z,t}$, we define 
\begin{equation} \label{xi_A_z_bar_general}
\bar{\xi}_{\A,t,\tau}^{\z} = \mathbb{E}\left[
 \partial_{\tau}\phi_{\z,t}(\mathcal{B}_{\z,t+1}, 
 \sqrt{\Lambda}\mathcal{B}_{\x,t+1}, \Lambda)
\right], 
\end{equation}
which corresponds to $\xi_{\A,t,\tau}^{\z}$ in (\ref{xi_A_z_general}). 
Using the definition of $\bar{\xi}_{\A,t,\tau}^{\z}$, 
we write an asymptotic alternative of $\boldsymbol{m}_{\z,t}$ in (\ref{mz}) as 
\begin{equation} \label{M_z}
M_{\z,t} = \phi_{\z,t} - \sum_{\tau=0}^{t}\bar{\xi}_{\A,t,\tau}^{\z}B_{\z,\tau},
\end{equation} 
with $\phi_{\z,t} = \phi_{\z,t}(\mathcal{B}_{\z,t+1}, 
\sqrt{\Lambda}\mathcal{B}_{\x,t+1}, \Lambda)$. In addition, we define 
$\tilde{M}=\sqrt{\Lambda}B_{\x,0}$ that corresponds to 
$\tilde{\boldsymbol{m}}=\boldsymbol{\Sigma}\boldsymbol{b}_{\x,0}$. 

Similarly, let 
\begin{equation} \label{xi_A_x_bar_general}
\bar{\xi}_{\A,t,\tau}^{\x}
= \delta_{t,\tau} + \mathbb{E}\left[
 \Lambda\partial_{\tau}\tilde{\phi}_{\x,t}(\sqrt{\Lambda}\mathcal{B}_{\x,t+1}, 
 \mathcal{B}_{\z,t+1}, \Lambda)
\right],
\end{equation}
which corresponds to $\xi_{\A,t,\tau}^{\x}$ in (\ref{xi_A_x_general}). 
We define an asymptotic alternative of $\boldsymbol{m}_{\x,t}$ in (\ref{mx}) as 
\begin{equation} \label{M_x}
M_{\x,t} = \sqrt{\Lambda}\tilde{\phi}_{\x,t}
(\sqrt{\Lambda}\mathcal{B}_{\x,t+1}, \mathcal{B}_{\z,t+1}, \Lambda). 
\end{equation}

We define the random variable $Q_{z,t+1}$ that corresponds to 
$\boldsymbol{q}_{\z,t+1}$ in (\ref{qz}). 
Let $\mathcal{H}_{\z,t+1}=\{H_{\z,\tau}\}_{\tau=0}^{t}$ denote 
zero-mean Gaussian random variables that correspond to 
$\{\boldsymbol{h}_{\z,\tau}\}_{\tau=0}^{t}$. They are independent of 
$W$ and have the covariance 
$\mathbb{E}[H_{\z,t}H_{\z,\tau}]=\mathbb{E}[M_{\z,t}M_{\z,\tau}]$ and 
$\mathbb{E}[\tilde{H}H_{\z,t}] = \mathbb{E}[\tilde{M}M_{\z,t}]$. 
Furthermore, let 
\begin{equation} \label{xi_B_z_bar_general}
\bar{\xi}_{\B,t,\tau}^{\z} 
= \mathbb{E}\left[
 \partial_{\tau}\psi_{\z,t}(\mathcal{H}_{\z,t+1}, \tilde{H}, W) 
\right], 
\end{equation}
\begin{equation} \label{eta_B_bar_general}
\bar{\eta}_{\B,t} 
= \mathbb{E}\left[
 \partial_{t+1}\psi_{\z,t}(\mathcal{H}_{\z,t+1}, \tilde{H}, W) 
\right], 
\end{equation}
which are asymptotic alternatives of $\xi_{\B,t,\tau}^{\z}$ and 
$\eta_{\B,t}$ in (\ref{xi_B_z_general}) and (\ref{eta_B_general}), 
respectively. Using these definitions, we define $Q_{\z,t+1}$ as 
\begin{equation} \label{Q_z}
Q_{\z,t+1} = \psi_{\z,t}(\mathcal{H}_{\z,t+1}, \tilde{H}, W)
- \sum_{\tau=0}^{t}\bar{\xi}_{\B,t,\tau}^{\z}H_{\z,\tau} - \bar{\eta}_{\B,t}\tilde{H}. 
\end{equation}

Finally, we present $\psi_{\x,t}(\boldsymbol{\Omega}_{t+1}, \boldsymbol{x})$, 
which corresponds to $\boldsymbol{q}_{\x,t+1}$ in (\ref{qx}).
 Let $\boldsymbol{\Omega}_{t+1}
=[\boldsymbol{\omega}_{0},\ldots,\boldsymbol{\omega}_{t}]
\in\mathbb{R}^{N\times(t+1)}$ denote a zero-mean Gaussian random matrix that is 
an asymptotic alternative of $\boldsymbol{H}_{\x,t+1}$. 
It is independent of $\boldsymbol{x}$ and has the covariance 
$\mathbb{E}[\boldsymbol{\omega}_{\tau}\boldsymbol{\omega}_{t}^{\mathrm{T}}]
=M^{-1}\mathbb{E}[M_{\x,t}M_{\x,\tau}]\boldsymbol{I}_{N}$. 
The quantity $\psi_{\x,t}(\boldsymbol{\Omega}_{t+1}, \boldsymbol{x})$ is given  
with these definitions. Furthermore, let 
\begin{equation} \label{xi_B_x_bar_general}
\bar{\xi}_{\B,t,\tau}^{\x}
= \lim_{N\to\infty}\frac{1}{M}\boldsymbol{1}^{\mathrm{T}}\mathbb{E}\left[
 \partial_{\tau}\psi_{\x,t}(\boldsymbol{\Omega}_{t+1}, \boldsymbol{x})
\right]
\end{equation}
in the sublinear sparsity limit, 
which corresponds to $\xi_{\B,t,\tau}^{\x}$. 

\begin{assumption} \label{assumption_inner_general} 
Let $\boldsymbol{\Omega}_{t+1,N}=[\boldsymbol{\omega}_{0,N},\ldots, 
\boldsymbol{\omega}_{t,N}]$ denote a zero-mean 
Gaussian random matrix that is independent of $\boldsymbol{x}$ and 
has the covariance $\mathbb{E}[\boldsymbol{\omega}_{\tau,N}
\boldsymbol{\omega}_{t,N}^{\mathrm{T}}]=N^{-1}\boldsymbol{m}_{\x,\tau}^{\mathrm{T}}
\boldsymbol{m}_{\x,t}\boldsymbol{I}_{N}$. Suppose that a random matrix 
$\boldsymbol{C}_{t+1}\in\mathbb{R}^{N\times(t+1)}$ satisfies the 
element-wise boundedness 
in probability of $\boldsymbol{C}_{t+1}^{\mathrm{T}}\boldsymbol{C}_{t+1}$ 
in the sublinear sparsity limit. 
\begin{itemize}
\item $\psi_{\x,t}$ in (\ref{psi_x}) is almost everywhere differentiable 
with respect to the first $(t+1)$ variables. Furthermore, the sublinear 
sparsity limit in (\ref{xi_B_x_bar_general}) exists. 

\item $\psi_{\x,\tau}(\boldsymbol{\Omega}_{\tau+1},
\boldsymbol{x})^{\mathrm{T}}\psi_{\x,t}(\boldsymbol{\Omega}_{t+1},
\boldsymbol{x})$ and $\boldsymbol{x}^{\mathrm{T}}
\psi_{\x,t}(\boldsymbol{\Omega}_{t+1},\boldsymbol{x})$ converge 
in probability to their expectation in the sublinear sparsity limit. 
 
\item 
$\|\psi_{\x,t}(\boldsymbol{C}_{t+1}\mathrm{diag}\{\boldsymbol{o}(1)\} 
+ \boldsymbol{\Omega}_{t+1,N}, \boldsymbol{x})
- \psi_{\x,t}(\boldsymbol{\Omega}_{t+1}, \boldsymbol{x})\|_{2}$ 
converges in probability to zero in the sublinear sparsity limit. 

\item $\boldsymbol{\omega}_{\tau}^{\mathrm{T}}
\psi_{\x,t}(\boldsymbol{C}_{t+1}\mathrm{diag}\{\boldsymbol{o}(1)\}
+ \boldsymbol{\Omega}_{t+1}, \boldsymbol{x}) 
- \mathbb{E}[\boldsymbol{\omega}_{\tau}^{\mathrm{T}}
\psi_{\x,t}(\boldsymbol{\Omega}_{t+1}$, $\boldsymbol{x})]$ converges  
in probability to zero in the sublinear sparsity limit. 
\end{itemize}
\end{assumption}

Assumption~\ref{assumption_inner_general} contains strong 
assumptions for the inner denoiser $\psi_{\x,t}$. 
Lemma~\ref{lemma_elimination} cannot be used to justify 
Assumption~\ref{assumption_inner_general} since the conditions required 
in Lemma~\ref{lemma_elimination} are not satisfied. 
Thus, Assumption~\ref{assumption_inner_general} needs to be justified 
for each inner denoiser in module~B. 

For instance, consider the memoryless 
function $\psi_{\x,t}=f_{\B}^{\x} - \boldsymbol{x}$ with a separable Bayesian 
estimator $f_{\B}^{\x}$ of $\boldsymbol{x}$ proposed in \cite{Takeuchi251}. 
In this case, \cite[Lemmas~1--4]{Takeuchi251} justifies 
Assumption~\ref{assumption_inner_general}. 
The function $\psi_{\x,t}$ has long-term memory when 
long-memory damping \cite{Liu22,Takeuchi22} is inserted just before the 
separable Bayesian estimator. Assumption~\ref{assumption_inner_general} may 
be justified in the same manner as in the memoryless case.

\begin{assumption} \label{assumption_SE_general}
Let $\tilde{\mathcal{M}}_{\z,t}
=\{M_{\z,0}, \ldots, M_{\z,t-1}, \tilde{M}\}$, 
$\mathcal{M}_{\x,t}$ 
$=\{M_{\x,0}, \ldots, M_{\x,t-1}\}$, and 
$\mathcal{Q}_{a,t+1} =\{Q_{a,0}, \ldots, Q_{a,t}\}$ for $a\in\{\z,\x\}$. 
There is some integer $\tau'\in\{0,\ldots,T\}$ such that the following 
conditions hold for all $\tau\in\{0,\ldots,\tau'\}$: 
\begin{itemize}
\item The variance $\mathbb{V}[M_{\z,0}]>0$ holds if the zero minimum MSE (MMSE) 
$\mathbb{E}[(M_{\z,0} - \mathbb{E}[M_{\z,0} | \tilde{M}])^{2}]=0$ is satisfied.  
Otherwise, $\mathbb{E}[(M_{\z,0} - \mathbb{E}[M_{\z,0} | \tilde{M}])^{2}]>0$ 
holds. For $\tau>0$, $\mathbb{E}[(M_{\z,\tau} - \mathbb{E}[M_{\z,\tau} 
| \tilde{\mathcal{M}}_{\z,\tau}])^{2}]>0$ holds. 

\item For $\tau>0$, $\mathbb{E}[(M_{\x,\tau} - \mathbb{E}[M_{\x,\tau} 
| \mathcal{M}_{\x,\tau}])^{2}]>0$ holds 
while $\mathbb{E}[M_{\x,0}^{2}]>0$ is satisfied.  

\item $\mathbb{E}[(Q_{\z,\tau+1} - \mathbb{E}[Q_{\z,\tau+1} 
| \mathcal{Q}_{\z,\tau+1}])^{2}]>0$ holds. 

\item Let $\{Q_{\x,t}\in\mathbb{R}\}_{t=0}^{T}$ denote zero-mean random 
variables satisfying $\mathbb{E}[Q_{\x,t}Q_{\x,t'}]=Q_{\x,t,t'}$. 
Then, $Q_{\x,\tau+1}$ satisfies $\mathbb{E}[(Q_{\x,\tau+1} 
- \mathbb{E}[Q_{\x,\tau+1} | \mathcal{Q}_{\x,\tau+1}])^{2}] >0$. 
\end{itemize}
\end{assumption}

The zero MMSE for $M_{\z,0}$ occurs when $\phi_{\z,0}$ 
in (\ref{phi_z}) is linear with respect to the first two variables and 
when $\Lambda=1$ holds with probability $1$. 
Assumption~\ref{assumption_SE_general} is broken once 
error-free signal reconstruction is achieved: 
Assumption~\ref{assumption_SE_general} is a technical assumption 
to terminate state evolution just before error-free signal reconstruction. 
In the last condition, the MMSE is invariant under the 
change of variables $\tilde{Q}_{\x,t}=Q_{\x,t}-\mu_{t}$ for any 
deterministic $\mu_{t}\in\mathbb{R}$. Thus, without loss of generality, 
we can make the zero-mean assumption for $Q_{\x,t}$. 

Assumption~\ref{assumption_SE_general} postulates 
nonlinear measurements since the outer sub-modules are not 
updated in OAMP for the linear measurement $g(z,w)=z+w$, as noted in 
Section~\ref{sec_OAMP}. When the linear measurement is considered, 
the vectors with subscript $\z$ should be eliminated from the general error 
model. Furthermore, Assumption~\ref{assumption_SE_general} should be 
replaced with that without assumptions on the outer sub-modules. In this sense, 
the main target of this paper is state evolution of message-passing algorithms for nonlinear measurements.   

We refer to the general error model with $\xi_{\B,t,\tau}^{\x}$ in 
(\ref{xi_B_x_general}) replaced by $\bar{\xi}_{\B,t,\tau}^{\x}$ in 
(\ref{xi_B_x_bar_general}) as the modified general error model. 
\begin{theorem} \label{theorem_SE_tech}
Suppose that Assumptions~\ref{assumption_w}, \ref{assumption_x}, 
\ref{assumption_A}, \ref{assumption_denoisers_general}, 
\ref{assumption_A_addition}, and 
\ref{assumption_inner_general} are 
satisfied. Then, in the initialization step for the modified general error 
model, the following results hold in the sublinear sparsity limit: 
\begin{enumerate}[start = 2, label* = I.\arabic*]
\item \label{I2}
The convergence in probability 
$M^{-1}\|\tilde{\boldsymbol{h}}\|_{2}^{2} \pto P$ holds. 

\item \label{I3}  
Assume $\phi, \psi\in\mathcal{L}_{2}$ in 
Definition~\ref{definition_integrability}. 
For $\tilde{\boldsymbol{m}}=\boldsymbol{U}^{\mathrm{T}}
\tilde{\boldsymbol{h}}=\boldsymbol{\Sigma}\boldsymbol{b}_{\x,0}$, we have 
\begin{equation} \label{I31}
\frac{\phi^{\mathrm{T}}(\tilde{\boldsymbol{m}}, \boldsymbol{\lambda})
\psi(\tilde{\boldsymbol{m}}, \boldsymbol{\lambda}) }{M}
\pto \mathbb{E}\left[
 \phi(\sqrt{\Lambda}B_{\x,0}, \Lambda)
 \psi(\sqrt{\Lambda}B_{\x,0}, \Lambda)
\right].  
\end{equation}

Suppose that $\phi: \mathbb{R}^{2}\to\mathbb{R}$ and 
$\psi: \mathbb{R}^{2}\to\mathbb{R}$ are piecewise 
Lipschitz-continuous functions. Then, we have 
\begin{equation} \label{I32}
\frac{1}{M}\phi^{\mathrm{T}}(\tilde{\boldsymbol{h}}, \boldsymbol{w})
\psi(\tilde{\boldsymbol{h}}, \boldsymbol{w})
\pto \mathbb{E}[\phi(\tilde{H}, W)\psi(\tilde{H}, W)],  
\end{equation}
\begin{equation} \label{I33}
\eta_{\B,-1} \pto \bar{\eta}_{\B,-1}. 
\end{equation}

\item \label{I4}
The asymptotic orthogonality 
$M^{-1}\tilde{\boldsymbol{h}}^{\mathrm{T}}\boldsymbol{q}_{\z,0}\pto 0$ holds. 

\item \label{I5}
The convergence in probability holds: 
$M^{-1}\|\tilde{\boldsymbol{m}}\|_{2}^{2}\pto Q_{\x,0,0}$, 
$M^{-1}\|\boldsymbol{q}_{\z,0}\|_{2}^{2} \pto \mathbb{E}[Q_{\z,0}^{2}]$, and 
$\|\boldsymbol{q}_{\x,0}\|_{2}^{2} \pto Q_{\x,0,0}$.
\end{enumerate}

For module~A in the modified general error model, 
the following results hold in the sublinear sparsity limit 
for all $\tau\in\{0,\ldots,T\}$ if Assumption~\ref{assumption_SE_general} 
with $\tau'=\tau-1$ is satisfied for $\tau>0$:  
\begin{enumerate}[label* = A.\arabic*]
\nonameditem
\begin{enumerate}[label* = .\submodules*]
\item \label{A1o}
For $\tau=0$, $\boldsymbol{b}_{\z,0}$ conditioned on $\Theta$ and 
$\mathfrak{E}_{0,0}^{\z}=\{\tilde{\boldsymbol{m}}, \tilde{\boldsymbol{h}}\}$ is 
distributed as 
\begin{equation}
\boldsymbol{b}_{\z,0} 
\sim o(1)\tilde{\boldsymbol{m}}
+ \{1 + o(1)\}\frac{\|\boldsymbol{q}_{\z,0}\|_{2}}
{\|\boldsymbol{\omega}_{\z,0}\|_{2}}\boldsymbol{\omega}_{\z,0},  
\end{equation}
where $\boldsymbol{\omega}_{\z,0}\sim\mathcal{N}(\boldsymbol{0},
\boldsymbol{I}_{M})$ is independent of $\{\Theta, \mathfrak{E}_{0,0}^{\z}\}$. 

For $\tau>0$, let $\boldsymbol{\beta}_{\z,\tau} 
= (\boldsymbol{Q}_{\z,\tau}^{\mathrm{T}}\boldsymbol{Q}_{\z,\tau})^{-1}
\boldsymbol{Q}_{\z,\tau}^{\mathrm{T}}\boldsymbol{q}_{\z,\tau}$, 
$\boldsymbol{q}_{\z,\tau}^{\perp}
= \boldsymbol{P}_{\boldsymbol{Q}_{\z,\tau}}^{\perp}\boldsymbol{q}_{\z,\tau}$. The 
vector $\boldsymbol{b}_{\z,\tau}$ conditioned on $\Theta$ and 
$\mathfrak{E}_{\tau,\tau}^{\z}$ is distributed as  
\begin{align}
\boldsymbol{b}_{\z,\tau} &\sim \boldsymbol{B}_{\z,\tau}\boldsymbol{\beta}_{\z,\tau} 
+ \boldsymbol{B}_{\z,\tau}\boldsymbol{o}(1) 
+ \tilde{\boldsymbol{M}}_{\z,\tau}\boldsymbol{o}(1) 
\nonumber \\
&+ \{1 + o(1)\}\frac{\|\boldsymbol{q}_{\z,\tau}^{\perp}\|_{2}}
{\|\boldsymbol{\omega}_{\z,\tau}\|_{2}}\boldsymbol{\omega}_{\z,\tau},
\end{align}
with $\boldsymbol{\omega}_{\z,\tau}\sim\mathcal{N}(\boldsymbol{0}, 
\boldsymbol{I}_{M})$ independent of $\Theta$ and $\mathfrak{E}_{\tau,\tau}^{\z}$. 

\item \label{A1i}
For $\tau>0$, let $\boldsymbol{\beta}_{\x,\tau} 
= (\boldsymbol{Q}_{\x,\tau}^{\mathrm{T}}\boldsymbol{Q}_{\x,\tau})^{-1}
\boldsymbol{Q}_{\x,\tau}^{\mathrm{T}}\boldsymbol{q}_{\x,\tau}$ 
and $\boldsymbol{q}_{\x,\tau}^{\perp} 
=\boldsymbol{P}_{\boldsymbol{Q}_{\x,\tau}}^{\perp}\boldsymbol{q}_{\x,\tau}$. The vector 
$\boldsymbol{b}_{\x,\tau}$ conditioned on $\Theta$ and 
$\mathfrak{E}_{\tau,\tau}^{\x}$ is distributed as 
\begin{align}
\boldsymbol{b}_{\x,\tau} &\sim \boldsymbol{B}_{\x,\tau}\boldsymbol{\beta}_{\x,\tau}
+ \boldsymbol{B}_{\x,\tau}\boldsymbol{o}(M/N) 
+ \boldsymbol{M}_{\x,\tau}\boldsymbol{o}(M/N) 
\nonumber \\
&+ \{1 + o(1)\}\frac{\|\boldsymbol{q}_{\x,\tau}^{\perp}\|_{2}}
{\|\boldsymbol{\omega}_{\x,\tau}\|_{2}}\boldsymbol{\omega}_{\x,\tau}, 
\end{align}
where $\boldsymbol{\omega}_{\x,\tau}\sim\mathcal{N}(\boldsymbol{0}, 
\boldsymbol{I}_{N})$ is independent of $\Theta$ and 
$\mathfrak{E}_{\tau,\tau}^{\x}$. 
\end{enumerate}

\item \label{A2}
For all $\tau'\in\{0,\ldots,\tau\}$, 
\begin{equation} \label{bzbz}
\frac{1}{M}\boldsymbol{b}_{\z,\tau'}^{\mathrm{T}}\boldsymbol{b}_{\z,\tau}
\pto \mathbb{E}[Q_{\z,\tau'}Q_{\z,\tau}],  
\end{equation}
\begin{equation} \label{bxbx}
\boldsymbol{b}_{\x,\tau'}^{\mathrm{T}}\boldsymbol{b}_{\x,\tau} 
\pto Q_{\x,\tau',\tau}. 
\end{equation}

\item \label{A3}
Assume $\phi, \psi\in\mathcal{L}_{2\tau+3}$ 
in Definition~\ref{definition_integrability}. Then, we have   
\begin{align}
&\frac{1}{M}\phi^{\mathrm{T}}(\boldsymbol{B}_{\z,\tau+1}, 
\boldsymbol{\Sigma}\boldsymbol{B}_{\x,\tau+1}, \boldsymbol{\lambda})
\psi(\boldsymbol{B}_{\z,\tau+1}, \boldsymbol{\Sigma}
\boldsymbol{B}_{\x,\tau+1}, \boldsymbol{\lambda})
\nonumber \\
&\pto \mathbb{E}[
\phi(\mathcal{B}_{\z,\tau+1}, \sqrt{\Lambda}\mathcal{B}_{\x,\tau+1}, 
\Lambda)
\psi(\mathcal{B}_{\z,\tau+1}, \sqrt{\Lambda}\mathcal{B}_{\x,\tau+1}, 
\Lambda)], 
\end{align}
\begin{equation}
\xi_{\A,\tau,\tau'}^{a} \pto \bar{\xi}_{\A,\tau,\tau'}^{a}
\end{equation}
for all $\tau'\in\{0,\ldots,\tau\}$, with $a\in\{\z,\x\}$. 

\nonameditem
\begin{enumerate}[label* = .\submodules*]
\item \label{A4o}
For all $\tau'\in\{0,\ldots,\tau\}$, 
\begin{equation} \label{zb}
\frac{1}{M}(\boldsymbol{\Sigma}\boldsymbol{b}_{\x,\tau'})^{\mathrm{T}}
\boldsymbol{b}_{\z,\tau} \pto 0, 
\end{equation}
\begin{equation} \label{bzmz}
\frac{1}{M}\boldsymbol{b}_{\z,\tau'}^{\mathrm{T}}\boldsymbol{m}_{\z,\tau}
\pto 0.
\end{equation}

\item \label{A4i}
For all $\tau'\in\{0,\ldots,\tau\}$, 
\begin{equation} \label{bxmx}
\boldsymbol{b}_{\x,\tau'}^{\mathrm{T}}\boldsymbol{m}_{\x,\tau} \pto 0. 
\end{equation}
\end{enumerate}

\nonameditem
\begin{enumerate}[label* = .\submodules*]
\item \label{A5o}
For all $\tau'\in\{0,\ldots,\tau\}$, 
the following convergence in probability holds: 
$M^{-1}\boldsymbol{m}_{\z,\tau'}^{\mathrm{T}}\boldsymbol{m}_{\z,\tau}
\pto \mathbb{E}[M_{\z,\tau'}M_{\z,\tau}]$ and 
$M^{-1}\tilde{\boldsymbol{m}}^{\mathrm{T}}\boldsymbol{m}_{\z,\tau}
\pto \mathbb{E}[\tilde{M}M_{\z,\tau}]$. Furthermore, 
there is some constant $C>0$ such that 
\begin{equation} \label{A5o_positivity}
\mathbb{P}\left(
 \lambda_{\mathrm{min}}\left(
  \frac{1}{M}\tilde{\boldsymbol{M}}_{\z,\tau+1}^{\mathrm{T}}
  \tilde{\boldsymbol{M}}_{\z,\tau+1}
 \right) > C
\right) \to 1 
\end{equation}
if the first condition in Assumption~\ref{assumption_SE_general} holds 
for $\tau'=\tau$. 

\item \label{A5i}
For all $\tau'\in\{0,\ldots,\tau\}$, the convergence in probability 
$(M/N)\boldsymbol{m}_{\x,\tau'}^{\mathrm{T}}\boldsymbol{m}_{\x,\tau}
\pto \mathbb{E}[M_{\x,\tau'}M_{\x,\tau}]$ holds. Furthermore, 
there is some constant $C>0$ such that 
\begin{equation} \label{A5i_positivity}
\mathbb{P}\left(
 \lambda_{\mathrm{min}}\left(
  \frac{M}{N}\boldsymbol{M}_{\x,\tau+1}^{\mathrm{T}}\boldsymbol{M}_{\x,\tau+1} 
 \right) > C
\right) \to 1 
\end{equation}
if the second condition in Assumption~\ref{assumption_SE_general} holds for $\tau'=\tau$. 
\end{enumerate}
\end{enumerate}

For module~B in the modified general error model, 
the following results hold in the sublinear sparsity limit 
for all $\tau\in\{0,\ldots,T\}$ if Assumption~\ref{assumption_SE_general} 
with $\tau'=\tau-1$ is satisfied for $\tau>0$ and if the first two conditions 
in Assumption~\ref{assumption_SE_general} with $\tau'=\tau$ is satisfied: 
\begin{enumerate}[label* = B.\arabic*]
\nonameditem
\begin{enumerate}[label* = .\submodules*]
\item \label{B1o} 
For $\tau=0$ the vector $\boldsymbol{h}_{\z,0}$ conditioned on $\Theta$ and 
$\mathfrak{E}_{0,1}^{\z}=\{\boldsymbol{q}_{\z,0}, \boldsymbol{b}_{\z,0}, 
\tilde{\boldsymbol{m}}, \tilde{\boldsymbol{h}}\}$ is distributed as 
\begin{align}
\boldsymbol{h}_{\z,0} 
&\sim \frac{\mathbb{E}[\tilde{M}M_{\z,0}]}{P}\tilde{\boldsymbol{h}}
+ o(1)\tilde{\boldsymbol{h}} + o(1)\boldsymbol{q}_{\z,0} 
\nonumber \\
&+ \{1 + o(1)\}\frac{(\boldsymbol{m}_{\z,0}^{\mathrm{T}}
\boldsymbol{P}_{\tilde{\boldsymbol{m}}}^{\perp}
\boldsymbol{m}_{\z,0})^{1/2}}
{\|\boldsymbol{\omega}_{\z,0}\|_{2}}\boldsymbol{\omega}_{\z,0}, 
\label{hz0}
\end{align}
where $\boldsymbol{\omega}_{\z,0}\sim\mathcal{N}(\boldsymbol{0}, 
\boldsymbol{I}_{M})$ is independent of $\Theta$ and $\mathfrak{E}_{0,1}^{\z}$. 

For $\tau>0$, let $\boldsymbol{\alpha}_{\z,\tau} 
= (\tilde{\boldsymbol{M}}_{\z,\tau}^{\mathrm{T}}
\tilde{\boldsymbol{M}}_{\z,\tau})^{-1}
\tilde{\boldsymbol{M}}_{\z,\tau}^{\mathrm{T}}
\boldsymbol{m}_{\z,\tau}$ 
and $\boldsymbol{m}_{\z,\tau}^{\perp}
= \boldsymbol{P}_{\tilde{\boldsymbol{M}}_{\z,\tau}}^{\perp}
\boldsymbol{m}_{\z,\tau}$. 
The vector $\boldsymbol{h}_{\z,\tau}$ conditioned on $\Theta$ and 
$\mathfrak{E}_{\tau,\tau+1}^{\z}$ is distributed as  
\begin{align}
&\boldsymbol{h}_{\z,\tau} 
\sim \tilde{\boldsymbol{H}}_{\z,\tau}\boldsymbol{\alpha}_{\z,\tau} 
+ \tilde{\boldsymbol{H}}_{\z,\tau}\boldsymbol{o}(1)
+ \boldsymbol{Q}_{\z,\tau+1}\boldsymbol{o}(1) 
\nonumber \\
&+ \{1 + o(1)\}\frac{\|\boldsymbol{m}_{\z,\tau}^{\perp}\|_{2}}
{\|\boldsymbol{\omega}_{\z,\tau}\|_{2}}\boldsymbol{\omega}_{\z,\tau},
\end{align}
with $\boldsymbol{\omega}_{\z,\tau}\sim\mathcal{N}(\boldsymbol{0}, 
\boldsymbol{I}_{M})$ independent of $\Theta$ and 
$\mathfrak{E}_{\tau,\tau+1}^{\z}$. 

\item \label{B1i} 
For $\tau=0$ the vector $\boldsymbol{h}_{\x,0}$ conditioned on $\Theta$ and 
$\mathfrak{E}_{0,1}^{\x}=\{\boldsymbol{q}_{\x,0}, \boldsymbol{b}_{\x,0}\}$ is 
distributed as 
\begin{equation} \label{hx0}
\boldsymbol{h}_{\x,0} \sim o(1)\boldsymbol{q}_{\x,0}
+ \{1 + o(1)\}\frac{\|\boldsymbol{m}_{\x,0}\|_{2}}
{\|\boldsymbol{\omega}_{\x,0}\|_{2}}\boldsymbol{\omega}_{\x,0}, 
\end{equation}
where $\boldsymbol{\omega}_{\x,0}\sim\mathcal{N}(\boldsymbol{0}, 
\boldsymbol{I}_{N})$ is independent of $\Theta$ and $\mathfrak{E}_{0,1}^{\x}$. 

For $\tau>0$, let $\boldsymbol{\alpha}_{\x,\tau} 
= (\boldsymbol{M}_{\x,\tau}^{\mathrm{T}}\boldsymbol{M}_{\x,\tau})^{-1}
\boldsymbol{M}_{\x,\tau}^{\mathrm{T}}\boldsymbol{m}_{\x,\tau}$ 
and $\boldsymbol{m}_{\x,\tau}^{\perp} 
=\boldsymbol{P}_{\boldsymbol{M}_{\x,\tau}}^{\perp}\boldsymbol{m}_{\x,\tau}$. The vector 
$\boldsymbol{h}_{\x,\tau}$ conditioned on $\Theta$ and 
$\mathfrak{E}_{\tau,\tau+1}^{\x}$ is distributed as 
\begin{align}
\boldsymbol{h}_{\x,\tau} &\sim \boldsymbol{H}_{\x,\tau}\boldsymbol{\alpha}_{\x,\tau}
+ \boldsymbol{H}_{\x,\tau}\boldsymbol{o}(1) 
+ \boldsymbol{Q}_{\x,\tau+1}\boldsymbol{o}(1) 
\nonumber \\
&+ \{1 + o(1)\}\frac{\|\boldsymbol{m}_{\x,\tau}^{\perp}\|_{2}}
{\|\boldsymbol{\omega}_{\x,\tau}\|_{2}}\boldsymbol{\omega}_{\x,\tau}, 
\end{align}
with $\boldsymbol{\omega}_{\x,\tau}\sim\mathcal{N}(\boldsymbol{0}, 
\boldsymbol{I}_{N})$ independent of $\Theta$ and 
$\mathfrak{E}_{\tau,\tau+1}^{\x}$. 
\end{enumerate}

\item \label{B2}
For all $\tau'\in\{0,\ldots,\tau\}$, 
\begin{equation}
\frac{1}{M}\boldsymbol{h}_{\z,\tau'}^{\mathrm{T}}\boldsymbol{h}_{\z,\tau}
\pto \mathbb{E}[M_{\z,\tau'}M_{\z,\tau}], 
\end{equation}
\begin{equation} \label{hz}
\frac{1}{M}\tilde{\boldsymbol{h}}^{\mathrm{T}}\boldsymbol{h}_{\z,\tau} 
\pto \mathbb{E}[\tilde{M}M_{\z,\tau}], 
\end{equation}
\begin{equation}
\frac{M}{N}\boldsymbol{h}_{\x,\tau'}^{\mathrm{T}}\boldsymbol{h}_{\x,\tau}
\pto \mathbb{E}[M_{\x,\tau'}M_{\x,\tau}]. 
\end{equation}

\nonameditem
\begin{enumerate}[label* = .\submodules*]
\item \label{B3o} 
For two piecewise Lipschitz-continuous functions 
$\phi: \mathbb{R}^{\tau+3}\to\mathbb{R}$ and 
$\psi: \mathbb{R}^{\tau+3}\to\mathbb{R}$,
\begin{align}
&\frac{1}{M}\phi^{\mathrm{T}}(\boldsymbol{H}_{\z,\tau+1}, \tilde{\boldsymbol{h}}, 
\boldsymbol{w}) 
\psi(\boldsymbol{H}_{\z,\tau+1}, \tilde{\boldsymbol{h}}, \boldsymbol{w}) 
\nonumber \\
&\pto \mathbb{E}\left[
 \phi(\mathcal{H}_{\z,\tau+1}, \tilde{H}, W)
 \psi(\mathcal{H}_{\z,\tau+1}, \tilde{H}, W)
\right],
\end{align}
\begin{equation}
\xi_{\B,\tau,\tau'}^{\z} \pto \bar{\xi}_{\B,\tau,\tau'}^{\z} 
\quad \eta_{\B,\tau} \pto \bar{\eta}_{\B,\tau}
\end{equation}
for all $\tau'\in\{0,\ldots,\tau\}$.

\item \label{B3i}
For all $\tau'\in\{0,\ldots,\tau\}$,  
\begin{align}
&\psi_{\x,\tau'}^{\mathrm{T}}(\boldsymbol{H}_{\x,\tau'+1}, \boldsymbol{x})
\psi_{\x,\tau}(\boldsymbol{H}_{\x,\tau+1}, \boldsymbol{x})
\nonumber \\
&- \mathbb{E}\left[
 \psi_{\x,\tau'}^{\mathrm{T}}(\boldsymbol{\Omega}_{\tau'+1}, \boldsymbol{x})
 \psi_{\x,\tau}(\boldsymbol{\Omega}_{\tau+1}, \boldsymbol{x})
\right] \pto 0, \label{psi_psi}
\end{align}
\begin{equation}
\boldsymbol{x}^{\mathrm{T}}
\psi_{\x,\tau}(\boldsymbol{H}_{\x,\tau+1}, \boldsymbol{x})
- \mathbb{E}\left[
 \boldsymbol{x}^{\mathrm{T}}
 \psi_{\x,\tau}(\boldsymbol{\Omega}_{\tau+1}, \boldsymbol{x})
\right] 
\pto 0, \label{x_psi}
\end{equation}
\begin{equation}
\boldsymbol{h}_{\x,\tau'}^{\mathrm{T}}
\psi_{\x,\tau}(\boldsymbol{H}_{\x,\tau+1}, \boldsymbol{x})
- \mathbb{E}\left[
 \boldsymbol{\omega}_{\tau'}^{\mathrm{T}}
 \psi_{\x,\tau}(\boldsymbol{\Omega}_{\tau+1}, \boldsymbol{x})
\right] 
\pto 0. \label{h_psi}
\end{equation}
\end{enumerate}

\nonameditem
\begin{enumerate}[label* = .\submodules*]
\item \label{B4o} 
For all $\tau'\in\{0,\ldots,\tau\}$, 
\begin{equation} \label{hzqz}
\frac{1}{M}\boldsymbol{h}_{\z,\tau'}^{\mathrm{T}}\boldsymbol{q}_{\z,\tau+1} \pto 0, 
\end{equation}
\begin{equation} \label{zq}
\frac{1}{M}\tilde{\boldsymbol{h}}^{\mathrm{T}}\boldsymbol{q}_{\z,\tau+1} 
\pto 0. 
\end{equation}

\item \label{B4i}
For all $\tau'\in\{0,\ldots,\tau\}$, 
\begin{equation} \label{hx}
\boldsymbol{x}^{\mathrm{T}}\boldsymbol{h}_{\x,\tau} \pto 0, 
\end{equation}
\begin{equation} \label{hxqx}
\boldsymbol{h}_{\x,\tau'}^{\mathrm{T}}\boldsymbol{q}_{\x,\tau+1} \pto 0.  
\end{equation}
\end{enumerate}

\nonameditem
\begin{enumerate}[label* = .\submodules*]
\item \label{B5o} 
For all $\tau'\in\{-1,\ldots,\tau\}$, the convergence in probability 
$M^{-1}\boldsymbol{q}_{\z,\tau'+1}^{\mathrm{T}}\boldsymbol{q}_{\z,\tau+1}\pto 
\mathbb{E}[Q_{\z,\tau'+1}Q_{\z,\tau+1}]$ holds. 
Furthermore, there is some constant $C>0$ such that 
\begin{equation} \label{B5o_positivity}
\mathbb{P}\left(
 \lambda_{\mathrm{min}}\left(
  \frac{1}{M}\boldsymbol{Q}_{\z,\tau+2}^{\mathrm{T}}\boldsymbol{Q}_{\z,\tau+2} 
 \right) > C
\right) \to 1 
\end{equation} 
if the third condition in Assumption~\ref{assumption_SE_general} holds 
for $\tau'=\tau$. 

\item \label{B5i}
For all $\tau'\in\{-1,\ldots,\tau\}$, the convergence in probability 
$\boldsymbol{q}_{\x,\tau'+1}^{\mathrm{T}}\boldsymbol{q}_{\x,\tau+1}\pto 
Q_{\x,\tau'+1,\tau+1}$ holds. 
Furthermore, there is some constant $C>0$ such that 
\begin{equation} \label{B5i_positivity}
\mathbb{P}\left(
 \lambda_{\mathrm{min}}\left(
  \boldsymbol{Q}_{\x,\tau+2}^{\mathrm{T}}\boldsymbol{Q}_{\x,\tau+2} 
 \right) > C
\right) \to 1
\end{equation}
if the last condition in Assumption~\ref{assumption_SE_general} holds 
for $\tau'=\tau$. 
\end{enumerate}
\end{enumerate}
\end{theorem}
\begin{IEEEproof}
The proof of Theorem~\ref{theorem_SE_tech} is by induction. The properties 
in the initialization step, modules~A, and B for $\tau=0$ are proved in 
Appendices~\ref{appen_initialization_step}, \ref{appen_module_A_0}, 
and \ref{appen_module_B_0}, respectively. 
For some $t\in\{1,\ldots,T\}$, suppose that Theorem~\ref{theorem_SE_tech} 
is correct for all $\tau<t$. Under this induction hypothesis, we prove 
the properties in modules~A and B for $\tau=t$ in 
Appendices~\ref{appen_module_A_t} and \ref{appen_module_B_t}. 
By induction, we arrive at Theorem~\ref{theorem_SE_tech}. 
\end{IEEEproof}

Properties~\ref{A3}, \ref{B3o}, and \ref{B3i} are asymptotic Gaussianity for module~A, the outer denoiser in module~B, and the inner denoiser in module~B, respectively. As originally proved in \cite{Takeuchi251}, the sublinear sparsity restricts the class of inner denoisers in module~B for justifying the asymptotic Gaussianity: For the sublinear sparsity the inner denoiser needs to satisfy the strong assumptions in Assumption~\ref{assumption_inner_general} while all Lipschitz-continuous inner denoisers are available for the linear sparsity.

\subsection{Proof of Theorem~\ref{theorem_SE}}
\subsubsection{Proof by Induction}
We first confirm Assumptions~\ref{assumption_denoisers_general}, 
\ref{assumption_A_addition}, and 
\ref{assumption_inner_general}. The functions 
$\phi_{\z,t}^{\mathrm{GO}}$ and $\phi_{\x,t}^{\mathrm{GO}}$ in (\ref{phi_z_GOAMP}) 
and (\ref{phi_x_GOAMP}) are linear with respect to the first two variables, 
so that $[\phi_{t}^{\mathrm{GO}}]_{m}$ and $[\phi_{\z,t}^{\mathrm{GO}}]_{m}$ are 
Lipschitz-continuous with the Lipschitz constants 
$L_{\x,t}(\lambda_{m}) = (\lambda_{m} + v_{\B\to\A}^{\z,t}/v_{\B\to\A}^{\x,t})^{-1}$ 
and $L_{\z,t}(\lambda_{m}) 
= \max\{v_{\B\to\A}^{\z,t}/v_{\B\to\A}^{\x,t}, \lambda_{m}\}L_{\x,t}(\lambda_{m})$, 
respectively. Since $L_{a,t}(\lambda_{m})$ is bounded and continuous, 
$\lambda_{m}L_{a,t}(\lambda_{m})$ is 
upper-bounded by a linear function of $\lambda_{m}$ for $a\in\{\z, x\}$. 
From the convergence in distribution and in mean 
under Assumption~\ref{assumption_A}, we justify  
Assumption~\ref{assumption_A_addition}. Furthermore, 
Assumption~\ref{assumption_denoisers} implies the piecewise 
Lipschitz-continuity of $\psi_{\z,t}$ in (\ref{psi_z_GOAMP}), so that 
Assumption~\ref{assumption_denoisers_general} holds. 
Assumptions~\ref{assumption_inner} implies 
Assumption~\ref{assumption_inner_general} for GOAMP. 

We prove that the state evolution 
for the modified general error model reduces to that for GOAMP and that 
Assumption~\ref{assumption_SE_general} with $\tau'=\tau$,  
$v_{\A\to\B}^{\z,\tau}\pto \bar{v}_{\A\to\B}^{\z,\tau}$,  
$v_{\A\to\B}^{\x,\tau}\pto \bar{v}_{\A\to\B}^{\x,\tau}$, 
$v_{\B\to\A}^{\z,\tau+1}\pto \bar{v}_{\B\to\A}^{\z,\tau+1}$, and 
$v_{\B\to\A}^{\x,\tau+1}\pto\bar{v}_{\B\to\A}^{\x,\tau+1}$ hold for all 
$\tau\in\{0,\ldots,T\}$. The proof is by induction. 
The proof for $\tau=0$ is omitted since it is similar to that for the general 
case. Suppose that there is some $t\in\{1,\ldots,T\}$ 
such that the statements are correct for all $\tau<t$. 
Under the induction hypothesis for Assumption~\ref{assumption_SE_general} 
we can use Properties~\ref{I2}--\ref{I5} and \ref{A1o}--\ref{A4i} 
in Theorem~\ref{theorem_SE_tech} for all $\tau\leq t$ while 
Assumption~\ref{assumption_SE_general} with $\tau'=t$ has to be proved 
to utilize Properties~\ref{A5o}--\ref{B5i} for $\tau=t$. 

\subsubsection{Outer Sub-Module in Module~A}
We first evaluate $\xi_{\A,t}^{\z}$ in (\ref{xi_A_z}). 
Comparing $\boldsymbol{m}_{\z,t}$ in (\ref{mz_GOAMP}) with 
$\phi_{\z,t}^{\mathrm{GO}}$ in (\ref{phi_z_GOAMP}) to $\boldsymbol{m}_{\z,t}$ in 
(\ref{mz}) with $\phi_{\z,t}$ in (\ref{phi_z}), and using the induction 
hypotheses $v_{\B\to\A}^{\z,t}\pto\bar{v}_{\B\to\A}^{\z,t}$ and 
$v_{\B\to\A}^{\x,t}\pto\bar{v}_{\B\to\A}^{\x,t}$, we define  
\begin{align} 
\phi_{\z,t}^{\mathrm{GO}}(B_{\z,t}, \sqrt{\Lambda}B_{\x,t}, \Lambda) 
&= \sqrt{\Lambda}B_{\x,t}
\nonumber \\
&+ \frac{\bar{v}_{\B\to\A}^{\x,t}\Lambda
(B_{\z,t} - \sqrt{\Lambda}B_{\x,t})}
{\bar{v}_{\B\to\A}^{\z,t} + \bar{v}_{\B\to\A}^{\x,t}\Lambda}.  
\label{phi_z_GOAMP_bar}
\end{align}
Using Property~\ref{A3} in Theorem~\ref{theorem_SE_tech} for $\tau=t$ and 
the definition of $\xi_{\A,t}^{\z}$ in (\ref{xi_A_z_GOAMP}), we find the 
reduction of $\xi_{\A,t}^{\z}$ to $\bar{\xi}_{\A,t}^{\z}$ in (\ref{xi_A_z_bar}): 
\begin{align} \label{xi_A_z_bar_tmp}
\xi_{\A,t}^{\z} 
&\pto \mathbb{E}\left[
 \partial_{0}\phi_{\z,t}^{\mathrm{GO}}(B_{\z,t}, \sqrt{\Lambda}B_{\x,t}, \Lambda)
\right]
\nonumber \\
&= \mathbb{E}\left[
 \frac{\bar{v}_{\B\to\A}^{\x,t}\Lambda} 
 {\bar{v}_{\B\to\A}^{\z,t} + \bar{v}_{\B\to\A}^{\x,t}\Lambda}
\right] \equiv \bar{\xi}_{\A,t}^{\z}. 
\end{align}

We next prove the first two conditions   
in Assumption~\ref{assumption_SE_general} for $\tau=t$. 
We only consider the random $\Lambda$ case: $\mathbb{P}(\Lambda=1)<1$ 
since the case $\mathbb{P}(\Lambda=1)=1$ can be treated in a similar manner.   
Comparing $\boldsymbol{m}_{\z,t}$ in (\ref{mz_GOAMP}) with  
$\boldsymbol{m}_{\z,t}$ in (\ref{mz}), we use the convergence in probability  
$\xi_{\A,t}^{\z}\pto\bar{\xi}_{\A,t}^{\z}$ to 
define $\phi_{\z,t}$ in the modified general error model as 
\begin{equation}
\phi_{\z,t}(B_{\z,t}, \sqrt{\Lambda}B_{\x,t}, \Lambda) 
= \frac{\phi_{\z,t}^{\mathrm{GO}}(B_{\z,t}, \sqrt{\Lambda}B_{\x,t}, \Lambda)}
{1 - \bar{\xi}_{\A,t}^{\z}},  
\end{equation}
with $\phi_{\z,t}^{\mathrm{GO}}$ in (\ref{phi_z_GOAMP_bar}). 
This implies that $M_{\z,t}$ in (\ref{M_z}) 
is linear with respect to $B_{\z,t}$ and $B_{\x,t}$ under the assumptions 
$\bar{v}_{\B\to\A}^{\z,t}>0$ and $\bar{v}_{\B\to\A}^{\x,t}>0$ 
in Assumption~\ref{assumption_SE}. We use the optimality of the conditional 
mean in terms of the MSE to obtain 
\begin{align}
&\mathbb{E}\left[
 (M_{\z,t} - \mathbb{E}[M_{\z,t}|\tilde{\mathcal{M}}_{\z,t}])^{2}
\right]
\nonumber \\
&\geq \mathbb{E}\left[
 (M_{\z,t} - \mathbb{E}[M_{\z,t}|\tilde{\mathcal{M}}_{\z,t}, 
 \mathcal{B}_{\z,t}, \mathcal{B}_{\x,t+1}, \Lambda])^{2}
\right]
\nonumber 
\end{align}
\begin{align}
&= \mathbb{E}\left[
 (M_{\z,t} - \mathbb{E}[M_{\z,t}| \mathcal{B}_{\z,t}, \mathcal{B}_{\x,t+1}, 
 \Lambda])^{2}
\right]
\nonumber \\
&= C\mathbb{E}\left[
 (B_{\z,t} - \mathbb{E}[B_{\z,t}| \mathcal{B}_{\z,t}, \Lambda])^{2}
\right] > 0
\end{align}
for some $C>0$. In the derivation of the first equality, we have used 
$\tilde{M}=\sqrt{\Lambda}B_{\x,0}$ and the fact that $M_{\z,\tau}$ is 
a deterministic function of $B_{\z,\tau}$, $B_{\x,\tau}$, and $\Lambda$. 
The last equality follows from the linearity of $M_{\z,t}$ with respect to 
$B_{\z,t}$ and $B_{\x,t}$, as well as 
the independence between $\mathcal{B}_{\z,t}$ and $\mathcal{B}_{\x,t+1}$. 
The strict positivity is due to $\bar{v}_{\B\to\A}^{\z,t}>0$ in 
Assumption~\ref{assumption_SE} and Property~\ref{B5o} in 
Theorem~\ref{theorem_SE_tech} for $\tau=t-1$ obtained from the induction 
hypothesis for Assumption~\ref{assumption_SE_general}. 
The strictly positive MMSE for $M_{\x,t}$ can be proved in a similar manner. 

We prove that $M^{-1}\|\boldsymbol{z}_{\A\to\B}^{t} 
- \boldsymbol{z}\|_{2}^{2}$ converges in probability to 
$\bar{v}_{\A\to\B}^{\z,t}$ in (\ref{moduleA_z_var_bar}). Since the first two 
conditions in Assumption~\ref{assumption_SE_general} have been 
proved for $\tau=t$, we can use Property~\ref{B2} in 
Theorem~\ref{theorem_SE_tech} for $\tau=t$, which implies 
$M^{-1}\|\boldsymbol{z}_{\A\to\B}^{t} - \boldsymbol{z}\|_{2}^{2}\pto 
\mathbb{E}[M_{\z,t}^{2}]\equiv\bar{v}_{\A\to\B}^{\z,t}$. 
Applying the definition of $\phi_{\z,t}^{\mathrm{GO}}$ in 
(\ref{phi_z_GOAMP_bar}) to $M_{\z,t}$ in (\ref{M_z}) yields 
\begin{align}
&(1 - \bar{\xi}_{\A,t}^{\z})^{2}\bar{v}_{\A\to\B}^{\z,t}
= \mathbb{E}\left[
 \left(
  \phi_{\z,t}^{\mathrm{GO}} - \bar{\xi}_{\A,t}^{\z}B_{\z,t} 
 \right)^{2}
\right]
\nonumber \\
&= \mathbb{E}\left[
 \{\phi_{\z,t}^{\mathrm{GO}}(B_{\z,t}, \sqrt{\Lambda}B_{\x,t}, \Lambda)\}^{2} 
\right]
- (\bar{\xi}_{\A,t}^{\z})^{2}\bar{v}_{\B\to\A}^{\z,t}, 
\end{align}
where the last equality follows from Lemma~\ref{lemma_Stein}, 
$\bar{\xi}_{\A,t}^{\z}$ in (\ref{xi_A_z_bar_tmp}), and 
$\mathbb{E}[B_{\z,t}^{2}]=\mathbb{E}[Q_{\z,t}^{2}]\equiv\bar{v}_{\B\to\A}^{\z,t}$. 
Using the definition of $\phi_{\z,t}^{\mathrm{GO}}$ in (\ref{phi_z_GOAMP_bar}), 
$\mathbb{E}[B_{\z,t}^{2}]=\bar{v}_{\B\to\A}^{\z,t}$,  
$\mathbb{E}[B_{\x,t}^{2}]=\bar{v}_{\B\to\A}^{\x,t}$, and 
the independence between $B_{\z,t}$ and $B_{\x,t}$, we have 
\begin{align}
&\mathbb{E}\left[
 \{\phi_{\z,t}^{\mathrm{GO}}\}^{2}
\right]
= \mathbb{E}\left[
 \left(
  \frac{\bar{v}_{\B\to\A}^{\x,t}\Lambda}
  {\bar{v}_{\B\to\A}^{\z,t} + \bar{v}_{\B\to\A}^{\x,t}\Lambda}
 \right)^{2}
\right]\bar{v}_{\B\to\A}^{\z,t}
\nonumber \\
&+ \mathbb{E}\left[
 \left(
  \frac{\bar{v}_{\B\to\A}^{\z,t}\sqrt{\Lambda}}
  {\bar{v}_{\B\to\A}^{\z,t} + \bar{v}_{\B\to\A}^{\x,t}\Lambda}
 \right)^{2}
\right]\bar{v}_{\B\to\A}^{\x,t}
= \bar{\xi}_{\A,t}^{\z}\bar{v}_{\B\to\A}^{\z,t},  
\end{align}
with $\bar{\xi}_{\A,t}^{\z}$ in (\ref{xi_A_z_bar_tmp}). 
Combining these results, we obtain $\bar{v}_{\A\to\B}^{\z,t}$ 
in (\ref{moduleA_z_var_bar}). 

Similarly, we use Property~\ref{B2} for $\tau=t$ and $\tilde{\boldsymbol{h}}=-\boldsymbol{z}$ to obtain the convergence in probability $-M^{-1}\boldsymbol{z}^{\mathrm{T}}(\boldsymbol{z}_{\A\to\B}^{t} - \boldsymbol{z})\pto\mathbb{E}[\tilde{H}H_{\z,t}]\equiv\bar{v}_{\A\to\B}^{\z,*,t}$. Repeating the evaluation of $\bar{v}_{\A\to\B}^{\z,t}$ with $\mathbb{E}[\tilde{H}H_{\z,t}]=\mathbb{E}[\tilde{M}M_{\z,t}]$ and 
$\tilde{M}=\sqrt{\Lambda}B_{\x,0}$ yields 
\begin{align}
&(1 - \bar{\xi}_{\A,t}^{\z})\bar{v}_{\A\to\B}^{\z,*,t}
= \mathbb{E}\left[
 \sqrt{\Lambda}B_{\x,0}\left(
  \phi_{\z,t}^{\mathrm{GO}} - \bar{\xi}_{\A,t}^{\z}B_{\z,t} 
 \right)
\right]
\nonumber \\
&= \mathbb{E}\left[
 \Lambda \left(
  1 - \frac{\bar{v}_{\B\to\A}^{\x,t}\Lambda}
  {\bar{v}_{\B\to\A}^{\z,t} + \bar{v}_{\B\to\A}^{\x,t}\Lambda}
 \right)
\right]\mathbb{E}[B_{\x,0}B_{\x,t}]
\nonumber \\
&= \mathbb{E}\left[
 \frac{\bar{v}_{\B\to\A}^{\z,t}\Lambda  }
 {\bar{v}_{\B\to\A}^{\z,t} + \bar{v}_{\B\to\A}^{\x,t}\Lambda}
\right]\mathbb{E}[B_{\x,0}B_{\x,t}]
= \frac{\bar{\xi}_{\A,t}^{\z}\bar{v}_{\B\to\A}^{\z,t}}{\bar{v}_{\B\to\A}^{\x,t}}
\bar{v}_{\B\to\A}^{\x,0,t},
\end{align}
with $\bar{v}_{\B\to\A}^{\x,0,t}=Q_{\x,0,t}$, 
where the last follows from $\bar{\xi}_{\A,t}^{\z}$ in 
(\ref{xi_A_z_bar}) and $\mathbb{E}[B_{\x,0}B_{\x,t}]=Q_{\x,0,t}$. 
Using $\bar{v}_{\A\to\B}^{\z,t}$ in (\ref{moduleA_z_var_bar}), 
we arrive at 
\begin{equation}
\bar{v}_{\A\to\B}^{\z,*,t}
= \frac{\bar{v}_{\B\to\A}^{\x,0,t}}{\bar{v}_{\B\to\A}^{\x,t}}
\bar{v}_{\A\to\B}^{\z,t}, 
\end{equation}
which is equivalent to $\bar{v}_{\A\to\B}^{\z,*,t}$ in (\ref{moduleA_z_var_bar}). 

Finally, we confirm the consistency of $v_{\A\to\B}^{\z,t}$. 
Using $v_{\A\to\B}^{\z,t}$ in (\ref{moduleA_z_mean}), $\xi_{\A,t}^{\z}\pto\bar{\xi}_{\A,t}^{\z}$, and the induction hypothesis $v_{\B\to\A}^{\z,t}\pto\bar{v}_{\B\to\A}^{\z,t}$, we find the consistency $v_{\A\to\B}^{\z,t}\pto\bar{v}_{\A\to\B}^{\z,t}$ in (\ref{moduleA_z_var_bar}).

\subsubsection{Inner Sub-Module in Module~A}
We first evaluate $\xi_{\A,t}^{\x}$. 
Comparing $\boldsymbol{m}_{\x,t}$ in (\ref{mz_GOAMP}) with 
$\phi_{\x,t}^{\mathrm{GO}}$ in (\ref{phi_x_GOAMP}) to $\boldsymbol{m}_{\x,t}$ in 
(\ref{mx}) with $\phi_{\x,t}$ in (\ref{phi_x}), and using the induction 
hypotheses $v_{\B\to\A}^{\z,t}\pto\bar{v}_{\B\to\A}^{\z,t}$ and 
$v_{\B\to\A}^{\x,t}\pto\bar{v}_{\B\to\A}^{\x,t}$, we define 
\begin{equation} \label{phi_t_GOAMP_bar}
\phi_{t}^{\mathrm{GO}}(\sqrt{\Lambda}B_{\x,t}, B_{\z,t}, \Lambda)
= \frac{\bar{v}_{\B\to\A}^{\x,t}(B_{\z,t} - \sqrt{\Lambda}B_{\x,t})}
{\bar{v}_{\B\to\A}^{\z,t} + \bar{v}_{\B\to\A}^{\x,t}\Lambda}. 
\end{equation}
Using Property~\ref{A3} in Theorem~\ref{theorem_SE_tech} and 
$\xi_{\A,t}^{\x}$ in (\ref{xi_A_x_GOAMP}) yields  
\begin{align}
&\xi_{\A,t}^{\x} 
= 1 + \mathbb{E}\left[
 \Lambda\partial_{0}\phi_{t}^{\mathrm{GO}}(\sqrt{\Lambda}B_{\x,t}, B_{\z,t}, \Lambda)
\right]
\nonumber \\
&= 1 - \mathbb{E}\left[
 \frac{\bar{v}_{\B\to\A}^{\x,t}\Lambda}
 {\bar{v}_{\B\to\A}^{\z,t} + \bar{v}_{\B\to\A}^{\x,t}\Lambda} 
\right]
= 1 - \bar{\xi}_{\A,t}^{\z}, 
\end{align}
with $\bar{\xi}_{\A,t}^{\z}$ in (\ref{xi_A_z_bar}), 
which is equal to $\bar{\xi}_{\A,t}^{\x}$ in (\ref{xi_A_x_bar}). 

We next prove that $(M/N)(\boldsymbol{x}_{\A\to\B}^{\tau} 
- \boldsymbol{x})^{\mathrm{T}}(\boldsymbol{x}_{\A\to\B}^{t} 
- \boldsymbol{x})$ converges in probability to 
$\bar{v}_{\A\to\B}^{\x,\tau,t}$ in (\ref{moduleA_x_var_bar}). 
Property~\ref{B2} in Theorem~\ref{theorem_SE_tech} for $\tau=t$ implies 
$(M/N)(\boldsymbol{x}_{\A\to\B}^{\tau} 
- \boldsymbol{x})^{\mathrm{T}}(\boldsymbol{x}_{\A\to\B}^{t} 
- \boldsymbol{x}) \pto\mathbb{E}[M_{\x,\tau}M_{\x,t}]
\equiv\bar{v}_{\A\to\B}^{\x,\tau,t}$. 
Substituting $\phi_{t}^{\mathrm{GO}}/(1 - \bar{\xi}_{\A,t}^{\x})$ into 
$\tilde{\phi}_{\x,t}$ in $M_{\x,t}$ given by (\ref{M_x}) yields 
\begin{equation} \label{M_x_GOAMP}
M_{\x,t} = \frac{\sqrt{\Lambda}\phi_{t}^{\mathrm{GO}}
(\sqrt{\Lambda}B_{\x,t}, B_{\z,t}, \Lambda)}
{1 - \bar{\xi}_{\A,t}^{\x}}. 
\end{equation}
We use $\phi_{t}^{\mathrm{GO}}$ in (\ref{phi_t_GOAMP_bar}), 
$\mathbb{E}[B_{\z,t}^{2}]=\bar{v}_{\B\to\A}^{\z,t}$,  
$\mathbb{E}[B_{\x,t}^{2}]=\bar{v}_{\B\to\A}^{\x,t}$, and 
the independence between $B_{\z,t}$ and $B_{\x,t}$ to obtain 
\begin{align}
&(1 - \bar{\xi}_{\A,\tau}^{\x})(1 - \bar{\xi}_{\A,t}^{\x})
\bar{v}_{\A\to\B}^{\x,\tau,t}
\nonumber \\
&= \mathbb{E}\left[
 \Lambda\phi_{\tau}^{\mathrm{GO}}(\sqrt{\Lambda}B_{\x,\tau}, B_{\z,\tau}, \Lambda)
 \phi_{t}^{\mathrm{GO}}(\sqrt{\Lambda}B_{\x,t}, B_{\z,t}, \Lambda)
\right]
\nonumber \\
&= \mathbb{E}\left[
 \frac{\bar{v}_{\B\to\A}^{\x,\tau}\bar{v}_{\B\to\A}^{\x,t}\Lambda}
 {(\bar{v}_{\B\to\A}^{\z,\tau} + \bar{v}_{\B\to\A}^{\x,\tau}\Lambda)
 (\bar{v}_{\B\to\A}^{\z,t} + \bar{v}_{\B\to\A}^{\x,t}\Lambda)}
\right]\bar{v}_{\B\to\A}^{\z,t} 
\nonumber \\
&+ \mathbb{E}\left[
 \frac{\bar{v}_{\B\to\A}^{\x,\tau}\bar{v}_{\B\to\A}^{\x,t}\Lambda^{2}}
 {(\bar{v}_{\B\to\A}^{\z,\tau} + \bar{v}_{\B\to\A}^{\x,\tau}\Lambda)
 (\bar{v}_{\B\to\A}^{\z,t} + \bar{v}_{\B\to\A}^{\x,t}\Lambda)}
\right]\bar{v}_{\B\to\A}^{\x,t}
\nonumber \\
&= \mathbb{E}\left[
 \frac{\bar{v}_{\B\to\A}^{\x,\tau}\Lambda}
 {\bar{v}_{\B\to\A}^{\z,\tau} + \bar{v}_{\B\to\A}^{\x,\tau}\Lambda}
\right]\bar{v}_{\B\to\A}^{\x,t}
= \bar{\xi}_{\A,\tau}^{\z}\bar{v}_{\B\to\A}^{\x,t},
\end{align}
with $\bar{\xi}_{\A,t}^{\z}$ in (\ref{xi_A_z_bar}), 
which reduces to (\ref{moduleA_x_var_bar}), due to $1 - \bar{\xi}_{\A,t}^{\x}=\bar{\xi}_{\A,t}^{\z}$. 

Finally, we prove the consistency $v_{\A\to\B}^{\x,t}\pto\bar{v}_{\A\to\B}^{\x,t,t}
\equiv \bar{v}_{\A\to\B}^{\x,t}$. 
Using the definitions of $v_{\A\to\B}^{\x,t}$ in (\ref{moduleA_x_mean}), 
$\boldsymbol{x}_{\A}^{t}$ in (\ref{moduleA_x_denoiser}), and repeating 
the proof of (\ref{mx_GOAMP_proof}) with $\boldsymbol{b}_{\x,t}
=\boldsymbol{V}^{\mathrm{T}}\boldsymbol{q}_{\x,t}$ in (\ref{b_GOAMP}), we obtain 
\begin{equation}
v_{\A\to\B}^{\x,t} = \frac{1}{MN}\frac{\|\boldsymbol{\Sigma}^{\mathrm{T}}
\phi_{t}^{\mathrm{GO}}(\boldsymbol{\Sigma}\boldsymbol{b}_{\x,t}, 
\boldsymbol{b}_{\z,t}, \boldsymbol{\lambda})\|_{2}^{2}}{(1 - \xi_{\A,t}^{\x})^{2}}
\pto \mathbb{E}[M_{\x,t}^{2}], 
\end{equation}
with $M_{\x,t}$ in (\ref{M_x_GOAMP}),  
where the last convergence follows from Property~\ref{A3} 
in Theorem~\ref{theorem_SE_tech} for $\tau=t$ 
and $\xi_{\A,t}^{\x}\pto\bar{\xi}_{\A,t}^{\x}$.  
Thus, we arrive at the consistency $v_{\A\to\B}^{\x,t}\pto\bar{v}_{\A\to\B}^{\x,t}$. 

\subsubsection{Outer Sub-Module in Module~B}
We first evaluate $\xi_{\B,t}^{\z}$. Comparing $\boldsymbol{q}_{\z,t+1}$ in (\ref{q_GOAMP}) with $\psi_{\z,t}^{\mathrm{GO}}$ in (\ref{psi_z_GOAMP}) to $\boldsymbol{q}_{\z,t+1}$ in (\ref{qz}) with $\phi_{\z,t}$ in (\ref{phi_z}), and using $v_{\A\to\B}^{\z,t}\pto\bar{v}_{\A\to\B}^{\z,t}$, we define  
\begin{equation} 
\psi_{\z,t}^{\mathrm{GO}}(H_{\z,t}, \tilde{H}, W) 
= f_{\B}^{\z}(-\tilde{H} + H_{\z,t}, g(-\tilde{H}, W); \bar{v}_{\A\to\B}^{\z,t})
+ \tilde{H}. 
\label{psi_z_GOAMP_bar}
\end{equation}
Using Property~\ref{B3o} in Theorem~\ref{theorem_SE_tech} for $\tau=t$ and 
the definition of $\xi_{\B,t}^{\z}$ in (\ref{xi_B_z_GOAMP}) yields  
\begin{equation}
\xi_{\B,t}^{\z} \pto \mathbb{E}\left[
 \partial_{0}f_{\B}^{\z}(-\tilde{H} + H_{\z,t}, g(-\tilde{H}, W); 
\bar{v}_{\A\to\B}^{\z,t}) 
\right] \equiv \bar{\xi}_{\B,t}^{\z}, 
\end{equation}
which is equivalent to $\bar{\xi}_{\B,t}^{\z}$ in (\ref{xi_B_z_bar}) because 
of $\tilde{H}=-Z$. Similarly, we use the definition of $\eta_{\B,t}$ in 
(\ref{eta_B_GOAMP}) to obtain 
\begin{align}
\bar{\eta}_{\B,t}
&\pto- \mathbb{E}\left[
 \left.
  \frac{\partial}{\partial z}f_{\B}^{\z}(-\tilde{H} + H_{\z,t}, g(z, W); 
  \bar{v}_{\A\to\B}^{\z,t}) 
 \right|_{z=-\tilde{H}}
\right]
\nonumber \\ 
&- \bar{\xi}_{\B,t}^{\z} + 1 
\equiv \bar{\eta}_{\B,t},
\end{align}
which is equal to $\bar{\eta}_{\B,t}$ in (\ref{eta_B_bar}) because 
of $\tilde{H}=-Z$. 

We prove the last two conditions in Assumption~\ref{assumption_SE_general} 
for $\tau=t$. The expression~(\ref{psi_z_GOAMP_bar}) implies that $Q_{\z,t+1}$ 
in (\ref{Q_z}) depends on $H_{\z,t}$ 
under the nonlinearity assumption in Assumption~\ref{assumption_denoisers}. 
Using the optimality of the conditional mean yields  
\begin{align}
&\mathbb{E}\left[
 (Q_{\z,t+1} - \mathbb{E}[Q_{\z,t+1} | \mathcal{Q}_{\z,t+1}])^{2}
\right]
\nonumber \\
&\geq \mathbb{E}\left[
 (Q_{\z,t+1} - \mathbb{E}[Q_{\z,t+1} | \mathcal{Q}_{\z,t+1}, 
 \mathcal{H}_{\z,t}, \tilde{H}, W])^{2}
\right]
\nonumber \\
&= \mathbb{E}\left[
 (Q_{\z,t+1} - \mathbb{E}[Q_{\z,t+1} | \mathcal{H}_{\z,t}, \tilde{H}, W])^{2}
\right] > 0.
\end{align}
The equality follows from the fact that $Q_{\z,\tau}$ 
is a deterministic function of $H_{\z,\tau-1}$, $\tilde{H}$, and $W$. 
The last positivity is due to the dependency of $Q_{\z,t+1}$ on $H_{\z,t}$ 
and the randomness of $H_{\z,t}$ conditioned on 
$\mathcal{H}_{\z,t}$, $\tilde{H}$, and $W$, of which the latter is due to 
$\bar{v}_{\A\to\B}^{\z,t}>0$ in Assumption~\ref{assumption_SE} and 
Property~\ref{A5o} in Theorem~\ref{theorem_SE_tech} for $\tau=t$. 
Similarly, the strictly positive MMSE for $Q_{\x,t+1}$ 
can be proved. Note that the nonlinearity of $f_{\B}^{\x}$ 
is not required because the Onsager correction in (\ref{moduleB_x_mean}) is 
negligibly small. 

We prove that $M^{-1}\|\boldsymbol{z}_{\B\to\A}^{t+1} - \boldsymbol{z}\|_{2}^{2}$ 
converges in probability to $\bar{v}_{\B\to\A}^{\z,t+1}$ in 
(\ref{moduleB_z_var_bar}). 
Property~\ref{B5o} in Theorem~\ref{theorem_SE_tech} for $\tau=t$ implies 
$M^{-1}\|\boldsymbol{z}_{\B\to\A}^{t+1} - \boldsymbol{z}\|_{2}^{2}\pto 
\mathbb{E}[Q_{\z,t+1}^{2}]\equiv \bar{v}_{\B\to\A}^{\z,t+1}$. 
Applying $\psi_{\z,t}^{\mathrm{GO}}/(1-\bar{\xi}_{\B,t}^{\z}-\bar{\eta}_{\B,t})$ 
in (\ref{psi_z_GOAMP_bar}) to 
$Q_{\z,t+1}$ in (\ref{Q_z}), and using Property~\ref{B4o} in Theorem~\ref{theorem_SE_tech} for $\tau=t$, we have 
\begin{align}
&(1 - \bar{\xi}_{\B,t}^{\z} - \bar{\eta}_{\B,t})^{2}
\bar{v}_{\B\to\A}^{\z,t+1}
\nonumber \\
&= \mathbb{E}\left[
 (f_{\B}^{\z} + \tilde{H})\left(
  f_{\B}^{\z} + \tilde{H}
  - \bar{\xi}_{\B,t}^{\z}H_{\z,t} - \bar{\eta}_{\B,t}\tilde{H}
 \right)
\right]
\nonumber \\
&= \mathbb{E}\left[
 (f_{\B}^{\z} + \tilde{H})^{2}
\right] - (\bar{\xi}_{\B,t}^{\z})^{2}\bar{v}_{\A\to\B}^{\z,t}
- \bar{\eta}_{\B,t}^{2}P
\nonumber \\
&- 2\bar{\xi}_{\B,t}^{\z}\bar{\eta}_{\B,t}\bar{v}_{\A\to\B}^{\z,*,t},
\end{align}
with $f_{\B}^{\z}=f_{\B}^{\z}(-\tilde{H} + H_{\z,t}, g(-\tilde{H}, W); 
\bar{v}_{\A\to\B}^{\z,t})$, 
where the last equality follows from Lemma~\ref{lemma_Stein}, 
$\bar{\xi}_{\B,t}^{\z}$ in (\ref{xi_B_z_bar}), $\bar{\eta}_{\B,t}$ in 
(\ref{eta_B_bar}), $\mathbb{E}[H_{\z,t}^{2}]=\mathbb{E}[M_{\z,t}^{2}]
=\bar{v}_{\A\to\B}^{\z,t}$, $\mathbb{E}[\tilde{H}^{2}]=P$, and 
$\mathbb{E}[\tilde{H}H_{\z,t}]=\bar{v}_{\A\to\B}^{\z,*,t}$. Thus, we arrive at 
$\bar{v}_{\B\to\A}^{\z,t+1}$ in (\ref{moduleB_z_var_bar}). 

Finally, we prove $v_{\B\to\A}^{\z,t+1}\pto\bar{v}_{\B\to\A}^{\z,t+1}$.  
Substituting the definition of $\boldsymbol{q}_{\z,t+1}
=\boldsymbol{z}_{\B\to\A}^{t+1} - \boldsymbol{z}$ with  
$\boldsymbol{z}=-\tilde{\boldsymbol{h}}$ into 
$v_{\B\to\A}^{\z,t+1}$ in (\ref{moduleB_z_mean}) and 
using Assumption~\ref{assumption_x}, we have 
\begin{equation}
v_{\B\to\A}^{\z,t+1} 
= \frac{\|\boldsymbol{q}_{\z,t+1} - \tilde{\boldsymbol{h}}\|_{2}^{2}}{M} 
- P 
\pto \mathbb{E}[Q_{\z,t+1}^{2}]= \bar{v}_{\B\to\A}^{\z,t+1},   
\end{equation}
where the last convergence follows from Properties~\ref{I2}, \ref{B4o}, and 
\ref{B5o} in Theorem~\ref{theorem_SE_tech} for $\tau=t$.

\subsubsection{Inner Sub-Module in Module~B} 
We prove that $\|\boldsymbol{x}_{\B\to\A}^{t+1} - \boldsymbol{x}\|_{2}^{2}$ and 
$-\boldsymbol{x}^{\mathrm{T}}(\boldsymbol{x}_{\B\to\A}^{t+1} 
- \boldsymbol{x})$ converge in probability to 
$\bar{v}_{\B\to\A}^{\x,t+1}$ and $\bar{v}_{\B\to\A}^{\x,0,t+1}$ in 
(\ref{moduleB_x_var_bar}) and (\ref{moduleB_x_cov_bar}), respectively. 
Property~\ref{B5i} in Theorem~\ref{theorem_SE_tech} for $\tau=t$ implies 
$\|\boldsymbol{x}_{\B\to\A}^{t+1} - \boldsymbol{x}\|_{2}^{2}\pto 
Q_{\x,t+1,t+1}\equiv \bar{v}_{\B\to\A}^{\x,t+1}$ and $-\boldsymbol{x}^{\mathrm{T}}
(\boldsymbol{x}_{\B\to\A}^{t+1} - \boldsymbol{x})\pto 
Q_{\x,0,t+1}\equiv \bar{v}_{\B\to\A}^{\x,0,t+1}$. 
Using $\psi_{\x,t}^{\mathrm{GO}}$ in (\ref{psi_x_GOAMP}) and  
$v_{\A\to\B}^{\x,t}\pto\bar{v}_{\A\to\B}^{\x,t}$, we have 
\begin{equation}
Q_{\x,t+1,t+1}
= \lim_{N\to\infty}\mathbb{E}\left[
 \|f_{\B}^{\x}(\boldsymbol{x} + \boldsymbol{\omega}_{t}; \bar{v}_{\A\to\B}^{\x,t}) 
 - \boldsymbol{x}\|_{2}^{2}
\right],
\end{equation}
\begin{equation}
Q_{\x,0,t+1}
= -\lim_{N\to\infty}\mathbb{E}\left[
 \boldsymbol{x}^{\mathrm{T}}
 \{f_{\B}^{\x}(\boldsymbol{x} + \boldsymbol{\omega}_{t}; \bar{v}_{\A\to\B}^{\x,t}) 
 - \boldsymbol{x}\}
\right],
\end{equation}
which are equal to $\bar{v}_{\B\to\A}^{\x,t+1}$ and $\bar{v}_{\B\to\A}^{\x,0,t+1}$ 
in (\ref{moduleB_x_var_bar}) and (\ref{moduleB_x_cov_bar}), respectively. 

We next confirm $\|\boldsymbol{x}_{\B}^{t+1} - \boldsymbol{x}\|_{2}^{2}
\pto\bar{v}_{\B\to\A}^{\x,t+1}$. Using the definition of 
$\boldsymbol{x}_{\B}^{t+1}$ in (\ref{moduleB_x_denoiser}), 
$v_{\A\to\B}^{\x,t}\pto\bar{v}_{\A\to\B}^{\x,t}$, and 
Property~\ref{B3i} in Theorem~\ref{theorem_SE_tech} for $\tau=t$ yields 
\begin{align}
\|\boldsymbol{x}_{\B}^{t+1} - \boldsymbol{x}\|_{2}^{2} 
\peq \|f_{\B}^{\x}(\boldsymbol{x} + \boldsymbol{h}_{\x,t}; 
\bar{v}_{\A\to\B}^{\x, t}) 
- \boldsymbol{x}\|_{2}^{2} + o(1)
\nonumber \\
\pto \lim_{N\to\infty}\mathbb{E}\left[
 \|f_{\B}^{\x}(\boldsymbol{x} + \boldsymbol{\omega}_{t}; 
\bar{v}_{\A\to\B}^{\x, t}) 
- \boldsymbol{x}\|_{2}^{2}
\right],
\end{align}
which is equal to $\bar{v}_{\B\to\A}^{\x,t+1}$ in (\ref{moduleB_x_var_bar}). 

To complete the proof of Theorem~\ref{theorem_SE}, we prove the consistency 
of $v_{\B\to\A}^{\x,t+1}$ in (\ref{moduleB_x_var}). Using $v_{\A\to\B}^{\x,t}
\pto\bar{v}_{\A\to\B}^{\x,t}$ and the replacement of $\xi_{\B,x}^{\x}$ with 
$\bar{\xi}_{\B,x}^{\x}$, we find $v_{\B\to\A}^{\x,t+1}\peq v_{\B}^{\x,t+1} + o(1) 
\pto \bar{v}_{\B\to\A}^{\x,t+1}$, in which the last convergence follows from 
the consistency assumption 
$\|\boldsymbol{x}_{\B}^{t+1} - \boldsymbol{x}\|_{2}^{2} 
- v_{\B}^{\x,t+1}\pto0$ and $\|\boldsymbol{x}_{\B}^{t+1} - \boldsymbol{x}\|_{2}^{2}
\pto\bar{v}_{\B\to\A}^{\x,t+1}$. Thus, Theorem~\ref{theorem_SE} holds. 

\section{Proof of Theorem~\ref{theorem_SE_Bayes}}
\label{proof_theorem_SE_Bayes}
We follow \cite[Lemmas~1--5]{Takeuchi251} under 
Assumption~\ref{assumption_inner_Bayes} to confirm 
Assumption~\ref{assumption_inner}. Furthermore, the posterior variance 
$v_{\B}^{\x,t+1}$ in (\ref{posterior_variance_Bayes}) is a consistent 
estimator of $\|\boldsymbol{x}_{\B}^{t+1} - \boldsymbol{x}\|_{2}^{2}$ 
in the sublinear sparsity limit. These observations allow us to use 
Theorem~\ref{theorem_SE} to justify the convergence in probability of 
the unnormalized square error 
$\|\boldsymbol{x}_{\B}^{t+1} - \boldsymbol{x}\|_{2}^{2}$ for modified 
Bayesian GOAMP with $\xi_{\B,t}^{\x}$ in (\ref{xi_B_x}) replaced 
by $\bar{\xi}_{\B,t}^{\x}$ in (\ref{xi_B_x_bar}). 
Thus, it is sufficient to prove that the state evolution 
recursion~(\ref{moduleB_z_var_bar}) reduces to 
(\ref{moduleB_z_var_bar_Bayes}), as well as the initial conditions.    

We first prove $\bar{\eta}_{\B,-1}=\bar{\xi}_{\B,-1}$ and $\bar{\eta}_{\B,t}=0$. 
Using \cite[Lemma~10]{Takeuchi242} for the Bayes-optimal outer 
denoiser~(\ref{outer_denoiser_Bayes}), we have the following identity: 
\begin{align}
&1 - \mathbb{E}\left[
 \partial_{0}f_{\B}^{\z}(Z_{t}, g(Z, W); \bar{v}_{\A\to\B}^{\z,t})
\right]
\nonumber \\
&= \mathbb{E}\left[
 \left.
  \frac{\partial}{\partial z}f_{\B}^{\z}(Z_{t}, g(z, W); 
  \bar{v}_{\A\to\B}^{\z,t}) 
 \right|_{z=Z}
\right],
\end{align}
which implies $\bar{\eta}_{\B,t}=0$ for (\ref{eta_B_bar}). 
We use the same identity 
with $Z_{t}=0$ for $\bar{\eta}_{\B,-1}$ in (\ref{eta_B0_bar}) to find 
$\bar{\eta}_{\B,-1}=\bar{\xi}_{\B,-1}$. 

We next prove the representation of $\bar{v}_{\B\to\A}^{\z,t+1}$ 
in (\ref{moduleB_z_var_bar_Bayes}). From $\bar{\eta}_{\B,t}=0$, we can 
simplify $\bar{v}_{\B\to\A}^{\z,t+1}$ in (\ref{moduleB_z_var_bar}) as 
\begin{align} 
\bar{v}_{\B\to\A}^{\z,t+1} 
= \frac{\mathbb{E}[\{f_{\B}^{\z}(Z_{t}, Y; \bar{v}_{\A\to\B}^{\z,t}) 
- Z\}^{2}] - (\bar{\xi}_{\B,t}^{\z})^{2}\bar{v}_{\A\to\B}^{\z,t}}
{(1 - \bar{\xi}_{\B,t}^{\z})^{2}}. 
\end{align}
To confirm the representation~(\ref{moduleB_z_var_bar_Bayes}), 
it is sufficient to prove 
\begin{equation} \label{MSE_outer_denoiser}
\mathbb{E}[\{f_{\B}^{\z}(Z_{t}, Y; \bar{v}_{\A\to\B}^{\z,t}) - Z\}^{2}] 
= \bar{\xi}_{\B,t}^{\z}\bar{v}_{\A\to\B}^{\z,t}. 
\end{equation}

We prove (\ref{MSE_outer_denoiser}). Using $Z\sim Z_{t}+N_{\z,t}$ yields 
\begin{align}
&\mathbb{E}[\{f_{\B}^{\z}(Z_{t}, Y; \bar{v}_{\A\to\B}^{\z,t}) - Z\}^{2}]
= \mathbb{E}[\{f_{\B}^{\z}(Z_{t}, Y; \bar{v}_{\A\to\B}^{\z,t}) - Z_{t}\}^{2}]
\nonumber \\
&- 2\mathbb{E}[\{f_{\B}^{\z}(Z_{t}, Y; \bar{v}_{\A\to\B}^{\z,t}) - Z_{t}\}N_{\z,t}]
+ \bar{v}_{\A\to\B}^{\z,t}. \label{MSE_outer_denoiser_tmp}
\end{align}
For the first term, we use \cite[Lemma 3]{Takeuchi242} 
for the Bayes-optimal outer denoiser to obtain  
\begin{equation}
\mathbb{E}\left[
 \left\{
  \frac{Z_{t} - f_{\B}^{\z}(Z_{t}, Y; \bar{v}_{\A\to\B}^{\z,t})}
  {\bar{v}_{\A\to\B}^{\z,t}}
 \right\}^{2}
\right]
= \frac{1 - \bar{\xi}_{\B,t}^{\z}}{\bar{v}_{\A\to\B}^{\z,t}}. 
\end{equation}
For the second term in (\ref{MSE_outer_denoiser_tmp}), we utilize 
Lemma~\ref{lemma_Stein} and $Y=g(Z,W)\sim g(Z_{t}+N_{\z,t}, W)$ to obtain 
\begin{align}
&\mathbb{E}[\{f_{\B}^{\z}(Z_{t}, Y; \bar{v}_{\A\to\B}^{\z,t}) - Z_{t}\}N_{\z,t}]
\nonumber \\
&= \bar{v}_{\A\to\B}^{\z,t}\mathbb{E}\left[
 \left.
  \frac{\partial}{\partial z}f_{\B}^{\z}(Z_{t}, g(z, W); \bar{v}_{\A\to\B}^{\z,t})
 \right|_{z=Z_{t} + N_{\z,t}}
\right]
\nonumber \\
&= (1 - \bar{\xi}_{\B,t}^{\z})\bar{v}_{\A\to\B}^{\z,t}, 
\end{align}
where the last equality follows from $\bar{\eta}_{\B,t}=0$ in 
(\ref{eta_B_bar}). Combining these results, we arrive at   
\begin{align}
&\mathbb{E}[\{f_{\B}^{\z}(Z_{t}, Y; \bar{v}_{\A\to\B}^{\z,t}) - Z\}^{2}]
= (1 - \bar{\xi}_{\B,t}^{\z})\bar{v}_{\A\to\B}^{\z,t} 
\nonumber \\
&- 2(1 - \bar{\xi}_{\B,t}^{\z})\bar{v}_{\A\to\B}^{\z,t}
+ \bar{v}_{\A\to\B}^{\z,t}
= \bar{\xi}_{\B,t}^{\z}\bar{v}_{\A\to\B}^{\z,t}. 
\end{align}
We repeat the same proof for $\bar{v}_{\B\to\A}^{\z,0}$ in 
(\ref{moduleB_z_var0_bar}) to find that 
the representation~(\ref{moduleB_z_var_bar_Bayes}) is valid for $t=-1$. 

\section{Proof of Theorem~\ref{theorem_linear}}
\label{proof_theorem_linear}
Let $y_{t}=\bar{v}_{\B\to\A}^{\x,t}\geq0$ and 
$x_{t}=2\bar{v}_{\A\to\B}^{\x,t}/\delta\geq0$. 
The state evolution recursion~(\ref{moduleB_x_var_bar_Bayes}) and 
(\ref{moduleA_x_var_bar_linear}) reduce to the dynamical system 
$y_{t+1} = \phi(x_{t})$ and $x_{t} = \psi(y_{t})$, with 
\begin{equation} \label{phi}
\phi(x) = \mathbb{E}\left[
 U^{2}1(U^{2}<x)
\right], 
\end{equation}
\begin{equation} \label{psi}
\psi(y) = \frac{2}{\delta}\left\{
 \mathbb{E}\left[
  \frac{\Lambda}{\sigma^{2} + y\Lambda}
 \right]
\right\}^{-1}. 
\end{equation}

We first investigate properties of the function $\psi$. Since 
$\tilde{\Lambda}(y)=\Lambda/(\sigma^{2}+y\Lambda)$ is a 
positive random variable with a compact support for all $y\geq0$ under 
Assumption~\ref{assumption_A}, we use $\tilde{\Lambda}'(y)
=-\tilde{\Lambda}^{2}(y)$ to evaluate the derivatives of $\psi$ as 
follows: 
\begin{equation} \label{psi_derivative}
\psi'(y) = \frac{2\mathbb{E}[\tilde{\Lambda}^{2}(y)]}
{\delta(\mathbb{E}[\tilde{\Lambda}(y)])^{2}} > 0, 
\end{equation}
\begin{equation}
\psi''(y) = \frac{4\{(\mathbb{E}[\tilde{\Lambda}^{2}(y)])^{2} 
- \mathbb{E}[\tilde{\Lambda}(y)]\mathbb{E}[\tilde{\Lambda}^{3}(y)]\}}
{\delta(\mathbb{E}[\tilde{\Lambda}(y)])^{3}} \leq0, 
\end{equation}
where the latter non-positivity follows from the positivity of 
$\tilde{\Lambda}(y)$ and the Cauchy-Schwarz inequality, 
\begin{equation}
(\mathbb{E}[\tilde{\Lambda}^{2}])^{2} 
= (\mathbb{E}[\tilde{\Lambda}^{1/2}\tilde{\Lambda}^{3/2}])^{2} 
\leq \mathbb{E}[|\tilde{\Lambda}|]\mathbb{E}[|\tilde{\Lambda}|^{3}]. 
\end{equation}
These results imply that the inverse function $\psi^{-1}(x)$ is monotonically 
increasing and convex for all $x\geq\psi(0)=2\sigma^{2}/\delta$. 

We prove that $y_{t}$ converges to zero for any 
distribution of $U$ as $t\to\infty$ if and only if $\delta>\delta_{*}$ holds. 
Consider the converse part $\delta\leq\delta_{*}$. 
The inequality $\delta\leq\delta_{*}$ is equivalent to 
\begin{equation}
\psi(u_{\mathrm{min}}^{2}) - u_{\mathrm{min}}^{2}
\geq \frac{2}{\delta_{*}}\left\{
 \mathbb{E}\left[
  \frac{\Lambda}{\sigma^{2} + u_{\mathrm{min}}^{2}\Lambda}
 \right]
\right\}^{-1} - u_{\mathrm{min}}^{2} = 0,  
\end{equation}
where the last equality follows from the definition of $\delta_{*}$ in 
(\ref{delta_threshold}). 
Thus, the curve $y=\psi^{-1}(x)$ passes through 
the point $(\psi(u_{\mathrm{min}}^{2}), u_{\mathrm{min}}^{2})$, which is below the 
straight line $y=x$ on the plane $(x, y)$. 
Since $\psi(y)$ is monotonically increasing and tends to infinity as 
$y\to\infty$, the inverse function $\psi^{-1}(x)$ is also monotonically 
increasing and diverges to infinity as $x\to\infty$. 

We focus on $\phi(x)$ for the constant-amplitude non-zero signals, i.e.\ 
$\mathbb{P}(|U|=u_{\mathrm{min}})=1$. For this $U$, we find that 
(\ref{phi}) reduces to $\phi(x)=u_{\mathrm{min}}^{2}1(u_{\mathrm{min}}^{2}<x)$. 
Thus, $y=\phi(x)$ and $y=\psi^{-1}(x)$ have a non-trivial intersection 
satisfying $y=u_{\mathrm{min}}^{2}$. These observations imply 
that $y_{t}$ cannot converge to zero for this $U$ 
if $\delta\leq\delta_{*}$ holds. 

Consider the direct part $\delta>\delta_{*}$. For this case, the inequality 
$\psi(u_{\mathrm{min}}^{2}) < u_{\mathrm{min}}^{2}$ holds. Thus, the point 
$(x, y)=(\psi(0), 0)$ is a fixed point, because of 
$u_{\mathrm{min}}^{2} - \psi(0)>\psi(u_{\mathrm{min}}^{2})-\psi(0)>0$.  

We evaluate the derivative of $\psi$ at $y=u_{\mathrm{min}}^{2}$. Using $\psi'$ 
in (\ref{psi_derivative}) and $\delta_{*}$ in (\ref{delta_threshold}) yields 
\begin{equation}
\psi'(u_{\mathrm{min}}^{2}) 
= \frac{\delta_{*}^{2}}{2\delta}
\mathbb{E}\left[
 u_{\mathrm{min}}^{4}\tilde{\Lambda}^{2}(u_{\mathrm{min}}^{2})
\right]. 
\end{equation}
Utilizing the trivial upper bound 
$u_{\mathrm{min}}^{2}\tilde{\Lambda}(u_{\mathrm{min}}^{2})<1$ yields 
\begin{equation}
\psi'(u_{\mathrm{min}}^{2}) 
< \frac{\delta_{*}^{2}}{2\delta}
\mathbb{E}\left[
 \frac{u_{\mathrm{min}}^{2}\Lambda}{\sigma^{2} + u_{\mathrm{min}}^{2}\Lambda}
\right] = \frac{\delta_{*}}{\delta} < 1, 
\end{equation}
where the equality follows from the definition of $\delta_{*}$ 
in (\ref{delta_threshold}). Since $\psi$ is monotonically increasing and 
concave, from the properties $\psi(u_{\mathrm{min}}^{2}) < u_{\mathrm{min}}^{2}$ and 
$\psi'(u_{\mathrm{min}}^{2})<1$ we find $\psi(y)<y$ for all 
$y\geq u_{\mathrm{min}}^{2}$. In other words, the curve $y=\psi^{-1}(x)$ is 
strictly above the straight line $y=x$ for all $x\geq \psi(u_{\mathrm{min}}^{2})$. 

\begin{algorithm}[t]
\caption{Bayesian GOAMP}
\label{alg1}
\begin{algorithmic}[1]
\State $v_{\A \to \B}^{\mathrm{z}, -1} = P$, 
$\boldsymbol{x}_{\B \to \A}^{0} = \boldsymbol{0}$, 
$v_{\B \to \A}^{\mathrm{x}, 0} = P$
\Comment{Initialization} 

\State $\boldsymbol{z}_{\B \to \A}^0 = f_{\B}^{\mathrm{z}}(\boldsymbol{0}, \boldsymbol{y}; v_{\A \to \B}^{\mathrm{z}, -1})$, $v_{\B \to \A}^{\mathrm{z}, 0} 
= \frac{1}{M}\|\boldsymbol{z}_{\B \to \A}^{0}\|_{2}^{2} - P$

\ForAll{$t\in\{0,\ldots, T\}$}

\Comment{Outer sub-module in module A}
\State $\tilde{\boldsymbol{z}}_{\B \to \A}^{t} 
= \boldsymbol{A}\boldsymbol{x}_{\B \to \A}^{t}$, 
$\varsigma_{t}^{2} = \frac{N v_{\B \to \A}^{\mathrm{z}, t}}
{v_{\B \to \A}^{\mathrm{x}, t}}$
\label{line4}

\State $\widehat{\Delta \boldsymbol{z}^{t}}
= \boldsymbol{A}\boldsymbol{A}^{\mathrm{T}}\left( 
 \varsigma_{t}^{2}\boldsymbol{I}_{M} 
 + \boldsymbol{A}\boldsymbol{A}^{\mathrm{T}} 
\right)^{-1}\left( 
 \boldsymbol{z}_{\B \to \A}^{t} - \tilde{\boldsymbol{z}}_{\B \to \A}^{t}
\right)$
\label{line8}

\State $\boldsymbol{z}_{\A}^{t} = 
\tilde{\boldsymbol{z}}_{\B \to \A}^{t} 
+ \widehat{\Delta \boldsymbol{z}^{t}}$, 
$\xi_{\A, t}^{\mathrm{z}} = \frac{1}{M}
\mathrm{div}_{\boldsymbol{z}_{\B \to \A}^t} \boldsymbol{z}_{\A}^{t}$

\State $\boldsymbol{z}_{\A \to \B}^{t} = \frac{\boldsymbol{z}_{\A}^{t} 
- \xi_{\A, t}^{\mathrm{z}} \boldsymbol{z}_{\B \to \A}^{t}}
{1 - \xi_{\A, t}^{\mathrm{z}}}$, 
$v_{\A \to \B}^{\mathrm{z}, t} 
= P - \frac{1}{M} \| \boldsymbol{z}_{\A \to \B}^t \|_{2}^{2}$
\label{line10}

\If{$v_{\A \to \B}^{\mathrm{z}, t} \leq 0$}
\State $v_{\A \to \B}^{\mathrm{z}, t} = \frac{\xi_{\A, t}^{\mathrm{z}} 
v_{\B \to \A}^{\mathrm{z}, t}}{1 - \xi_{\A, t}^{\mathrm{z}}}$
\EndIf

\Comment{Inner sub-module in module A}
\State $\widehat{\Delta \boldsymbol{x}^{t}} 
= \boldsymbol{A}^{\mathrm{T}}\left( 
 \varsigma_{t}^{2}\boldsymbol{I}_{M} 
 + \boldsymbol{A}\boldsymbol{A}^{\mathrm{T}} 
\right)^{-1}\left( 
 \boldsymbol{z}_{\B \to \A}^{t} - \tilde{\boldsymbol{z}}_{\B \to \A}^{t}
\right)$
\label{line11}

\State $\boldsymbol{x}_{\A}^{t} = \boldsymbol{x}_{\B \to \A}^{t} 
+ \frac{N}{M}\widehat{\Delta \boldsymbol{x}^{t}}$, 
$\xi_{\A, t}^{\mathrm{x}} = \frac{1}{N}\mathrm{div}_{\boldsymbol{x}_{\B \to \A}^{t}}
\boldsymbol{x}_{\A}^{t}$

\State $\boldsymbol{x}_{\A \to \B}^{t} = \frac{\boldsymbol{x}_{\A}^{t} 
- \xi_{\A, t}^{\mathrm{x}}\boldsymbol{x}_{\B \to \A}^{t}}
{1 - \xi_{\A, t}^{\mathrm{x}}}$, 
$v_{\A \to \B}^{\mathrm{x}, t} = \frac{M \| \boldsymbol{x}_{\A}^{t} 
- \boldsymbol{x}_{\B \to \A}^{t} \|_{2}^{2}}{N (1 - \xi_{\A, t}^{\mathrm{x}})^{2}}$

\Comment{Outer sub-module in module B}

\State $\boldsymbol{z}_{\B}^{t+1} 
= f_{\B}^{\mathrm{z}}(\boldsymbol{z}_{\A \to \B}^{t}, \boldsymbol{y}; 
v_{\A \to \B}^{\mathrm{z}, t})$, $\xi_{\B, t}^{\mathrm{z}} 
= \frac{1}{M} \mathrm{div}_{\boldsymbol{z}_{\A \to \B}^{t}} \boldsymbol{z}_{\B}^{t+1}$
\label{line16}

\State $\boldsymbol{z}_{\B \to \A}^{t+1} = \frac{\boldsymbol{z}_{\B}^{t+1} 
- \xi_{\B, t}^{\mathrm{z}} \boldsymbol{z}_{\A \to \B}^t}
{1 - \xi_{\B, t}^{\mathrm{z}}}$, 
$v_{\B \to \A}^{\mathrm{z}, t+1} = \frac{\xi_{\B, t}^{\mathrm{z}} 
v_{\A \to \B}^{\mathrm{z}, t}}{1 - \xi_{\B, t}^{\mathrm{z}}}$

\Comment{Inner sub-module in module B}

\State $\boldsymbol{x}_{\B}^{t+1} 
= f_{\B}^{\mathrm{x}}(\boldsymbol{x}_{\A \to \B}^{t}; 
v_{\A \to \B}^{\mathrm{x}, t})$, $\xi_{\B, t}^{\mathrm{x}} 
= \frac{1}{M}\mathrm{div}_{\boldsymbol{x}_{\A \to \B}^{t}}\boldsymbol{x}_{\B}^{t+1}$
\label{line14}

\State \!$\boldsymbol{x}_{\B \to \A}^{t+1} 
= \frac{\boldsymbol{x}_{\B}^{t+1} - (M/N)\xi_{\B, t}^{\mathrm{x}} 
\boldsymbol{x}_{\A \to \B}^{t}}{1 - (M/N)\xi_{\B, t}^{\mathrm{x}}}, v_{\B \to \A}^{\mathrm{x}, t+1} = \frac{v_{\A \to \B}^{\mathrm{x}, t} 
\xi_{\B, t}^{\mathrm{x}}}{1 - (M/N)\xi_{\B, t}^{\mathrm{x}}}$

\If{$t>0$}
\Comment{Damping}
\State $\boldsymbol{z}_{\B\to\A}^{t+1}:= \theta_{\z}\boldsymbol{z}_{\B\to\A}^{t+1} 
+ (1 - \theta_{\z})\boldsymbol{z}_{\B\to\A}^{t}$ 
\State $v_{\B\to\A}^{\z,t+1} := \theta_{\z}v_{\B\to\A}^{\z,t+1} 
+ (1 - \theta_{\z})v_{\B\to\A}^{\z,t}$

\State $\boldsymbol{x}_{\B\to\A}^{t+1}:= \theta_{\x}\boldsymbol{x}_{\B\to\A}^{t+1} 
+ (1 - \theta_{\x})\boldsymbol{x}_{\B\to\A}^{t}$
\State $v_{\B\to\A}^{\x,t+1} := \theta_{\x}v_{\B\to\A}^{\x,t+1} 
+ (1 - \theta_{\x})v_{\B\to\A}^{\x,t}$
\EndIf
\EndFor

\State \Return $\boldsymbol{x}_{\B}^{T+1}$
\end{algorithmic}
\end{algorithm}

We next evaluate $\phi(x)$ for any distribution of $U$. Applying the upper 
bound $U^{2}1(U^{2}<x)\leq x$ for all $x\geq0$ to $\phi$ in (\ref{phi}), 
we obtain $\phi(x)\leq x$. Thus, the curve $y=\phi(x)$ is below the 
straight line $y=x$. From $\phi(x)=0$ for all $x\leq\psi(u_{\mathrm{min}}^{2})
< u_{\mathrm{min}}^{2}$, the point $(\psi(0), 0)$ is the 
unique fixed point of the dynamical system for $\delta>\delta_{*}$. 
This implies that $y_{t}$ converges to zero as $t\to\infty$ if 
$\delta>\delta_{*}$ holds.  

\begin{algorithm}[t]
\caption{State Evolution for Bayesian GOAMP}
\label{alg2}
\begin{algorithmic}[1]
\State $\bar{v}_{\A \to \B}^{\z, -1} = P$, $\bar{v}_{\B \to \A}^{\x, 0} = P$
\Comment{Initialization}

\State $\bar{\xi}_{\B,-1}^{\z} 
= \mathbb{E}[\partial_{0} f_{\B}^{\z}(0, Y; \bar{v}_{\A \to \B}^{\z,-1})]$

\State $\bar{v}_{\B \to \A}^{\z, 0} = \frac{\bar{\xi}_{\B,-1}^{\z} 
\bar{v}_{\A \to \B}^{\z, -1}}{1 - \bar{\xi}_{\B,-1}^{\z}}$

\ForAll{$t \in \{0, \ldots, T\}$}

\Comment{Outer sub-module in module~A}
\State $\bar{\xi}_{\A,t}^{\z} = \mathbb{E}\left[ 
 \frac{\bar{v}_{\B \to \A}^{\x, t} \Lambda}
 {\bar{v}_{\B \to \A}^{\z, t} + \bar{v}_{\B \to \A}^{\x, t} \Lambda} 
\right]$
\label{GOAMP_line5}

\State $\bar{v}_{\A \to \B}^{\z, t} 
= \frac{\bar{\xi}_{\A,t}^{\z} \bar{v}_{\B \to \A}^{\z, t}}{1 - \bar{\xi}_{\A,t}^{\z}}$

\State $\bar{\xi}_{\A,t}^{\x} = 1 - \bar{\xi}_{\A,t}^{\z}$
\Comment{Inner sub-module in module~A}

\State $\bar{v}_{\A \to \B}^{\x, t} = 
\frac{\bar{v}_{\B \to \A}^{\x, t}}{1 - \bar{\xi}_{\A,t}^{\x}}$

\Comment{Outer sub-module in module~B}

\State $\bar{\xi}_{\B,t}^{\z} = \mathbb{E} \left[ 
 \partial_{0} f_{\B}^{\z}(Z + H_{\z,t}, g(Z, W); \bar{v}_{\A \to \B}^{\z, t}) 
\right]$
\label{line9_GOAMP}

\State $\bar{v}_{\B \to \A}^{\z, t+1} 
= \frac{\bar{\xi}_{\B,t}^\z \bar{v}_{\A \to \B}^{\z, t}}{1 - \bar{\xi}_{\B,t}^\z}$

\Comment{Inner sub-module in module~B}

\State $\bar{v}_{\B \to \A}^{\x, t+1} = \mathbb{E} \left[ 
 U^{2}1\left( 
  U^{2} < \frac{2 \bar{v}_{\A \to \B}^{\x, t}}{\delta} 
 \right) 
\right]$
\label{GOAMP_line11}

\EndFor
\State \Return $\bar{v}_{\B \to \mathrm{A}}^{\x, T+1}$
\end{algorithmic}
\end{algorithm}

\section{Pseudocodes} \label{appen_pseudocodes}
Algorithm~\ref{alg1} shows a pseudocode for Bayesian GOAMP with damping. 
In line~\ref{line10}, the update rule~(\ref{moduleA_z_mean}) for 
$v_{\A\to\B}^{\z, t}$ is replaced with the 
modification~(\ref{moduleA_z_var_modify}). 
In lines~\ref{line16} and \ref{line14}, $f_{\B}^{\mathrm{z}}$ and 
$f_{\B}^{\mathrm{x}}$ represent the Bayes-optimal outer 
denoiser~(\ref{outer_denoiser_Bayes}) and the Bayesian inner denoiser 
$f_{\B}^{\mathrm{x}}(\boldsymbol{x}_{t}; \bar{v}_{\A \to \B}^{\mathrm{x}, t})
=\mathbb{E}[\boldsymbol{x} | \boldsymbol{x}_{t}]$ with 
$\boldsymbol{x}_{t}$ in (\ref{virtual_AWGN}), respectively. 
In this case, the update rule~(\ref{moduleB_x_var}) for 
$v_{\B\to\A}^{\x,t+1}$ is simplified to (\ref{moduleB_x_var_modify}). 
Furthermore, we use $\eta_{\B,t}=0$ to simplify the update 
rule~(\ref{moduleB_z_mean}) for $\boldsymbol{z}_{\B\to\A}^{t+1}$. 
For simplicity, the update rule~(\ref{moduleB_z_var_Bayes}) 
for $v_{\B\to\A}^{\z, t+1}$ is used instead of (\ref{moduleB_z_mean}). 

The SVD of $\boldsymbol{A}$ should be pre-computed to reduce the 
per-iteration complexity due to the matrix inversion in lines~\ref{line8} 
and \ref{line11}. As a result, the vector 
$\widehat{\Delta \boldsymbol{z}^{t}}$ in line~\ref{line8} can be computed 
efficiently via direct computation, rather than the matrix-vector 
multiplication $\widehat{\Delta \boldsymbol{z}^{t}}
=\boldsymbol{A}\widehat{\Delta\boldsymbol{x}^{t}}$. The per-iteration 
complexity of Bayesian GOAMP is dominated by the two matrix-vector 
multiplication operations in lines~\ref{line4} and \ref{line11}. Note that 
the overall complexity of Bayesian GOAMP is dominated by the pre-computation 
of the SVD.   

\begin{algorithm}[t]
\caption{State Evolution for Bayesian OAMP}
\label{alg3}
\begin{algorithmic}[1]
\State $\bar{v}_{\B \to \A}^{\x, 0} = P$
\Comment{Initialization}

\ForAll{$t \in \{0, \ldots, T\}$}
\State $\bar{v}_{\A\to\B}^{\x,t} = \left\{
 \mathbb{E}\left[
  \frac{\Lambda}{\sigma^{2} + \bar{v}_{\B\to\A}^{\x,t}\Lambda}
 \right]
\right\}^{-1}$
\label{OAMP_line3}
\Comment{Module A}

\State $\bar{v}_{\B \to \A}^{\x, t+1} = \mathbb{E} \left[ 
 U^{2}1\left( 
  U^{2} < \frac{2 \bar{v}_{\A \to \B}^{\x, t}}{\delta} 
 \right) 
\right]$
\label{OAMP_line4}
\Comment{Module B}
\EndFor
\State \Return $\bar{v}_{\B \to \mathrm{A}}^{\x, T+1}$
\end{algorithmic}
\end{algorithm}

Algorithms~\ref{alg2} and \ref{alg3} show pseudocodes for the state evolution 
recursion of Bayesian GOAMP and Bayesian OAMP, respectively. 
The random variable $\Lambda$ represents the asymptotic eigenvalue 
distribution of $\boldsymbol{\Lambda}$ in Assumption~\ref{assumption_A}
while $U$ is defined in Assumption~\ref{assumption_x}. The function  
$f_{\B}^{\z}$ in Algorithm~\ref{alg2} represents the Bayes-optimal outer 
denoiser~(\ref{outer_denoiser_Bayes}). 
See (\ref{virtual_outer_measurement_model}) and Assumption~\ref{assumption_w} 
for the definitions of random variables related to $f_{\B}^{\z}$. 
The expectation in line~\ref{line9_GOAMP} for GOAMP can be computed via 
numerical integration or numerical sampling, in which we can use 
the representation 
$Z\sim Z_{t} + N_{\z,t}$ with $N_{\z,t}\sim\mathcal{N}(0, \bar{v}_{\A\to\B}^{\z,t})$ 
independent of $Z_{t}=Z+H_{\z,t}\sim\mathcal{N}(0,P - \bar{v}_{\A\to\B}^{\z,t})$ 
and $W$. 

In practical implements, different definitions may be used to compute 
the other expectation in Algorithms~\ref{alg2} and \ref{alg3}.  
When $\boldsymbol{\Lambda}=N^{-1}\boldsymbol{\Sigma}
\boldsymbol{\Sigma}^{\mathrm{T}}$ in Assumption~\ref{assumption_A} is available, 
line~\ref{GOAMP_line5} for GOAMP and line~\ref{OAMP_line3} for OAMP 
can be replaced with 
\begin{equation}
\bar{\xi}_{\A,t}^{\z} 
= \frac{1}{M}\sum_{m=1}^{M}\frac{\bar{v}_{\B \to \A}^{\x, t} \lambda_{m}}
{\bar{v}_{\B \to \A}^{\z, t} + \bar{v}_{\B \to \A}^{\x, t} \lambda_{m}}, 
\end{equation}
\begin{equation}
\bar{v}_{\A\to\B}^{\x,t} = \left\{
 \frac{1}{M}\sum_{m=1}^{M}\frac{\lambda_{m}}
 {\sigma^{2} + \bar{v}_{\B\to\A}^{\x,t}\lambda_{m}}
\right\}^{-1}, 
\end{equation}
respectively. To improve the accuracy of $\bar{v}_{\B\to\A}^{\x,t+1}$ in 
lines~\ref{GOAMP_line11} and \ref{OAMP_line4} for GOAMP and OAMP, 
the following may be used: 
\begin{equation} \label{moduleB_x_bar_var_finite}
\bar{v}_{\B\to\A}^{\x,t+1} 
= \mathbb{E}\left[
 \|f_{\B}^{\x}(\boldsymbol{x} + \boldsymbol{\omega}_{t}; \bar{v}_{\A\to\B}^{\x,t}) 
 - \boldsymbol{x}\|_{2}^{2}
\right].
\end{equation}
In (\ref{moduleB_x_bar_var_finite}), $f_{\B}^{\x}$ denotes the Bayesian inner denoiser proposed in \cite{Takeuchi251} while $\boldsymbol{\omega}_{t}$ is defined in the virtual Gaussian measurement~(\ref{virtual_AWGN}). Since finite-length analysis of (\ref{moduleB_x_bar_var_finite}) is not available, $\{\boldsymbol{x}, \boldsymbol{\omega}_{t}\}$ should be sampled to evaluate the expectation.

\section{Proof of Theorem~\ref{theorem_SE_tech}}
\label{proof_theorem_SE_tech} 
\subsection{Initialization Step} 
\label{appen_initialization_step}
\begin{IEEEproof}[Proof of (\ref{I31}) for Lipschitz-continuous $\phi$ and 
$\psi$]
We first assume that $\phi, \psi\in\mathcal{L}_{2}$ 
are everywhere Lipschitz-continuous 
with respect to the first variable. Using the initial condition 
$\boldsymbol{q}_{\x,0}=-\boldsymbol{x}$ and (\ref{Qx_0}), 
we confirm $\|\boldsymbol{q}_{\x,0}\|_{2}^{2}\pto Q_{\x,0,0}$. 
For Lipschitz-continuous
$\phi: \mathbb{R}^{2}\to\mathbb{R}$ and $\psi: \mathbb{R}^{2}\to\mathbb{R}$ 
with respect to the first variable, we use 
$\tilde{\boldsymbol{m}}=\boldsymbol{\Sigma}\boldsymbol{b}_{\x,0}$ and 
Lemma~\ref{lemma_representation} for 
$\boldsymbol{b}_{\x,0}=\boldsymbol{V}^{\mathrm{T}}
\boldsymbol{q}_{\x,0}$ in (\ref{b}) under Assumption~\ref{assumption_A} to have 
\begin{equation}
\frac{\phi^{\mathrm{T}}(\tilde{\boldsymbol{m}}, \boldsymbol{\lambda})
\psi(\tilde{\boldsymbol{m}}, \boldsymbol{\lambda})}{M}
\sim \boldsymbol{\phi}^{\mathrm{T}}\left(
 \frac{\|\boldsymbol{q}_{\x,0}\|_{2}\boldsymbol{\omega}_{\x}}
 {\|\boldsymbol{\omega}_{\x}\|_{2}}
\right)\boldsymbol{\psi}\left(
 \frac{\|\boldsymbol{q}_{\x,0}\|_{2}\boldsymbol{\omega}_{\x}}
 {\|\boldsymbol{\omega}_{\x}\|_{2}}
\right), 
\end{equation}
where $\boldsymbol{\omega}_{\x}\sim\mathcal{N}(\boldsymbol{0}, 
\boldsymbol{I}_{N})$ is independent of $\boldsymbol{q}_{\x,0}$ and 
$\boldsymbol{\Sigma}$, with $\phi_{n}=\psi_{n}=0$ for all $n>M$ and otherwise 
\begin{equation}
\phi_{n}(\boldsymbol{\omega}) 
= \frac{\phi(\sigma_{n}\omega_{n}, \lambda_{n})}{\sqrt{M}}, \quad
\psi_{n}(\boldsymbol{\omega}) 
= \frac{\psi(\sigma_{n}\omega_{n}, \lambda_{n})}{\sqrt{M}}.
\end{equation}

To utilize Lemma~\ref{lemma_elimination} with $f(N)=N^{-1/2}$, we investigate 
properties of the separable functions $\boldsymbol{\phi}$ and 
$\boldsymbol{\psi}$. 
For all $m>M$, $\phi_{m}$ and $\psi_{m}$ have the Lipschitz constant $L_{m}=0$. 
For $m\leq M$, we use the Lipschitz continuity of 
$\phi, \psi\in\mathcal{L}_{2}$ in Definition~\ref{definition_integrability} 
with respect to the first variable to obtain 
\begin{align}
\sqrt{M}|\phi_{m}(x) - \phi_{m}(y)|
&= \left|
 \phi\left(
  \sigma_{m}x, \lambda_{m}
 \right)
 -  \phi\left(
  \sigma_{m}y, \lambda_{m}
 \right)
\right|
\nonumber \\
&\leq L_{\phi}(\lambda_{m})\sigma_{m}|x - y|,
\end{align}
\begin{equation}
\sqrt{M}|\psi_{m}(x) - \psi_{m}(y)|
\leq L_{\psi}(\lambda_{m})\sigma_{m}|x - y|,
\end{equation}
where $L_{a}(\lambda_{m})$ is some almost everywhere continuous function of 
$\lambda_{m}$ with uniformly integrable 
$M^{-1}\sum_{m=1}^{M}\lambda_{m}L_{a}^{2}(\lambda_{m})$ for $a\in\{\phi, \psi\}$. 
Let $L(\lambda)=\max\{L_{\phi}(\lambda), L_{\psi}(\lambda)\}$, which satisfies 
the same properties as $L_{\phi}$ and $L_{\psi}$.  
Thus, $\phi_{m}$ and $\psi_{m}$ are Lipschitz-continuous 
with the common Lipschitz constant $L_{m}=M^{-1/2}\sigma_{m}L(\lambda_{m})$ for 
$m\leq M$. 

We confirm all conditions in Lemma~\ref{lemma_elimination} with 
$f(N)=N^{-1/2}$. From $\tilde{a}_{N}=\|\boldsymbol{q}_{\x,0}\|_{2}^{2}\pto 
Q_{\x,0,0}$ we use the definition $N^{-1}\sigma_{m}^{2}=\lambda_{m}$ to evaluate 
\begin{align}
\left\|
 \boldsymbol{\phi}\left(
  \frac{Q_{\x,0,0}^{1/2}}{\sqrt{N}}\boldsymbol{\omega}_{\x}
 \right)
\right\|_{2}^{2}
&= \frac{1}{M}\sum_{m=1}^{M}\phi^{2}\left(
 \sqrt{\lambda_{m}}Q_{\x,0,0}^{1/2}\omega_{\x,m}, \lambda_{m}
\right)
\nonumber \\
&\pto \mathbb{E}\left[
 \phi^{2}(\sqrt{\Lambda}B_{\x,0}, \Lambda)
\right]
\end{align}
with $B_{\x,0}\sim\mathcal{N}(0, Q_{\x,0,0})$, where the last convergence 
follows from the weak law of large numbers for $\boldsymbol{\omega}_{\x}$,  
Assumption~\ref{assumption_A}, and the uniform integrability of 
$M^{-1}\sum_{m=1}^{M}\lambda_{m}L^{2}(\lambda_{m})$. 
Similarly, we find the boundedness 
in probability of $\|\boldsymbol{\psi}(N^{-1/2}Q_{\x,0,0}^{1/2}
\boldsymbol{\omega}_{\x})\|_{2}^{2}$. Furthermore, 
we use the definition $N^{-1}\sigma_{m}^{2}=\lambda_{m}$, 
Assumption~\ref{assumption_A}, and the uniform integrability to find 
\begin{equation} 
f^{2}(N)\sum_{n=1}^{N}L_{n}^{2} 
= \frac{1}{M}\sum_{m=1}^{M}\lambda_{m}L^{2}(\lambda_{m})
\pto \mathbb{E}[\Lambda L^{2}(\Lambda)]   
\end{equation}
for $f(N)=N^{-1/2}$. 

We have confirmed all conditions in 
Lemma~\ref{lemma_elimination} with 
$f(N)=N^{-1/2}$.
Thus, we can use Lemma~\ref{lemma_elimination} to obtain 
\begin{align}
&\boldsymbol{\phi}^{\mathrm{T}}\left(
 \frac{\|\boldsymbol{q}_{\x,0}\|_{2}\boldsymbol{\omega}_{\x}}
 {\|\boldsymbol{\omega}_{\x}\|_{2}}
\right)\boldsymbol{\psi}\left(
 \frac{\|\boldsymbol{q}_{\x,0}\|_{2}\boldsymbol{\omega}_{\x}}
 {\|\boldsymbol{\omega}_{\x}\|_{2}}
\right)
\nonumber \\
&\peq \frac{1}{M}\sum_{m=1}^{M}\phi\left(
 \sqrt{\lambda_{m}}Q_{\x,0,0}^{1/2}\omega_{\x,m}, \lambda_{m}
\right)
\nonumber \\
&\cdot\psi\left(
 \sqrt{\lambda_{m}}Q_{\x,0,0}^{1/2}\omega_{\x,m}, \lambda_{m}
\right) + o(1)
\nonumber \\
&\pto \mathbb{E}\left[
 \phi(\sqrt{\Lambda}B_{\x,0}, \Lambda)
 \psi(\sqrt{\Lambda}B_{\x,0}, \Lambda)
\right],
\end{align}
where the last convergence follows from the weak law of large numbers 
for $\boldsymbol{\omega}_{\x}$, Assumption~\ref{assumption_A}, and  
the uniform integrability of 
$M^{-1}\sum_{m=1}^{M}\lambda_{m}L^{2}(\lambda_{m})$. 
Thus, (\ref{I31}) holds for everywhere Lipschitz-continuous 
$\phi, \psi\in\mathcal{L}_{2}$ with respect to the first variable. 
\end{IEEEproof}

\begin{IEEEproof}[Proof of (\ref{I31}) for piecewise Lipschitz-continuous 
$\phi$ and~$\psi$]
We next prove (\ref{I31}) for piecewise Lipschitz-continuous functions 
$\phi, \psi\in\mathcal{L}_{2}$ with respect to the first variable. 
We have proved that (\ref{I31}) is correct for all Lipschitz-continuous 
functions $\phi, \psi\in\mathcal{L}_{2}$ with respect to the first variable. 
This implies the convergence in probability of the empirical distribution 
$\rho_{M}$ in the sense of the weak convergence 
for Lemma~\ref{lemma_generalization} 
with $\boldsymbol{X}_{1,M} = \tilde{\boldsymbol{m}}$ and 
$\boldsymbol{X}_{2,M} = \boldsymbol{\lambda}$. 

We confirm all conditions in Lemma~\ref{lemma_generalization}. 
Since the functions $\phi$ and $\psi$ are 
piecewise Lipschitz-continuous with respect to the first variable, 
$f(\tilde{m}, \lambda_{m})=\phi(\tilde{m}, \lambda_{m})
\psi(\tilde{m}, \lambda_m)$ 
is an almost everywhere continuous function of $\tilde{m}$ for fixed 
$\lambda_{m}$. Thus, the first condition in Lemma~\ref{lemma_generalization} 
is satisfied. For the second condition, we use (\ref{I31}) for the 
Lipschitz-continuous functions $\phi(\tilde{m}, \lambda_{m}) 
= \psi(\tilde{m}, \lambda_{m}) = \tilde{m}$ to obtain 
\begin{equation} \label{z_tilde2}
\frac{1}{M}\|\tilde{\boldsymbol{m}}\|_{2}^{2} 
\pto \mathbb{E}[\Lambda B_{\x,0}^{2}] 
= \mathbb{E}[\tilde{M}^{2}], \quad 
\tilde{M}=\sqrt{\Lambda}B_{\x,0}. 
\end{equation}
The third condition in Lemma~\ref{lemma_generalization}  
$\langle \mathbb{E}[f(\sqrt{\Lambda}B_{\x,0}, 
\boldsymbol{\lambda})| \tilde{\boldsymbol{m}}, \boldsymbol{\lambda}]\rangle
= M^{-1}\sum_{m=1}^{M}\mathbb{E}[f(\sqrt{\Lambda}B_{\x,0}, 
\lambda_{m}) | \boldsymbol{\lambda}]
\pto \mathbb{E}[
f(\sqrt{\Lambda}B_{\x,0}, \Lambda)]$ 
follows from $\phi, \psi\in\mathcal{L}_{2}$ under 
Assumption~\ref{assumption_A}. 
Finally, from the piecewise 
Lipschitz-continuity of $\phi$ and $\psi$ with respect to the first variable 
we find that there is some almost everywhere continuous function 
$L(\lambda_{m})>0$ of $\lambda_{m}$ with the uniform integrability of 
$M^{-1}\sum_{m=1}^{M}\lambda_{m}L^{2}(\lambda_{m})$  
such that $|\phi(\tilde{m}, \lambda_{m})|\leq L(\lambda_{m})(1 + |\tilde{m}|)$ 
and $|\psi(\tilde{m}, \lambda_{m})|\leq L(\lambda_{m})(1 + |\tilde{m}|)$ hold 
for all $\tilde{m}$ and $\lambda_{m}$. These upper bounds result in 
$|f(\tilde{m},\lambda_{m})|\leq L^{2}(\lambda_{m})(1 + |\tilde{m}|)^{2}
\leq 2L^{2}(\lambda_{m})(1 + \tilde{m}^{2})$. The last condition in 
Lemma~\ref{lemma_generalization} follows from 
Assumption~\ref{assumption_A} and the properties of $L(\lambda)$. 
Thus, we can use Lemma~\ref{lemma_generalization} to prove (\ref{I31}) 
for all piecewise Lipschitz-continuous functions 
$\phi, \psi\in\mathcal{L}_{2}$ with respect to the first variable. 
\end{IEEEproof}

\begin{IEEEproof}[Proof of (\ref{I32})]
We first prove (\ref{I32}) for everywhere Lipschitz-continuous  
$\phi: \mathbb{R}^{2}\to\mathbb{R}$ and $\psi: \mathbb{R}^{2}\to\mathbb{R}$. 
Using Lemma~\ref{lemma_representation} for $\tilde{\boldsymbol{h}}
=\boldsymbol{U}\tilde{\boldsymbol{m}}$ under 
Assumption~\ref{assumption_A} yields  
\begin{equation}
\frac{\phi^{\mathrm{T}}(\tilde{\boldsymbol{h}}, \boldsymbol{w})
\psi(\tilde{\boldsymbol{h}}, \boldsymbol{w})}{M}
\sim \frac{1}{M}\phi^{\mathrm{T}}
\left(
 \frac{a_{M}\boldsymbol{\omega}_{\z}}{\|\boldsymbol{\omega}_{\z}\|_{2}}, 
 \boldsymbol{w} 
\right)\psi
\left(
 \frac{a_{M}\boldsymbol{\omega}_{\z}}{\|\boldsymbol{\omega}_{\z}\|_{2}}, 
 \boldsymbol{w} 
\right), 
\end{equation}
with $a_{M}=\|\tilde{\boldsymbol{m}}\|_{2}$, where 
$\boldsymbol{\omega}_{\z}\sim\mathcal{N}(\boldsymbol{0}, \boldsymbol{I}_{M})$ 
is independent of $\tilde{\boldsymbol{m}}$ and $\boldsymbol{w}$. 

To utilize Lemma~\ref{lemma_elimination} with $f(M)=1$, 
we use (\ref{z_tilde2}) to obtain 
\begin{equation} \label{Sbx_0}
\frac{1}{M}a_{M}^{2} \pto \mathbb{E}[\tilde{M}^{2}] = P. 
\end{equation} 
For fixed $w_{m}$, let $\phi_{m}(\tilde{h})=M^{-1/2}\phi(\tilde{h}, w_{m})$ and 
$\psi_{m}(\tilde{h})=M^{-1/2}\psi(\tilde{h}, w_{m})$. 
Since $\phi$ and $\psi$ 
are Lipschitz-continuous, $\phi_{m}(\tilde{h})$ and $\psi_{m}(\tilde{h})$ 
have a common Lipschitz constant $L_{m}=L/\sqrt{M}$ for some $L>0$. 
This implies that the boundedness $\sum_{m=1}^{M}L_{m}^{2}=L<\infty$ holds. 
Furthermore, we use the weak law of large numbers for 
$\boldsymbol{\omega}_{\z}$ and Assumption~\ref{assumption_w} to have 
$\|\boldsymbol{\phi}(\sqrt{P}\boldsymbol{\omega}_{\z})\|_{2}^{2}
=M^{-1}\|\phi(\sqrt{P}\boldsymbol{\omega}_{\z}, 
\boldsymbol{w})\|_{2}^{2}\pto \mathbb{E}[\phi^{2}(\tilde{H}, W)]$ 
with $\tilde{H}\sim\mathcal{N}(0, P)$. Similarly, we find the boundedness in 
probability of $\|\boldsymbol{\psi}(\sqrt{P}\boldsymbol{\omega}_{\z})\|_{2}^{2}$. 
Thus, we can use Lemma~\ref{lemma_elimination} to obtain 
\begin{align}
&\frac{1}{M}\phi^{\mathrm{T}}
\left(
 \frac{a_{M}\boldsymbol{\omega}_{\z}}{\|\boldsymbol{\omega}_{\z}\|_{2}}, 
 \boldsymbol{w} 
\right)\psi
\left(
 \frac{a_{M}\boldsymbol{\omega}_{\z}}{\|\boldsymbol{\omega}_{\z}\|_{2}}, 
 \boldsymbol{w} 
\right)
\nonumber \\
&\peq \frac{1}{M}\phi^{\mathrm{T}}(\sqrt{P}\boldsymbol{\omega}_{\z}, 
\boldsymbol{w})
\psi(\sqrt{P}\boldsymbol{\omega}_{\z}, \boldsymbol{w})  + o(1)
\nonumber \\
&\pto \mathbb{E}[\phi(\tilde{H}, W)\psi(\tilde{H}, W)], 
\end{align}
where the last convergence follows from the weak law of large numbers for 
$\boldsymbol{\omega}_{\z}$ and Assumption~\ref{assumption_w}.  

We next consider piecewise Lipschitz-continuous $\phi$ and $\psi$. 
We have already proved the convergence in probability of 
the empirical distribution $\rho_{M}$ in Lemma~\ref{lemma_generalization} for 
$\boldsymbol{X}_{1,M}=[\tilde{\boldsymbol{h}}, \boldsymbol{w}]$, 
$\boldsymbol{X}_{2,M}=\boldsymbol{0}$, 
$f(\tilde{h}, w, 0) = \phi(\tilde{h}, w)\psi(\tilde{h}, w)$, 
$t=3$, and $t_{0}=2$. By definition, we have 
\begin{equation}
|f(\tilde{h}, w, 0)| \leq \tilde{L}^{2}\left(
 1 + \sqrt{\tilde{h}^{2} + w^{2}}
\right)^{2}\leq 2\tilde{L}^{2}\left(
 1 + \tilde{h}^{2} + w^{2}
\right)
\end{equation}
for some $\tilde{L}>0$. Since $\phi$ and $\psi$ are almost 
everywhere continuous, the first condition in 
Lemma~\ref{lemma_generalization} is satisfied. The second condition 
$M^{-1}\|\tilde{\boldsymbol{h}}\|_{2}^{2} + M^{-1}\|\boldsymbol{w}\|_{2}^{2}
\pto \mathbb{E}[\tilde{H}^{2} + W^{2}]$ follows from (\ref{I32}) for everywhere 
Lipschitz-continuous $\phi$ and $\psi$. The other two conditions are trivial 
for $\boldsymbol{X}_{2,M}=\boldsymbol{0}$. Thus, we can use 
Lemma~\ref{lemma_generalization} to arrive at (\ref{I32}) for 
piecewise Lipschitz-continuous $\phi$ and $\psi$. 
\end{IEEEproof}

\begin{IEEEproof}[Proof of (\ref{I33})]
Let $\boldsymbol{X}_{1,M}=[\tilde{\boldsymbol{h}}, \boldsymbol{w}]$,  
$\boldsymbol{X}_{2,M}=\boldsymbol{0}$, $t=3$, and $t_{0}=2$ 
in Lemma~\ref{lemma_generalization}. 
The convergence~(\ref{I32}) implies the convergence in probability 
of the empirical distribution $\rho_{M}$ in Lemma~\ref{lemma_generalization} 
in the sense of the weak convergence. Furthermore, it indicates the second 
and third conditions in Lemma~\ref{lemma_generalization}. 
Since any piecewise Lipschitz-continuous function 
is almost everywhere differentiable, 
Assumption~\ref{assumption_denoisers_general} 
implies that $\partial_{0}\psi_{\z,-1}$ is almost everywhere continuous. Thus, 
the first condition in Lemma~\ref{lemma_generalization} is satisfied. 
Furthermore, we use the piecewise Lipschitz-continuity of $\psi_{\z,-1}$ to 
find that the boundedness $|\partial_{0}\psi_{\z,-1}|\leq L$ for some $L>0$, 
so that the last condition in Lemma~\ref{lemma_generalization} is trivial. 
Thus, we can use 
Lemma~\ref{lemma_generalization} to arrive at
$\eta_{\B,-1}\pto\bar{\eta}_{\B,-1}$. 
\end{IEEEproof}

\begin{IEEEproof}[Proof of \ref{I2}]
We use (\ref{I32}) to obtain 
\begin{equation} \label{zz}
\frac{1}{M}\|\tilde{\boldsymbol{h}}\|_{2}^{2}
\pto \mathbb{E}[\tilde{H}^{2}] = P.  
\end{equation}
Thus, Property~\ref{I2} holds. 
\end{IEEEproof}

\begin{IEEEproof}[Proof of \ref{I4}]
Since Assumption~\ref{assumption_denoisers_general} implies the piecewise 
Lipschitz-continuity of 
$\boldsymbol{q}_{\z,0}=\psi_{\z,-1}(\tilde{\boldsymbol{h}}, \boldsymbol{w}) 
- \eta_{\B,-1}\tilde{\boldsymbol{h}}$ in (\ref{initial_condition_general}), 
we can use (\ref{I32}) and (\ref{I33}) 
for $\tilde{\boldsymbol{h}}^{\mathrm{T}}\boldsymbol{q}_{\z,0}$ to obtain  
\begin{equation}
\frac{1}{M}\tilde{\boldsymbol{h}}^{\mathrm{T}}\boldsymbol{q}_{\z,0} 
\pto \mathbb{E}[\tilde{H}\psi_{\z,-1}(\tilde{H}, W)] 
- \bar{\eta}_{\B,-1}\mathbb{E}[\tilde{H}^{2}]= 0,
\end{equation}
where the last equality follows from Lemma~\ref{lemma_Stein} and 
the definition of $\bar{\eta}_{\B,-1}$ in (\ref{eta_B_0}). 
Thus, Property~\ref{I4} holds.   
\end{IEEEproof}

\begin{IEEEproof}[Proof of \ref{I5}]
We have already proved the first and last convergence. 
For the second convergence, using (\ref{I32}) and (\ref{I33}) 
for the initial condition $\boldsymbol{q}_{\z,0}
=\psi_{\z,-1}(\tilde{\boldsymbol{h}},\boldsymbol{w}) 
- \eta_{\B,-1}\tilde{\boldsymbol{h}}$ in (\ref{initial_condition_general}), 
we have 
\begin{equation} \label{qz_00}
\frac{\|\boldsymbol{q}_{\z,0}\|_{2}^{2}}{M}
\pto \mathbb{E}[\{\psi_{\z,-1}(\tilde{H}, W) 
- \bar{\eta}_{\B,-1}\tilde{H}\}^{2}]
= \mathbb{E}[Q_{\z,0}^{2}]. 
\end{equation}
Thus, Property~\ref{I5} holds. 
\end{IEEEproof}

\subsection{Module~A for $\tau=0$}
\label{appen_module_A_0}
\begin{IEEEproof}[Proof of \ref{A1o}]
From Assumption~\ref{assumption_A} and Property~\ref{I2} 
we can use Lemma~\ref{lemma_conditioning} for the constraint 
$\tilde{\boldsymbol{h}}=\boldsymbol{U}\tilde{\boldsymbol{m}}$ yields 
\begin{equation}
\boldsymbol{U}\sim 
\|\tilde{\boldsymbol{h}}\|_{2}^{-2}
\tilde{\boldsymbol{h}}\tilde{\boldsymbol{m}}^{\mathrm{T}} 
+ \boldsymbol{\Phi}_{\tilde{\boldsymbol{h}}}^{\perp}
\tilde{\boldsymbol{U}}
(\boldsymbol{\Phi}_{\tilde{\boldsymbol{m}}}^{\perp})^{\mathrm{T}} 
\end{equation}
conditioned on $\Theta$ and $\mathfrak{E}_{0,0}^{\z}$, where 
$\tilde{\boldsymbol{U}}$ is a Haar-distributed orthogonal matrix independent 
of $\{\Theta, \mathfrak{E}_{0,0}^{\z}\}$. 
Substituting this representation into the definition of 
$\boldsymbol{b}_{\z,0}$ in (\ref{b}), we arrive at  
\begin{align}
\boldsymbol{b}_{\z,0} 
&\sim \frac{\tilde{\boldsymbol{h}}^{\mathrm{T}}\boldsymbol{q}_{\z,0}}
{ \|\tilde{\boldsymbol{h}}\|_{2}^{2}}\tilde{\boldsymbol{m}}
+ \boldsymbol{\Phi}_{\tilde{\boldsymbol{m}}}^{\perp}
\tilde{\boldsymbol{U}}^{\mathrm{T}}
(\boldsymbol{\Phi}_{\tilde{\boldsymbol{h}}}^{\perp})^{\mathrm{T}} 
\boldsymbol{q}_{\z,0}
\nonumber \\
&\peq o(1)\tilde{\boldsymbol{m}}
+ \boldsymbol{\Phi}_{\tilde{\boldsymbol{m}}}^{\perp}
\tilde{\boldsymbol{U}}^{\mathrm{T}}
\boldsymbol{q}_{\z,0}^{\perp}, 
\end{align}
with $\boldsymbol{q}_{\z,0}^{\perp}
=(\boldsymbol{\Phi}_{\tilde{\boldsymbol{h}}}^{\perp})^{\mathrm{T}}\boldsymbol{q}_{\z,0}$, 
where the last equality follows from Properties~\ref{I2} and \ref{I4}. 

To utilize Lemma~\ref{lemma_modified_representation}, we confirm 
\begin{equation}
\frac{\|\boldsymbol{q}_{\z,0}^{\perp}\|_{2}^{2}}{M} 
= \frac{\boldsymbol{q}_{\z,0}^{\mathrm{T}}\boldsymbol{P}_{\tilde{\boldsymbol{h}}}^{\perp}
\boldsymbol{q}_{\z,0}}{M}
= \frac{\|\boldsymbol{q}_{\z,0}\|_{2}^{2}}{M} 
- \frac{(\tilde{\boldsymbol{h}}^{\mathrm{T}}\boldsymbol{q}_{\z,0})^{2}}
{M\|\tilde{\boldsymbol{h}}\|_{2}^{2}}
\pto \mathbb{E}[Q_{\z,0}^{2}], \label{qz_perp_00}
\end{equation}
where the last convergence follows from Properties~\ref{I2}, \ref{I4} and 
\ref{I5}. Combining this result with Property~\ref{I5}, 
we can use Lemma~\ref{lemma_modified_representation} to arrive at
\begin{equation}
\boldsymbol{b}_{\z,0} 
\sim o(1)\tilde{\boldsymbol{m}}
+ \{1 + o(1)\}\frac{\|\boldsymbol{q}_{\z,0}\|_{2}}
{\|\boldsymbol{\omega}_{\z,0}\|_{2}}\boldsymbol{\omega}_{\z,0},
\end{equation}
with $\boldsymbol{\omega}_{\z,0}
\sim\mathcal{N}(\boldsymbol{0},\boldsymbol{I}_{M})$ independent of 
$\{\Theta, \mathfrak{E}_{0,0}^{\z}\}$. 
Thus, Property~\ref{A1o} holds for $\tau=0$. 
\end{IEEEproof}

\begin{IEEEproof}[Proof of \ref{A2}]
For (\ref{bzbz}) we use the definition of $\boldsymbol{b}_{\z,0}$ in (\ref{b}) 
to obtain 
\begin{equation}
\frac{\|\boldsymbol{b}_{\z,0}\|_{2}^{2}}{M}
= \frac{\|\boldsymbol{q}_{\z,0}\|_{2}^{2} }{M}
\pto \mathbb{E}[Q_{\z,0}^{2}], 
\end{equation}
where the last convergence follows from Property~\ref{I5}. 
The convergence~(\ref{bxbx}) for $\tau=0$ follows from 
$\boldsymbol{b}_{\x,0}$ in (\ref{b}) and Property~\ref{I5}. Thus, 
Property~\ref{A2} holds for $\tau=0$. 
\end{IEEEproof}

\begin{IEEEproof}[Proof of \ref{A3}]
We first prove that the former convergence for $\tau=0$ implies the latter 
convergence for $\tau=0$. We only evaluate $\xi_{\A,0,0}^{\z}$ since 
$\xi_{\A,0,0}^{\x}$ can be evaluated in the same manner. 

Let $\boldsymbol{X}_{1,M}
=[\boldsymbol{b}_{\z,0},\boldsymbol{\Sigma}\boldsymbol{b}_{\x,0}]$ and 
$\boldsymbol{X}_{2,M}=\boldsymbol{\lambda}$ in Lemma~\ref{lemma_generalization}. 
The former convergence for $\tau=0$ implies the convergence in probability 
of the empirical distribution $\rho_{M}$ in Lemma~\ref{lemma_generalization} 
in the sense of the weak convergence. Furthermore, it indicates the second 
condition in Lemma~\ref{lemma_generalization}. 
Since any piecewise Lipschitz-continuous function 
is almost everywhere differentiable, 
Assumption~\ref{assumption_denoisers_general} 
implies that $\partial_{0}\phi_{\z,0}(\cdot, \cdot, \lambda_{m})$ is 
almost everywhere continuous for fixed $\lambda_{m}$. Thus, the first 
condition in Lemma~\ref{lemma_generalization} is satisfied. 
From Assumption~\ref{assumption_A_addition}, 
for each piece~$\mathcal{D}_{i}\subset\mathbb{R}$ in the domain of the 
piecewise Lipschitz function $\phi_{\z,0}(\cdot, \sigma_{m}b_{\x}, \lambda_{m})$, 
there is some almost everywhere continuous function $L_{i}(\lambda_{m})$ of 
$\lambda_{m}$ such that 
the following inequality holds for all $b_{\z}\in\mathcal{D}_{i}$: 
\begin{equation}
|\phi_{\z,0}(b_{\z}, \sigma_{m}b_{\x}, \lambda_{m})
- \phi_{\z,0}(b_{\z}', \sigma_{m}b_{\x}, \lambda_{m})|
\leq L_{i}(\lambda_{m})|b_{\z} - b_{\z}'|,  
\end{equation} 
which implies the upper bound 
$|\partial_{0}\phi_{\z,0}(b_{\z}, \sigma_{m}b_{\x}, \lambda_{m})|
\leq \max_{i}L_{i}(\lambda_{m})$ for all $b_{\z}\in\mathbb{R}$.
The last two conditions in Lemma~\ref{lemma_generalization} follows from 
Assumptions~\ref{assumption_A} and \ref{assumption_A_addition}. 
Thus, we can use 
Lemma~\ref{lemma_generalization} to arrive at the latter convergence 
$\xi_{\A,0,0}^{\z}\pto\bar{\xi}_{\A,0,0}^{\z}$ for $\tau=0$. 

We next prove the former convergence for $\tau=0$. We only consider 
everywhere Lipschitz-continuous functions $\phi, \psi\in\mathcal{L}_{3}$ 
in Definition~\ref{definition_integrability} with respect to 
the first two variables since the generalization to the piecewise case can 
be proved in a similar manner to that for the proof of (\ref{I31}) 
in Property~\ref{I3}. 

Suppose that two functions 
$\phi(\cdot, \cdot, \lambda_{m}): \mathbb{R}^{2}\to\mathbb{R}$ and 
$\psi(\cdot, \cdot, \lambda_{m}): \mathbb{R}^{2}\to\mathbb{R}$ are 
Lipschitz-continuous with a common Lipschitz constant $L(\lambda_{m})$, 
which is some almost everywhere continuous function of $\lambda_{m}$ with 
uniformly integrable $M^{-1}\sum_{m=1}^{M}\lambda_{m}L^{2}(\lambda_{m})$. Let 
$a_{M}=\{1 + o(1)\}\|\boldsymbol{q}_{\z,0}\|_{2}$. 
We use $\tilde{\boldsymbol{m}}=\boldsymbol{\Sigma}
\boldsymbol{b}_{\x,0}$ and Property~\ref{A1o} for $\tau=0$ to obtain   
\begin{align}
&\phi^{\mathrm{T}}(\boldsymbol{b}_{\z,0}, \tilde{\boldsymbol{m}}, 
\boldsymbol{\lambda})\psi(\boldsymbol{b}_{\z,0}, \tilde{\boldsymbol{m}}, 
\boldsymbol{\lambda})
\nonumber \\
&\sim \phi^{\mathrm{T}}\left(
 o(1)\tilde{\boldsymbol{m}}
 + \frac{a_{M}\boldsymbol{\omega}}{\|\boldsymbol{\omega}\|_{2}}, 
 \tilde{\boldsymbol{m}}, \boldsymbol{\lambda}
\right)
\psi\left(
 o(1)\tilde{\boldsymbol{m}}
 + \frac{a_{M}\boldsymbol{\omega}}{\|\boldsymbol{\omega}\|_{2}}, 
 \tilde{\boldsymbol{m}}, \boldsymbol{\lambda}
\right)
\end{align}
conditioned on $\Theta$ and $\mathfrak{E}_{0,0}^{\z}$,  
with $\boldsymbol{\omega}\sim\mathcal{N}(\boldsymbol{0}, 
\boldsymbol{I}_{M})$ independent of $\{\Theta, \mathfrak{E}_{0,0}^{\z}\}$.   

Let $\boldsymbol{a}_{2}=a_{M}\boldsymbol{\omega}/\|\boldsymbol{\omega}\|_{2}$, 
$\boldsymbol{a}_{0}=\boldsymbol{a}_{1}=\tilde{\boldsymbol{m}}$, and 
$f_{m}=\phi(o(1)a_{m,0} + a_{m,2}, a_{m,1}, \lambda_{m})
\psi(o(1)a_{m,0} + a_{m,2}, a_{m,1}, \lambda_{m})$ 
in Lemma~\ref{lemma_bias_elimination}. Since $\phi, \psi\in\mathcal{L}_{3}$ 
are Lipschitz-continuous with respect to the first two variables, 
$f_{m}$ is second-order pseudo-Lipschitz with a Lipschitz constant 
$\tilde{L}(\lambda_{m})$, which is some almost everywhere continuous function 
of $\lambda_{m}$ with uniformly integrable 
$M^{-1}\sum_{m=1}^{M}\lambda_{m}\tilde{L}(\lambda_{m})$.  
We use the weak law of large numbers 
$M^{-1}\|\boldsymbol{\omega}\|_{2}^{2}\pto1$ and Property~\ref{I5} to find 
\begin{equation}
\frac{a_{M}^{2}}{\|\boldsymbol{\omega}\|_{2}^{2}}\pto 
\mathbb{E}[Q_{\z,0}^{2}]. 
\end{equation}
Thus, from Property~\ref{I3} we find that 
all conditions in Lemma~\ref{lemma_bias_elimination} are satisfied. 
Thus, we can use Lemma~\ref{lemma_bias_elimination} to obtain 
\begin{align}
&\frac{1}{M}\phi^{\mathrm{T}}(\boldsymbol{b}_{\z,0}, \tilde{\boldsymbol{m}}, 
\boldsymbol{\lambda})\psi(\boldsymbol{b}_{\z,0}, \tilde{\boldsymbol{m}}, 
\boldsymbol{\lambda})
\nonumber \\
&\sim \frac{1}{M}\phi^{\mathrm{T}}\left(
 \frac{a_{M}\boldsymbol{\omega}}{\|\boldsymbol{\omega}\|_{2}}, 
 \tilde{\boldsymbol{m}}, \boldsymbol{\lambda}
\right)
\psi\left(
 \frac{a_{M}\boldsymbol{\omega}}{\|\boldsymbol{\omega}\|_{2}}, 
 \tilde{\boldsymbol{m}}, \boldsymbol{\lambda}
\right) + o(1). 
\end{align}

To utilize Lemma~\ref{lemma_elimination} with $f(M)=1$, we consider 
$\phi_{m}(x) = M^{-1/2}\phi(x, \tilde{m}_{m}, \lambda_{m})$ and 
$\psi_{m}(x) = M^{-1/2}\psi(x, \tilde{m}_{m}, \lambda_{m})$ for fixed 
$\tilde{m}_{m}$ and $\lambda_{m}$. 
Since $\phi, \psi\in\mathcal{L}_{3}$ are Lipschitz-continuous with respect to 
the first two variables, $\phi_{m}$ and $\psi_{m}$ are Lipschitz-continuous  
with a common Lipschitz constant $L_{m}=M^{-1/2}L(\lambda_{m})$ for some 
almost everywhere continuous function $L(\lambda_{m})$ of $\lambda_{m}$ with 
uniformly integrable $M^{-1}\sum_{m=1}^{M}\lambda_{m}L^{2}(\lambda_{m})$. 
This implies $\sum_{m=1}^{M}L_{m}^{2}=M^{-1}\sum_{m=1}^{M}L^{2}(\lambda_{m})
\pto \mathbb{E}[L^{2}(\Lambda)]$ under Assumption~\ref{assumption_A}.  
Furthermore, from $a_{M}^{2}/M\pto\mathbb{E}[Q_{\z,0}^{2}]\equiv a^{2}$ 
we use the weak law of large numbers for $\boldsymbol{\omega}$ and 
Property~\ref{I3} to have 
\begin{equation}
\left\|
 \boldsymbol{\phi}\left(
  a\boldsymbol{\omega}
 \right)
\right\|_{2}^{2}
= \frac{\|\phi(a\boldsymbol{\omega}, \tilde{\boldsymbol{m}}, 
\boldsymbol{\lambda})\|_{2}^{2}}{M}
\pto \mathbb{E}[\phi^{2}(B_{\z,0}, \sqrt{\Lambda}B_{\x,0}, \Lambda)],
\end{equation}
with $B_{\z,0}\sim\mathcal{N}(0, \mathbb{E}[Q_{\z,0}^{2}])$ independent of 
$B_{\x,0}$ and $\Lambda$. Similarly, 
we confirm the boundedness in probability of 
$\|\boldsymbol{\psi}(a\boldsymbol{\omega})\|_{2}^{2}$. 
Thus, we can use Lemma~\ref{lemma_elimination} to arrive at 
\begin{align}
&\frac{1}{M}\phi^{\mathrm{T}}(\boldsymbol{b}_{\z,0}, \tilde{\boldsymbol{m}}, 
\boldsymbol{\lambda}) 
\phi(\boldsymbol{b}_{\z,0}, \tilde{\boldsymbol{m}}, \boldsymbol{\lambda}) 
\nonumber \\
&\sim \frac{1}{M}\phi\left(
 a\boldsymbol{\omega}, 
 \tilde{\boldsymbol{m}}, \boldsymbol{\lambda}
\right)
\psi\left(
 a\boldsymbol{\omega}, 
 \tilde{\boldsymbol{m}}, \boldsymbol{\lambda}
\right)
+ o(1)
\nonumber \\
&\pto \mathbb{E}[\phi(B_{\z,0}, \sqrt{\Lambda}B_{\x,0}, \Lambda)
\psi(B_{\z,0}, \sqrt{\Lambda}B_{\x,0}, \Lambda)], 
\end{align}
where the last convergence follows from the weak law of large numbers for 
$\boldsymbol{\omega}$ and Property~\ref{I3}. 
Thus, Property~\ref{A3} holds for $\tau=0$. 
\end{IEEEproof}

\begin{IEEEproof}[Proof of \ref{A4o}]
We first confirm (\ref{zb}) for $\tau=0$, which is equivalent to 
Property~\ref{I4}:  
\begin{equation}
\frac{1}{M}(\boldsymbol{\Sigma}\boldsymbol{b}_{\x,0})^{\mathrm{T}}
\boldsymbol{b}_{\z,0}
= \frac{1}{M}\tilde{\boldsymbol{h}}^{\mathrm{T}}\boldsymbol{q}_{\z,0}
\pto 0,
\end{equation}
where the first equality follows from 
$\tilde{\boldsymbol{m}}=\boldsymbol{\Sigma}\boldsymbol{b}_{\x,0}$, 
$\tilde{\boldsymbol{m}} = \boldsymbol{U}^{\mathrm{T}}\tilde{\boldsymbol{h}}$, 
and the definition of $\boldsymbol{b}_{\z,0}$ in (\ref{b}).  

We next prove (\ref{bzmz}) for $\tau=0$. 
Using $\boldsymbol{m}_{\z,0}$ in (\ref{mz}) and 
Property~\ref{A3} for $\tau=0$ under 
Assumptions~\ref{assumption_denoisers_general} and 
\ref{assumption_A_addition} yields 
\begin{align} 
\frac{\boldsymbol{b}_{\z,0}^{\mathrm{T}}\boldsymbol{m}_{\z,0}}{M}
&\pto \mathbb{E}[B_{\z,0}\phi_{\z,0}(B_{\z,0}, \sqrt{\Lambda}B_{\x,0}, \Lambda)] 
- \bar{\xi}_{\A,0,0}^{\z}\mathbb{E}[B_{\z,0}^{2}]
\label{bm0}
\nonumber \\
&= 0,  
\end{align}
where the last equality follows from Lemma~\ref{lemma_Stein} and the 
definition of $\bar{\xi}_{\A,0,0}^{\z}$ in (\ref{xi_A_z_bar_general}). 
\end{IEEEproof}

\begin{IEEEproof}[Proof of \ref{A4i}]
Using the definition of $\boldsymbol{m}_{\x,0}$ in (\ref{mx}), 
Property~\ref{A2} for $\tau=0$, and the latter convergence in 
Property~\ref{A3} for $a=\x$ and $\tau=0$ yields  
\begin{align}
&\boldsymbol{b}_{\x,0}^{\mathrm{T}}\boldsymbol{m}_{\x,0} 
\peq \boldsymbol{b}_{\x,0}^{\mathrm{T}}\phi_{\x,0} 
- \bar{\xi}_{\A,0,0}^{\x}Q_{\x,0,0} + o(1)
\nonumber \\
&\peq  \mathbb{E}\left[
 \sqrt{\Lambda}B_{\x,0}\tilde{\phi}_{\x,0}(\sqrt{\Lambda}B_{\x,0}, B_{\z,0}, \Lambda)
\right]
- (\bar{\xi}_{\A,0,0}^{\x} - 1)Q_{\x,0,0} 
\nonumber \\
&+ o(1),
\label{bxmx_0}
\end{align}
where the last equality follows from $\phi_{\x,t}$ in (\ref{phi_x}) and 
Property~\ref{A3} for $\tau=0$ under 
Assumptions~\ref{assumption_denoisers_general} and 
\ref{assumption_A_addition}. 
Utilizing Lemma~\ref{lemma_Stein}, $\mathbb{E}[ B_{\x,0}^{2}]=Q_{\x,0,0}$, and 
$\bar{\xi}_{\A,0,0}^{\x}$ in (\ref{xi_A_x_bar_general}), 
we arrive at (\ref{bxmx}) for $\tau=0$. 
\end{IEEEproof}

\begin{IEEEproof}[Proof of \ref{A5o}]
We prove properties for $[\boldsymbol{m}_{\z,0}, \tilde{\boldsymbol{m}}]$. 
Using the definition of $\boldsymbol{m}_{\z,0}$ in (\ref{mz}) and 
Property~\ref{A3} for $\tau=0$ under 
Assumptions~\ref{assumption_denoisers_general} and 
\ref{assumption_A_addition}, we confirm 
$M^{-1}\|\boldsymbol{m}_{\z,0}\|_{2}^{2}\pto \mathbb{E}[M_{\z,0}^{2}]$. 
Similarly, we use $\tilde{\boldsymbol{m}}
=\boldsymbol{\Sigma}\boldsymbol{b}_{\x,0}$, Property~\ref{I5}, and 
Property~\ref{A3} for $\tau=0$ to obtain  
\begin{equation} \label{M_tilde_M_tilde_0}
\frac{1}{M}[\boldsymbol{m}_{\z,0}, \tilde{\boldsymbol{m}}]^{\mathrm{T}}
[\boldsymbol{m}_{\z,0}, \tilde{\boldsymbol{m}}]
\pto 
\begin{bmatrix}
\mathbb{E}[M_{\z,0}^{2}] & \mathbb{E}[M_{\z,0}\tilde{M}] \\
\mathbb{E}[\tilde{M}M_{\z,0}] & \mathbb{E}[\tilde{M}^{2}]
\end{bmatrix}, 
\end{equation}
with $\tilde{M}=\sqrt{\Lambda}B_{\x,0}$. 

The first condition in Assumption~\ref{assumption_SE_general} for $\tau'=0$ 
implies (\ref{A5o_positivity}) for 
$\tilde{\boldsymbol{M}}_{\z,1}=\boldsymbol{m}_{\z,0}$ 
when $\tilde{\boldsymbol{m}}$ and $\boldsymbol{m}_{\z,0}$ are linearly 
dependent. Assume that $\tilde{\boldsymbol{m}}$ and $\boldsymbol{m}_{\z,0}$ 
are linearly independent.
We know that (\ref{A5o_positivity}) holds for $\tau=0$ 
if $M^{-1}\|\boldsymbol{P}_{\tilde{\boldsymbol{m}}}^{\perp}
\boldsymbol{m}_{\z,0}\|_{2}^{2}$ is strictly positive in probability 
in the sublinear sparsity limit~\cite[Lemmas~8 and 9]{Bayati11}. 
Using (\ref{M_tilde_M_tilde_0}) yields 
\begin{align}
&\frac{\|\boldsymbol{P}_{\tilde{\boldsymbol{m}}}^{\perp}
\boldsymbol{m}_{\z,0}\|_{2}^{2}}{M}
\pto \mathbb{E}[M_{\z,0}^{2}]
- \frac{(\mathbb{E}[\tilde{M}M_{\z,0}])^{2}}
{\mathbb{E}[\tilde{M}^{2}]} \nonumber \\
&\geq \mathbb{E}\left[
 (M_{\z,0} - \mathbb{E}[M_{\z,0} | \tilde{M}])^{2} 
\right] > 0. 
\end{align}
In the derivation of the lower bound, we have used 
Lemma~\ref{lemma_positive_definite}. The last strict positivity is due to 
The first condition in Assumption~\ref{assumption_SE_general} for $\tau'=0$. 
Thus, (\ref{A5o_positivity}) holds for $\tau=0$. 
\end{IEEEproof}

\begin{IEEEproof}[Proof of \ref{A5i}]
We prove properties for $\boldsymbol{m}_{\x,0}$. Using 
$\boldsymbol{m}_{\x,0}$ in (\ref{mx}), $\phi_{\x,0}$ in (\ref{phi_x}), 
Property~\ref{A3} for $\tau=0$ under 
Assumptions~\ref{assumption_denoisers_general} and 
\ref{assumption_A_addition}, Property~\ref{A4i} for $\tau=0$, and 
$\boldsymbol{\Lambda}=N^{-1}\boldsymbol{\Sigma}\boldsymbol{\Sigma}^{\mathrm{T}}$, 
we obtain 
\begin{align}
&\frac{M\|\boldsymbol{m}_{\x,0}\|_{2}^{2}}{N}
\peq \frac{1}{N}\boldsymbol{m}_{\x,0}^{\mathrm{T}}\boldsymbol{\Sigma}^{\mathrm{T}}
\tilde{\phi}_{\x,0} + o(1) 
\nonumber \\
&\pto \mathbb{E}[\Lambda
\tilde{\phi}_{\x,0}^{2}(\sqrt{\Lambda}B_{\x,0}, B_{\z,0}, \Lambda)] 
= \mathbb{E}[M_{\x,0}^{2}], 
\end{align}
which also implies (\ref{A5i_positivity}) for $\tau=0$ 
under the second condition in Assumption~\ref{assumption_SE_general} 
for $\tau'=0$. 
\end{IEEEproof}

\subsection{Module~B for $\tau=0$} 
\label{appen_module_B_0}
\begin{IEEEproof}[Proof of \ref{B1o}]
From Properties~\ref{I2}, \ref{I4}, and \ref{I5}, as well as 
Assumption~\ref{assumption_A}, we can use 
Lemma~\ref{lemma_conditioning} for the constraint 
\begin{equation}
[\boldsymbol{q}_{\z,0}, \tilde{\boldsymbol{h}}]
= \boldsymbol{U}[\boldsymbol{b}_{\z,0}, \tilde{\boldsymbol{m}}] 
\end{equation}
to have 
\begin{align}
\boldsymbol{U} &\sim [\boldsymbol{q}_{\z,0}, \tilde{\boldsymbol{h}}]
\begin{bmatrix}
\|\boldsymbol{q}_{\z,0}\|_{2}^{2} & o(M) \\
o(M) & \|\tilde{\boldsymbol{h}}\|_{2}^{2}
\end{bmatrix}^{-1}
[\boldsymbol{b}_{\z,0}, \tilde{\boldsymbol{m}}]^{\mathrm{T}}
\nonumber \\
&+ \boldsymbol{\Phi}_{[\boldsymbol{q}_{\z,0}, \tilde{\boldsymbol{h}}]}^{\perp}
\tilde{\boldsymbol{U}}
(\boldsymbol{\Phi}_{[\boldsymbol{b}_{\z,0}, \tilde{\boldsymbol{m}}]}^{\perp})^{\mathrm{T}} 
\end{align}
conditioned on $\Theta$ and $\mathfrak{E}_{0,1}^{\z}$, 
where $\tilde{\boldsymbol{U}}$ is a Haar-distributed orthogonal matrix 
independent of $\Theta$ and $\mathfrak{E}_{0,1}^{\z}$. 
Applying this expression to the definition of $\boldsymbol{h}_{\z,0}$ 
in (\ref{h}), and using Property~\ref{A4o} for $\tau=0$, we have 
\begin{align}
\boldsymbol{h}_{\z,0} 
&\sim \frac{\tilde{\boldsymbol{m}}^{\mathrm{T}}\boldsymbol{m}_{\z,0}}
{\|\tilde{\boldsymbol{h}}\|_{2}^{2}}\tilde{\boldsymbol{h}}
+ o(1)\tilde{\boldsymbol{h}} + o(1)\boldsymbol{q}_{\z,0}
+ \boldsymbol{\Phi}_{[\boldsymbol{q}_{\z,0}, \tilde{\boldsymbol{h}}]}^{\perp}
\tilde{\boldsymbol{U}}\boldsymbol{a}
\nonumber \\
&\peq \frac{\mathbb{E}[\tilde{M}M_{\z,0}]}{P}\tilde{\boldsymbol{h}}
+ o(1)\tilde{\boldsymbol{h}} + o(1)\boldsymbol{q}_{\z,0} 
+ \boldsymbol{\Phi}_{[\boldsymbol{q}_{\z,0}, \tilde{\boldsymbol{h}}]}^{\perp}
\tilde{\boldsymbol{U}}\boldsymbol{a},  \label{hz0_tmp}
\end{align}
with $\boldsymbol{a}
=(\boldsymbol{\Phi}_{[\boldsymbol{b}_{\z,0},\tilde{\boldsymbol{m}}]}^{\perp})^{\mathrm{T}}
\boldsymbol{m}_{\z,0}$, where the last equality follows from 
Properties~\ref{I2} and \ref{A5o} for $\tau=0$. The first three terms 
in (\ref{hz0_tmp}) are equal to those in (\ref{hz0}). 

To use Lemma~\ref{lemma_modified_representation} for the last term 
in (\ref{hz0_tmp}), we use Property~\ref{A4o} for $\tau=0$,  
Properties~\ref{I2}, \ref{I4}, and \ref{I5} to confirm 
\begin{equation}
\frac{\|\boldsymbol{a}\|_{2}^{2}}{M}
= \frac{\boldsymbol{m}_{\z,0}^{\mathrm{T}}
\boldsymbol{P}_{[\boldsymbol{b}_{\z,0},\tilde{\boldsymbol{m}}]}^{\perp}
\boldsymbol{m}_{\z,0}}{M}
\peq \frac{\boldsymbol{m}_{\z,0}^{\mathrm{T}}
\boldsymbol{P}_{\tilde{\boldsymbol{m}}}^{\perp}
\boldsymbol{m}_{\z,0}}{M}
+ o(1), 
\end{equation}
\begin{equation}
\frac{1}{M}[\boldsymbol{q}_{\z,0}, \tilde{\boldsymbol{h}}]^{\mathrm{T}}
[\boldsymbol{q}_{\z,0}, \tilde{\boldsymbol{h}}]
\pto \begin{bmatrix}
\mathbb{E}[Q_{\z,0}^{2}] & 0 \\
0 & P
\end{bmatrix},
\end{equation}
which imply that all conditions in Lemma~\ref{lemma_modified_representation} 
are satisfied. Thus, we can use 
Lemma~\ref{lemma_modified_representation} to obtain 
\begin{equation}
\boldsymbol{\Phi}_{[\boldsymbol{q}_{\z,0}, \tilde{\boldsymbol{h}}]}^{\perp}
\tilde{\boldsymbol{U}}\boldsymbol{a}
\sim o(1)\tilde{\boldsymbol{h}} + o(1)\boldsymbol{q}_{\z,0}
+ \{1 + o(1)\}\frac{\|\boldsymbol{a}\|_{2}}
{\|\boldsymbol{\omega}_{\z,0}\|_{2}}\boldsymbol{\omega}_{\z,0},
\end{equation}
with $\boldsymbol{\omega}_{\z,0}\sim\mathcal{N}(\boldsymbol{0}, 
\boldsymbol{I}_{M})$ independent of $\Theta$ and $\mathfrak{E}_{0,1}^{\z}$. 
Combining these results, we arrive at (\ref{hz0}). 
\end{IEEEproof}

\begin{IEEEproof}[Proof of \ref{B1i}]
From Property~\ref{I5} and Assumption~\ref{assumption_A} 
we can use Lemma~\ref{lemma_conditioning} for the constraint 
$\boldsymbol{q}_{\x,0}=\boldsymbol{V}\boldsymbol{b}_{\x,0}$ to have 
\begin{equation}
\boldsymbol{V}\sim \frac{\boldsymbol{q}_{\x,0}}
{\|\boldsymbol{q}_{\x,0}\|_{2}^{2}}\boldsymbol{b}_{\x,0}^{\mathrm{T}} 
+ \boldsymbol{\Phi}_{\boldsymbol{q}_{\x,0}}^{\perp}\tilde{\boldsymbol{V}}
(\boldsymbol{\Phi}_{\boldsymbol{b}_{\x,0}}^{\perp})^{\mathrm{T}} 
\end{equation}
conditioned on $\Theta$ and $\mathfrak{E}_{0,1}^{\x}$, 
where $\tilde{\boldsymbol{V}}$ is a Haar-distributed orthogonal matrix 
independent of $\Theta$ and $\mathfrak{E}_{0,1}^{\x}$. 
Applying this expression to the definition of $\boldsymbol{h}_{\x,0}$ 
in (\ref{h}) yields 
\begin{equation}
\boldsymbol{h}_{\x,0} 
\sim o(1)\boldsymbol{q}_{\x,0}
+ \boldsymbol{\Phi}_{\boldsymbol{q}_{\x,0}}^{\perp}
\tilde{\boldsymbol{V}}\boldsymbol{a}, 
\end{equation}
with $\boldsymbol{a}=(\boldsymbol{\Phi}_{\boldsymbol{b}_{\x,0}}^{\perp})^{\mathrm{T}}
\boldsymbol{m}_{\x,0}$, 
where we have used Properties~\ref{I5} and \ref{A4i} for $\tau=0$ 
in evaluation of the first term. 

To utilize Lemma~\ref{lemma_modified_representation} for the second term, 
we use Property~\ref{A4i} for $\tau=0$ and Property~\ref{I5} to confirm 
\begin{equation}
\frac{\|\boldsymbol{a}\|_{2}^{2}}{N\|\boldsymbol{q}_{\x,0}\|_{2}^{2}}
= \frac{\boldsymbol{m}_{\x,0}^{\mathrm{T}}
\boldsymbol{P}_{\boldsymbol{b}_{\x,0}}^{\perp}\boldsymbol{m}_{\x,0}}
{N\|\boldsymbol{q}_{\x,0}\|_{2}^{2}}
\peq \frac{\|\boldsymbol{m}_{\x,0}\|_{2}^{2}}{NQ_{\x,0,0}} + o(1)
\pto 0,  
\end{equation}
because of Property~\ref{A5i} for $\tau=0$. Thus, 
we can use Lemma~\ref{lemma_modified_representation} to obtain 
\begin{equation}
\boldsymbol{\Phi}_{\boldsymbol{q}_{\x,0}}^{\perp}
\tilde{\boldsymbol{V}}\boldsymbol{a}
\sim o(1)\boldsymbol{q}_{\x,0} + \{1 + o(1)\}
\frac{\|\boldsymbol{m}_{\x,0}\|_{2}}
{\|\boldsymbol{\omega}_{\x,0}\|_{2}}\boldsymbol{\omega}_{\x,0}, 
\end{equation}
with $\boldsymbol{\omega}_{\x,0}\sim\mathcal{N}(\boldsymbol{0}, 
\boldsymbol{I}_{N})$ independent of $\Theta$ and $\mathfrak{E}_{0,1}^{\x}$. 
Combining these results, we arrive at (\ref{hx0}). 
\end{IEEEproof}

\begin{IEEEproof}[Proof of \ref{B2}]
The first and last convergence in Property~\ref{B2} for $\tau=0$ follows 
from the definitions of $\boldsymbol{h}_{\z,0}$ and $\boldsymbol{h}_{\x,0}$ 
in (\ref{h}), as well as Properties~\ref{A5o} and \ref{A5i} for $\tau=0$.  
For the second convergence, we use $\tilde{\boldsymbol{h}}
=\boldsymbol{U}\tilde{\boldsymbol{m}}$ and the definition of 
$\boldsymbol{h}_{\z,0}$ in (\ref{h}) to obtain  
\begin{equation}
\frac{1}{M}\tilde{\boldsymbol{h}}^{\mathrm{T}}\boldsymbol{h}_{\z,0}
= \frac{1}{M}\tilde{\boldsymbol{m}}^{\mathrm{T}}\boldsymbol{m}_{\z,0}
\pto \mathbb{E}[\tilde{M}M_{\z,0}],
\end{equation} 
where the last convergence follows from Property~\ref{A5o} for $\tau=0$. 
Thus, Property~\ref{B2} holds for $\tau=0$. 
\end{IEEEproof}

\begin{IEEEproof}[Proof of \ref{B3o}]
Let $\boldsymbol{X}_{1,M}=[\boldsymbol{h}_{\z,0}, \tilde{\boldsymbol{h}}, 
\boldsymbol{w}]$, $\boldsymbol{X}_{2,M}=\boldsymbol{0}$, $t=4$, and $t_{0}=3$ 
in Lemma~\ref{lemma_generalization}. Repeating the proof of the latter 
convergence in Property~\ref{A3} for $\tau=0$, 
we can prove that the former convergence in Property~\ref{B3o} for $\tau=0$ 
implies the latter convergence in Property~\ref{B3o} for $\tau=0$. 
   
We prove the former convergence for $\tau=0$. We only consider everywhere 
Lipschitz-continuous $\phi$ and $\psi$ since the generalization to the 
piecewise Lipschitz-continuous case can be proved with 
Lemma~\ref{lemma_generalization}. 

To use Lemma~\ref{lemma_bias_elimination} for the separable function 
$\boldsymbol{f}(\boldsymbol{h}_{\z,0}, \tilde{\boldsymbol{h}}, \boldsymbol{w})$ 
satisfying 
$f_{m}=[\phi(\boldsymbol{h}_{\z,0}, \tilde{\boldsymbol{h}}, \boldsymbol{w})]_{m}
[\psi(\boldsymbol{h}_{\z,0}, \tilde{\boldsymbol{h}}, \boldsymbol{w})]_{m}$, 
we confirm all conditions in Lemma~\ref{lemma_bias_elimination}. The 
Lipschitz continuity of $\phi$ and $\psi$ implies that the $m$th element 
function $f_{m}$ is second-order pseudo-Lipschitz with some $m$-independent 
Lipschitz constant. From Properties~\ref{I3} and \ref{I5} we can confirm 
the other conditions in Lemma~\ref{lemma_bias_elimination}. 
Applying Property~\ref{B1o} for $\tau=0$ to $\boldsymbol{h}_{\z,0}$ and 
using Lemma~\ref{lemma_bias_elimination} repeatedly, we have 
\begin{align}
&\frac{1}{M}\phi^{\mathrm{T}}(\boldsymbol{h}_{\z,0}, \tilde{\boldsymbol{h}}, 
\boldsymbol{w})\psi(\boldsymbol{h}_{\z,0}, \tilde{\boldsymbol{h}}, 
\boldsymbol{w})
\nonumber \\
&\sim \left\langle\boldsymbol{f}\left(
 \frac{\mathbb{E}[\tilde{M}M_{\z,0}]}{P}\tilde{\boldsymbol{h}} 
 + \frac{a_{M}}{\|\boldsymbol{\omega}_{\z,0}\|_{2}}
 \boldsymbol{\omega}_{\z,0}, \tilde{\boldsymbol{h}}, \boldsymbol{w}
\right)
\right\rangle
+ o(1)
\end{align}
conditioned on $\Theta$ and $\mathfrak{E}_{0,1}^{\z}$, 
with $a_{M}=\{1 + o(1)\}(\boldsymbol{m}_{\z,0}^{\mathrm{T}}
\boldsymbol{P}_{\tilde{\boldsymbol{m}}}^{\perp}$ 
$\cdot\boldsymbol{m}_{\z,0})^{1/2}$, 
where $\boldsymbol{\omega}_{\z,0}\sim\mathcal{N}(\boldsymbol{0}, 
\boldsymbol{I}_{M})$ is independent of $\Theta$ and $\mathfrak{E}_{0,1}^{\z}$. 

To utilize Lemma~\ref{lemma_elimination} with $f(M)=1$, we let 
$\phi_{m}(h_{\z}) = M^{-1/2}\phi( P^{-1}\mathbb{E}[\tilde{M}M_{\z,0}]\tilde{h}_{m} 
+ h_{\z}, \tilde{h}_{m}, w_{m})$ and $\psi_{m}(h_{\z}) 
= M^{-1/2}\psi(P^{-1}\mathbb{E}[\tilde{M}M_{\z,0}]\tilde{h}_{m} + h_{\z}, 
\tilde{h}_{m}, w_{m})$ for fixed $\tilde{h}_{m}$ and 
$w_{m}$. The Lipschitz continuity of $\phi$ and $\psi$ implies that $\phi_{m}$ 
and $\psi_{m}$ are Lipschitz-continuous with a common Lipschitz constant 
$L_{m}=M^{-1/2}L$ for some $L>0$, which results in $\sum_{m=1}^{M}L_{m}^{2}=L^{2}$. 
From Property~\ref{A5o} for $\tau=0$ we find 
$M^{-1}a_{M}^{2}\pto a^{2}$ for some $a>0$. Furthermore, 
we use the weak law of large numbers for $\boldsymbol{\omega}_{\z,0}$ and 
Property~\ref{I3} to obtain 
\begin{align}
&\|\boldsymbol{\phi}(a\boldsymbol{\omega}_{\z,0})\|_{2}^{2}
= \frac{1}{M}\left\|
 \phi\left(
  \frac{\mathbb{E}[\tilde{M}M_{\z,0}]}{P}\tilde{\boldsymbol{h}} 
  + a\boldsymbol{\omega}_{\z,0}, \tilde{\boldsymbol{h}}, 
  \boldsymbol{w}
 \right)
\right\|_{2}^{2}
\nonumber \\
&\pto \mathbb{E}\left[
 \phi(H_{\z,0}, \tilde{H}, W)
\right],
\end{align}
where the zero-mean Gaussian random variable $H_{\z,0}$ is independent of $W$ 
and has the covariance 
$\mathbb{E}[\tilde{H}H_{\z,0}] = \mathbb{E}[\tilde{M}M_{\z,0}]$ and  
$\mathbb{E}[H_{\z,0}^{2}] = \mathbb{E}[M_{\z,0}^{2}]$ due to 
Property~\ref{B2} for $\tau=0$. 
Similarly, we confirm the boundedness in probability of 
$\|\boldsymbol{\psi}(a\boldsymbol{\omega}_{\z,0})\|_{2}^{2}$. Thus, we can use 
Lemma~\ref{lemma_elimination} to obtain 
\begin{align}
&\left\langle\boldsymbol{f}\left(
 \frac{\mathbb{E}[\tilde{M}M_{\z,0}]}{P}\tilde{\boldsymbol{h}} 
 + \frac{a_{M}}{\|\boldsymbol{\omega}_{\z,0}\|_{2}}
 \boldsymbol{\omega}_{\z,0}, \tilde{\boldsymbol{h}}, \boldsymbol{w}
\right)
\right\rangle
\nonumber \\
&\peq \left\langle\boldsymbol{f}\left(
 \frac{\mathbb{E}[\tilde{M}M_{\z,0}]}{P}\tilde{\boldsymbol{h}} 
 + a\boldsymbol{\omega}_{\z,0}, \tilde{\boldsymbol{h}}, \boldsymbol{w}
\right)
\right\rangle + o(1)
\nonumber \\
&\pto \mathbb{E}\left[
 \phi(H_{\z,0}, \tilde{H}, W)\psi(H_{\z,0}, \tilde{H}, W)
\right],
\end{align}
where the last convergence follows from the weak law of large 
numbers for $\boldsymbol{\omega}_{\z,0}$ and 
Property~\ref{I3}. Combining these results, 
we arrive at Property~\ref{B3o} for $\tau=0$. 
\end{IEEEproof}

\begin{IEEEproof}[Proof of \ref{B3i}]
We first prove (\ref{psi_psi}) for $\tau=0$. 
Define $a_{N}=\{1 + o(1)\}\|\boldsymbol{m}_{\x,0}\|_{2}$ and 
$\tilde{a}_{N}=(M/N)^{1/2}a_{N}$. 
Property~\ref{A5i} for $\tau=0$ implies 
$\tilde{a}_{N}\pto\{\mathbb{E}[M_{\x,0}^{2}]\}^{1/2}\equiv a$. 
Using Property~\ref{B1i} for $\tau=0$ yields 
\begin{align}
&\psi_{\x,0}(\boldsymbol{h}_{\x,0}, \boldsymbol{x})
\sim \psi_{\x,0}\left(
 o(1)\boldsymbol{q}_{\x,0} + \frac{a_{N}\boldsymbol{\omega}_{\x,0}}
 {\|\boldsymbol{\omega}_{\x,0}\|_{2}}, \boldsymbol{x}
\right)
\nonumber \\
&\peq \psi_{\x,0}\left(
 o(1)\boldsymbol{q}_{\x,0} + o(1)\boldsymbol{a}
 + \frac{a_{N}\boldsymbol{\omega}_{\x,0}}{\sqrt{N}}, \boldsymbol{x}
\right) \label{psi_0}
\end{align}
conditioned on $\Theta$ and $\mathfrak{E}_{0,1}^{\x}$, 
with $\boldsymbol{a}=N^{-1/2}\boldsymbol{\omega}_{\x,0}$ and 
$\boldsymbol{\omega}_{\x,0}\sim\mathcal{N}(\boldsymbol{0}, 
\boldsymbol{I}_{N})$ independent of $\{\Theta, \mathfrak{E}_{0,1}^{\x}\}$. 
In the derivation of the last equality we have used 
Proposition~\ref{proposition1}. From Property~\ref{I5} and the weak law of 
large numbers $\|\boldsymbol{a}\|_{2}^{2}=N^{-1}\|\boldsymbol{w}_{\x,0}\|_{2}^{2}
\pto 1$, we can use Assumption~\ref{assumption_inner_general} to arrive at  
\begin{equation}
\|\psi_{\x,0}(\boldsymbol{h}_{\x,0}, \boldsymbol{x})
- \psi_{\x,0}(\boldsymbol{\omega}_{0}, \boldsymbol{x})\|_{2}
\pto 0, 
\end{equation}
with $\boldsymbol{\omega}_{0}=M^{-1/2}a\boldsymbol{\omega}_{\x,0}$. 
Combining this result with the convergence in probability 
$\|\psi_{\x,0}(\boldsymbol{\omega}_{0}, \boldsymbol{x})\|_{2}^{2}
\pto \mathbb{E}[\|\psi_{\x,0}(\boldsymbol{\omega}_{0}, \boldsymbol{x})\|_{2}^{2}]$ 
in Assumption~\ref{assumption_inner_general}, we arrive at (\ref{psi_psi}) 
for $\tau=0$. 

We next prove (\ref{x_psi}) for $\tau=0$. To utilize 
Lemma~\ref{lemma_elimination} for $\boldsymbol{\phi}(\boldsymbol{\omega})
=\boldsymbol{x}$, $\boldsymbol{\psi}(\boldsymbol{\omega})
=\psi_{\x,0}(o(1)\boldsymbol{q}_{\x,0} + \boldsymbol{\omega}, 
\boldsymbol{x})$, and $f(N)=M^{-1/2}$, with $M$ regarded as a function of $N$, 
we confirm all conditions in Lemma~\ref{lemma_elimination}. 
Since $\boldsymbol{\phi}(\boldsymbol{\omega})$ 
does not depend on $\boldsymbol{\omega}$, 
$\|\Delta\boldsymbol{\phi}(\boldsymbol{\omega})\|_{2}=0$ is trivial.   
We have already observed $\tilde{a}_{N}\pto a$, 
the boundedness in probability of 
$\|\boldsymbol{\psi}(M^{-1/2}a\boldsymbol{\omega})\|_{2}$, and 
$\|\Delta\boldsymbol{\psi}(\boldsymbol{\omega})\|_{2}\pto0$. 
Using (\ref{Qx_0}) yields $\|\boldsymbol{\phi}(\boldsymbol{\omega})\|_{2}^{2}
=\|\boldsymbol{x}\|_{2}^{2}\pto P$. Thus, we can use 
Lemma~\ref{lemma_elimination} with (\ref{psi_0}) to obtain 
\begin{align}
\boldsymbol{x}^{\mathrm{T}}
\psi_{\x,0}(\boldsymbol{h}_{\x,0}, \boldsymbol{x})
&\sim \boldsymbol{x}^{\mathrm{T}}
\psi_{\x,0}(o(1)\boldsymbol{q}_{\x,0} + o(1)\boldsymbol{a} 
+ \boldsymbol{\omega}_{0}, \boldsymbol{x}) + o(1)
\nonumber \\
&= \boldsymbol{x}^{\mathrm{T}}
\psi_{\x,0}(\boldsymbol{\omega}_{0}, \boldsymbol{x}) + o(1)
\nonumber \\
&\peq \mathbb{E}[\boldsymbol{x}^{\mathrm{T}}
\psi_{\x,0}(\boldsymbol{\omega}_{0}, \boldsymbol{x})] + o(1).
\end{align}
In the derivation of the second equality, we have used 
Lemma~\ref{lemma_bias_elimination_inner}. The last equality follows from 
Assumption~\ref{assumption_inner_general}.

Finally, We prove (\ref{h_psi}) for $\tau=0$. Repeating the proof of 
(\ref{x_psi}) for $\tau=0$ yields 
\begin{align}
\boldsymbol{h}_{\x,0}^{\mathrm{T}}\psi_{\x,0}(\boldsymbol{h}_{\x,0}, \boldsymbol{x}) 
&\sim \boldsymbol{\omega}_{0}^{\mathrm{T}}\psi_{\x,0}\left(
 o(1)\boldsymbol{q}_{\x,0} + o(1)\boldsymbol{a} + \boldsymbol{\omega}_{0}
\right) + o(1)
\nonumber  \\
&\peq \mathbb{E}\left[
 \boldsymbol{\omega}_{0}^{\mathrm{T}}\psi_{\x,0}(\boldsymbol{\omega}_{0}) 
\right] + o(1) 
\end{align}
conditioned on $\Theta$ and $\mathfrak{E}_{0,1}^{\x}$, where the first 
equivalence follows from Lemmas~\ref{lemma_bias_elimination_inner} and 
\ref{lemma_elimination} while 
the last equality is due to Assumption~\ref{assumption_inner_general}. 
Thus, Property~\ref{B3i} holds for $\tau=0$. 
\end{IEEEproof}

\begin{IEEEproof}[Proof of \ref{B4o}]
We first prove (\ref{hzqz}) for $\tau=0$. Using the definition of 
$\boldsymbol{q}_{\z,1}$ in (\ref{qz}) and Property~\ref{B3o} 
for $\tau=0$ under Assumption~\ref{assumption_denoisers_general}, we obtain 
\begin{align}
&\frac{1}{M}\boldsymbol{h}_{\z,0}^{\mathrm{T}}\boldsymbol{q}_{\z,1} 
\nonumber \\
&\pto\mathbb{E}[H_{\z,0}\psi_{\z,0}(H_{\z,0}, \tilde{H}, W)]
- \bar{\xi}_{\B,0,0}^{\z}\mathbb{E}[H_{\z,0}^{2}] 
- \bar{\eta}_{\B,0}\mathbb{E}[H_{\z,0}\tilde{H}] 
\nonumber \\
&= \mathbb{E}[H_{\z,0}^{2}]\mathbb{E}[\partial_{0}\psi_{\z,0}]
+ \mathbb{E}[H_{\z,0}\tilde{H}]\mathbb{E}[\partial_{1}\psi_{\z,0}]
\nonumber \\
&- \bar{\xi}_{\B,0,0}^{\z}\mathbb{E}[H_{\z,0}^{2}]
- \bar{\eta}_{\B,0}\mathbb{E}[H_{\z,0}\tilde{H}]
= 0.
\end{align}
In the derivation of the second last equality, we have used 
Lemma~\ref{lemma_Stein}. The last equality is obtained from the definitions  
of $\bar{\xi}_{\B,0,0}^{\z}$ in (\ref{xi_B_z_bar_general}) and 
$\bar{\eta}_{\B,0}$ in (\ref{eta_B_bar_general}).  

We next prove (\ref{zq}) for $\tau=0$. Repeating the proof of 
(\ref{hzqz}) for $\tau=0$, we have  
\begin{align}
\frac{1}{M}\tilde{\boldsymbol{h}}^{\mathrm{T}}\boldsymbol{q}_{\z,1} 
&\pto \mathbb{E}[\tilde{H}\psi_{\z,0}(H_{\z,0}, \tilde{H}, W)]
- \bar{\xi}_{\B,0,0}^{\z}\mathbb{E}[\tilde{H}H_{\z,0}]
\nonumber \\
&- \bar{\eta}_{\B,0}\mathbb{E}[\tilde{H}^{2}] 
= 0, 
\end{align}
where the last follows from Lemma~\ref{lemma_Stein}, 
$\bar{\xi}_{\B,0,0}^{\z}$ in (\ref{xi_B_z_bar_general}),  
and $\bar{\eta}_{\B,0}$ in (\ref{eta_B_bar_general}). 
Thus, Property~\ref{B4o} holds for $\tau=0$. 
\end{IEEEproof}

\begin{IEEEproof}[Proof of \ref{B4i}]
We first prove (\ref{hx}) for $\tau=0$. Using Property~\ref{B1i} for $\tau=0$, 
the initial condition $\boldsymbol{q}_{\x,0}=-\boldsymbol{x}$, 
and (\ref{Qx_0}) yields 
\begin{equation}
\boldsymbol{x}^{\mathrm{T}}\boldsymbol{h}_{\x,0} 
\sim o(1) + \{1 + o(1)\}\frac{\|\boldsymbol{m}_{\x,0}\|_{2}}
{\|\boldsymbol{\omega}_{\x,0}\|_{2}}
\boldsymbol{x}^{\mathrm{T}}\boldsymbol{\omega}_{\x,0}
\end{equation}
conditioned on $\Theta$ and $\mathfrak{E}_{0,1}^{\x}$, where 
$\boldsymbol{\omega}_{\x,0}\sim\mathcal{N}(\boldsymbol{0}, \boldsymbol{I}_{N})$ 
is independent of $\Theta$ and $\mathfrak{E}_{0,1}^{\x}$. 
From Property~\ref{A5i} for $\tau=0$ we have 
$\|\boldsymbol{m}_{\x,0}\|_{2}^{2}/\|\boldsymbol{\omega}_{\x,0}\|_{2}^{2}\peq 
{\cal O}(M^{-1})$. Furthermore, we use Assumption~\ref{assumption_x} to find 
that $\boldsymbol{x}^{\mathrm{T}}\boldsymbol{\omega}_{\x,0}$ converges in 
distribution to a zero-mean Gaussian random variable with variance $P$ 
in the sublinear sparsity limit. These observations imply 
(\ref{hx}) for $\tau=0$. 

We next prove (\ref{hxqx}) for $\tau=0$. Using the definition of 
$\boldsymbol{q}_{\x,1}$ in (\ref{qx}), with $\xi_{\B,0,0}^{\x}$ replaced by 
$\bar{\xi}_{\B,0,0}^{\x}$, and Property~\ref{B2} for $\tau=0$, we have 
\begin{equation} 
\boldsymbol{h}_{\x,0}^{\mathrm{T}}\boldsymbol{q}_{\x,1}
\peq \boldsymbol{h}_{\x,0}^{\mathrm{T}}\psi_{\x,0}(\boldsymbol{h}_{\x,0}, 
\boldsymbol{x}) 
- \bar{\xi}_{\B,0,0}^{\x}\mathbb{E}[M_{\x,0}^{2}] + o(1).  
\end{equation}
Applying Property~\ref{B3i} for $\tau=0$ to the first term yields 
\begin{align}
&\boldsymbol{h}_{\x,0}^{\mathrm{T}}\boldsymbol{q}_{\x,1}
\peq \mathbb{E}[\boldsymbol{\omega}_{0}^{\mathrm{T}}
\psi_{\x,0}(\boldsymbol{\omega}_{0}, \boldsymbol{x})] 
- \bar{\xi}_{\B,0,0}^{\x}\mathbb{E}[M_{\x,0}^{2}] + o(1)
\nonumber \\
&= \sum_{n=1}^{N}\mathbb{E}[\partial_{0}\psi_{\x,0}(\omega_{n,0}, x_{n})]
\mathbb{E}[\omega_{n,0}^{2}]
- \bar{\xi}_{\B,0,0}^{\x}\mathbb{E}[M_{\x,0}^{2}] + o(1)
\nonumber \\
&\to o(1), 
\end{align}
with $\boldsymbol{\omega}_{0}\sim\mathcal{N}(\boldsymbol{0}, 
M^{-1}\mathbb{E}[M_{\x,0}^{2}]\boldsymbol{I}_{N})$ independent of 
$\boldsymbol{x}$. In the derivation of the second equality, we have used 
Lemma~\ref{lemma_Stein} under Assumption~\ref{assumption_inner_general}. 
The last convergence follows from 
$\mathbb{E}[\omega_{n,0}^{2}]=M^{-1}\mathbb{E}[M_{\x,0}^{2}]$ and 
the definition of $\bar{\xi}_{\B,0,0}^{\x}$ in (\ref{xi_B_x_bar_general}). 
Thus, Property~\ref{B4i} holds for $\tau=0$.   
\end{IEEEproof}

\begin{IEEEproof}[Proof of \ref{B5o}]
We prove properties for $\boldsymbol{q}_{\z,1}$. 
Using the definition of $\boldsymbol{q}_{\z,1}$ in (\ref{qz}) and 
Property~\ref{B3o} for $\tau=0$ under 
Assumption~\ref{assumption_denoisers_general}, we have 
$M^{-1}\boldsymbol{q}_{\z,\tau'}^{\mathrm{T}}\boldsymbol{q}_{\z,1}
\pto\mathbb{E}[Q_{\z,\tau'}Q_{\z,1}]$ for all $\tau'\in\{0, 1\}$. 
Repeat the proof of (\ref{A5o_positivity}) for $\tau=0$ to prove 
(\ref{B5o_positivity}) for $\tau=0$ under the third condition 
in Assumption~\ref{assumption_SE_general} for $\tau'=0$. 
\end{IEEEproof}

\begin{IEEEproof}[Proof of \ref{B5i}]
We prove properties for $\boldsymbol{q}_{\x,1}$. Since 
$\xi_{\B,0,0}^{\x}$ has been replaced with $\bar{\xi}_{\B,0,0}$, 
we use the definition of $\boldsymbol{q}_{\x,1}$ in (\ref{qx}) and 
Property~\ref{B4i} for $\tau=0$, we have 
\begin{align}
\boldsymbol{q}_{\x,0}^{\mathrm{T}}\boldsymbol{q}_{\x,1}
&\peq -\boldsymbol{x}^{\mathrm{T}}\psi_{\x,0} + o(1)
\nonumber \\
&\pto -\mathbb{E}\left[
 \boldsymbol{x}^{\mathrm{T}}\psi_{\x,0}(\boldsymbol{\omega}_{0}, \boldsymbol{x})
\right] = Q_{\x,0,1}, 
\end{align}
where the convergence in probability follows from Property~\ref{B3i} 
for $\tau=0$. Similarly, we have 
\begin{align}
&\|\boldsymbol{q}_{\x,1}\|_{2}^{2}
\peq \boldsymbol{q}_{\x,1}^{\mathrm{T}}\psi_{\x,0} + o(1)
\nonumber \\
&\peq \|\psi_{\x,0}\|_{2}^{2} 
- \frac{M}{N}\bar{\xi}_{\B,0,0}^{\x}\boldsymbol{h}_{\x,0}^{\mathrm{T}}
\psi_{\x,0} + o(1)
\nonumber \\
&\pto \mathbb{E}\left[
 \|\psi_{\x,0}(\boldsymbol{\omega}_{0}, \boldsymbol{x})\|_{2}^{2}
\right] = Q_{\x,1,1}. 
\end{align} 
For the proof of (\ref{B5i_positivity}) for $\tau=0$, 
repeat the proof of (\ref{A5o_positivity}) for $\tau=0$ under the last 
condition in Assumption~\ref{assumption_SE_general} for $\tau'=0$. 
\end{IEEEproof}

\subsection{Module~A for $\tau=t$}
\label{appen_module_A_t}
\begin{IEEEproof}[Proof of \ref{A1o}]
From the induction hypotheses~\ref{A5o}, \ref{B4o}, and \ref{B5o} for all 
$\tau<t$, as well as Assumption~\ref{assumption_A},  
we can use Lemma~\ref{lemma_conditioning} for the constraints 
\begin{equation}
[\boldsymbol{Q}_{\z,t}, \tilde{\boldsymbol{H}}_{\z,t}] 
= \boldsymbol{U}[\boldsymbol{B}_{\z,t}, \tilde{\boldsymbol{M}}_{\z,t}].  
\end{equation}
Using the definition of $\boldsymbol{b}_{\z,t}$ in (\ref{b}) and 
the induction hypothesis~\ref{B4o} for all $\tau<t$ yields 
\begin{equation}
\boldsymbol{b}_{\z,t} \sim \boldsymbol{B}_{\z,t}\boldsymbol{\beta}_{\z,t} 
+ \boldsymbol{B}_{\z,t}\boldsymbol{o}(1) 
+ \tilde{\boldsymbol{M}}_{\z,t}\boldsymbol{o}(1) 
+ \boldsymbol{\Phi}_{[\boldsymbol{B}_{\z,t}, \tilde{\boldsymbol{M}}_{\z,t}]}^{\perp}
\tilde{\boldsymbol{U}}^{\mathrm{T}}\boldsymbol{a}
\end{equation}
conditioned on $\Theta$ and $\mathfrak{E}_{t,t}^{\z}$, 
with $\boldsymbol{\beta}_{\z,t} = (\boldsymbol{Q}_{\z,t}^{\mathrm{T}}
\boldsymbol{Q}_{\z,t})^{-1}\boldsymbol{Q}_{\z,t}^{\mathrm{T}}\boldsymbol{q}_{\z,t}$ 
and $\boldsymbol{a}=(\boldsymbol{\Phi}_{[\boldsymbol{Q}_{\z,t}, 
\tilde{\boldsymbol{H}}_{\z,t}]}^{\perp})^{\mathrm{T}}\boldsymbol{q}_{\z,t}$, 
where $\tilde{\boldsymbol{U}}$ is a Haar-distributed orthogonal matrix 
independent of $\Theta$ and $\mathfrak{E}_{t,t}^{\z}$. 

To exploit Lemma~\ref{lemma_modified_representation}, we use 
the induction hypothesis~\ref{A4o} for all $\tau<t$ to confirm 
\begin{align}
&\lambda_{\mathrm{min}}\left(
 \frac{1}{M}[ \boldsymbol{B}_{\z,t}, \tilde{\boldsymbol{M}}_{\z,t}]^{\mathrm{T}}
 [ \boldsymbol{B}_{\z,t}, \tilde{\boldsymbol{M}}_{\z,t}]
\right)
\nonumber \\
&\peq \lambda_{\mathrm{min}}\left(
 \mathrm{diag}\left\{
  \frac{\boldsymbol{B}_{\z,t}^{\mathrm{T}}\boldsymbol{B}_{\z,t}}{M},  
  \frac{\tilde{\boldsymbol{M}}_{\z,t}^{\mathrm{T}}
  \tilde{\boldsymbol{M}}_{\z,t}}{M} 
 \right\}
\right) + o(1),
\end{align}
which is strictly positive in probability, because of 
the induction hypotheses~\ref{A2}, \ref{A5o}, and \ref{B5o} for all $\tau<t$.  
Similarly, using the induction hypothesis~\ref{B4o} for $\tau=t-1$ yields 
\begin{equation}
\frac{\|\boldsymbol{a}\|_{2}^{2} }{M}
= \frac{\boldsymbol{q}_{\z,t}^{\mathrm{T}}
\boldsymbol{P}_{[\boldsymbol{Q}_{\z,t}, \tilde{\boldsymbol{H}}_{\z,t}]}^{\perp}
\boldsymbol{q}_{\z,t}}{M}
\peq \frac{\boldsymbol{q}_{\z,t}^{\mathrm{T}}
\boldsymbol{P}_{\boldsymbol{Q}_{\z,t}}^{\perp}\boldsymbol{q}_{\z,t}}{M} + o(1). 
\end{equation}
Thus, we can use Lemma~\ref{lemma_modified_representation} to obtain 
\begin{align}
\boldsymbol{\Phi}_{[\boldsymbol{B}_{\z,t}, \tilde{\boldsymbol{M}}_{\z,t}]}^{\perp}
\tilde{\boldsymbol{U}}^{\mathrm{T}}\boldsymbol{a}
&\sim \boldsymbol{B}_{\z,t}\boldsymbol{o}(1) 
+ \tilde{\boldsymbol{M}}_{\z,t}\boldsymbol{o}(1) 
\nonumber \\
&+ \{1 + o(1)\}\frac{\|\boldsymbol{q}_{\z,t}^{\perp}\|_{2}}
{\|\boldsymbol{\omega}_{\z,t}\|_{2}}\boldsymbol{\omega}_{\z,t}, 
\end{align}
with $\boldsymbol{q}_{\z,t}^{\perp}
= \boldsymbol{P}_{\boldsymbol{Q}_{\z,t}}^{\perp}\boldsymbol{q}_{\z,t}$, where 
$\boldsymbol{\omega}_{\z,t}\sim\mathcal{N}(\boldsymbol{0},\boldsymbol{I}_{M})$ 
is independent of $\{\Theta, \mathfrak{E}_{t,t}^{\z}\}$. 
Thus, Property~\ref{A1o} holds for $\tau=t$. 
\end{IEEEproof}

\begin{IEEEproof}[Proof of \ref{A1i}]
From the induction hypotheses~\ref{A5i}, \ref{B4i}, and \ref{B5i} 
for all $\tau<t$, as well as Assumption~\ref{assumption_A}, 
we can use Lemma~\ref{lemma_conditioning} for the constraints 
\begin{equation}
[\boldsymbol{Q}_{\x,t}, \tilde{\boldsymbol{H}}_{\x,t}] 
= \boldsymbol{V}[\boldsymbol{B}_{\x,t}, (M/N)^{1/2}\boldsymbol{M}_{\x,t}],   
\end{equation}
with $\tilde{\boldsymbol{H}}_{\x,t}=(M/N)^{1/2}\boldsymbol{H}_{\x,t}$. 
In applying Lemma~\ref{lemma_conditioning}, we utilize the induction 
hypotheses~\ref{B2}, \ref{B4i}, and \ref{B5i} for all $\tau<t$ to find 
\begin{align}
&\begin{bmatrix}
\boldsymbol{Q}_{\x,t}^{\mathrm{T}}\boldsymbol{Q}_{\x,t} & 
\boldsymbol{Q}_{\x,t}^{\mathrm{T}}\tilde{\boldsymbol{H}}_{\x,t} \\
\tilde{\boldsymbol{H}}_{\x,t}^{\mathrm{T}}\boldsymbol{Q}_{\x,t} &
\tilde{\boldsymbol{H}}_{\x,t}^{\mathrm{T}}\tilde{\boldsymbol{H}}_{\x,t}
\end{bmatrix}^{-1}
\nonumber \\
&= \begin{bmatrix}
(\boldsymbol{Q}_{\x,t}^{\mathrm{T}}\boldsymbol{Q}_{\x,t})^{-1} 
+ \boldsymbol{o}(M/N) & \boldsymbol{o}(\sqrt{M/N}) \\
\boldsymbol{o}(\sqrt{M/N}) & 
(\tilde{\boldsymbol{H}}_{\x,t}^{\mathrm{T}}\tilde{\boldsymbol{H}}_{\x,t})^{-1}
+ \boldsymbol{o}(M/N)
\end{bmatrix}. 
\end{align}
From this observation we use the same induction hypotheses to obtain 
\begin{align}
&\begin{bmatrix}
\boldsymbol{Q}_{\x,t}^{\mathrm{T}}\boldsymbol{Q}_{\x,t} & 
\boldsymbol{Q}_{\x,t}^{\mathrm{T}}\tilde{\boldsymbol{H}}_{\x,t} \\
\tilde{\boldsymbol{H}}_{\x,t}^{\mathrm{T}}\boldsymbol{Q}_{\x,t} &
\tilde{\boldsymbol{H}}_{\x,t}^{\mathrm{T}}\tilde{\boldsymbol{H}}_{\x,t}
\end{bmatrix}^{-1}
\begin{bmatrix}
\boldsymbol{Q}_{\x,t}^{\mathrm{T}}\boldsymbol{q}_{\x,t} \\
\tilde{\boldsymbol{H}}_{\x,t}^{\mathrm{T}}\boldsymbol{q}_{\x,t} 
\end{bmatrix}
\nonumber \\
&= \begin{bmatrix}
\boldsymbol{\beta}_{\x,t} + \boldsymbol{o}(M/N) \\
\boldsymbol{o}(\sqrt{M/N})
\end{bmatrix}, 
\end{align}
with $\boldsymbol{\beta}_{\x,t} = (\boldsymbol{Q}_{\x,t}^{\mathrm{T}}
\boldsymbol{Q}_{\x,t})^{-1}\boldsymbol{Q}_{\x,t}^{\mathrm{T}}\boldsymbol{q}_{\x,t}$.  
Thus, applying Lemma~\ref{lemma_conditioning} and the definition of 
$\boldsymbol{b}_{\x,t}$ in (\ref{b}) yields 
\begin{align}
&\boldsymbol{b}_{\x,t} \sim 
\boldsymbol{B}_{\x,t}\boldsymbol{\beta}_{\x,t}
+ \boldsymbol{B}_{\x,t}\boldsymbol{o}(M/N) 
+ \boldsymbol{M}_{\x,t}\boldsymbol{o}(M/N)
\nonumber \\
&+ \boldsymbol{\Phi}_{[\boldsymbol{B}_{\x,t}, (M/N)^{1/2}\boldsymbol{M}_{\x,t}]}^{\perp}
\tilde{\boldsymbol{V}}^{\mathrm{T}}
(\boldsymbol{\Phi}_{[\boldsymbol{Q}_{\x,t}, 
\tilde{\boldsymbol{H}}_{\x,t}]}^{\perp})^{\mathrm{T}}\boldsymbol{q}_{\x,t} 
\end{align}
conditioned on $\Theta$ and $\mathfrak{E}_{t,t}^{\x}$, where 
$\tilde{\boldsymbol{V}}$ is a Haar-distributed orthogonal matrix 
independent of $\Theta$ and $\mathfrak{E}_{t,t}^{\x}$. 

To complete the proof of Property~\ref{A1i} for $\tau=t$, we let 
$\boldsymbol{a}=(\boldsymbol{\Phi}_{[\boldsymbol{Q}_{\x,t}, 
\tilde{\boldsymbol{H}}_{\x,t}]}^{\perp})^{\mathrm{T}}\boldsymbol{q}_{\x,t}$ and 
confirm the conditions in Lemma~\ref{lemma_modified_representation} 
for $\boldsymbol{\Phi}_{[\boldsymbol{B}_{\x,t}, (M/N)^{1/2}\boldsymbol{M}_{\x,t}]}^{\perp}
\tilde{\boldsymbol{V}}^{\mathrm{T}}\boldsymbol{a}$. 
Using the induction hypothesis~\ref{B4i} for all $\tau<t$ yields 
\begin{equation}
\|\boldsymbol{a}\|_{2}^{2} = \boldsymbol{q}_{\x,t}^{\mathrm{T}}
\boldsymbol{P}_{[\boldsymbol{Q}_{\x,t}, \tilde{\boldsymbol{H}}_{\x,t}]}^{\perp}
\boldsymbol{q}_{\x,t}
\peq \|\boldsymbol{q}_{\x,t}^{\perp}\|_{2}^{2} + o(1), 
\end{equation}
with $\boldsymbol{q}_{\x,t}^{\perp} 
=\boldsymbol{P}_{\boldsymbol{Q}_{\x,t}}^{\perp}\boldsymbol{q}_{\x,t}$. 
On the other hand, we use the induction hypothesis~\ref{A4i} 
for all $\tau<t$ to obtain  
\begin{align}
&\lambda_{\mathrm{min}}\left(
 [\boldsymbol{B}_{\x,t}, (M/N)^{1/2}\boldsymbol{M}_{\x,t}]^{\mathrm{T}}
 [\boldsymbol{B}_{\x,t}, (M/N)^{1/2}\boldsymbol{M}_{\x,t}]
\right)
\nonumber \\
&\peq \lambda_{\mathrm{min}}\left(
 \mathrm{diag}\left\{
  \boldsymbol{B}_{\x,t}^{\mathrm{T}}\boldsymbol{B}_{\x,t},
  (M/N)\boldsymbol{M}_{\x,t}^{\mathrm{T}}\boldsymbol{M}_{\x,t} 
 \right\}
\right) + o(1). 
\end{align}
The induction hypotheses~\ref{A2}, \ref{A5i}, and \ref{B5i} for all $\tau<t$ 
imply that the minimum eigenvalue is strictly positive. Thus, we can use 
Lemma~\ref{lemma_modified_representation} to obtain 
\begin{align}
&\boldsymbol{\Phi}_{[\boldsymbol{B}_{\x,t}, (M/N)^{1/2}\boldsymbol{M}_{\x,t}]}^{\perp}
\tilde{\boldsymbol{V}}^{\mathrm{T}}\boldsymbol{a}
\sim \boldsymbol{B}_{\x,t}\boldsymbol{o}(N^{-1}) 
\nonumber \\
&+ \boldsymbol{M}_{\x,t}\boldsymbol{o}(\sqrt{M/N^{3}}) 
+ \{1 + o(1)\}\frac{\|\boldsymbol{q}_{\x,t}^{\perp}\|_{2}}
{\|\boldsymbol{\omega}_{\x,t}\|_{2}}\boldsymbol{\omega}_{\x,t}, 
\end{align}
with $\boldsymbol{\omega}_{\x,t}\sim\mathcal{N}(\boldsymbol{0}, 
\boldsymbol{I}_{N})$ independent of $\Theta$ and $\mathfrak{E}_{t,t}^{\x}$. 
Combining these results, we arrive at Property~\ref{A1i} for $\tau=t$. 
\end{IEEEproof}

\begin{IEEEproof}[Proof of \ref{A2}]
We only prove (\ref{bxbx}) for $\tau=t$ because (\ref{bzbz}) for $\tau=t$ 
can be proved in a similar manner. We first consider the case $\tau'<t$. 
Using Property~\ref{A1i} for $\tau=t$, as well as the induction 
hypotheses~\ref{A2} and \ref{A4i} for all $\tau<t$, we have 
\begin{equation}
\boldsymbol{b}_{\x,\tau'}^{\mathrm{T}}\boldsymbol{b}_{\x,t} 
\sim \boldsymbol{b}_{\x,\tau'}^{\mathrm{T}}\boldsymbol{B}_{\x,t}
\boldsymbol{\beta}_{\x,t} + o(1) + \{1 + o(1)\}
\frac{\|\boldsymbol{q}_{\x,t}^{\perp}\|_{2}}
{\|\boldsymbol{\omega}_{\x,t}\|_{2}}
\boldsymbol{b}_{\x,\tau'}^{\mathrm{T}}\boldsymbol{\omega}_{\x,t} 
\end{equation}
conditioned on $\Theta$ and $\mathfrak{E}_{t,t}^{\x}$, with 
$\boldsymbol{\omega}_{\x,t}\sim\mathcal{N}(\boldsymbol{0}, 
\boldsymbol{I}_{N})$ independent of $\Theta$ and $\mathfrak{E}_{t,t}^{\x}$. 
It is straightforward to confirm that the last term converges in probability 
to zero, because of the weak law of large numbers. Using the induction 
hypotheses~\ref{A2} and \ref{B5i} for all $\tau<t$, 
we find that the first term reduces to 
\begin{align}
&\boldsymbol{b}_{\x,\tau'}^{\mathrm{T}}\boldsymbol{B}_{\x,t}
\boldsymbol{\beta}_{\x,t}
\peq \boldsymbol{q}_{\x,\tau'}^{\mathrm{T}}\boldsymbol{Q}_{\x,t}
(\boldsymbol{Q}_{\x,t}^{\mathrm{T}}\boldsymbol{Q}_{\x,t})^{-1}
\boldsymbol{Q}_{\x,t}^{\mathrm{T}}\boldsymbol{q}_{\x,t} + o(1)
\nonumber \\
&= \boldsymbol{q}_{\x,\tau'}^{\mathrm{T}}\boldsymbol{P}_{\boldsymbol{Q}_{\x,t}}^{\parallel}
\boldsymbol{q}_{\x,t} + o(1)
= \boldsymbol{q}_{\x,\tau'}^{\mathrm{T}}\boldsymbol{q}_{\x,t} + o(1)
\nonumber \\
&\pto Q_{\x,\tau',t}  
\end{align}
for $\tau'<t$. Thus, (\ref{bxbx}) holds for $\tau=t$ and all $\tau'<t$.  

We next consider the case $\tau'=t$. Using Property~\ref{A1i} for $\tau=t$ 
and the weak law of large numbers, as well as the induction 
hypotheses~\ref{A2}, \ref{A4i}, and \ref{A5i} for all $\tau<t$, we have 
\begin{equation}
\|\boldsymbol{b}_{\x,t}\|_{2}^{2} 
\sim \boldsymbol{\beta}_{\x,t}^{\mathrm{T}}\boldsymbol{B}_{\x,t}^{\mathrm{T}}
\boldsymbol{B}_{\x,t}\boldsymbol{\beta}_{\x,t} + o(1) 
+ \{1 + o(1)\}\|\boldsymbol{q}_{\x,t}^{\perp}\|_{2}^{2} 
\end{equation}
conditioned on $\Theta$ and $\mathfrak{E}_{t,t}^{\x}$. 
For the first term, we utilize the induction hypotheses~\ref{A2} and 
\ref{B5i} for all $\tau<t$ to obtain 
\begin{align}
\boldsymbol{\beta}_{\x,t}^{\mathrm{T}}\boldsymbol{B}_{\x,t}^{\mathrm{T}}
\boldsymbol{B}_{\x,t}\boldsymbol{\beta}_{\x,t}
&\peq \boldsymbol{q}_{\x,t}^{\mathrm{T}}\boldsymbol{Q}_{\x,t}
(\boldsymbol{Q}_{\x,t}^{\mathrm{T}}\boldsymbol{Q}_{\x,t})^{-1}
\boldsymbol{Q}_{\x,t}^{\mathrm{T}}\boldsymbol{q}_{\x,t} + o(1)
\nonumber \\
&= \|\boldsymbol{q}_{\x,t}^{\parallel}\|_{2}^{2} + o(1), 
\end{align}
with $\boldsymbol{q}_{\x,t}^{\parallel}=\boldsymbol{P}_{\boldsymbol{Q}_{\x,t}}^{\parallel}
\boldsymbol{q}_{\x,t}$. Combining these results, we arrive at 
\begin{equation}
\|\boldsymbol{b}_{\x,t}\|_{2}^{2} \sim
\|\boldsymbol{q}_{\x,t}^{\parallel}\|_{2}^{2} 
+ \|\boldsymbol{q}_{\x,t}^{\perp}\|_{2}^{2} + o(1) 
= Q_{\x,t,t} + o(1),  
\end{equation}
where the last follows from the induction hypothesis~\ref{B5i} for 
all $\tau<t$. Thus, Property~\ref{A2} holds for $\tau=t$. 
\end{IEEEproof}

\begin{IEEEproof}[Proof of \ref{A3}]
The latter convergence in Property~\ref{A3} for $\tau=t$ can be proved with 
the former convergence for $\tau=t$ by repeating the proof of the 
latter convergence in Property~\ref{A3} for $\tau=0$. 
Furthermore, the piecewise case in the former convergence can be treated 
in a similar manner to that in the proof of Property~\ref{I3}. 
Thus, we only prove the former convergence in Property~\ref{A3} for $\tau=t$ 
when $\phi, \psi\in\mathcal{L}_{2t+3}$ are everywhere Lipschitz-continuous 
with respect to the first $(2t+2)$ variables. 

Define the separable function 
$\boldsymbol{f}: \mathbb{R}^{M\times(2t+3)}\to\mathbb{R}^{M}$ with 
$f_{m}=[\phi(\boldsymbol{B}_{\z,\tau+1}, 
\boldsymbol{\Sigma}\boldsymbol{B}_{\x,\tau+1}, \boldsymbol{\lambda})]_{m}
[\psi(\boldsymbol{B}_{\z,\tau+1}, \boldsymbol{\Sigma}
\boldsymbol{B}_{\x,\tau+1}$, 
$\boldsymbol{\lambda})]_{m}$. 
Using Properties~\ref{A1o} and \ref{A1i} for $\tau=t$ yields 
\begin{align}
&\frac{1}{M}\phi^{\mathrm{T}}(\boldsymbol{B}_{\z,t+1}, 
\boldsymbol{\Sigma}\boldsymbol{B}_{\x,t+1}, \boldsymbol{\lambda})
\psi(\boldsymbol{B}_{\z,t+1}, 
\boldsymbol{\Sigma}\boldsymbol{B}_{\x,t+1}, \boldsymbol{\lambda}) 
\nonumber \\ 
&=\left\langle
 \boldsymbol{f}(\boldsymbol{B}_{\z,t+1}, 
 \boldsymbol{\Sigma}\boldsymbol{B}_{\x,t+1}, \boldsymbol{\lambda})
\right\rangle
\nonumber \\
&\sim \left\langle
 \boldsymbol{f}(\boldsymbol{B}_{\z,t}, 
 \tilde{\boldsymbol{b}}_{\z,t} + o(1)\boldsymbol{m}_{\z,t-1}, 
 \boldsymbol{\Sigma}\boldsymbol{B}_{\x,t}, 
 \boldsymbol{\Sigma}\boldsymbol{b}_{\x,t}, \boldsymbol{\lambda})
\right\rangle 
\end{align}
conditioned on $\Theta$, $\mathfrak{E}_{t,t}^{\z}$, and 
$\mathfrak{E}_{t,t}^{\x}$, with 
\begin{align}
\tilde{\boldsymbol{b}}_{\z,t} 
&= \boldsymbol{B}_{\z,t}\boldsymbol{\beta}_{\z,t}
+ \boldsymbol{B}_{\z,t}\boldsymbol{o}(1) 
+ \tilde{\boldsymbol{M}}_{\z,t-1}\boldsymbol{o}(1)
\nonumber \\
&+ \{1 + o(1)\}\frac{\|\boldsymbol{q}_{\z,t}^{\perp}\|_{2}}
{\|\boldsymbol{\omega}_{\z,t}\|_{2}}\boldsymbol{\omega}_{\z,t}, 
\end{align}
\begin{align}
\boldsymbol{b}_{\x,t} &= \boldsymbol{B}_{\x,t}\boldsymbol{\beta}_{\x,t}
+ \boldsymbol{B}_{\x,t}\boldsymbol{o}(M/N) 
+ \boldsymbol{M}_{\x,t}\boldsymbol{o}(M/N) 
\nonumber \\
&+ \{1 + o(1)\}\frac{\|\boldsymbol{q}_{\x,t}^{\perp}\|_{2}}
{\|\boldsymbol{\omega}_{\x,t}\|_{2}}\boldsymbol{\omega}_{\x,t}, 
\end{align}
where $\boldsymbol{\omega}_{\z,t}\sim\mathcal{N}(\boldsymbol{0}, 
\boldsymbol{I}_{M})$ independent of $\Theta$, $\mathfrak{E}_{t,t}^{\z}$, 
$\mathfrak{E}_{t,t}^{\x}$, and $\boldsymbol{\omega}_{\x,t}$ while 
$\boldsymbol{\omega}_{\x,t}\sim\mathcal{N}(\boldsymbol{0}, 
\boldsymbol{I}_{N})$ is independent of $\Theta$ and 
$\mathfrak{E}_{t,t}^{\x}$. 

To utilize Lemma~\ref{lemma_bias_elimination}, we confirm all conditions 
in Lemma~\ref{lemma_bias_elimination}. Since 
$\phi, \psi\in\mathcal{L}_{2t+3}$ in 
Definition~\ref{definition_integrability} are 
Lipschitz-continuous with respect to the first $(2t+2)$ variables, 
$f_{m}$ is second-order pseudo-Lipschitz with respect to the same variables. 
Let $L_{m}=\tilde{L}(\lambda_{m})>0$ denote a 
Lipschitz constant of $f_{m}(\cdots, \lambda_{m})$ for fixed $\lambda_{m}$, 
in which $\tilde{L}(\lambda_{m})$ is some almost everywhere continuous 
function of $\lambda_{m}$ with uniformly integrable 
$M^{-1}\sum_{m=1}^{M}\lambda_{m}\tilde{L}(\lambda_{m})$. 
All conditions follow from 
\begin{equation}
\frac{1}{M}\sum_{m=1}^{M}L_{m} \pto \mathbb{E}[\tilde{L}(\Lambda)] <\infty, 
\end{equation}
\begin{equation}
\frac{1}{M}\sum_{m=1}^{M}L_{m}b_{\z,m,\tau}^{2} \pto 
\mathbb{E}[\tilde{L}(\Lambda)B_{\z,\tau}^{2}] < \infty,
\end{equation}
\begin{equation}
\frac{1}{M}\sum_{m=1}^{M}L_{m}m_{\z,m,\tau}^{2} \pto 
\mathbb{E}[\tilde{L}(\Lambda)M_{\z,\tau}^{2}] < \infty,
\end{equation}
\begin{equation}
\frac{1}{M}\sum_{m=1}^{M}L_{m}\omega_{\z,m,t}^{2} \pto 
\mathbb{E}[\tilde{L}(\Lambda)] < \infty, 
\end{equation}
\begin{equation}
\frac{1}{M}\sum_{m=1}^{M}L_{m}
\{[\boldsymbol{\Sigma}\boldsymbol{b}_{\x,\tau}]_{m}\}^{2} 
\pto \mathbb{E}[\Lambda \tilde{L}(\Lambda) B_{\x,\tau}^{2}] < \infty,
\end{equation}
\begin{equation}
\frac{1}{M}\sum_{m=1}^{M}L_{m}\{(M/N)\sigma_{m}m_{\x,m,\tau}\}^{2} 
\pto \mathbb{E}[\Lambda \tilde{L}(\Lambda)M_{\x,\tau}^{2}] < \infty, 
\end{equation}
\begin{equation}
\frac{1}{M}\sum_{m=1}^{M}L_{m}(N^{-1/2}\sigma_{m}\omega_{\x,m,t})^{2} \pto 
\mathbb{E}[\Lambda \tilde{L}(\Lambda)] < \infty 
\end{equation}
for all $\tau<t$, obtained from Assumption~\ref{assumption_A}, the 
uniform integrability of $M^{-1}\sum_{m=1}^{M}\lambda_{m}\tilde{L}(\lambda_{m})$, 
the induction hypothesis~\ref{A3} for $\tau<t$, and 
the weak law of large numbers. 

We can use Lemma~\ref{lemma_bias_elimination} to obtain 
\begin{align}
&\left\langle
 \boldsymbol{f}(\boldsymbol{B}_{\z,t+1}, 
 \boldsymbol{\Sigma}\boldsymbol{B}_{\x,t+1}, \boldsymbol{\lambda})
\right\rangle
\nonumber \\
&\peq \langle \boldsymbol{f}(\boldsymbol{B}_{\z,t}, 
\tilde{\boldsymbol{b}}_{\z,t}, \boldsymbol{\Sigma}\boldsymbol{B}_{\x,t}, 
\boldsymbol{\Sigma}\boldsymbol{b}_{\x,t}, \boldsymbol{\lambda}) \rangle + o(1).  
\end{align}
Repeating the use of Lemma~\ref{lemma_bias_elimination}, we obtain 
\begin{align}
&\langle \boldsymbol{f}(\boldsymbol{B}_{\z,t+1}, 
\boldsymbol{\Sigma}\boldsymbol{B}_{\x,t+1}, \boldsymbol{\lambda}) \rangle 
\peq o(1)
\nonumber \\
&+\left\langle 
 \boldsymbol{f}(\boldsymbol{B}_{\z,t}, 
 \boldsymbol{B}_{\z,t}\boldsymbol{\beta}_{\z,t}
 + \frac{a_{\z, t, M}\boldsymbol{\omega}_{\z,t}}
 {\|\boldsymbol{\omega}_{\z,t}\|_{2}}, 
 \boldsymbol{\Sigma}\boldsymbol{B}_{\x,t}, 
 \boldsymbol{\Sigma}\tilde{\boldsymbol{b}}_{\x,t}, \boldsymbol{\lambda} 
\right\rangle,   
\end{align}
with $a_{\alpha, t, M} = \{1 + o(1)\}\|\boldsymbol{q}_{\alpha,t}^{\perp}\|_{2}$ 
for $\alpha\in\{\z, \x\}$ and 
\begin{equation}
\tilde{\boldsymbol{b}}_{\x,t} = \boldsymbol{B}_{\x,t}\boldsymbol{\beta}_{\x,t}
+ \frac{a_{\x,t,M}}
{\|\boldsymbol{\omega}_{\x,t}\|_{2}}\boldsymbol{\omega}_{\x,t}. 
\end{equation}

To utilize Lemma~\ref{lemma_elimination} with $f(M)=1$, we define 
$\boldsymbol{\phi}$ and $\boldsymbol{\psi}$ as 
\begin{equation}
\boldsymbol{\phi}(\boldsymbol{\omega})
= \frac{1}{\sqrt{M}}\phi(\boldsymbol{B}_{\z,t}, 
\boldsymbol{B}_{\z,t}\boldsymbol{\beta}_{\z,t} + \boldsymbol{\omega}, 
\boldsymbol{\Sigma}\boldsymbol{B}_{\x,t}, 
\boldsymbol{\Sigma}\tilde{\boldsymbol{b}}_{\x,t}, \boldsymbol{\lambda}), 
\end{equation}
\begin{equation}
\boldsymbol{\psi}(\boldsymbol{\omega}) 
= \frac{1}{\sqrt{M}}\psi(\boldsymbol{B}_{\z,t}, 
\boldsymbol{B}_{\z,t}\boldsymbol{\beta}_{\z,t} + \boldsymbol{\omega}, 
\boldsymbol{\Sigma}\boldsymbol{B}_{\x,t}, 
\boldsymbol{\Sigma}\tilde{\boldsymbol{b}}_{\x,t}, \boldsymbol{\lambda}).
\end{equation}
The functions $\phi_{m}, \psi_{m}\in\mathcal{L}_{2t+3}$ 
are Lipschitz-continuous with a common Lipschitz constant 
$L_{m}=M^{-1/2}L(\lambda_{m})$, in which $L(\lambda_{m})$ is some almost 
everywhere continuous function of $\lambda_{m}$ with uniformly integrable 
$M^{-1}\sum_{m=1}^{M}\lambda_{m}L^{2}(\lambda_{m})$. 
We use Assumption~\ref{assumption_A} and the uniform integrability to find 
$f^{2}(M)\sum_{m=1}^{M}L_{m}^{2}
=M^{-1}\sum_{m=1}^{M}L^{2}(\lambda_{m})\pto\mathbb{E}[L^{2}(\Lambda)]$. 
From the induction hypothesis~\ref{B5o} for $\tau=t-1$ we confirm that 
$a_{\z,t,M}^{2}/M$ converges 
in probability to $a_{\z,t}^{2}$ for some constant $a_{\z,t}>0$. Furthermore, 
we use the induction hypothesis~\ref{A3} for $\tau=t-1$ and the weak law of 
large numbers to find the boundedness in probability of 
$\|\boldsymbol{\phi}(a_{\z,t}\boldsymbol{\omega})\|_{2}^{2}$ and 
$\|\boldsymbol{\psi}(a_{\z,t}\boldsymbol{\omega})\|_{2}^{2}$. 
Thus, we can use Lemma~\ref{lemma_elimination} to obtain 
\begin{align}
&\langle \boldsymbol{f}(\boldsymbol{B}_{\z,t+1}, 
\boldsymbol{\Sigma}\boldsymbol{B}_{\x,t+1}, \boldsymbol{\lambda}) \rangle 
\peq o(1)
\nonumber \\
&+ \left\langle 
 \boldsymbol{f}(\boldsymbol{B}_{\z,t}, 
 \boldsymbol{B}_{\z,t}\boldsymbol{\beta}_{\z,t}
 + a_{\z,t}\boldsymbol{\omega}_{\z,t}, 
 \boldsymbol{\Sigma}\boldsymbol{B}_{\x,t}, 
 \boldsymbol{\Sigma}\tilde{\boldsymbol{b}}_{\x,t}, \boldsymbol{\lambda}
\right\rangle.    
\end{align}

We next utilize Lemma~\ref{lemma_elimination} with $f(N)=N^{-1/2}$. 
Using the induction hypothesis~\ref{B5i} for $\tau=t-1$, we confirm that 
$a_{\x,t,M}^{2}$ converges in probability to $a_{\x,t}^{2}$ 
in the sublinear sparsity limit for some constant $a_{\x,t}>0$. 
Define a function $\boldsymbol{\phi}(\boldsymbol{\omega}): 
\mathbb{R}^{N}\to\mathbb{R}^{N}$ with 
$\phi_{n}= M^{-1/2}[\phi(\boldsymbol{B}_{\z,t}, 
\boldsymbol{B}_{\z,t}\boldsymbol{\beta}_{\z,t} + a_{\z,t}\boldsymbol{\omega}_{\z,t}, 
\boldsymbol{\Sigma}\boldsymbol{B}_{\x,t}, 
\boldsymbol{\Sigma}\boldsymbol{B}_{\x,t}\boldsymbol{\beta}_{\x,t}
+ \boldsymbol{\Sigma}\boldsymbol{\omega}, \boldsymbol{\lambda})]_{n}$ 
for all $n\leq M$ and $\phi_{n}=0$ for all $n>M$.  
Similarly, we define $\boldsymbol{\psi}(\boldsymbol{\omega})$. 
The functions $\phi_{n}, \psi_{n}\in\mathcal{L}_{2t+3}$ are 
Lipschitz-continuous with a Lipschitz constant $L_{n}=0$ for $n>M$ and 
$L_{n}=M^{-1/2}\sigma_{n}L(\lambda_{n})$ for $n\leq M$, in which $L(\lambda_{m})$ 
is some almost everywhere continuous function of $\lambda_{m}$ with 
uniformly integrable $M^{-1}\sum_{m=1}^{M}\lambda_{m}L^{2}(\lambda_{m})$. 
We use the definition $\lambda_{m}=N^{-1}\sigma_{m}^{2}$, 
Assumption~\ref{assumption_A}, and the uniform integrability to find 
\begin{equation}
f^{2}(N)\sum_{n=1}^{N}L_{n}^{2} 
= \frac{1}{M}\sum_{m=1}^{M}\lambda_{m}L^{2}(\lambda_{m})
\pto \mathbb{E}\left[
 \Lambda L^{2}(\Lambda)
\right]. 
\end{equation}
Furthermore, we use the induction hypothesis~\ref{A3} for $\tau=t-1$ and 
the weak law of large numbers to find the boundedness in probability of 
$\|\boldsymbol{\phi}(a_{\x,t}\boldsymbol{\omega})\|_{2}^{2}$ and 
$\|\boldsymbol{\psi}(a_{\x,t}\boldsymbol{\omega})\|_{2}^{2}$. 
Thus, we can use Lemma~\ref{lemma_elimination} with $f(N)=N^{-1/2}$ to have 
\begin{align}
&\langle \boldsymbol{f}(\boldsymbol{B}_{\z,t+1}, 
\boldsymbol{\Sigma}\boldsymbol{B}_{\x,t+1}, \boldsymbol{\lambda}) \rangle 
\peq \langle \boldsymbol{f}(\boldsymbol{B}_{\z,t}, 
\boldsymbol{B}_{\z,t}\boldsymbol{\beta}_{\z,t}
+ a_{\z,t}\boldsymbol{\omega}_{\z,t}, 
\nonumber \\
&\left.
 \boldsymbol{\Sigma}\boldsymbol{B}_{\x,t}, 
 \boldsymbol{\Sigma}(\boldsymbol{B}_{\x,t}\boldsymbol{\beta}_{\x,t}
 + N^{-1/2}a_{\x,t}\boldsymbol{\omega}_{\x,t}), 
 \boldsymbol{\lambda}) 
\right\rangle + o(1). \label{A3_tmp}
\end{align}

Define $\boldsymbol{b}_{\z,0}^{\mathrm{G}} = a_{\z,0}\boldsymbol{\omega}_{\z,0}$,  
$\boldsymbol{b}_{\x,0}^{\mathrm{G}} = N^{-1/2}a_{\x,0}\boldsymbol{\omega}_{\x,0}$, and 
\begin{equation}
\boldsymbol{b}_{\z,\tau}^{\mathrm{G}} 
= \boldsymbol{B}_{\z,\tau}^{\mathrm{G}}\boldsymbol{\beta}_{\z,\tau} 
+ a_{\z,\tau}\boldsymbol{\omega}_{\z,\tau}, 
\end{equation}
\begin{equation}
\boldsymbol{b}_{\x,\tau}^{\mathrm{G}} 
= \boldsymbol{B}_{\x,\tau}^{\mathrm{G}}\boldsymbol{\beta}_{\x,\tau} 
+ N^{-1/2}a_{\x,\tau}\boldsymbol{\omega}_{\x,\tau} 
\end{equation}
for $\tau\in\{1,\ldots, t\}$. By definition, 
$\boldsymbol{B}_{\z,t+1}^{\mathrm{G}}$ and $\boldsymbol{B}_{\x,t+1}^{\mathrm{G}}$ 
have zero-mean i.i.d. Gaussian rows, which follow the same distribution  
as that of $\mathcal{B}_{\z,t+1}$ and $\mathcal{B}_{\x,t+1}$, respectively.  
Repeating the derivation of (\ref{A3_tmp}), we arrive at 
\begin{align}
&\langle \boldsymbol{f}(\boldsymbol{B}_{\z,t+1}, 
\boldsymbol{\Sigma}\boldsymbol{B}_{\x,t+1}, \boldsymbol{\lambda}) \rangle 
\peq \langle \boldsymbol{f}(\boldsymbol{B}_{\z,t+1}^{\mathrm{G}}, 
\boldsymbol{\Sigma}\boldsymbol{B}_{\x,t+1}^{\mathrm{G}}, 
\boldsymbol{\lambda}) \rangle + o(1)
\nonumber \\
&\pto \mathbb{E}\left[
 \boldsymbol{f}(\mathcal{B}_{\z,t+1}, 
 \sqrt{\Lambda}\mathcal{B}_{\x,t+1}, \Lambda)
\right], 
\end{align}
where the last convergence follows from the weak law of large numbers, 
Assumption~\ref{assumption_A}, and $\phi, \psi\in\mathcal{L}_{2t+3}$. 
\end{IEEEproof}

\begin{IEEEproof}[Proof of \ref{A4o}]
We first prove (\ref{zb}) for $\tau=t$. Using Property~\ref{A3} for $\tau=t$ 
yields 
\begin{equation}
\frac{1}{M}(\boldsymbol{\Sigma}\boldsymbol{b}_{\x,\tau'})^{\mathrm{T}}
\boldsymbol{b}_{\z,t} 
\pto \mathbb{E}\left[
 \sqrt{\Lambda}B_{\x,\tau'}B_{\z,t}
\right] = 0,
\end{equation}
where the last equality follows from the independence between the zero-mean 
random variables $B_{\x,\tau'}$ and $B_{\z,t}$. 

We next prove (\ref{bzmz}) for $\tau=t$. Using the definition of 
$\boldsymbol{m}_{\z,t}$ in (\ref{mz}) and Property~\ref{A3} for $\tau=t$ 
under Assumptions~\ref{assumption_denoisers_general} and 
\ref{assumption_A_addition}, we have 
\begin{align}
&\frac{1}{M}\boldsymbol{b}_{\z,\tau'}^{\mathrm{T}}\boldsymbol{m}_{\z,t}
\pto \mathbb{E}[B_{\z,\tau'}
\phi_{\z,t}(\mathcal{B}_{\z,t+1}, \sqrt{\Lambda}\mathcal{B}_{\x,t+1}, \Lambda)]
\nonumber \\
&- \sum_{\tau=0}^{t}\bar{\xi}_{\A,t,\tau}^{\z}\mathbb{E}[B_{\z,\tau'}B_{\z,\tau}]
= 0, 
\end{align}
where the last equality follows from Lemma~\ref{lemma_Stein} and 
the definition of $\bar{\xi}_{\A,t,\tau}^{\z}$ in (\ref{xi_A_z_bar_general}). 
\end{IEEEproof}

\begin{IEEEproof}[Proof of \ref{A4i}]
Applying the definition of $\boldsymbol{m}_{\x,t}$ in (\ref{mx}),  
Property~\ref{A2}, and Property~\ref{A3} for all $\tau\leq t$ yields 
\begin{equation}
\boldsymbol{b}_{\x,\tau'}^{\mathrm{T}}\boldsymbol{m}_{\x,t}
\peq \boldsymbol{b}_{\x,\tau'}^{\mathrm{T}}\phi_{\x,t} 
- \sum_{\tau=0}^{t}\bar{\xi}_{\A,t,\tau}^{\x}Q_{\x,\tau',\tau} + o(1). 
\end{equation}
We utilize the definition of $\phi_{\x,t}$ in (\ref{phi_x}), Property~\ref{A2} 
for $\tau=t$, and Property~\ref{A3} for $\tau=t$ under 
Assumptions~\ref{assumption_denoisers_general} and 
\ref{assumption_A_addition} to obtain 
\begin{equation}
\boldsymbol{b}_{\x,\tau'}^{\mathrm{T}}\phi_{\x,t} 
\pto Q_{\x,\tau',t} + \mathbb{E}\left[
 \sqrt{\Lambda}B_{\x,\tau'}\tilde{\phi}_{\x,t}(\sqrt{\Lambda}\mathcal{B}_{\x,t+1}, 
 \mathcal{B}_{\z,t+1}, \Lambda)
\right]. 
\end{equation}
Using Lemma~\ref{lemma_Stein}, $\mathbb{E}[B_{x,\tau'}B_{\x,\tau}]
= Q_{\x,\tau',\tau}$, and the definition of $\bar{\xi}_{\A,t,\tau}^{\x}$ 
in (\ref{xi_A_x_bar_general}) yields  
\begin{align}
\boldsymbol{b}_{\x,\tau'}^{\mathrm{T}}\phi_{\x,t} 
&\pto Q_{\x,\tau',t} + \sum_{\tau=0}^{t}(\bar{\xi}_{\A,t,\tau}^{\x} - \delta_{t,\tau})
Q_{\x,\tau',\tau}
\nonumber \\
&= \sum_{\tau=0}^{t}\bar{\xi}_{\A,t,\tau}^{\x}Q_{\x,\tau',\tau}.
\end{align}
Combining these results, we find that Property~\ref{A4i} holds 
for $\tau=t$. 
\end{IEEEproof}

\begin{IEEEproof}[Proof of \ref{A5o}]
We prove properties for $\tilde{\boldsymbol{M}}_{\z,t+1}$. 
Using the definition of $\boldsymbol{m}_{\z,t}$ in (\ref{mz}) 
and Property~\ref{A3} for $\tau=t$ under 
Assumptions~\ref{assumption_denoisers_general} and 
\ref{assumption_A_addition} yields 
\begin{align}
\frac{1}{M}\tilde{\boldsymbol{m}}^{\mathrm{T}}\boldsymbol{m}_{\z,\tau}
&\pto \mathbb{E}\left[
 \sqrt{\Lambda}B_{\x,0} \left(
  \phi_{\z,\tau} - \sum_{\tau'=0}^{\tau}\bar{\xi}_{\A,\tau,\tau'}^{\z}B_{\z,\tau'}
 \right)
\right]
\nonumber \\
&= \mathbb{E}[\tilde{M}M_{\z,\tau}] \label{m_tilde_m_t}
\end{align}
for all $\tau\leq t$. Similarly, we have  
\begin{align}
\frac{1}{M}&\boldsymbol{m}_{\z,\tau'}^{\mathrm{T}}\boldsymbol{m}_{\z,t} 
\pto \mathbb{E}\left[
 \left(
  \phi_{\z,\tau'} - \sum_{\tau=0}^{\tau'}\bar{\xi}_{\A,\tau',\tau}^{\z}B_{\z,\tau}
 \right)
\right. \nonumber \\
&\left.
 \cdot\left(
  \phi_{\z,t} - \sum_{\tau=0}^{t}\bar{\xi}_{\A,t,\tau}^{\z}B_{\z,\tau}
 \right)
\right] 
= \mathbb{E}[M_{\z,\tau'}M_{\z,t}]. \label{m_tau_m_t}
\end{align} 

To prove (\ref{A5o_positivity}) for $\tau=t$, it is sufficient to confirm that 
$M^{-1}\|\boldsymbol{P}_{\tilde{\boldsymbol{M}}_{\z,t}}^{\perp}
\boldsymbol{m}_{\z,t}\|_{2}^{2}$ is strictly positive in the sublinear 
sparsity limit~\cite[Lemmas~8 and 9]{Bayati11}. 
Using (\ref{m_tilde_m_t}) and (\ref{m_tau_m_t}) under the induction 
hypothesis~\ref{A5o} for $\tau=t-1$ yields  
\begin{align}
&\frac{1}{M}\|\boldsymbol{P}_{\tilde{\boldsymbol{M}}_{\z,t}}^{\perp}
\boldsymbol{m}_{\z,t}\|_{2}^{2}
\pto \mathbb{E}[M_{\z,t}^{2}] 
\nonumber \\
&- \mathbb{E}\left[
 M_{\z,t}\bar{\boldsymbol{M}}_{\z,t}^{\mathrm{T}}
\right]\left(
 \mathbb{E}[\bar{\boldsymbol{M}}_{\z,t}
 \bar{\boldsymbol{M}}_{\z,t}^{\mathrm{T}}]
\right)^{-1}
\mathbb{E}[ \bar{\boldsymbol{M}}_{\z,t}M_{\z,t} ], 
\end{align}
where we define 
$\bar{\boldsymbol{M}}_{\z,t}=(M_{\z,0},\ldots,M_{\z,t-1},
\tilde{M})^{\mathrm{T}}$ if $\tilde{\boldsymbol{m}}$ and $\boldsymbol{m}_{\z,0}$ 
are linearly independent. Otherwise, we define 
$\bar{\boldsymbol{M}}_{\z,t}=(M_{\z,0},\ldots,M_{\z,t-1})^{\mathrm{T}}$. 
We utilize Lemma~\ref{lemma_positive_definite} to obtain 
\begin{align}
&\frac{1}{M}\|\boldsymbol{P}_{\tilde{\boldsymbol{M}}_{\z,t}}^{\perp}
\boldsymbol{m}_{\z,t}\|_{2}^{2}
\nonumber \\
&\geq \mathbb{E}\left[
 (M_{\z,t} - \mathbb{E}[M_{\z,t} | \bar{\boldsymbol{M}}_{\z,t}])^{2} 
\right] \quad \hbox{in probability,}
\end{align}
which is strictly positive under the first condition in 
Assumption~\ref{assumption_SE_general} for $\tau'=t$. 
\end{IEEEproof}

\begin{IEEEproof}[Proof of \ref{A5i}]
We prove properties for $\boldsymbol{M}_{\x,t+1}$. 
Using the definition of $\boldsymbol{m}_{\x,t}$ in (\ref{mx}), 
Property~\ref{A3}, and Property~\ref{A4i} for all $\tau\leq t$, we have 
\begin{align}
&\frac{M}{N}\boldsymbol{m}_{\x,\tau'}^{\mathrm{T}}\boldsymbol{m}_{\x,t}
\peq \frac{M}{N}\phi_{\x,\tau'}^{\mathrm{T}}\left(
 \phi_{\x,t} - \sum_{\tau=0}^{t}\bar{\xi}_{\A,t,\tau}^{\x}\boldsymbol{b}_{\x,\tau}
\right) + o(1)
\nonumber \\
&\peq \frac{1}{N}\tilde{\phi}_{\x,\tau'}^{\mathrm{T}}\boldsymbol{\Sigma}\left(
 \phi_{\x,t} - \sum_{\tau=0}^{t}\bar{\xi}_{\A,t,\tau}^{\x}\boldsymbol{b}_{\x,\tau}
\right) + o(1)
\nonumber \\
&\peq \frac{1}{M}\tilde{\phi}_{\x,\tau'}^{\mathrm{T}}
\boldsymbol{\Lambda}\tilde{\phi}_{\x,t} + o(1), 
\end{align}
which converges to $\mathbb{E}[M_{\x,\tau'}M_{\x,t}]$, because of 
Property~\ref{A3} for $\tau=t$ under 
Assumptions~\ref{assumption_denoisers_general} and 
\ref{assumption_A_addition}. 
In the derivation of the second equality, we have used the definition of 
$\phi_{\x,t}$ in (\ref{phi_x}) and Property~\ref{A4i} for $\tau=t$. 
The last equality follows from Property~\ref{A3} for $\tau=t$ and 
the definition $\boldsymbol{\Lambda}=
N^{-1}\boldsymbol{\Sigma}\boldsymbol{\Sigma}^{\mathrm{T}}$. 
Repeat the proof of (\ref{A5o_positivity}) for $\tau=t$ to prove 
(\ref{A5i_positivity}) for 
$\tau=t$ under the second condition in Assumption~\ref{assumption_SE_general} 
for $\tau'=t$. 
\end{IEEEproof}

\subsection{Module~B for $\tau=t$}
\label{appen_module_B_t}
\begin{IEEEproof}[Proofs of \ref{B1o} and \ref{B1i}]
Repeat the proofs of Properties~\ref{A1o} and \ref{A1i} for $\tau=t$ 
to prove Properties~\ref{B1o} and \ref{B1i} for $\tau=t$, respectively.   
\end{IEEEproof}

\begin{IEEEproof}[Proof of \ref{B2}]
Property~\ref{B2} for $\tau=t$ can be proved by 
repeating the proof of Property~\ref{A2} for $\tau=t$. 
\end{IEEEproof}

\begin{IEEEproof}[Proof of \ref{B3o}]
Repeat the proof of Property~\ref{A3} for $\tau=t$. 
\end{IEEEproof}

\begin{IEEEproof}[Proof of \ref{B3i}]
Let $\{\boldsymbol{\omega}_{\x,\tau}\in\mathbb{R}^{N}\}_{\tau=0}^{t}$ denote 
independent standard Gaussian random vectors. Each 
$\boldsymbol{\omega}_{\x,\tau}$ is independent of $\Theta$ and 
$\mathfrak{E}_{\tau,\tau+1}^{\x}$. Define 
$\boldsymbol{h}_{\x,0}^{\mathrm{G}}=N^{-1/2}a_{0,N}\boldsymbol{\omega}_{\x,0}$ and 
\begin{equation}
\boldsymbol{h}_{\x,\tau}^{\mathrm{G}} 
= \boldsymbol{H}_{\x,\tau}^{\mathrm{G}}\boldsymbol{\alpha}_{\x,\tau}
+ \frac{a_{\tau,N}}{\sqrt{N}}\boldsymbol{\omega}_{\x,\tau}  
\end{equation}
for $\tau\in\{1,\ldots,t\}$, with 
$a_{0,N}=\{1 + o(1)\}\|\boldsymbol{m}_{\x,0}\|_{2}$ and 
$a_{\tau,N}=\{1 + o(1)\}\|\boldsymbol{m}_{\x,\tau}^{\perp}\|_{2}$ for $\tau>0$. 
By definition, we have $\boldsymbol{H}_{\x,t+1}^{\mathrm{G}}
\sim\boldsymbol{\Omega}_{t+1,N}$ in Assumption~\ref{assumption_inner_general}. 
We first prove 
$\|\boldsymbol{h}_{\x,\tau} - \boldsymbol{h}_{\x,\tau}^{\mathrm{G}}\|_{2}\pto0$ 
for all $\tau\in\{0,\ldots,t\}$. 

The proof is by induction. For $\tau=0$, we use Property~\ref{B1i} for 
$\tau=0$, Proposition~\ref{proposition1}, and Property~\ref{A5i} for $\tau=0$ 
to obtain 
\begin{align}
\|\boldsymbol{h}_{\x,0} - \boldsymbol{h}_{\x,0}^{\mathrm{G}}\|_{2}
\sim \left\|
 o(1)\boldsymbol{q}_{\x,0}
 + {\cal O}(M^{-1/2})\frac{\boldsymbol{\omega}_{\x,0}}{\sqrt{N}}
\right\|_{2}
\nonumber \\
\leq o(1)\|\boldsymbol{q}_{\x,0}\|_{2} + {\cal O}(M^{-1/2})
\frac{\|\boldsymbol{\omega}_{\x,0}\|_{2}}{\sqrt{N}} \pto 0
\end{align}
conditioned on $\Theta$ and $\mathfrak{E}_{0,1}^{\x}$. In the derivation of 
the inequality, we have used the triangle inequality. The last convergence 
follows from Property~\ref{I5} and the weak law of large numbers. 

For some $\tau_{0}\in\{1,\ldots,t\}$, suppose that $\|\boldsymbol{h}_{\x,\tau} 
- \boldsymbol{h}_{\x,\tau}^{\mathrm{G}}\|_{2}\pto0$ holds  
for all $\tau\in\{0,\ldots,\tau_{0}-1\}$. We prove 
$\|\boldsymbol{h}_{\x,\tau_{0}} 
- \boldsymbol{h}_{\x,\tau_{0}}^{\mathrm{G}}\|_{2}\pto0$. 
Using Property~\ref{B1i} for $\tau=\tau_{0}$, Proposition~\ref{proposition1}, 
and Property~\ref{A5i} for $\tau=\tau_{0}$ yields 
\begin{align}
\|\boldsymbol{h}_{\x,\tau_{0}} &- \boldsymbol{h}_{\x,\tau_{0}}^{\mathrm{G}}\|_{2}
\sim \Big\|(\boldsymbol{H}_{\x,\tau_{0}} - \boldsymbol{H}_{\x,\tau_{0}}^{\mathrm{G}})
\boldsymbol{\alpha}_{\x,\tau_{0}}
+ \boldsymbol{H}_{\x,\tau_{0}}\boldsymbol{o}(1) 
\nonumber \\
&+ \left.
 \boldsymbol{Q}_{\x,\tau_{0}+1}\boldsymbol{o}(1) 
 + {\cal O}(M^{-1/2})\frac{\boldsymbol{\omega}_{\x,\tau_{0}}}{\sqrt{N}}
\right\|_{2} \pto 0 
\end{align}
conditioned on $\Theta$ and $\mathfrak{E}_{\tau_{0},\tau_{0}+1}^{\x}$, where the last 
convergence follows from the triangle inequality, the induction hypothesis 
$\|\boldsymbol{h}_{\x,\tau} - \boldsymbol{h}_{\x,\tau}^{\mathrm{G}}\|_{2}\pto0$ 
for all $\tau\in\{0,\ldots,\tau_{0}-1\}$, Property~\ref{B2} 
for $\tau=\tau_{0}-1$, the induction hypothesis~\ref{B5i} for $\tau=\tau_{0}-1$, 
and the weak law of large numbers 
for $\|\boldsymbol{\omega}_{\x,\tau_{0}}\|_{2}/\sqrt{N}$. Thus, 
$\|\boldsymbol{h}_{\x,\tau} - \boldsymbol{h}_{\x,\tau}^{\mathrm{G}}\|_{2}\pto0$ 
holds for all $\tau\in\{0,\ldots,t\}$. 

We only prove (\ref{psi_psi}) for $\tau=t$ since (\ref{x_psi}) and 
(\ref{h_psi}) for $\tau=t$ can be proved in a similar manner. To utilize 
Lemma~\ref{lemma_bias_elimination_inner} for 
$\boldsymbol{\phi}=\psi_{\x,\tau'}(\boldsymbol{\Omega}_{\tau'+1}, 
\boldsymbol{x})$, $\tilde{\boldsymbol{\phi}}
=\psi_{\x,\tau'}(\boldsymbol{H}_{\x,\tau'+1}, \boldsymbol{x})$, 
$\boldsymbol{\psi}=\psi_{\x,t}(\boldsymbol{\Omega}_{t+1}, 
\boldsymbol{x})$, and 
$\tilde{\boldsymbol{\psi}}=\psi_{\x,t}(\boldsymbol{H}_{\x,t+1}, \boldsymbol{x})$, 
we confirm all conditions in Lemma~\ref{lemma_bias_elimination_inner}. 
Assumption~\ref{assumption_inner_general} implies 
the boundedness in probability of $\|\boldsymbol{\phi}\|_{2}$ and 
$\|\boldsymbol{\psi}\|_{2}$. Furthermore, we use 
$\|\boldsymbol{h}_{\x,\tau} - \boldsymbol{h}_{\x,\tau}^{\mathrm{G}}\|_{2}\pto0$ 
for all $\tau\in\{0,\ldots,t\}$, $\boldsymbol{H}_{\x,t+1}^{\mathrm{G}}
\sim\boldsymbol{\Omega}_{t+1,N}$, and 
Assumption~\ref{assumption_inner_general} to find 
$\|\Delta\boldsymbol{\phi}\|_{2}\pto0$ and 
$\|\Delta\boldsymbol{\psi}\|_{2}\pto0$. Thus, we can use  
Lemma~\ref{lemma_bias_elimination_inner} to obtain 
\begin{align}
&\psi_{\x,\tau'}^{\mathrm{T}}(\boldsymbol{H}_{\x,\tau'+1}, \boldsymbol{x})
\psi_{\x,t}(\boldsymbol{H}_{\x,t+1}, \boldsymbol{x})
\nonumber \\
&\peq \psi_{\x,\tau'}^{\mathrm{T}}(\boldsymbol{\Omega}_{\tau'+1}, \boldsymbol{x})
\psi_{\x,t}(\boldsymbol{\Omega}_{t+1}, \boldsymbol{x}) + o(1)
\nonumber \\
&\peq \mathbb{E}\left[
 \psi_{\x,\tau'}^{\mathrm{T}}(\boldsymbol{\Omega}_{\tau'+1}, \boldsymbol{x})
 \psi_{\x,t}(\boldsymbol{\Omega}_{t+1}, \boldsymbol{x})
\right] + o(1),
\end{align}
where the last equality follows from 
Assumption~\ref{assumption_inner_general}. 
Thus, Property~\ref{B3i} holds for $\tau=t$. 
\end{IEEEproof}

\begin{IEEEproof}[Proof of \ref{B4o}]
We first prove (\ref{hzqz}) for $\tau=t$. Using the definition of 
$\boldsymbol{q}_{\z,t+1}$ in (\ref{qz}) and Property~\ref{B3o} for $\tau=t$ 
under Assumption~\ref{assumption_denoisers_general} yields 
\begin{align}
&\frac{\boldsymbol{h}_{\z,\tau'}^{\mathrm{T}}\boldsymbol{q}_{\z,t+1}}{M}
\pto \mathbb{E}\left[
 H_{\z,\tau'}\psi_{\z,t}(\mathcal{H}_{\z,t+1}, Z, W)
\right] 
\nonumber \\
&- \sum_{\tau=0}^{t}\bar{\xi}_{\B,t,\tau}^{\z}\mathbb{E}[H_{\z,\tau'}H_{\z,\tau}] 
- \bar{\eta}_{\B,t}\mathbb{E}[H_{\z,\tau'}\tilde{H}]
= 0,  
\end{align}
where the last equality follows from Lemma~\ref{lemma_Stein}, 
$\bar{\xi}_{\B,t,\tau}^{\z}$ in (\ref{xi_B_z_bar_general}), and 
$\bar{\eta}_{\B,t}$ in (\ref{eta_B_bar_general}). 

We next prove (\ref{zq}) for $\tau=t$. Using the definition of 
$\boldsymbol{q}_{\z,t+1}$ in (\ref{qz}) and  
Property~\ref{B3o} for $\tau=t$ under 
Assumption~\ref{assumption_denoisers_general} yields 
\begin{align}
\frac{\tilde{\boldsymbol{h}}^{\mathrm{T}}\boldsymbol{q}_{\z,t+1}}{M} 
&\pto \mathbb{E}\left[
 \tilde{H}\psi_{\z,t}(\mathcal{H}_{\z,t+1}, \tilde{H}, W)
\right]
\nonumber \\
&- \sum_{\tau=0}^{t}\bar{\xi}_{\B,t,\tau}^{\z}\mathbb{E}[\tilde{H}H_{\z,\tau}]
- \bar{\eta}_{\B,t}\mathbb{E}[\tilde{H}^{2}]
= 0,
\end{align}
where the last follows from Lemma~\ref{lemma_Stein}, 
$\bar{\xi}_{\B,t,\tau}^{\z}$ in (\ref{xi_B_z_bar_general}), and 
$\bar{\eta}_{\B,t}$ in (\ref{eta_B_bar_general}). Thus, 
Property~\ref{B4o} holds for $\tau=t$. 
\end{IEEEproof}

\begin{IEEEproof}[Proof of \ref{B4i}]
We first prove (\ref{hx}) for $\tau=t$. Using Property~\ref{B1i} for 
$\tau=t$, as well as the induction hypotheses~\ref{B4i} and \ref{B5i} 
for all $\tau<t$, we have 
\begin{equation}
\boldsymbol{x}^{\mathrm{T}}\boldsymbol{h}_{\x,t} 
\sim o(1) + \{1 + o(1)\}\frac{\|\boldsymbol{m}_{\x,t}^{\perp}\|_{2}}
{\|\boldsymbol{\omega}_{\x,t}\|_{2}}\boldsymbol{x}^{\mathrm{T}}
\boldsymbol{\omega}_{\x,t}  
\end{equation}
conditioned on $\Theta$ and $\mathfrak{E}_{t,t+1}^{\x}$, 
with $\boldsymbol{\omega}_{\x,t}\sim\mathcal{N}(\boldsymbol{0}, 
\boldsymbol{I}_{N})$ independent of $\Theta$ and 
$\mathfrak{E}_{t,t+1}^{\x}$. Repeating the proof of (\ref{hx}) for $\tau=0$, 
we arrive at $\boldsymbol{x}^{\mathrm{T}}\boldsymbol{h}_{\x,t}\pto 0$. 

We next prove (\ref{hxqx}) for $\tau=t$. 
Using the definition of $\boldsymbol{q}_{\x,t+1}$ in (\ref{qx}) with 
$\xi_{\B,t,\tau}^{\x}$ replaced by $\bar{\xi}_{\B,t,\tau}^{\x}$, as well 
as Properties~\ref{B2} for all $\tau\leq t$, we have  
\begin{equation}
\boldsymbol{h}_{\x,\tau'}^{\mathrm{T}}\boldsymbol{q}_{\x,t+1}
\peq \boldsymbol{h}_{\x,\tau'}^{\mathrm{T}}\psi_{\x,t} 
- \sum_{\tau=0}^{t}\bar{\xi}_{\B,t,\tau}^{\x}\mathbb{E}[M_{\x,\tau'}M_{\x,\tau}] 
+ o(1).  
\end{equation}
Utilizing Property~\ref{B3i} for $\tau=t$ yields  
\begin{align}
&\boldsymbol{h}_{\x,\tau'}^{\mathrm{T}}\psi_{\x,t}
\peq \sum_{n=1}^{N}\mathbb{E}\left[
 \omega_{n,\tau'}\psi_{\x,t}(\omega_{n,0},\ldots,\omega_{n,t}, x_{n})
\right] + o(1)
\nonumber \\
&= \sum_{\tau=0}^{t}\bar{\xi}_{\B,t,\tau}^{\x}\mathbb{E}[M_{\x,\tau'}M_{\x,\tau}]
+ o(1), 
\end{align}
where the last equality follows from Lemma~\ref{lemma_Stein} under 
Assumption~\ref{assumption_inner_general}, 
$\mathbb{E}[\omega_{n,\tau'}\omega_{n,\tau}]
=M^{-1}\mathbb{E}[M_{\x,\tau'}M_{\x,\tau}]$, and the definition of 
$\bar{\xi}_{\B,t,\tau}^{\x}$ in (\ref{xi_B_x_bar_general}). 
Combining these results, 
we arrive at (\ref{hxqx}) for $\tau=t$. 
\end{IEEEproof}

\begin{IEEEproof}[Proof of \ref{B5o}]
We evaluate 
$M^{-1}\boldsymbol{q}_{\z,\tau'+1}^{\mathrm{T}}\boldsymbol{q}_{\z,t+1}$ 
for $\tau'>0$. 
Using the definition of $\boldsymbol{q}_{\z,t+1}$ in (\ref{qz}) and 
Property~\ref{B3o} for $\tau=t$ under 
Assumption~\ref{assumption_denoisers_general} yields 
\begin{align}
&\frac{\boldsymbol{q}_{\z,\tau'+1}^{\mathrm{T}}\boldsymbol{q}_{\z,t+1}}{M} 
\pto \mathbb{E}\left[
 \left(
  \psi_{\z,\tau'} - \sum_{\tau=0}^{\tau'}\bar{\xi}_{\B,\tau',\tau}^{\z}H_{\z,\tau}
  - \bar{\eta}_{\B,\tau'}\tilde{H}
 \right)
\right.
\nonumber \\
&\cdot \left. 
 \left(
  \psi_{\z,t} - \sum_{\tau=0}^{t}\bar{\xi}_{\B,t,\tau}^{\z}H_{\z,\tau}
  - \bar{\eta}_{\B,t}\tilde{H}
 \right)
\right],
\end{align}
which is equal to $\mathbb{E}[Q_{\z,\tau'+1}Q_{\z,t+1}]$. 
Repeat the proof of (\ref{A5o_positivity}) for $\tau=t$ to prove 
(\ref{B5o_positivity}) for $\tau=t$ under the third condition in 
Assumption~\ref{assumption_SE_general} for $\tau'=t$. 
\end{IEEEproof}

\begin{IEEEproof}[Proof of \ref{B5i}]
We evaluate $\boldsymbol{q}_{\x,\tau'+1}\boldsymbol{q}_{\x,t+1}$ only 
for $\tau'>0$. Since $\xi_{\B,t,\tau}^{\x}$ has been replaced with 
$\bar{\xi}_{\B,t,\tau}^{\x}$, we use the definition of 
$\boldsymbol{q}_{\x,t+1}$ in (\ref{qx}) and 
Property~\ref{B4i} for $\tau=t$ to obtain  
\begin{align}
&\boldsymbol{q}_{\x,\tau'+1}^{\mathrm{T}}\boldsymbol{q}_{\x,t+1}
\peq \psi_{\x,\tau'}^{\mathrm{T}}\left(
 \psi_{\x,t} - \frac{M}{N}\sum_{\tau=0}^{t}\bar{\xi}_{\B,t,\tau}^{\x}
 \boldsymbol{h}_{\x,\tau}
\right) + o(1)
\nonumber \\
&\to \mathbb{E}\left[
 \psi_{\x,\tau'}^{\mathrm{T}}(\boldsymbol{\Omega}_{\x,\tau'+1}, \boldsymbol{x})
 \psi_{\x,t}(\boldsymbol{\Omega}_{\x,t+1}, \boldsymbol{x})
\right], 
\end{align}
which is equal to $Q_{\x,\tau'+1,t+1}$, 
where the convergence in probability follows from Property~\ref{B3i} 
for $\tau=t$. Repeat the proof of (\ref{A5o_positivity}) for $\tau=t$ to prove 
(\ref{B5i_positivity}) for $\tau=t$ under the last condition in 
Assumption~\ref{assumption_SE_general} for $\tau'=t$. 
\end{IEEEproof}

\section{Proofs of Technical Results}
\subsection{Proof of Lemma~\ref{lemma_bias_elimination}}
\label{proof_lemma_bias_elimination}
We use the second-order pseudo-Lipschitz assumption for 
$\boldsymbol{f}$ and the upper bound $\|\cdot\|_{2}\leq\|\cdot\|_{1}$ 
to obtain 
\begin{align}
|\langle\Delta\boldsymbol{f}\rangle|  
&\leq \frac{|\delta_{M}|}{M}\sum_{m=1}^{M}L_{m}|a_{m,0}|
+ \frac{\delta_{M}^{2}}{M}\sum_{m=1}^{M}L_{m}a_{m,0}^{2}
\nonumber \\
&+ \frac{|\delta_{M}|}{M}\sum_{\tau=1}^{t}\sum_{m=1}^{M}L_{m}|a_{m,0}a_{m,\tau}|. 
\end{align}
Since $\lim_{M\to\infty}M^{-1}\sum_{m=1}^{M}L_{m}a_{m,0}^{2}$ is bounded 
in probability, the assumption $\delta_{M}\pto0$ implies that 
the second term converges to zero. 

For the first term, we use the Cauchy-Schwarz inequality to have 
\begin{align}
&\frac{|\delta_{M}|}{M}\sum_{m=1}^{M}L_{m}|a_{m,0}|
= \frac{|\delta_{M}|}{M}\sum_{m=1}^{M}L_{m}^{1/2}(L_{m}a_{m,0}^{2})^{1/2} 
\nonumber \\
&\leq |\delta_{M}|\left(
 \frac{1}{M}\sum_{m=1}^{M}L_{m}\frac{1}{M}\sum_{m=1}^{M}L_{m}a_{m,0}^{2} 
\right)^{1/2}
\pto 0,
\end{align}
where the last convergence follows from $\delta_{M}\pto0$, as well as 
the boundedness assumption of $M^{-1}\sum_{m=1}^{M}L_{m}$ and 
$M^{-1}\sum_{m=1}^{M}L_{m}a_{m,0}^{2}$ in the limit $M\to\infty$. 

For the last term, similarly, using the Cauchy-Schwarz inequality yields 
\begin{align}
&\frac{|\delta_{M}|}{M}\sum_{m=1}^{M}L_{m}|a_{m,0}a_{m,\tau}|
\nonumber \\
&\leq |\delta_{M}|\left(
 \frac{1}{M}\sum_{m=1}^{M}L_{m}a_{m,0}^{2}
 \frac{1}{M}\sum_{m=1}^{M}L_{m}a_{m,\tau}^{2}
\right)^{1/2}\pto 0 
\end{align}
for all $\tau\in\{1, \ldots, t\}$. 
Thus, Lemma~\ref{lemma_bias_elimination} holds.

\subsection{Proof of Lemma~\ref{lemma_modified_representation}}
\label{proof_lemma_modified_representation}
We use Lemma~\ref{lemma_representation} for $\boldsymbol{V}\boldsymbol{a}$ 
to represent 
$\boldsymbol{\Phi}_{\boldsymbol{M}}^{\perp}\boldsymbol{V}\boldsymbol{a}$ as 
\begin{equation}
\boldsymbol{\Phi}_{\boldsymbol{M}}^{\perp}
\boldsymbol{V}\boldsymbol{a}
\sim \frac{\|\boldsymbol{a}\|_{2}}{\|\boldsymbol{\omega}_{N-t}\|_{2}}
\boldsymbol{\Phi}_{\boldsymbol{M}}^{\perp}\boldsymbol{\omega}_{N-t},
\end{equation}
where $\boldsymbol{\omega}_{N-t}\sim\mathcal{N}(\boldsymbol{0}, 
\boldsymbol{I}_{N-t})$ is independent of $\boldsymbol{a}$ and $\boldsymbol{M}$. 
Let $\boldsymbol{\omega}_{0}=[\omega_{N-t+1},\ldots,\omega_{N}]^{\mathrm{T}}
\sim\mathcal{N}(\boldsymbol{0}, \boldsymbol{I}_{t})$ denote a standard 
Gaussian random vector that is independent of 
$\boldsymbol{\omega}_{N-t}$, $\boldsymbol{a}$, and $\boldsymbol{M}$. 
Then, we have 
$\boldsymbol{\omega}_{N}\sim\mathcal{N}(\boldsymbol{0}, \boldsymbol{I}_{N})$ 
independent of $\boldsymbol{a}$ and $\boldsymbol{M}$. 
For the orthogonal matrix 
$\boldsymbol{\Phi}_{\boldsymbol{M}}=[\boldsymbol{\Phi}_{\boldsymbol{M}}^{\perp}, 
\boldsymbol{\Phi}_{\boldsymbol{M}}^{\parallel}]$, 
\begin{align}
\frac{\|\boldsymbol{a}\|_{2}}{\|\boldsymbol{\omega}_{N-t}\|_{2}}
\boldsymbol{\Phi}_{\boldsymbol{M}}^{\perp}\boldsymbol{\omega}_{N-t}
&= \frac{\|\boldsymbol{a}\|_{2}}{\|\boldsymbol{\omega}_{N-t}\|_{2}}\left(
 \boldsymbol{\Phi}_{\boldsymbol{M}}\boldsymbol{\omega}_{N} 
 - \boldsymbol{\Phi}_{\boldsymbol{M}}^{\parallel}\boldsymbol{\omega}_{0}
\right)
\nonumber \\
&= \boldsymbol{M}\boldsymbol{v}
+ \frac{\|\boldsymbol{a}\|_{2}}{\|\boldsymbol{\omega}_{N-t}\|_{2}}
\boldsymbol{\Phi}_{\boldsymbol{M}}\boldsymbol{\omega}_{N},
\label{decomposition}
\end{align}
with
\begin{equation}
\boldsymbol{v} = -\frac{\|\boldsymbol{a}\|_{2}}{\|\boldsymbol{\omega}_{N-t}\|_{2}}
(\boldsymbol{M}^{\mathrm{T}}\boldsymbol{M})^{-1}
\boldsymbol{M}^{\mathrm{T}}\boldsymbol{\Phi}_{\boldsymbol{M}}^{\parallel}
\boldsymbol{\omega}_{0}, 
\end{equation}
where the representation of $\boldsymbol{v}$ follows from 
$(\boldsymbol{\Phi}_{\boldsymbol{M}}^{\parallel})^{\mathrm{T}}
\boldsymbol{\Phi}_{\boldsymbol{M}}^{\parallel}=\boldsymbol{I}_{t}$ and 
$\boldsymbol{\Phi}_{\boldsymbol{M}}^{\parallel}
(\boldsymbol{\Phi}_{\boldsymbol{M}}^{\parallel})^{\mathrm{T}}
=\boldsymbol{P}_{\boldsymbol{M}}^{\parallel}
=\boldsymbol{M}(\boldsymbol{M}^{\mathrm{T}}\boldsymbol{M})^{-1}
\boldsymbol{M}^{\mathrm{T}}$. 

For the first term in (\ref{decomposition}), we evaluate the expectation of 
$\|\boldsymbol{v}\|_{2}^{2}$ over $\boldsymbol{\omega}_{0}$ as 
\begin{align}
&\mathbb{E}_{\boldsymbol{\omega}_{0}}\left[
 \|\boldsymbol{v}\|_{2}^{2}
\right]
= \frac{\|\boldsymbol{a}\|_{2}^{2}}{\|\boldsymbol{\omega}_{N-t}\|_{2}^{2}}
\mathrm{Tr}\left\{
 (\boldsymbol{M}^{\mathrm{T}}\boldsymbol{M})^{-1}
\right\}
\nonumber \\
&\leq \frac{t\|\boldsymbol{a}\|_{2}^{2}}
{\|\boldsymbol{\omega}_{N-t}\|_{2}^{2}\lambda_{\mathrm{min}}
(\boldsymbol{M}^{\mathrm{T}}\boldsymbol{M})}
\peq \frac{t\|\boldsymbol{a}\|_{2}^{2}}
{N\lambda_{\mathrm{min}}
(\boldsymbol{M}^{\mathrm{T}}\boldsymbol{M})} + o(1),
\end{align}
where the last equality follows from the weak law of large 
numbers $N^{-1}\|\boldsymbol{\omega}_{N-t}\|_{2}^{2}\pto1$. In particular, 
we have $\boldsymbol{v}\peq\boldsymbol{o}(1)$ if $\|\boldsymbol{a}\|_{2}^{2}/
\{N\lambda_{\mathrm{min}}(\boldsymbol{M}^{\mathrm{T}}\boldsymbol{M})\}\pto0$
holds. 

The second term in (\ref{decomposition}) reduces to 
\begin{align}
\frac{\|\boldsymbol{a}\|_{2}}{\|\boldsymbol{\omega}_{N-t}\|_{2}}
\boldsymbol{\Phi}_{\boldsymbol{M}}\boldsymbol{\omega}_{N} 
&= \left(
 1 - \frac{\|\boldsymbol{\omega}_{0}\|_{2}^{2}}
 {\|\boldsymbol{\omega}_{N}\|_{2}^{2}}
\right)^{-1/2}\frac{\|\boldsymbol{a}\|_{2}}{\|\boldsymbol{\omega}_{N}\|_{2}}
\boldsymbol{\Phi}_{\boldsymbol{M}}\boldsymbol{\omega}_{N} 
\nonumber \\
&\sim \{1 + o(1)\} 
\frac{\|\boldsymbol{a}\|_{2}}{\|\boldsymbol{\omega}_{N}\|_{2}}
\boldsymbol{\omega}_{N},
\end{align}
where we have used $\|\boldsymbol{\omega}_{N}\|_{2}^{-1}
\boldsymbol{\Phi}_{\boldsymbol{M}}\boldsymbol{\omega}_{N}
\sim\|\boldsymbol{\omega}_{N}\|_{2}^{-1}\boldsymbol{\omega}_{N}$ and 
$\|\boldsymbol{\omega}_{0}\|_{2}^{2}
/\|\boldsymbol{\omega}_{N}\|_{2}^{2}\pto 0$. Combining these results, 
we arrive at Lemma~\ref{lemma_modified_representation}. 

\subsection{Proof of Proposition~\ref{proposition1}}
\label{proof_proposition1}
By definition we have 
\begin{align}
&\mathbb{E}\left[
 \left|
  \frac{1}{\|\boldsymbol{w}\|_{2}} - \frac{1}{\sqrt{N}}
 \right|^{p}
\right]
= \mathbb{E}\left[
 \frac{|\sqrt{N} - \|\boldsymbol{w}\|_{2}|^{p}}{N^{p/2}\|\boldsymbol{w}\|_{2}^{p}}
\right]
\nonumber \\
&= \mathbb{E}\left[
 \frac{|N - \|\boldsymbol{w}\|_{2}^{2}|^{p}}
 {N^{p/2}\|\boldsymbol{w}\|_{2}^{p}(\sqrt{N} + \|\boldsymbol{w}\|_{2})^{p}}
\right]
\leq \mathbb{E}\left[
 \frac{|N - \|\boldsymbol{w}\|_{2}^{2}|^{p}}{N^{p}\|\boldsymbol{w}\|_{2}^{p}}
\right], 
\end{align}
where the last inequality follows from $\|\boldsymbol{w}\|_{2}\geq0$. 
Using the Cauchy-Schwarz inequality yields
\begin{align}
\mathbb{E}\left[
 \frac{|N - \|\boldsymbol{w}\|_{2}^{2}|^{p}}{\|\boldsymbol{w}\|_{2}^{p}}
\right]
&\leq \left(
 \mathbb{E}\left[
  |N - \|\boldsymbol{w}\|_{2}^{2}|^{2p}
 \right]\mathbb{E}\left[
  \|\boldsymbol{w}\|_{2}^{-2p}
 \right]
\right)^{1/2}
\nonumber \\
&= {\cal O}(1), 
\end{align} 
where the last equality follows from \cite[Propositions~3 and 4]{Takeuchi20}. 
Combining these results, we arrive at Proposition~\ref{proposition1}.

\subsection{Proof of Lemma~\ref{lemma_elimination}}
\label{proof_lemma_elimination} 
We formulate 
$\boldsymbol{\phi}^{\mathrm{T}}(a_{N}\boldsymbol{\omega}
/\|\boldsymbol{\omega}\|_{2})\boldsymbol{\psi}(a_{N}\boldsymbol{\omega}
/\|\boldsymbol{\omega}\|_{2})$ as 
\begin{align}
&\boldsymbol{\phi}^{\mathrm{T}}\left(
 \frac{a_{N}\boldsymbol{\omega}}{\|\boldsymbol{\omega}\|_{2}}
\right)
\boldsymbol{\psi}\left(
 \frac{a_{N}\boldsymbol{\omega}}{\|\boldsymbol{\omega}\|_{2}}
\right)
= \boldsymbol{\phi}^{\mathrm{T}}(af(N)\boldsymbol{\omega})
\boldsymbol{\psi}(af(N)\boldsymbol{\omega})
\nonumber \\
&+ \Delta\boldsymbol{\phi}^{\mathrm{T}}(\boldsymbol{\omega})  
\boldsymbol{\psi}(af(N)\boldsymbol{\omega})
+ \boldsymbol{\phi}^{\mathrm{T}}(af(N)\boldsymbol{\omega})  
\Delta\boldsymbol{\psi}(\boldsymbol{\omega})
\nonumber \\
&+ \Delta\boldsymbol{\phi}^{\mathrm{T}}(\boldsymbol{\omega})
\Delta\boldsymbol{\psi}(\boldsymbol{\omega}). 
\end{align}
Since $\|\boldsymbol{\phi}(af(N)\boldsymbol{\omega})\|_{2}$ and 
$\|\boldsymbol{\psi}(af(N)\boldsymbol{\omega})\|_{2}$ have been 
assumed to be bounded in probability, we use the Cauchy-Schwarz inequality 
for the last three terms to find that 
$\|\Delta\boldsymbol{\phi}(\boldsymbol{\omega})\|_{2}\pto 0$ and 
$\|\Delta\boldsymbol{\psi}(\boldsymbol{\omega})\|_{2}\pto 0$ 
imply Lemma~\ref{lemma_elimination}. 

We prove $\|\Delta\boldsymbol{\phi}(\boldsymbol{\omega})\|_{2}\pto 0$ and 
$\|\Delta\boldsymbol{\psi}(\boldsymbol{\omega})\|_{2}\pto 0$. We only prove 
$\|\Delta\boldsymbol{\phi}(\boldsymbol{\omega})\|_{2}\pto 0$ 
since $\|\Delta\boldsymbol{\psi}(\boldsymbol{\omega})\|_{2}\pto 0$ can be 
proved in the same manner. Let 
\begin{equation}
\Delta_{1}\boldsymbol{\phi}(\boldsymbol{\omega}) 
=\boldsymbol{\phi}\left(
  \frac{a_{N}}{\|\boldsymbol{\omega}\|_{2}}\boldsymbol{\omega}
 \right)
- \boldsymbol{\phi}\left(
 \frac{a_{N}}{\sqrt{N}}\boldsymbol{\omega}
\right),
\end{equation}
\begin{equation}
\Delta_{2}\boldsymbol{\phi}(\boldsymbol{\omega}) 
=\boldsymbol{\phi}\left(
  \frac{a_{N}}{\sqrt{N}}\boldsymbol{\omega}
 \right)
- \boldsymbol{\phi}\left(
 af(N)\boldsymbol{\omega}
\right), 
\end{equation}
which imply $\Delta\boldsymbol{\phi}(\boldsymbol{\omega}) 
=\Delta_{1}\boldsymbol{\phi}(\boldsymbol{\omega}) 
+\Delta_{2}\boldsymbol{\phi}(\boldsymbol{\omega})$. It is sufficient to prove 
$\|\Delta_{1}\boldsymbol{\phi}(\boldsymbol{\omega})\|_{2}\pto0$ and  
$\|\Delta_{2}\boldsymbol{\phi}(\boldsymbol{\omega})\|_{2}\pto0$. 

We use the Lipschitz continuity of 
$\boldsymbol{\phi}$ and the independence between $a_{N}$ and 
$\boldsymbol{\omega}$ to evaluate 
$\mathbb{E}[\|\Delta_{1}\boldsymbol{\phi}(\boldsymbol{\omega})\|_{2}^{2}
| \tilde{a}_{N}]$ with $\tilde{a}_{N}=a_{N}/\{\sqrt{N}f(N)\}$ as  
\begin{align}
&\mathbb{E}\left[
 \left.
  \|\Delta_{1}\boldsymbol{\phi}(\boldsymbol{\omega})\|_{2}^{2}
 \right| \tilde{a}_{N}
\right]
\nonumber \\
&\leq Nf^{2}(N)\tilde{a}_{N}^{2}\mathbb{E}\left[
 \left|
  \frac{1}{\|\boldsymbol{\omega}\|_{2}} - \frac{1}{\sqrt{N}} 
 \right|^{2}\omega_{1}^{2}
\right]\sum_{n=1}^{N}L_{n}^{2}. 
\end{align}  
Applying the Cauchy-Schwarz inequality yields  
\begin{align}
\left\{
 \mathbb{E}\left[
  \left|
   \frac{1}{\|\boldsymbol{\omega}\|_{2}} - \frac{1}{\sqrt{N}} 
  \right|^{2}\omega_{1}^{2}
 \right]
\right\}^{2}
&= \mathbb{E}\left[
 \left|
  \frac{1}{\|\boldsymbol{\omega}\|_{2}} - \frac{1}{\sqrt{N}} 
 \right|^{4}
\right]
\mathbb{E}\left[
 \omega_{1}^{4}
\right]
\nonumber \\
&= {\cal O}(N^{-4})
\end{align}
as $N\to\infty$, where the last equality follows from 
Proposition~\ref{proposition1}. 
Combining these results, as well as the convergence in probability 
$\tilde{a}_{N}\pto a$ and the boundedness assumption in probability  
$f^{2}(N)\sum_{n=1}^{N}L_{n}^{2}<\infty$, we obtain 
$\mathbb{E}[\|\Delta_{1}\boldsymbol{\phi}(\boldsymbol{\omega})\|_{2}^{2}\to0$. 

We next prove $\|\Delta_{2}\boldsymbol{\phi}(\boldsymbol{\omega})\|_{2}\pto 0$. 
Without loss of generality, the symmetry 
$-\boldsymbol{\omega}\sim\boldsymbol{\omega}$ allows us to assume $a\geq0$. 
For some $u>a$, we consider two events 
$\{\tilde{a}_{N}\in[0, u)\}$ and $\{\tilde{a}_{N}\notin[0, u)\}$ to have 
\begin{align}
\|\Delta_{2}\boldsymbol{\phi}(\boldsymbol{\omega})\|_{2}
&= \|\Delta_{2}\boldsymbol{\phi}(\boldsymbol{\omega})\|_{2}
1(\tilde{a}_{N}\in[0, u)) 
\nonumber \\
&+ \|\Delta_{2}\boldsymbol{\phi}(\boldsymbol{\omega})\|_{2}
1(\tilde{a}_{N}\notin[0, u)). 
\end{align}
Since the assumption $\tilde{a}_{N}\pto a\in[0, u)$ implies 
$\mathbb{P}(\tilde{a}_{N}\notin[0, u))\to0$, we find the convergence 
in probability for the second term: For any $\epsilon>0$ 
\begin{align}
&\mathbb{P}(\|\Delta_{2}\boldsymbol{\phi}(\boldsymbol{\omega})\|_{2}
1(\tilde{a}_{N}\notin[0, u)) > \epsilon)
\nonumber \\
&= \mathbb{P}(\|\Delta_{2}\boldsymbol{\phi}(\boldsymbol{\omega})\|_{2}
> \epsilon | \tilde{a}_{N}\notin[0, u))
\mathbb{P}(\tilde{a}_{N}\notin[0, u))\to 0. 
\end{align}
Thus, it is sufficient to prove 
$\|\Delta_{2}\boldsymbol{\phi}(\boldsymbol{\omega})\|_{2}^{2}
1(\tilde{a}_{N}\in[0, u))\pto 0$. 

We evaluate $\mathbb{E}[\|\Delta_{2}\boldsymbol{\phi}
(\boldsymbol{\omega})\|_{2}^{2}1(\tilde{a}_{N}\in[0, u))]$. 
Applying the Lipschitz continuity of $\boldsymbol{\phi}$ yields 
\begin{align}
&\mathbb{E}\left[
 \|\Delta_{2}\boldsymbol{\phi}(\boldsymbol{\omega})\|_{2}^{2}
 1(\tilde{a}_{N}\in[0, u))
\right]
\nonumber \\
&\leq f^{2}(N)\mathbb{E}\left[
 (\tilde{a}_{N} - a)^{2}1(\tilde{a}_{N}\in[0, u))
\right]\sum_{n=1}^{N}L_{n}^{2}, \label{probability_Delta_phi}
\end{align}
where we have used the independence between $\tilde{a}_{N}$ and 
$\boldsymbol{\omega}$, as well as $\mathbb{E}[\omega_{n}^{2}]=1$. 
Since $(\tilde{a}_{N} - a)^{2}1(\tilde{a}_{N}\in[0, u))$ is bounded, 
the assumption $\tilde{a}_{N}\pto a$ implies 
$\mathbb{E}[(\tilde{a}_{N} - a)^{2}1(\tilde{a}_{N}\in[0, u))]\to0$. 
Combining this result, as well as the boundedness assumption of  
$f^{2}(N)\sum_{n=1}^{N}L_{n}^{2}$, we arrive at 
$\mathbb{E}[\|\Delta_{2}\boldsymbol{\phi}(\boldsymbol{\omega})\|_{2}^{2}
1(\tilde{a}_{N}\in[0, u))]\to0$. Thus, Lemma~\ref{lemma_elimination} holds. 

\subsection{Proof of Lemma~\ref{lemma_generalization}}
\label{proof_lemma_generalization}
We follow \cite[Lemmas~4 and 5]{Bayati11} to prove 
Lemma~\ref{lemma_generalization}. 
From the convergence assumption on the empirical distribution $\rho_{M}$, 
we use Skorokhod's theorem to find that there are some random vectors
$\tilde{\boldsymbol{X}}_{M}=[\tilde{X}_{1,M},\ldots,\tilde{X}_{t,M}]^{\mathrm{T}}$ 
and 
$\tilde{\boldsymbol{X}}_{*}=[\tilde{X}_{1,*},\ldots,\tilde{X}_{t,*}]^{\mathrm{T}}$ 
on a common probability space $(\Omega, \mathscr{F}, \mathbb{P})$ such that 
$\tilde{\boldsymbol{X}}_{M}\sim\rho_{M}$ holds, as well as   
$\tilde{\boldsymbol{X}}_{*}\sim\rho_{*}$, and that 
$\tilde{\boldsymbol{X}}_{M}$ converges in probability to 
$\tilde{\boldsymbol{X}}_{*}$ with respect to this probability measure 
$\mathbb{P}$ as $M\to\infty$. 
We use the assumption $\mathbb{P}(\tilde{\boldsymbol{X}}_{1,*}
\in\mathcal{C}_{f}(\tilde{\boldsymbol{X}}_{2,M})| \boldsymbol{X}_{M})=1$ 
to obtain 
\begin{equation}
\mathbb{E}\left[
 \left. 
  f(\tilde{\boldsymbol{X}}_{M}) 
 \right| \boldsymbol{X}_{M}
\right]
= \tilde{f}(\boldsymbol{X}_{M}), 
\end{equation}
with 
\begin{equation}
\tilde{f}(\boldsymbol{X}_{M}) 
= \mathbb{E}\left[
 \left.
  f(\tilde{\boldsymbol{X}}_{M})1\left(
   \tilde{\boldsymbol{X}}_{1,*}
   \in\mathcal{C}_{f}(\tilde{\boldsymbol{X}}_{2,M})
  \right)
 \right| \boldsymbol{X}_{M}
\right]. \label{empirical_average_tmp}
\end{equation}

We use the standard truncation technique to evaluate 
(\ref{empirical_average_tmp}). For $a\in\{\mathrm{L}, \mathrm{U}\}$ we define  
\begin{equation}
\tilde{f}_{a}(\boldsymbol{X}_{M}) 
= \mathbb{E}\left[
 \left.
  f_{a}(\tilde{\boldsymbol{X}}_{M})
  1\left(
   \tilde{\boldsymbol{X}}_{1,*}\in\mathcal{C}_{f}
   (\tilde{\boldsymbol{X}}_{2,M})
  \right)
 \right| \boldsymbol{X}_{M}
\right], \label{empirical_average_tmp2}
\end{equation}
with $f_{\mathrm{L}}(\tilde{\boldsymbol{X}}_{M})
=f(\tilde{\boldsymbol{X}}_{M})1(|f(\tilde{\boldsymbol{X}}_{M})|\leq T)$ 
and $f_{\mathrm{U}}(\tilde{\boldsymbol{X}}_{M})
=f(\tilde{\boldsymbol{X}}_{M})1(|f(\tilde{\boldsymbol{X}}_{M})|> T)$  
for some threshold $T>0$. By definition we have 
$\tilde{f}(\boldsymbol{X}_{M}) = \tilde{f}_{\mathrm{L}}(\boldsymbol{X}_{M}) 
+ \tilde{f}_{\mathrm{U}}(\boldsymbol{X}_{M})$. 
Since the convergence in probability 
$\tilde{\boldsymbol{X}}_{1,M}\pto\tilde{\boldsymbol{X}}_{1,*}$ implies 
$f(\tilde{\boldsymbol{X}}_{M})\pto f(\tilde{\boldsymbol{X}}_{1,*}, 
\tilde{\boldsymbol{X}}_{2,M})$ conditioned on $\tilde{\boldsymbol{X}}_{2,M}$  
for $\tilde{\boldsymbol{X}}_{1,*}\in\mathcal{C}_{f}
(\tilde{\boldsymbol{X}}_{2,M})$, 
we use the boundedness of $f_{\mathrm{L}}(\tilde{\boldsymbol{X}}_{M})
1(\tilde{\boldsymbol{X}}_{1,*}
\in\mathcal{C}_{f}(\tilde{\boldsymbol{X}}_{2,M}))$ to find that 
(\ref{empirical_average_tmp2}) for $a=\mathrm{L}$ reduces to 
\begin{align}
&\lim_{M\to\infty}\tilde{f}_{\mathrm{L}}(\boldsymbol{X}_{M}) 
\nonumber \\
&= \lim_{M\to\infty}\mathbb{E}\left[
 \left.
  f_{\mathrm{L}}(\tilde{\boldsymbol{X}}_{1,*}, \tilde{\boldsymbol{X}}_{2,M})
  1\left(
   \tilde{\boldsymbol{X}}_{1,*}
   \in\mathcal{C}_{f}(\tilde{\boldsymbol{X}}_{2,M})
  \right)
 \right| \boldsymbol{X}_{M}
\right]
\nonumber \\
&= \lim_{M\to\infty}\mathbb{E}\left[
 \left. 
  f_{\mathrm{L}}(\tilde{\boldsymbol{X}}_{1,*}, \tilde{\boldsymbol{X}}_{2,M})
 \right| \boldsymbol{X}_{M}
\right]
\nonumber \\
&= \lim_{M\to\infty}\mathbb{E}\left[
 \left.
  f(\tilde{\boldsymbol{X}}_{1,*}, \tilde{\boldsymbol{X}}_{2,M})
 \right| \boldsymbol{X}_{M}
\right] \quad 
\hbox{as $T\to\infty$,} 
\end{align} 
which converges in probability to 
$\mathbb{E}[f(\tilde{\boldsymbol{X}}_{*})]$, because of the third 
assumption in Lemma~\ref{lemma_generalization}. 
In the derivation of the second equality, we have used the assumption 
$\mathbb{P}(\tilde{\boldsymbol{X}}_{1,*}
\in\mathcal{C}_{f}(\tilde{\boldsymbol{X}}_{2,M}) 
| \boldsymbol{X}_{M})=1$. Thus, 
it is sufficient to prove that (\ref{empirical_average_tmp2}) for 
$a=\mathrm{U}$ converges in probability to zero. 

We evaluate (\ref{empirical_average_tmp2}) 
for $a=\mathrm{U}$. Using the trivial upper bound 
$1(\tilde{\boldsymbol{X}}_{1,*}
\in\mathcal{C}_{f}(\tilde{\boldsymbol{X}}_{2,M}))\leq1$ yields 
\begin{align}
&\left|
 \tilde{f}_{\mathrm{U}}(\boldsymbol{X}_{M})
\right|
\leq \mathbb{E}\left[
 \left.
  |f_{\mathrm{U}}(\tilde{\boldsymbol{X}}_{M})|
 \right| \boldsymbol{X}_{M}
\right]
\nonumber \\
&\leq \mathbb{E}\left[
 \left. 
  L(\tilde{\boldsymbol{X}}_{2,M})
  h(L(\tilde{\boldsymbol{X}}_{2,M}), \boldsymbol{X}_{M}) 
 \right| \boldsymbol{X}_{M}
\right], 
\end{align}
with 
\begin{align}
&h(L, \boldsymbol{X}_{M}) 
= \mathbb{E}\Bigg[(1 + \|\tilde{\boldsymbol{X}}_{1,M}\|_{2}^{2})
\nonumber \\
&\cdot\left.
 \left.
  \left\{
   1 - 1\left(
    \|\tilde{\boldsymbol{X}}_{1,M}\|_{2}^{2} 
    \leq \frac{T}{L} - 1
   \right)
  \right\}
 \right| \boldsymbol{X}_{M}
\right]. \label{h_function} 
\end{align}
The last inequality follows from 
the assumption $|f(\tilde{\boldsymbol{X}}_{M})|
\leq L(\tilde{\boldsymbol{X}}_{2,M})
(1 + \|\tilde{\boldsymbol{X}}_{1,M}\|_{2}^{2})$. 

We use the truncation technique for $L(\tilde{\boldsymbol{X}}_{2,M})$ with 
some threshold $L_{0}>0$ to obtain 
\begin{align}
&\mathbb{E}\left[
 \left. 
  L(\tilde{\boldsymbol{X}}_{2,M})
  h(L(\tilde{\boldsymbol{X}}_{2,M}), \boldsymbol{X}_{M}) 
 \right| \boldsymbol{X}_{M}
\right]
\nonumber \\
&= \mathbb{E}\left[
 \left.
  L_{\mathrm{L}}(\tilde{\boldsymbol{X}}_{2,M}) 
  h(L(\tilde{\boldsymbol{X}}_{2,M}), \boldsymbol{X}_{M}) 
 \right| \boldsymbol{X}_{M}
\right]
\nonumber \\
&+ \mathbb{E}\left[
 \left. 
  L_{\mathrm{U}}(\tilde{\boldsymbol{X}}_{2,M})
  h(L(\tilde{\boldsymbol{X}}_{2,M}), \boldsymbol{X}_{M}) 
 \right| \boldsymbol{X}_{M}
\right], \label{second_term_bound}
\end{align}
with the truncation $L_{\mathrm{L}}(\tilde{\boldsymbol{X}}_{2,M})
= L(\tilde{\boldsymbol{X}}_{2,M}) 1(L(\tilde{\boldsymbol{X}}_{2,M})\leq L_{0})$ 
and $L_{\mathrm{U}}(\tilde{\boldsymbol{X}}_{2,M})
=L(\tilde{\boldsymbol{X}}_{2,M})1(L(\tilde{\boldsymbol{X}}_{2,M})> L_{0})$. 

For the first term in (\ref{second_term_bound}), we 
use the non-decreasing property of $h(L, \boldsymbol{X}_{M})$ 
with respect to $L$ and the upper bound 
$L_{\mathrm{L}}(\tilde{\boldsymbol{X}}_{2,M})\leq L_{0}$ to obtain 
\begin{equation}
\mathbb{E}\left[
 \left.
  L_{\mathrm{L}}(\tilde{\boldsymbol{X}}_{2,M})
  h(L(\tilde{\boldsymbol{X}}_{2,M}), \boldsymbol{X}_{M}) 
 \right| \boldsymbol{X}_{M}
\right]
\leq L_{0}h(L_{0}, \boldsymbol{X}_{M}). 
\end{equation}
From the convergence 
in probability $\tilde{\boldsymbol{X}}_{1,M}\pto\tilde{\boldsymbol{X}}_{1,*}$ 
and the second assumption $\mathbb{E}[\|\tilde{\boldsymbol{X}}_{1,M}\|_{2}^{2}
| \boldsymbol{X}_{M}]\pto\mathbb{E}[\|\tilde{\boldsymbol{X}}_{1,*}\|_{2}^{2}]$,    
we use the boundedness of $(1+\|\tilde{\boldsymbol{X}}_{1,M}\|_{2}^{2})
1(\|\tilde{\boldsymbol{X}}_{1,M}\|_{2}^{2} \leq TL_{0}^{-1} - 1)$ 
in (\ref{h_function}) to have 
\begin{align}
&\lim_{M\to\infty}h(L_{0}, \boldsymbol{X}_{M})
\nonumber \\
&\peq \mathbb{E}\left[
 (1 + \|\tilde{\boldsymbol{X}}_{1,*}\|_{2}^{2})
 \left\{
  1 - 1\left(
   \|\tilde{\boldsymbol{X}}_{1,*}\|_{2}^{2} 
   \leq \frac{T}{L_{0}} - 1
  \right)
 \right\}
\right]
\nonumber \\
&= \mathbb{E}\left[
 (1 + \|\tilde{\boldsymbol{X}}_{1,*}\|_{2}^{2})
 1\left(
  \|\tilde{\boldsymbol{X}}_{1,*}\|_{2}^{2} >
  \frac{T}{L_{0}} - 1
 \right)
\right]\to 0
\end{align}
as $T\to\infty$, where the last convergence follows from the fact that 
we can use the dominated convergence theorem from 
the $T$-independent upper bound 
$(1 + \|\tilde{\boldsymbol{X}}_{1,*}\|_{2}^{2})
1(\|\tilde{\boldsymbol{X}}_{1,*}\|_{2}^{2} > TL_{0}^{-1} - 1)\leq 
1 + \|\tilde{\boldsymbol{X}}_{1,*}\|_{2}^{2}$ and the boundedness assumption 
$\mathbb{E}[\|\tilde{\boldsymbol{X}}_{1,*}\|_{2}^{2} ]<\infty$. 

For the second term in (\ref{second_term_bound}), 
we have the upper bound 
$h(L(\tilde{\boldsymbol{X}}_{2,M}), \boldsymbol{X}_{M})
\leq 1 + \mathbb{E}[ \|\tilde{\boldsymbol{X}}_{1,M}\|_{2}^{2} 
| \boldsymbol{X}_{M}]$, which implies  
\begin{align}
&\lim_{M\to\infty}\mathbb{E}\left[
 \left.
  L_{\mathrm{U}}(\tilde{\boldsymbol{X}}_{2,M})
  h(L(\tilde{\boldsymbol{X}}_{2,M}), \boldsymbol{X}_{M})
 \right| \boldsymbol{X}_{M} 
\right]
\nonumber \\
&\leq \lim_{M\to\infty}\left(
 1 + \mathbb{E}[ \|\tilde{\boldsymbol{X}}_{1,M}\|_{2}^{2} | \boldsymbol{X}_{M}]
\right)
\nonumber \\
&\cdot\mathbb{E}\left[
 \left.
L(\tilde{\boldsymbol{X}}_{2,M})
  1(L(\tilde{\boldsymbol{X}}_{2,M})> L_{0})
 \right| \boldsymbol{X}_{M} 
\right]
\pto 0 
\end{align}
as $L_{0}\to\infty$, where the last convergence follows from the second 
assumption and the uniform-integrability assumption of 
$L(\tilde{\boldsymbol{X}}_{2,M})$ in Lemma~\ref{lemma_generalization}. 
Combining these results, we arrive at Lemma~\ref{lemma_generalization}. 

\subsection{Proof of Lemma~\ref{lemma_positive_definite}}
\label{proof_lemma_positive_definite}
It is sufficient to prove the following inequality: 
\begin{equation}
\mathbb{E}[X_{t}\boldsymbol{X}_{t}^{\mathrm{T}}]
(\mathbb{E}[\boldsymbol{X}_{t}\boldsymbol{X}_{t}^{\mathrm{T}}])^{-1}
\mathbb{E}[\boldsymbol{X}_{t}X_{t}] 
\leq \mathbb{E}\left[
 \left(
  \mathbb{E}[X_{t} | \boldsymbol{X}_{t}] 
 \right)^{2}
\right]. 
\end{equation} 
Let $Y = X_{t} - \boldsymbol{X}_{t}^{\mathrm{T}}
(\mathbb{E}[\boldsymbol{X}_{t}\boldsymbol{X}_{t}^{\mathrm{T}}])^{-1}
\mathbb{E}[\boldsymbol{X}_{t}X_{t}]$. 
Evaluating $\mathbb{E}[\boldsymbol{X}_{t}Y]$ yields 
$\mathbb{E}[\boldsymbol{X}_{t}Y] 
= \mathbb{E}[\boldsymbol{X}_{t}X_{t}] 
- \mathbb{E}[\boldsymbol{X}_{t}X_{t}]
= \boldsymbol{0}$,
which implies 
$\mathbb{E}[ \mathbb{E}[Y|\boldsymbol{X}_{t}]\boldsymbol{X}_{t}^{\mathrm{T}} ]
= \mathbb{E}[Y\boldsymbol{X}_{t}^{\mathrm{T}}] = \boldsymbol{0}$. 
Using the representation $X_{t} = Y + \boldsymbol{X}_{t}^{\mathrm{T}}
(\mathbb{E}[\boldsymbol{X}_{t}\boldsymbol{X}_{t}^{\mathrm{T}}])^{-1}
\mathbb{E}[\boldsymbol{X}_{t}X_{t}]$ and this result, we arrive at
\begin{align}
&\mathbb{E}\left[
 \left(
  \mathbb{E}[X_{t} | \boldsymbol{X}_{t}] 
 \right)^{2}
\right]
\nonumber \\
&= \mathbb{E}\left[
 \left\{
  \mathbb{E}[Y | \boldsymbol{X}_{t}] + \boldsymbol{X}_{t}^{\mathrm{T}}
  (\mathbb{E}[\boldsymbol{X}_{t}\boldsymbol{X}_{t}^{\mathrm{T}}])^{-1}
  \mathbb{E}[\boldsymbol{X}_{t}X_{t}]
 \right\}^{2}
\right] \nonumber \\
&= \mathbb{E}\left[
 \left(
  \mathbb{E}[Y | \boldsymbol{X}_{t}]
 \right)^{2}
\right]
+ \mathbb{E}[X_{t}\boldsymbol{X}_{t}^{\mathrm{T}}]
(\mathbb{E}[\boldsymbol{X}_{t}\boldsymbol{X}_{t}^{\mathrm{T}}])^{-1}
\mathbb{E}[\boldsymbol{X}_{t}X_{t}]
\nonumber \\
&\geq  \mathbb{E}[X_{t}\boldsymbol{X}_{t}^{\mathrm{T}}]
(\mathbb{E}[\boldsymbol{X}_{t}\boldsymbol{X}_{t}^{\mathrm{T}}])^{-1}
\mathbb{E}[\boldsymbol{X}_{t}X_{t}],
\end{align}
where the last inequality follows from the non-negativity of  
$\mathbb{E}[(\mathbb{E}[Y | \boldsymbol{X}_{t}])^{2}]$. Thus, 
Lemma~\ref{lemma_positive_definite} holds.  


\balance

\bibliographystyle{IEEEtran}
\bibliography{IEEEabrv,kt-it2025_1}






\end{document}